\newcommand*{\authors}{Markus F. Weber and Erwin Frey}  
\newcommand*{\addresses}{Arnold Sommerfeld Center for Theoretical Physics and Center for NanoScience, Department of Physics, Ludwig-Maximilians-Universit\"at M\"unchen, Theresienstra\ss e 37, 80333 M\"unchen, Germany.}
\newcommand*{\contactemail}{frey@lmu.de}

\newcommand*{\shortTitle}{Master equations and the theory of stochastic path integrals}  
\newcommand*{\longTitle}{Master equations and the theory of\\ stochastic path integrals}

\newcommand*{\keywords}{Stochastic processes, Markov processes, master equations, path integrals, path summation, spectral analysis, rare event probabilities}
\newcommand*{\pacsNumbers}{02.50.-r, 02.50.Ey, 02.50.Ga, 02.70.Hm, 05.10.Gg, 05.10.Cc, 05.40.-a}

\documentclass[10pt]{iopart}

\makeatletter% 
  \newcommand*{\coloneqq}{%
    \mathrel{%
      \rlap{\raisebox{0.37ex}{$\m@th\cdot$}}%
      \raisebox{-0.37ex}{$\m@th\cdot$}%
    }%
  \hspace{-0.06em}=}%
\makeatother%
\makeatletter%
  \newcommand*{\eqqcolon}{%
     =\hspace{-0.06em}%
     \mathrel{%
       \rlap{\raisebox{0.37ex}{$\m@th\cdot$}}%
       \raisebox{-0.37ex}{$\m@th\cdot$}%
    }%
  }%
\makeatother%

\newcommand{\aref}[1]{appendix~\ref{#1}}

\makeatletter%
  \def\@fnsymbol#1{\ensuremath{\ifcase#1 
  \or \textsuperscript{1} \or \textsuperscript{2} \or \textsuperscript{3} \or \textsuperscript{4} \or \textsuperscript{5}%
  \or \textsuperscript{6} \or \textsuperscript{7} \or \textsuperscript{8} \or \textsuperscript{9} \or \textsuperscript{10}%
  \or \textsuperscript{11}%
  \else\@ctrerr\fi}}%
\makeatother%
\usepackage{bm}
\renewcommand*{\bi}[1]{\bm{#1}}

\usepackage[OT2,T1]{fontenc}
\DeclareSymbolFont{cyrletters}{OT2}{wncyr}{m}{n}
\DeclareMathSymbol{\sha}{\mathalpha}{cyrletters}{"58}

\usepackage{amssymb}
\usepackage[british]{babel}
\usepackage{graphicx}
\usepackage{esint}%  path integral signs
\usepackage{dsfont}%  unit matrix

\usepackage{cite}

\PassOptionsToPackage{hyphens}{url}% Correct line breaks of URLs
\usepackage{hyperxmp}
\usepackage{hyperref}
\hypersetup{%
  pdftitle={\shortTitle{}},%
  pdfauthor={\authors{}},%
  pdfsubject={\shortTitle{}},%
  pdfkeywords={\keywords{}},%
  pdfcontactemail={\contactemail{}},%
  pdflang={en}%
}
\usepackage{url}

\usepackage{tikz}
\usetikzlibrary{positioning}
\usetikzlibrary{arrows.meta}
\usetikzlibrary{decorations.markings}
\usetikzlibrary{calc}
\tikzset{
  linePlain/.style={draw=black, semithick},
  lineWithArrow1/.style={draw=black, semithick, postaction={decorate},decoration={markings,mark=at position .6 with {\arrow[scale=1.75]{stealth}}}},
  lineWithArrow2/.style={draw=black, semithick, postaction={decorate},decoration={markings,mark=at position .55 with {\arrow[scale=1.75]{stealth}}}},
  lineWithArrowInline/.style={draw=black, semithick, postaction={decorate},decoration={markings,mark=at position .7 with {\arrow[scale=1.75]{stealth}}}},
  vertex/.style={draw, shape=circle, fill=black, minimum size=1.1mm, inner sep=0mm, outer sep=0mm},
}
\newcommand{\upperloop}[3][]{%
  \draw[#1] let \p1 = ($(#2)-(#3)$) in (#2) arc (0:180:({0.5*veclen(\x1,\y1)});)
} 
\newcommand{\lowerloop}[3][]{%
  \draw[#1] let \p1 = ($(#2)-(#3)$) in (#2) arc (360:180:({0.5*veclen(\x1,\y1)});)
}

\newcommand*{\mathtext}[1]{\textnormal{#1}}
\newcommand*{\mathtextit}[1]{\textit{#1}}

\newcommand*{\diff}{\mathop{}\!\mathrm{d}}
\newcommand*{\pathintegral}[2]{\fint_{#1}^{#2}\!}
\newcommand*{\binomCoeff}[2]{{#1 \choose #2}}
 
\newcommand*{\numberOperator}{\mathcal{N}}
\newcommand*{\cre}{c}  
\newcommand*{\ann}{a}  

\newcommand*{\LLangle}{\langle\kern-2\nulldelimiterspace\langle}
\newcommand*{\bigLangle}{\big\langle}
\newcommand*{\bigLLangle}{\bigLangle\kern-2.5\nulldelimiterspace\bigLangle}
\newcommand*{\BigLangle}{\Big\langle}
\newcommand*{\BigLLangle}{\BigLangle\kern-3\nulldelimiterspace\BigLangle}

\newcommand*{\RRangle}{\rangle\kern-2\nulldelimiterspace\rangle}
\newcommand*{\bigRangle}{\big\rangle}
\newcommand*{\bigRRangle}{\bigRangle\kern-2.5\nulldelimiterspace\bigRangle}
\newcommand*{\BigRangle}{\Big\rangle}
\newcommand*{\BigRRangle}{\BigRangle\kern-3\nulldelimiterspace\BigRangle}

\newcommand*{\braA}[1]{{\langle #1 |}}
\newcommand*{\ketA}[1]{{| #1 \rangle}}
\newcommand*{\braketA}[2]{\langle #1 | #2 \rangle}

\newcommand*{\braB}[1]{\LLangle #1 |}
\newcommand*{\ketB}[1]{| #1 \RRangle}
\newcommand*{\braketB}[2]{\LLangle #1 | #2 \RRangle}

\newcommand*{\braketBA}[2]{\LLangle #1 | #2 \rangle}
\newcommand*{\braketAB}[2]{\langle #1 | #2 \RRangle}

\newcommand*{\reals}{\mathbb{R}}
\newcommand*{\complex}{\mathbb{C}}
\newcommand*{\naturals}{\mathbb{N}}
\newcommand*{\integers}{\mathbb{Z}}
\newcommand*{\lattice}{\mathbb{L}}
\newcommand*{\neighbors}{\mathsf{N}}

\newcommand*{\prob}{p}  
\newcommand*{\gen}{g}  

\newcommand*{\waitingTimeDist}{\mathcal{W}}
\newcommand*{\survivalProb}{\mathcal{S}}  
\newcommand*{\gaussianDistribution}{\mathcal{G}}

\newcommand*{\vect}[1]{\bi{#1}}
\newcommand*{\bigO}{\mathord{\mathrm{O}}}
\newcommand*{\transpose}{\intercal}
\newcommand*{\laplaceOp}{\Delta}
\newcommand*{\discreteLaplaceOp}{\underline{\Delta}}
\newcommand*{\unitMatrix}{\mathds{1}}
\newcommand*{\hermite}{\mathit{He}}  
\newcommand*{\legendreTransform}{\mathcal{L}}  
\newcommand*{\complexPath}{\mathcal{C}}  
\newcommand*{\hldots}{\dots}  
\newcommand*{\HeavisideStep}{\Theta}  
\newcommand*{\HeavisideStepDiscrete}{\Theta}  
\newcommand*{\propagator}{G}

\newcommand*{\ee}{\mathrm{e}}% or \rme 
\newcommand*{\ii}{\mathrm{i}}% or \rmi

\newcommand*{\tVar}{\tau}  
\newcommand*{\sVar}{s}  

\newcommand*{\x}{x}
\newcommand*{\xAd}{\x}
\newcommand*{\X}{X}
\newcommand*{\XAd}{\X}
\newcommand*{\tx}{\tilde{x}}
\newcommand*{\txAd}{\tx}
\newcommand*{\q}{q}
\newcommand*{\qAd}{\q}
\newcommand*{\Q}{Q}
\newcommand*{\QAd}{\Q}
\newcommand*{\tq}{\tilde{q}}
\newcommand*{\tqAd}{\tq}
\newcommand*{\transEvoOp}{\widetilde{\mathcal{Q}}}
\newcommand*{\transEvoOpAd}{\transEvoOp^\dagger}
\newcommand*{\transitionOp}{\mathcal{Q}}
\newcommand*{\transitionOpAd}{\transitionOp^\dagger}
\newcommand*{\evolutionOp}{\mathcal{E}}
\newcommand*{\evolutionOpAd}{\evolutionOp}
\newcommand*{\perturbationOp}{\mathcal{P}}
\newcommand*{\perturbationOpAd}{\perturbationOp^\dagger}
\newcommand*{\action}{\mathcal{S}}
\newcommand*{\actionAd}{\action^\dagger}
\newcommand*{\generator}{\mathcal{L}}
\newcommand*{\generatorAd}{\generator^\dagger}
\newcommand*{\fieldGenFct}{\mathcal{Z}}
\newcommand*{\fieldGenFctAd}{\fieldGenFct}

\newcommand*{\stochPath}{\mathcal{P}}
\newcommand*{\wienerProcess}{W}  
\newcommand*{\qMatrix}{Q}
\newcommand*{\transitionRate}{w}
\newcommand*{\transitionRateJumpSize}{\kappa}
\newcommand*{\exitRate}{w}
\newcommand*{\observable}{A}
\newcommand*{\jumpMoment}[1]{\mathcal{M}^{(#1)}}

\newcommand*{\rateCoeffGeneric}{\gamma}

\newcommand*{\rateCoeffSpontGrowth}{\gamma}  
\newcommand*{\rateCoeffLinGrowth}{\gamma}  

\newcommand*{\rateCoeffLinDecay}{\mu}
\newcommand*{\rateCoeffAnnihilation}{\mu}
\newcommand*{\rateCoeffCoagulation}{\mu}

\newcommand*{\rateCoeffDiffCont}{D}  
\newcommand*{\rateCoeffDiffDisc}{\varepsilon}  

\newcommand*{\basisPrefactor}{\zeta}

\begin{document}

\review[\shortTitle{}]{\longTitle{}}

\author{\authors{}}

\address{\addresses{}}
\ead{\contactemail{}}

\begin{abstract}
  This review provides a pedagogic and self-contained introduction to master equations and to their representation by path integrals. Since the 1930s, master equations have served as a fundamental tool to understand the role of fluctuations in complex biological, chemical, and physical systems. Despite their simple appearance, analyses of masters equations most often rely on low-noise approximations such as the Kramers-Moyal or the system size expansion, or require ad-hoc closure schemes for the derivation of low-order moment equations. We focus on numerical and analytical methods going beyond the low-noise limit and provide a unified framework for the study of master equations. After deriving the forward and backward master equations from the Chapman-Kolmogorov equation, we show how the two master equations can be cast into either of four linear partial differential equations (PDEs). Three of these PDEs are discussed in detail. The first PDE governs the time evolution of a generalized probability generating function whose basis depends on the stochastic process under consideration. Spectral methods, WKB approximations, and a variational approach have been proposed for the analysis of the PDE. The second PDE is novel and is obeyed by a distribution that is marginalized over an initial state. It proves useful for the computation of mean extinction times. The third PDE describes the time evolution of a ``generating functional'', which generalizes the so-called Poisson representation. Subsequently, the solutions of the PDEs are expressed in terms of two path integrals: a ``forward'' and a ``backward'' path integral. Combined with inverse transformations, one obtains two distinct path integral representations of the conditional probability distribution solving the master equations. We exemplify both path integrals in analysing elementary chemical reactions. Moreover, we show how a well-known path integral representation of averaged observables can be recovered from them. Upon expanding the forward and the backward path integrals around stationary paths, we then discuss and extend a recent method for the computation of rare event probabilities. Besides, we also derive path integral representations for processes with continuous state spaces whose forward and backward master equations admit Kramers-Moyal expansions. A truncation of the backward expansion at the level of a diffusion approximation recovers a classic path integral representation of the (backward) Fokker-Planck equation. One can rewrite this path integral in terms of an Onsager-Machlup function and, for purely diffusive Brownian motion, it simplifies to the path integral of Wiener. To make this review accessible to a broad community, we have used the language of probability theory rather than quantum (field) theory and do not assume any knowledge of the latter. The probabilistic structures underpinning various technical concepts, such as coherent states, the Doi-shift, and normal-ordered observables, are thereby made explicit.
\end{abstract}

\pacs{\pacsNumbers{}}
\vspace{2pc}

\ \\\ \\\ \\
\noindent \textbf{Preprint of the article:}\\
M.~F. Weber and E.~Frey.
Master equations and the theory of stochastic path integrals.
{\em Rep. Prog. Phys.}, 80(4):046601, 2017.
\href {http://dx.doi.org/10.1088/1361-6633/aa5ae2}
{\path{doi:10.1088/1361-6633/aa5ae2}}.

%\submitto{\submittedTo{}}

\maketitle

% Table of contents
\tableofcontents
\thispagestyle{empty}
\newpage

\ioptwocol

\markboth{\shortTitle{}}{\shortTitle{}}

%\linenumbers

\section{Introduction}\label{sec:Intro}  

  \subsection{Scope of this review}\label{subsec:Intro_Scope}  

    The theory of continuous-time Markov processes is largely built on two equations: the Fokker-Planck~\cite{Einstein:1905,Fokker:1914,Planck:1917,Kolmogoroff:1931} and the master equation~\cite{Kolmogoroff:1931,Nordsieck:1940}. Both equations assume that the future of a system depends only on its current state, memories of its past having been wiped out by randomizing forces. This \textit{Markov} assumption is sufficient to derive either of the two equations. Whereas the Fokker-Planck equation describes systems that evolve continuously from one state to another, the master equation models systems that perform jumps in state space.
        
    Path integral representations of the master equation were first derived around 1980~\cite{Doi:1976a,Doi:1976b,Zeldovich:1978,Rose:1979,Grassberger:1980,Mikhailov:1981a,Mikhailov:1981b,Goldenfeld:1984,Mikhailov:1985,Peliti:1985}, shortly after such representations had been derived for the Fokker-Planck equation~\cite{Martin:1973,deDominicis:1976,Janssen:1976,Bausch:1976}. Both approaches were heavily influenced by quantum theory, introducing such concepts as the Fock space~\cite{Fock:1932} with its ``bras'' and ``kets''~\cite{Dirac:1937},  coherent states~\cite{Schroedinger:1926b,Sudarshan:1963,Glauber:1963b}, and ``normal-ordering''~\cite{Wick:1950} into non-equilibrium theory. Some of these concepts are now well established and the original ``bosonic'' path integral representation has been complemented with a ``fermionic'' counterpart~\cite{Sandow:1993,Patzlaff:1994,Bares:1999,Brunel:2000,Schulz:2005,Silva:2008}. Nevertheless, we feel that the theory of these ``stochastic'' path integrals may benefit from a step back and a closer look at the probabilistic structures behind the integrals. Therefore, the objects imported from quantum theory make place for their counterparts from probability theory in this review. For example, the coherent states give way to the Poisson distribution. Moreover, we use the bras and kets as particular basis functionals and functions whose choice depends on the stochastic process at hand (a functional maps functions to numbers). Upon choosing the basis functions as Poisson distributions, one can thereby recover both a classic path integral representation of averaged observables as well as the Poisson representation of Gardiner and Chaturvedi~\cite{Gardiner:1977,Chaturvedi:1978}. The framework presented in this review integrates a variety of different approaches to the master equation. Besides the Poisson representation, these approaches include a spectral method for the computation of stationary probability distributions~\cite{Walczak:2009}, WKB approximations and other ``semi-classical'' methods for the computation of rare event probabilities~\cite{Elgart:2004,Assaf:2006a,Assaf:2006b,Assaf:2008}, and a variational approach that was proposed in the context of stochastic gene expression~\cite{Sasai:2003}. All of these approaches can be treated within a unified framework. Knowledge about this common framework makes it possible to systematically search for new ways of solving the master equation. 
    
    Before outlining the organization of this review, let us note that by focusing on the above path integral representations of master and Fokker-Planck equations, we neglect several other ``stochastic'' or ``statistical'' path integrals that have been developed. These include Edwards path integral approach to turbulence~\cite{Edwards:1964,Sreenivasan:2005}, a path integral representation of Haken~\cite{Haken:1976}, path integral representations of non-Markov processes~\cite{Wodkiewicz:1981,Pesquera:1983,Fox:1986a,Fox:1986b,Luciani:1987,Forster:1988,Luciani:1988,Bray:1989,Haenggi:1989,McKane:1989,McKane:1990,Bray:1990,Luckock:1990,Venkatesh:1993,Einchcomb:1996,Mahanta:2000} and of polymers~\cite{Wiegel:1986,Doi:1988,Vilgis:2000,Kleinert:2009}, and a representation of ``hybrid'' processes~\cite{Bressloff:2014,Bressloff:2014b,Bressloff:2015}. The dynamics of these stochastic hybrid processes are piecewise-deterministic. Moreover, we do not discuss the application of renormalization group techniques, despite their significant importance. Excellent texts exploring these techniques in the context of non-equilibrium critical phenomena~\cite{Hinrichsen:2000,Odor:2004,Henkel:2008} are provided by the review of T\"auber, Howard, and Vollmayr-Lee~\cite{Taeuber:2005} as well as the book by T\"auber~\cite{Taeuber:2014}. Our main interest lies in a mathematical framework unifying the different approaches from the previous paragraph and in two path integrals that are based on this framework. Both of these path integrals provide exact representations of the conditional probability distribution solving the master equation. We exemplify the use of the path integrals for elementary processes, which we choose for their pedagogic value. Most of these processes do not involve spatial degrees of freedom but the application of the presented methods to processes on spatial lattices or networks is straightforward. A process with diffusion and linear decay serves as an example of how path integrals can be evaluated perturbatively using Feynman diagrams. The particles' linear decay is treated as a perturbation to their free diffusion. The procedure readily extends to more complex processes. Moreover, we show how the two path integrals can be used for the computation of rare event probabilities. Let us emphasize that we only consider Markov processes obeying the Chapman-Kolmogorov equation and associated master equations~\cite{Kolmogoroff:1931}. It may be interesting to extend the discussed methods to ``generalized'' or ``physical'' master equations with memory kernels~\cite{Zwanzig:1961,Zwanzig:1964,Mori:1965,Oppenheim:1965}.
    
    \begin{figure*}[tb] 
      \centering
      \includegraphics{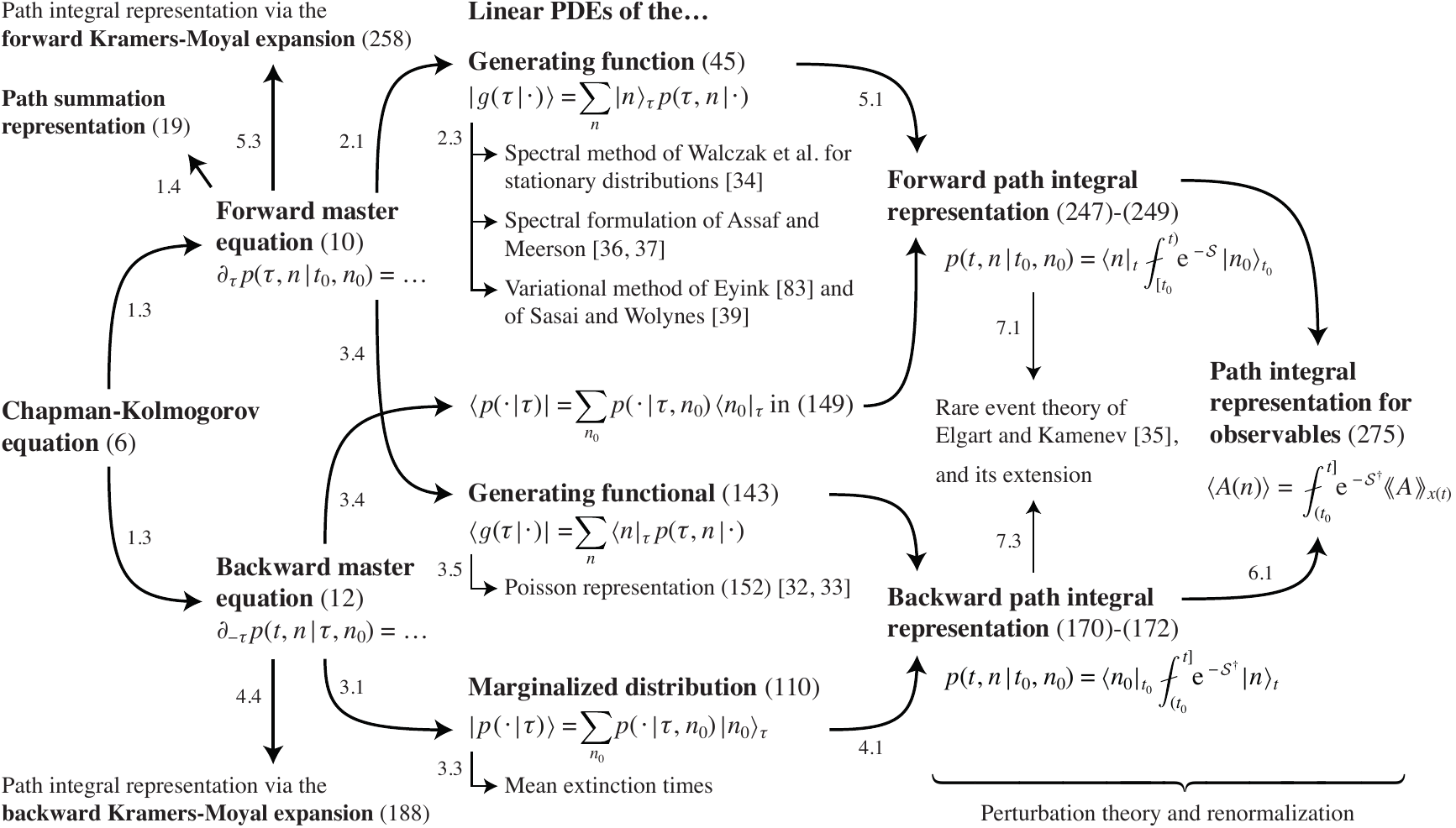}
      \caption{\label{fig:Outline}
        Roadmap and summary of the methods considered in this review. The arrows represent possible routes for derivations. Labelled arrows represent derivations that are explicitly treated in the respective sections. For example, the forward and backward master equations are derived from the Chapman-Kolmogorov equation in \sref{subsec:Intro_Mesoscopic}. In this section, we also discuss the path summation representation~\eref{eq:Intro_Mescoscopic_Solution} of the conditional probability distribution $\prob(\tVar, n | t_0, n_0)$. This representation can be derived by examining the stochastic simulation algorithm (SSA) of Gillespie~\cite{Gillespie:1976,Gillespie:1977,Gillespie:1992,Gillespie:2007} or by performing a Laplace transformation of the forward master equation~\eref{eq:Intro_Mescoscopic_MasterEq} (cf.~\aref{sec:A_PathSummationProof}). In sections~\ref{sec:GenFct} and~\ref{sec:GenFctnl}, the forward and the backward master equations are cast into four linear PDEs, also called ``flow equations''. These equations are obeyed by a probability generating function, a probability generating functional, a marginalized distribution, and a further series expansion. The flow equations can be solved in terms of a forward and a backward path integral as shown in sections~\ref{sec:PathInts_Backward} and~\ref{sec:PathInts_Forward}. Upon performing inverse transformations, the path integrals provide two distinct representations of the conditional probability distribution solving the master equations. Moreover, they can be used to represent averaged observables as explained in \sref{sec:PathInts_Observables}. Besides the methods illustrated in the figure, we discuss path integral representations of processes with continuous state spaces whose master equations admit Kramers-Moyal expansions (sections~\ref{subsec:PathInts_BackwardKramersMoyal} and~\ref{subsec:PathInts_ForwardKramersMoyal}). A truncation of the backward Kramers-Moyal expansion at the level of a diffusion approximation results in a path integral representation of the (backward) Fokker-Planck equation whose original development goes back to works of Martin, Siggia, and Rose~\cite{Martin:1973}, de Dominicis~\cite{deDominicis:1976}, Janssen~\cite{Janssen:1976,Bausch:1976}, and Bausch, Janssen, and Wagner~\cite{Bausch:1976}. The representation can be rewritten in terms of an Onsager-Machlup function~\cite{Onsager:1953}, and it simplifies to Wiener's path integral~\cite{Wiener:1921a,Wiener:1921b} for purely diffusive Brownian motion~\cite{Brown:1827}. Renormalization group techniques are not considered in this review. Information on these techniques can be found in~\cite{Taeuber:2005, Taeuber:2014}.
      }
    \end{figure*}    
    
  \subsection{Organization of this review}\label{subsec:Intro_Organization}  

    The organization of this review is summarized in \fref{fig:Outline} and is as follows. In the next \sref{subsec:Intro_Mesoscopic}, we introduce the basic concepts of the theory of continuous-time Markov processes. After discussing the roles of the forward and backward Fokker-Planck equations for processes with continuous sample paths, we turn to processes with discontinuous sample paths. The probability of finding such a ``jump process'' in a generic state $n$ at time $\tVar > t_0$, given that the process has been in state $n_0$ at time $t_0$, is represented by the conditional probability distribution $\prob(\tVar, n | t_0, n_0)$. Whereas the forward master equation evolves this probability distribution forward in time, starting out at time $\tVar=t_0$, the backward master equation evolves the distribution $\prob(t, n | \tVar, n_0)$ backward in time, starting out at time $\tVar = t$. Both master equations can be derived from the Chapman-Kolmogorov equation (cf.\ left side of \fref{fig:Outline}). In \sref{subsec:Intro_MasterEqSolution}, we discuss two explicit representations of the conditional probability distribution solving the two master equations. Moreover, we comment on various numerical methods for the approximation of this distribution and for the generation of sample paths. Afterwards, \sref{subsec:Intro_PathInts} provides a brief historical overview of contributions to the development of stochastic path integrals. 
    
    The main part of this review begins with \sref{sec:GenFct}. We first exemplify how a generalized probability generating function can be used to determine the stationary probability distribution of an elementary chemical reaction. This example introduces the bra-ket notation used in this review. In \sref{subsec:GenFct_Flow}, we formulate conditions under which a general forward master equation can be transformed into a linear partial differential equation (PDE) obeyed by the generating function. This function is defined as the sum of the conditional probability distribution $\prob(\tVar, n | t_0, n_0)$ over a set of basis functions $\{\ketA{n}\}$, the ``kets'' (cf.\ middle column of \fref{fig:Outline}). The explicit choice of the basis functions depends on the process being studied. We discuss different choices of the basis functions in \sref{subsec:GenFct_ForwardBases}, first for a random walk, afterwards for chemical reactions and for processes whose particles locally exclude one another. Several methods have recently been proposed for the analysis of the PDE obeyed by the generating function. These methods include the variational method of Eyink~\cite{Eyink:1996} and of Sasai and Wolynes~\cite{Sasai:2003}, the WKB approximations~\cite{Bender:1999} and spectral methods of Elgart and Kamenev~\cite{Elgart:2004} and of Assaf and Meerson~\cite{Assaf:2006a,Assaf:2006b,Assaf:2008}, and the spectral method of Walczak, Mugler, and Wiggins~\cite{Walczak:2009}. We comment on these methods in \sref{subsec:GenFct_Spectral}. 
    
    In \sref{subsec:GenFctnl_MarginalizedDist}, we formulate conditions under which a general backward master equation can be transformed into a novel, backward-time PDE obeyed by a ``marginalized distribution''. This object is defined as the sum of the conditional probability distribution $\prob(t, n | \tVar, n_0)$ over a set of basis functions $\{\ketA{n_0}\}$ (cf.\ middle column of \fref{fig:Outline}). If the basis function $\ketA{n_0}$ is chosen as a probability distribution, the marginalized distribution also constitutes a true probability distribution. Different choices of the basis function are considered in \sref{subsec:GenFctnl_BackwardBases}. In \sref{subsec:GenFctnl_ExtTimes}, the use of the marginalized distribution is exemplified in the calculation of mean extinction times. Afterwards, in \sref{subsec:GenFctnl_Flow}, we derive yet another linear PDE, which is obeyed by a ``probability generating functional''. This functional is defined as the sum of the conditional probability distribution $\prob(\tVar, n | t_0, n_0)$ over a set of basis functionals $\{\braA{n}\}$, the ``bras''. In \sref{subsec:GenFctnl_PoissonRep}, we show that the way in which the generating functional ``generates'' probabilities generalizes the Poisson representation of Gardiner and Chaturvedi~\cite{Gardiner:1977,Chaturvedi:1978}.
    
    Sections~\ref{sec:PathInts_Backward} and~\ref{sec:PathInts_Forward} share the goal of representing the master equations' solution by path integrals. In \sref{subsec:PathInts_Backward_Derivation}, we first derive a novel \textit{backward path integral representation} from the PDE obeyed by the marginalized distribution (cf.\ right side of \fref{fig:Outline}). Its use is exemplified in sections~\ref{subsec:PathInts_SimpleGrowthLinearDecay}, \ref{subsec:PathInts_BackwardAlongPaths}, and \ref{subsec:PathInts_BackwardSolutions} in which we solve several elementary processes. Although we do not discuss the application of renormalization group techniques, \sref{subsubsec:PathInts_BackwardSolutions_DiffusionAndDecay} includes a discussion of how the backward path integral representation can be evaluated in terms of a perturbation expansion. The summands of the expansion are expressed by Feynman diagrams. Besides, we derive a path integral representation for Markov processes with continuous state spaces in the ``intermezzo'' \sref{subsec:PathInts_BackwardKramersMoyal}. This representation is obtained by performing a Kramers-Moyal expansion of the backward master equation and it comprises a classic path integral representation~\cite{Martin:1973,deDominicis:1976,Janssen:1976,Bausch:1976} of the (backward) Fokker-Planck equation as a special case. One can rewrite the representation of the Fokker-Planck equation in terms of an Onsager-Machlup function~\cite{Onsager:1953} and, for purely diffusive Brownian motion~\cite{Brown:1827}, the representation simplifies to the path integral of Wiener~\cite{Wiener:1921a,Wiener:1921b}. Moreover, we recover a Feynman-Kac like formula~\cite{Kac:1949}, which solves the (backward) Fokker-Planck equation in terms of an average over the paths of an It\^{o} stochastic differential equation~\cite{Ito:1944,Ito:1946,Ito:1950} (or of a Langevin equation~\cite{Lemons:1997}).
        
    In \sref{sec:PathInts_Forward}, we complement the backward path integral representation with a \textit{forward path integral representation}. Its derivation in \sref{subsec:PathInts_Forward_Derivation} starts out from the PDE obeyed by the generalized generating function (cf.\ right side of \fref{fig:Outline}). The forward path integral representation can, for example, be used to compute the generating function of generic linear processes as we demonstrate in \sref{subsec:PathInts_LinearProcesses}. Besides, we briefly outline how a Kramers-Moyal expansion of the (forward) master equation can be employed to derive a path integral representation for processes with continuous state spaces in \sref{subsec:PathInts_ForwardKramersMoyal}. This path integral can be expressed in terms of an average over the paths of an SDE proceeding backward in time. Its potential use remains to be explored.
    
    Before proceeding, let us briefly point out some properties of the forward and backward path integral representations. First, the paths along which these path integrals proceed are described by real variables and all integrations are performed over the real line. Grassmann path integrals~\cite{Sandow:1993,Patzlaff:1994,Bares:1999,Brunel:2000,Schulz:2005,Silva:2008} for systems whose particles locally exclude one another are not considered. It is, however, explained in \sref{subsec:GenFct_ForwardBases_Exclusion} how such systems can be treated without the need for Grassmann variables, based on a method recently proposed by van Wijland~\cite{Wijland:2001}. Second, transformations of the path integral variables such as the ``Doi-shift''~\cite{Cardy:2008} are implemented on the level of the basis functions and functionals. Third, our derivations of the forward and backward path integral representations do not involve coherent states or combinatoric relations for the commutation of exponentiated operators. Last, the path integrals allow for time-dependent rate coefficients of the stochastic processes.
    
    In \sref{sec:PathInts_Observables}, we derive a path integral representation of averaged observables (cf.\ right side of \fref{fig:Outline}). This representation can be derived both from the backward and forward path integral representations (cf.\ \sref{subsec:PathInts_Observables_Derivation}), and by representing the forward master equation in terms of the eigenvectors of creation and annihilation matrices (``coherent states''; cf.\ \sref{subsec:PathInts_Observables_Algebraic}). The duality between these two approaches resembles the duality between the wave~\cite{Schroedinger:1926c} and matrix~\cite{Heisenberg:1925,Born:1925,Born:1926} formulations of quantum mechanics. Let us note that our resulting path integral does not involve a ``second-quantized'' or ``normal-ordered'' observable~\cite{Cardy:1998}. In fact, we show that this object agrees with the average of an observable over a Poisson distribution. In \sref{subsec:PathInts_Observables_Perturbation}, we then explain how the path integral can be evaluated perturbatively using Feynman diagrams. Such an evaluation is demonstrated for the coagulation reaction $2\, A \to A$ in \sref{subsec:PathInts_Observables_Binary}, restricting ourselves to the ``tree level'' of the diagrams.
      
    In \sref{sec:StationaryPaths}, we review and extend a recent method of Elgart and Kamenev for the computation of rare event probabilities~\cite{Elgart:2004}. As explained in \sref{subsec:StationaryPaths_GenFct}, this method evaluates a probability distribution by expanding the forward path integral representation from \sref{sec:PathInts_Forward} around ``stationary'', or ``extremal'', paths. In a first step, one thereby acquires an approximation of the ordinary probability generating function. In a second step, this generating function is transformed back into the underlying probability distribution. The evaluation of this back transformation typically involves an additional saddle-point approximation. In \sref{subsec:StationaryPaths_GenFct_BinaryAnnihilation}, we demonstrate both of the steps for the binary annihilation reaction $2\, A \to \emptyset$, improving an earlier approximation of the process by Elgart and Kamenev~\cite{Elgart:2004} by terms of sub-leading order. In \sref{subsec:StationaryPaths_GenFctnl}, we then extend the ``stationary path method'' to the backward path integral representation from \sref{sec:PathInts_Backward}. The backward path integral provides direct access to a probability distribution without requiring an auxiliary saddle-point approximation.  However, the leading order term of its expansion is not normalized. We demonstrate the procedure for the binary annihilation reaction in~\sref{subsec:StationaryPaths_GenFctnl_BinaryAnnihilation}.
    
    Finally, \sref{sec:Summary} closes with a summary of the different approaches discussed in this review and outlines open challenges and  promising directions for future research.

  \subsection{Continuous-time Markov processes and the forward and backward master equations}\label{subsec:Intro_Mesoscopic}  
  
    Our main interest lies in a special class of stochastic processes, namely in the class of continuous-time Markov processes with discontinuous sample paths. These processes are also called ``jump processes''. In the following, we outline the mathematical theory of jump processes and derive the central equations obeyed by them: the forward and the backward master equation. Before going into the mathematical details, let us explain when a system's time evolution can be modelled as a continuous-time Markov process with discontinuous sample paths and what that phrase actually means. 
    
    First of all, if the evolution of a system is to be modelled as a continuous-time Markov process, it must be possible to describe the system's state by some variable $n$. In fact, it must be possible to do so at every point $\tVar$ in time throughout an observation period $[t_0,t]$. A variable $n\in\integers$ could, for example, represent the position of a molecular motor along a cytoskeletal filament, or a variable $n\in\reals_{\geq 0}$ the price of a stock between the opening and closing times of an exchange. The assumption of a continuous time parameter $\tVar$ is rather natural and conforms to our everyday experience. Still, a discrete time parameter may sometimes be preferred, for example, to denote individual generations of an evolving population~\cite{Sinervo:1996}. By allowing $\tVar$ to take on any value between the initial time $t_0$ and the final time $t$, we can choose it to be arbitrarily close to one of those times. Below, this possibility will allow us to describe the evolution of the process in terms of a differential equation.
    
    The (unconditional) probability of finding the system in state $n$ at time $\tVar$ is represented by the ``single-time'' probability distribution $\prob(\tVar, n)$. Upon demanding that the system has visited some state $n_0$ at an earlier time $t_0<\tVar$, the probability of observing state $n$ at time $\tVar$ is instead encoded by the conditional probability distribution $\prob(\tVar, n | t_0, n_0)$. If the conditional probability distribution is known, the single-time distribution can be inferred from any given initial distribution $\prob(t_0, n_0)$ via $\prob(\tVar, n) = \sum_{n_0} \prob(\tVar, n | t_0, n_0) \prob(t_0, n_0)$. A stochastic process is said to be \textit{Markovian} if a distribution conditioned on multiple points in time actually depends only on the state that was realized most recently. In other words, a conditional distribution $\prob(t, n | \tVar_k, m_k; \cdots ; \tVar_1, m_1; t_0, n_0)$ must agree with $\prob(t, n | \tVar_k, m_k)$ whenever $t > \tVar_k > \tVar_{k-1},\hldots, t_0$.\footnote{Note that we do not distinguish between random variables and their outcomes. Moreover, we stick to the physicists' convention of ordering times in descending order. In the mathematical literature, the reverse order is more common, see e.g.~\cite{Kolmogoroff:1931}.} Therefore, a Markov process is fully characterized by the single-time distribution $\prob(t_0, n_0)$ and the conditional distribution $\prob(\tVar, n | t_0, n_0)$. The latter function is commonly referred to as the ``transition probability'' for Markov processes~\cite{vanKampen:2007}.
    
    The stochastic dynamics of a system can be modelled in terms of a Markov process if the system has no memory. Let us explain this requirement with the example of a Brownian particle suspended in a fluid~\cite{Brown:1827}. Over a very short time scale, the motion of such a particle is ballistic and its velocity highly auto-correlated~\cite{Huang:2011}. But as the particle collides with molecules of the fluid, that memory fades away. A significant move of the particle due to fluctuations in the isotropy of its molecular bombardment then appears to be completely uncorrelated from previous moves (provided that the observer does not look too closely~\cite{Franosch:2011}). Thus, on a sufficiently coarse time scale, the motion of the particle looks diffusive and can be modelled as a Markov process. However, the validity of the Markov assumption does not extend beyond the coarse time scale.
    
    The Brownian particle exemplifies only two of the properties that we are looking for: its position is well-defined at every time $\tVar$ and its movement is effectively memoryless on the coarse time scale. But the paths of the Brownian particle are continuous, meaning that it does not spontaneously vanish and then reappear at another place. If the friction of the fluid surrounding the Brownian particle is high (over-damped motion), the probability of observing the particle at a particular place can be described by the Smoluchowski equation~\cite{vonSmoluchowski:1906}. This equation coincides with the simple diffusion equation when the particle is not subject to an external force. In the general case, the probability of observing the particle at a particular place with a particular velocity obeys the Klein-Kramers equation~\cite{Klein:1921,Kramers:1940} (the book of Risken~\cite{Risken:1996} provides a pedagogic introduction to these equations). From a mathematical point of view, all of these equations constitute special cases of the (forward) Fokker-Planck equation~\cite{Einstein:1905,Fokker:1914,Planck:1917,Kolmogoroff:1931,Risken:1996}. For a single random variable $\xAd\in\reals$, e.g.\ the position of the Brownian particle, this equation has the generic form
    \begin{equation}  
      \partial_\tVar \prob(\tVar, \xAd | t_0, \xAd_0)      \label{eq:Intro_Mescoscopic_FP}
        =   - \partial_\xAd \bigl[\alpha_\tVar(\xAd) \prob \bigr]
          +\frac{1}{2}\partial_\xAd^2 \bigl[ \beta_\tVar(\xAd) \prob \bigr] \,.
    \end{equation} 
    The initial condition of this equation is given by the Dirac delta distribution (or generalized function) $\prob(t_0, \xAd | t_0, \xAd_0) = \delta(\xAd-\xAd_0)$. Here we used the letter $\xAd$ for the random variable because the letter $n$ would suggest a discrete state space. The function $\alpha_\tVar$ is often called a drift coefficient and $\beta_\tVar$ a diffusion coefficient (note, however, that in the context of population genetics, $\beta_\tVar$ describes the strength of random genetic ``drift''~\cite{Kimura:1957,deVladar:2011}). For reasons addressed below, the diffusion coefficient must be non-negative at every point in time for every value of $\xAd$ (for a multivariate process, $\beta_\tVar$ represents a positive-semidefinite matrix). In the mathematical community,  the Fokker-Planck equation is better known as the Kolmogorov forward equation~\cite{Kimura:1957}, honouring Kolmogorov's fundamental contributions to the theory of continuous-time Markov processes~\cite{Kolmogoroff:1931}. Whereas the above Fokker-Planck equation evolves the conditional probability distribution forward in time, one can also evolve this distribution backward in time, starting out from the final condition $\prob(t, \xAd | t, \xAd_0) = \delta(\xAd-\xAd_0)$. The corresponding equation is called the Kolmogorov backward or backward Fokker-Planck equation. It has the generic form
    \begin{equation}  
      \partial_{-\tVar} \prob(t, \xAd | \tVar, \xAd_0)        \label{eq:Intro_Mescoscopic_FP_BW}
        =   \alpha_\tVar(\xAd_0) \partial_{\xAd_0}\prob
          +\frac{1}{2} \beta_\tVar(\xAd_0)\partial_{\xAd_0}^2 \prob \,.
    \end{equation} 

    The forward and backward Fokker-Planck equations provide information about the conditional probability distribution but not about the individual paths of a Brownian particle. The general theory of how partial differential equations connect to the individual sample paths of a stochastic process goes back to works of Feynman and Kac~\cite{Feynman:2005,Kac:1949}.\footnote{Due to the central importance of the Feynman-Kac formula, we provide a brief proof of it in \aref{sec:A_FeynmanKacProof}. We also encounter the formula in \sref{sec:PathInts_Backward} in evaluating a path integral representation of the (backward) master equation.}
    Their theory allows us to write the solution of the backward Fokker-Planck equation~\eref{eq:Intro_Mescoscopic_FP_BW} in terms of the following Kolmogorov formula, which constitutes a special case of the Feynman-Kac formula~\cite{Kloeden:1992,Majumdar:2005,Gardiner:2009}:
    \begin{equation}
      \prob(t, \xAd | \tVar, \xAd_0)    \label{eq:Intro_Mescoscopic_FP_BW_Solution}
      = \bigLLangle \delta \bigl(\xAd - \xAd(t)\bigr) \bigRRangle_\wienerProcess  \,.
    \end{equation}
    The brackets $\LLangle \cdot \RRangle_{\wienerProcess}$ represent an average over realizations of a Wiener process $\wienerProcess$, which evolves through uncorrelated Gaussian increments $\diff{W}$. The Wiener process drives the evolution of the sample path $\xAd(\sVar)$ from $\xAd(\tVar) = \xAd_0$ to $\xAd(t)$ via the It\^{o} stochastic differential equation (SDE)~\cite{Ito:1944,Ito:1946,Ito:1950}
    \begin{equation}
      \diff{\xAd}(\sVar)     \label{eq:Intro_Mescoscopic_Ito}
      = \alpha_\sVar(\xAd(\sVar)) \diff{\sVar}  + \sqrt{\beta_\sVar(\xAd(\sVar))}\diff{W}(\sVar)   \,.
    \end{equation} 
    The diffusion coefficient $\beta_\sVar$ must be non-negative because $\xAd(\sVar)$ describes the position of a real particle. Otherwise, the sample path heads off into imaginary space (for a multivariate process, $\sqrt{\beta_\sVar}$ may be chosen as the unique symmetric and positive-semidefinite square root of $\beta_\sVar$~\cite{Harville:1997}). Algorithms for the numerical solution of SDEs are provided in~\cite{Kloeden:1992}. In the physical sciences, SDEs are often written as Langevin equations~\cite{Lemons:1997}.\footnote{The Langevin equation corresponding to the SDE~\eref{eq:Intro_Mescoscopic_Ito} reads $\partial_\sVar \xAd(\sVar)  = \alpha_\sVar(\xAd(\sVar))  + \sqrt{\beta_\sVar(\xAd(\sVar))} \eta(\sVar)$, with the Gaussian white noise $\eta(\sVar)$ having zero mean and the auto-correlation function $\langle \eta(\sVar)\eta(\sVar^\prime)\rangle = \delta(\sVar-\sVar^\prime)$.} For a discussion of stochastic differential equations the reader may refer to a recent report on progress \cite{Volpe:2016}. 
    
    After this brief detour to continuous-time Markov processes with continuous sample paths, let us return to jump processes, whose sample paths are discontinuous. A system that can be modelled as such a process are motor proteins on cytoskeletal filaments~\cite{Lipowsky:2001,Parmeggiani:2003,Parmeggiani:2004}. The uni-directional walk of a molecular motor such as myosin, kinesin, or dynein along an actin filament or a microtubule is driven by the hydrolysis of adenosine triphosphate (ATP) and is intrinsically stochastic~\cite{HowardJ:1997}. Once a sufficient amount of energy is available, one of the two ``heads'' of the motor unbinds from its current binding site on the filament and moves to the next binding site. Each binding site can only be occupied by a single head. On a coarse-grained level, the state of the system at time $\tVar$ is therefore characterized by the occupation of its binding sites. With only a single cytoskeletal filament whose binding sites are labelled as $\{0,1,2,\hldots \}\eqqcolon\lattice$, the variable $n \equiv (n_0, n_1, n_2,\hldots)\in\{0,1\}^{|\lattice|}$ can be used to represent the occupied and unoccupied binding sites. Here, $|\lattice|$ denotes the total number of binding sites along the filament and $n_i=1$ signifies that the $i$-th binding site is occupied. Since the state space of all the binding site configurations is discrete, a change in the binding site configuration involves a ``jump'' in state space. Provided that the jumps are uncorrelated from one another (which needs to be verified experimentally), the dynamics of the system can be described by a continuous-time Markov process with discontinuous sample paths. Before addressing further systems for which this is the case, let us derive the fundamental equations obeyed by these processes: the forward and the backward master equation.
    
    In his classic textbook~\cite{Gardiner:2009}, Gardiner presents a succinct derivation of both the (forward) master and the (forward) Fokker-Planck equation by distinguishing between discontinuous and continuous contributions to sample paths. In the following, we are only interested in the master equation, which governs the evolution of systems whose states change discontinuously. To prevent the occurrence of continuous changes, we assume that the state of our system is represented by a discrete variable $n$ and that the space of all states is countable. With the state space chosen as the set of integers $\integers$, $n$ could, for example, represent the position of a molecular motor along a cytoskeletal filament. On the other hand, $n\in\naturals_0$ could represent the number of molecules in a chemical reaction. The minimal jump size is one in both cases. By keeping the explicit role of $n$ unspecified, the following considerations also apply to systems harbouring different kind of molecules (e.g.\ $n \equiv (n^{A}, n^{B}, n^{C}) \in \naturals_0^3$), and to systems whose molecules perform random walks in a (discrete) space (e.g.\ $n \equiv \{n_i\in \naturals_0\}_{i\in\integers}$). 
    
    To derive the master equation, we start out by marginalizing the joint conditional distribution $\prob(t, n ; \tVar, m | t_0, n_0)$ over the state $m$ at the intermediate time $\tVar$ ($t \geq \tVar \geq t_0$), resulting in
    \begin{equation}
      \prob(t, n | t_0, n_0) = \sum_{ m } \prob(t, n ; \tVar, m | t_0, n_0)   \,.
    \end{equation}
    Whenever the range of a sum is not specified, it shall cover the whole state space of its summation variable. The above equation holds for arbitrary stochastic processes. But for a Markov process, one can employ the relation between joint and conditional distributions to turn the equation into the Chapman-Kolmogorov equation
    \begin{equation} 
      \prob(t, n | t_0, n_0)              
      =  \sum_{ m }   \prob(t, n | \tVar, m) \prob(\tVar, m | t_0, n_0)          \label{eq:Intro_Mescoscopic_ChapmanKolmogorov}  \,.
    \end{equation} 
    Letting $\prob(t | t_0)$ denote the matrix with elements $\prob(t, n | t_0, n_0)$, the Chapman-Kolmogorov equation can also be written as $\prob(t | t_0) = \prob(t | \tVar) \prob(\tVar | t_0)$ (semigroup property). Note that the matrix notation requires a mapping between the state space of $n$ and $n_0$ and an appropriate index set $I\subset\naturals$. However, we also make use of this notation when the state space is countably infinite.
    
    To derive the (forward) master equation from the Chapman-Kolmogorov equation~\eref{eq:Intro_Mescoscopic_ChapmanKolmogorov}, we define
    \begin{equation} 
      \qMatrix_{\tVar,\Delta t}(n, m) \label{eq:Intro_Mescoscopic_RateOfProbChange_Discrete}
        \coloneqq
        \frac{\prob(\tVar + \Delta t, n | \tVar, m) - \delta_{n,m}}{\Delta t}       
    \end{equation} 
    for all values of $n$ and $m$ and assume the existence and finiteness of the limits
    \begin{equation} 
      \qMatrix_\tVar(n, m) 
        \coloneqq \lim_{\Delta t \to 0} \qMatrix_{\tVar,\Delta t}(n, m)       \,.
        \label{eq:Intro_Mescoscopic_RateOfProbChange}    
    \end{equation} 
    These are the elements of the transition (rate) matrix $\qMatrix_\tVar$, which is also called the infinitesimal generator of the Markov process or is simply referred to as the $\qMatrix_\tVar$-matrix. Its off-diagonal elements $\transitionRate_\tVar(n, m)   \coloneqq \qMatrix_\tVar(n, m)$ denote the rates at which probability flows from a state $m$ to a state $n\neq m$. The ``exit rates'' $\exitRate_{\tVar}(m) \coloneqq -\qMatrix_{\tVar}(m,m)$, on the other hand, describe the rates at which probability leaves state~$m$. Both $\transitionRate_\tVar(n, m)$ and $\exitRate_{\tVar}(m)$ are non-negative for all $n$ and $m$. All of the processes considered here shall conserve the total probability, requiring that $\sum_{n} \qMatrix_\tVar(n, m) = 0$ or, equivalently, $\exitRate_{\tVar}(m) = \sum_{n} \transitionRate_\tVar(n, m)$ (with $\transitionRate_\tVar(m, m) \coloneqq 0$). The finiteness of the exit rate $\exitRate_{\tVar}(m)$ and the conservation of total probability imply that we consider a \textit{stable} and \textit{conservative} Markov process~\cite{Anderson:1991}. In the natural sciences, the master equation is commonly written in terms of $\exitRate_{\tVar}$, but most mathematicians prefer $\qMatrix_\tVar$. These matrices can be converted into one another by employing
    \begin{equation} 
      \qMatrix_\tVar(n, m)   \label{eq:Intro_Mescoscopic_TransitionRate}
      = \transitionRate_\tVar(n, m) - \delta_{n, m} \exitRate_{\tVar}(m) \,.
    \end{equation} 
    We refer to both of the matrices as transition (rate) matrices and to their (identical) off-diagonal elements as transition rates. The transition rates fully specify the stochastic process.
    
    Assuming that the limit in~\eref{eq:Intro_Mescoscopic_RateOfProbChange} interchanges with a sum over the state $m$, the (forward) master equation now follows from the Chapman-Kolmogorov equation~\eref{eq:Intro_Mescoscopic_ChapmanKolmogorov} as
    \begin{eqnarray} 
      \partial_\tVar \prob(\tVar, n | t_0, n_0 )    \label{eq:Intro_Mescoscopic_MasterEq}  \\
      = \lim_{\Delta t \to 0} \frac{\prob(\tVar+\Delta t, n | t_0, n_0 )-\prob(\tVar, n | t_0, n_0 )}{\Delta t} \nonumber\\
      = \lim_{\Delta t \to 0} \sum_{m} \frac{\prob(\tVar+\Delta t, n | \tVar, m )-\delta_{n,m}}{\Delta t} \prob(\tVar, m | t_0, n_0 )  \nonumber\\
      = \sum_{m} \qMatrix_\tVar(n, m) \prob(\tVar, m | t_0, n_0 )    \nonumber  \,.
    \end{eqnarray} 
    Thus, the master equation constitutes a set of coupled, linear, first-order ordinary differential equations (ODEs). The time evolution of the distribution starts out from $\prob(t_0, n | t_0, n_0 ) = \delta_{n,n_0}$. In matrix notation, the equation can be written as $\partial_\tVar \prob(\tVar | t_0) = \qMatrix_\tVar \prob(\tVar | t_0)$. In terms of $\transitionRate_\tVar$, it assumes the intuitive gain-loss form
    \begin{eqnarray}
      \partial_\tVar \prob(\tVar, n | t_0, n_0 )      \label{eq:Intro_Mescoscopic_MasterEqGainLoss}  \\
      = \sum_{m} \bigl[
          \transitionRate_\tVar(n, m) \prob(\tVar, m | \cdot )   
           - \transitionRate_\tVar(m, n) \prob(\tVar,n | \cdot )      
        \bigr]    \,.  \nonumber  
    \end{eqnarray} 
    The dot inside the probability distribution's argument abbreviates the initial parameters $t_0$ and $n_0$, which are of secondary concern right here. That will change below in the derivation of the backward master equation. An omission of the parameters also makes it impossible to distinguish the conditional distribution $\prob(\tVar, n | t_0, n_0)$ from the single-time distribution $\prob(\tVar,n)$. The single-time distribution obeys the master equation as well, as can inferred directly from the relation $\prob(\tVar, n) = \sum_{n_0} \prob(\tVar, n | t_0, n_0) \prob(t_0, n_0)$ or by summing the above master equation over an initial distribution~$\prob(t_0,n_0)$. In fact, the single-time distribution would even obey the master equation if the process was not Markovian, but without providing a complete characterization of the process~\cite{Oppenheim:1965,Taeuber:2014}. The master equation~\eref{eq:Intro_Mescoscopic_MasterEq} or~\eref{eq:Intro_Mescoscopic_MasterEqGainLoss} is particularly interesting for transition rates causing an imbalance between forward and backward transitions along closed cycles of states, i.e.\ for rates violating Kolmogorov's criterion~\cite{Kolmogoroff:1936} for detailed balance~\cite{Taeuber:2014}. Such systems are truly out of thermal equilibrium. If detailed balance is instead fulfilled, the system eventually converges to a stationary Boltzmann-Gibbs distribution with vanishing probability currents between states~\cite{Taeuber:2014}. Whether or not detailed balance is actually fulfilled is, however, not relevant for the methods discussed in this review. Information on the existence and uniqueness of an asymptotic stationary distribution of the master equation can be found in~\cite{Anderson:1991}.
    
    The name ``master equation'' was originally coined by Nordsieck, Lamb, and Uhlenbeck~\cite{Nordsieck:1940} in their study of the Furry model of cosmic rain showers~\cite{Furry:1937}. Shortly before, Feller applied an equation of the same structure to the growth of populations~\cite{Feller:1939} and Delbr\"uck to well-mixed, auto-catalytic chemical reactions~\cite{Delbrueck:1940}. Delbr\"uck's line of research was followed by several others~\cite{Singer:1953,Bartholomay:1958,Krieger:1960,Ishida:1964}, most notably by McQuarrie~\cite{McQuarrie:1963,McQuarrie:1964,McQuarrie:1967} (see also the books~\cite{Nicolis:1977,vanKampen:2007,Gardiner:2009}). In these articles, several elementary chemical reactions are solved by methods that also appear later in this review. When the particles engaging in a reaction can also diffuse in space, their density may exhibit dynamics that are not expected from observations made in well-mixed environments. Hence, reaction-diffusion master equations have been the focus of intense research and have been analysed using path integrals (see, for example, \cite{Lee:1994b,Lee:1995,Cardy:1996,Cardy:1998,Janssen:2005} and the references in \sref{subsec:Intro_PathInts}). Master equations, and simulations algorithms based on master equations, are now being used in numerous fields of research. They are being applied in the contexts of spin dynamics~\cite{Glauber:1963a,Kawasaki:1966a,Kawasaki:1966b,Kawasaki:1966c}, gene regulatory networks~\cite{Elowitz:2000,Thattai:2001,Rao:2002,Karlebach:2008,Shahrezaei:2008,Walczak:2009,Tsimring:2014}, the spreading of diseases~\cite{Bailey:1950,Keeling:2008,Black:2011,Rock:2014}, epidermal homeostasis~\cite{Clayton:2007}, nucleosome repositioning~\cite{Chou:2007}, ecological~\cite{Volkov:2003,McKane:2004,McKane:2005,Mobilia:2006,Reichenbach:2006a,Volkov:2007,Butler:2009a,Butler:2009b,Park:2010,Noble:2011,Dobrinevski:2012,Black:2012a,Taeuber:2012,Shih:2014} and bacterial dynamics~\cite{Park:2010,Cremer:2012,Reiter:2014,Weber:2014,Wienand:2015}, evolutionary game theory~\cite{Reichenbach:2007a,Reichenbach:2007b,Reichenbach:2008,Mobilia:2010b,Assaf:2010c,Frey:2010,Melbinger:2010,Cremer:2011,Traulsen:2012,Black:2012b,Knebel:2015,Melbinger:2015}, surface growth~\cite{Krug:1991}, and social and economic processes~\cite{Weidlich:1991,Weidlich:1992,Lux:1995,Castellano:2009}. Queuing processes are also often modelled in terms of master equations, but in this context, the equations are typically referred to as Kolmogorov equations~\cite{Gross:2008}. Moreover, master equations and the SSA have helped to understand the formation of traffic jams on highways~\cite{Nagel:1992,Helbing:2001}, the walks of molecular motors along cytoskeletal filaments~\cite{Lipowsky:2001,Parmeggiani:2003,Parmeggiani:2004,Reichenbach:2006b,Kriecherbauer:2010,Chou:2011}, and the condensation of bosons in driven-dissipative quantum systems~\cite{Vorberg:2013,Knebel:2015,Vorberg:2015}. The master equation that was found to describe the coarse-grained dynamics of these bosons coincides with the master equation of the (asymmetric) inclusion process~\cite{Evans:2014,Grosskinsky:2011,Giardina:2010,Giardina:2007}. Transport processes are commonly modelled in terms of the (totally) asymmetric simple exclusion process (ASEP or TASEP)~\cite{Spitzer:1970,Spohn:1991,Liggett:1999,Schuetz:2000}. The ASEP describes the biased hopping of particles along a one-dimensional lattice, with each lattice site providing space for at most one particle. The ASEP and the TASEP are regarded as paradigmatic models in the field of non-equilibrium statistical mechanics, with many exact mathematical results having been established~\cite{Janowsky:1992,Derrida:1992,Derrida:1993,Schuetz:1993,Schuetz:1997,Johansson:2000,Priezzhev:2003,Priezzhev:2005,Tracy:2008a,Tracy:2008b,Tracy:2009a,Tracy:2009b}. Some of these results were established by applying the Bethe ansatz to the master equation of the ASEP~\cite{Schuetz:1997,Tracy:2008a}. The review of Chou, Mallick, and Zia provides a comprehensive account of the ASEP and of its variants~\cite{Chou:2011}. The master equation of the TASEP with Langmuir kinetics was recently used to understand the length regulation of microtubules~\cite{Melbinger:2012}.
    
    Unlike deterministic models, the master equations describing the dynamics of the above systems take into account that discrete and finite populations are prone to ``demographic fluctuations''. The populations of the above systems consist of genes or proteins, infected persons, bacteria or cars and they are typically small, at least compared to the number of molecules in a mole of gas. For example, the copy number of low abundance proteins in \textit{Escherichia coli} cytosol was found to be in the tens to hundreds~\cite{Ishihama:2008}. Therefore, the presence or absence of a single protein is much more important than the presence or absence of an individual molecule in a mole of gas. A demographic fluctuation may even be fatal for a system, for example, when the copy number of an auto-catalytic reactant drops to zero. The master equation~\eref{eq:Intro_Mescoscopic_MasterEq} provides a useful tool to describe such an effect.
            
    Up to this point, we have only considered the forward master equation. But just as the (forward) Fokker-Planck equation~\eref{eq:Intro_Mescoscopic_FP} is complemented by the backward Fokker-Planck equation~\eref{eq:Intro_Mescoscopic_FP_BW}, the (forward) master equation~\eref{eq:Intro_Mescoscopic_MasterEq} is complemented by a backward master equation. This equation can be derived from the Chapman-Kolmogorov equation~\eref{eq:Intro_Mescoscopic_ChapmanKolmogorov} as
    \begin{eqnarray} 
      \partial_{-\tVar} \prob(t, n | \tVar, n_0 )    \label{eq:Intro_Mescoscopic_BackwardMasterEq}\\
      = \lim_{\Delta t \to 0} \frac{\prob(t, n | \tVar-\Delta t, n_0 )-\prob(t, n | \tVar, n_0 )}{\Delta t}  \nonumber\\
      = \lim_{\Delta t \to 0} \sum_{m}\prob(t, n | \tVar, m )  \frac{\prob(\tVar, m | \tVar - \Delta t, n_0 )-\delta_{m,n_0}}{\Delta t}  \nonumber\\
      = \sum_{m}\prob(t, n | \tVar, m ) \qMatrix_\tVar(m, n_0)
          \nonumber  \,.
    \end{eqnarray} 
    Here, the transition rate is obtained in the limit $\lim_{\Delta t \to 0} \qMatrix_{\tVar-\Delta t,\Delta t}(m, n_0)$ (cf.\ \eref{eq:Intro_Mescoscopic_RateOfProbChange_Discrete}). In matrix notation, the backward master equation reads $\partial_{-\tVar} \prob(t | \tVar) = \prob(t | \tVar) \qMatrix_\tVar$. In terms of $\transitionRate_\tVar$, it assumes the form (cf.~\eref{eq:Intro_Mescoscopic_TransitionRate})
    \begin{eqnarray} 
       \partial_{-\tVar} \prob(t, n | \tVar, n_0 )   \label{eq:Intro_Mescoscopic_BackwardMasterEqGainLoss}  \\
       = \sum_{m} \bigl[
          \prob(\cdot | \tVar, m) - \prob(\cdot | \tVar, n_0)      
        \bigr] \transitionRate_\tVar(m, n_0)    \,.  \nonumber
    \end{eqnarray} 
    In this equation, the dots abbreviate the final parameters $t$ and $n$. The backward master equation evolves the conditional probability distribution backward in time, starting out from the final condition $\prob(t, n | t, n_0 ) = \delta_{n,n_0}$. Just as the backward Fokker-Planck equation, the backward master equation proves useful for the computation of mean extinction and first passage times (see~\cite{Doering:2005,Gardiner:2009} and \sref{subsec:GenFctnl_ExtTimes}). Furthermore, it follows from the backward master equation~\eref{eq:Intro_Mescoscopic_BackwardMasterEq} that the (conditional) average $\langle \observable \rangle(t | \tVar, n_0) \coloneqq \sum_{n} A(n) \prob(t, n | \tVar, n_0)$ of an observable $A$ fulfils an equation of just the same form, namely
    \begin{equation} 
      \partial_{-\tVar} \langle \observable \rangle(t | \tVar, n_0)   \label{eq:Intro_Mescoscopic_BackwardMasterEqObservable}
      = \sum_{m}\langle\observable \rangle(t | \tVar, m) \qMatrix_\tVar(m, n_0)     \,.
    \end{equation} 
    The final condition of the equation is given by $\langle \observable \rangle(t | t, n_0) = A(n_0)$. The validity of equation~\eref{eq:Intro_Mescoscopic_BackwardMasterEqObservable} is the reason why we later employ a ``backward'' path integral to represent the average $\langle \observable \rangle$ (cf.\ \sref{sec:PathInts_Observables}).
    
  \subsection{Analytical and numerical methods for the solution of master equations}\label{subsec:Intro_MasterEqSolution}  
    
     If the dynamics of a system are restricted to a finite number of states and if its transition rates are independent of time, both the forward master equation $\partial_\tVar \prob(\tVar | t_0) = \qMatrix \prob(\tVar | t_0)$ and the backward master equation $\partial_{-\tVar_0} \prob(t | \tVar_0) = \prob(t | \tVar_0) \qMatrix$ are solved by~\cite{Norris:1998}
     \begin{equation}
       \prob(\tVar| \tVar_0) = \ee^{\qMatrix(\tVar-\tVar_0)}\unitMatrix    
       \label{eq:Intro_MasterEqSolution_MatrixExp}
     \end{equation}
     (recall that $\prob(\tVar | \tVar_0)$ is the matrix with elements $\prob(\tVar, n | \tVar_0, n_0)$). The Chapman-Kolmogorov equation~\eref{eq:Intro_Mescoscopic_ChapmanKolmogorov} is also solved by the distribution. Although the matrix exponential inside this solution can in principle be evaluated in terms of the (convergent) Taylor expansion $\sum_{k=0}^\infty\frac{(\tVar-\tVar_0)^k}{k!}\qMatrix^k$, the actual calculation of this series is typically infeasible for non-trivial processes, both analytically and numerically (a truncation of the Taylor series may induce severe round-off errors and serves as a lower bound on the performance of algorithms in~\cite{Moler:2003}). Consequently, alternative numerical algorithms have been developed to evaluate the matrix exponential. Moler and Van Loan reviewed ``nineteen dubious ways'' of computing the exponential in~\cite{Moler:1978,Moler:2003}. Algorithms that can deal with very large state spaces are considered in~\cite{Munsky:2006,Burrage:2006}. For time-dependent transition rates, the matrix exponential generalizes to a Magnus expansion~\cite{Magnus:1954,Blanes:2009}.
       
    In the previous paragraph, we restricted the dynamics to a finite state space to ensure the existence of the matrix exponential in~\eref{eq:Intro_MasterEqSolution_MatrixExp}. Provided that the supremum $\sup_{m} |\qMatrix(m,m)|$ of all the exit rates is finite (uniformly bounded $\qMatrix$-matrix), the validity of the above solution extends to state spaces comprising a countable number of states~\cite{Klenke:2014}. To see that, 
    we define the left stochastic matrix $P\coloneqq \unitMatrix + \lambda^{-1} \qMatrix$, with the parameter $\lambda$ being larger than the above supremum. Writing $\ee^{\qMatrix \Delta t} = \ee^{-\lambda\Delta t} \ee^{\lambda \Delta t P}$ with $\Delta t \coloneqq \tVar-\tVar_0$, the matrix exponential can be evaluated in terms of the convergent Taylor series
    \begin{equation}
      \prob(\tVar| \tVar_0)   \label{eq:Intro_MasterEqSolution_Uniform}
      = \sum_{k=0}^\infty \frac{\ee^{-\lambda \Delta t} (\lambda \Delta t)^k}{k!} P^k     \,.
    \end{equation} 
    Effectively, one has thereby decomposed the continuous-time Markov process with transition matrix $\qMatrix$ into a discrete-time Markov chain with transition matrix $P$, subordinated to a continuous-time Poisson process with rate coefficient $\lambda$ (the Poisson process acts as a ``clock'' with sufficiently high ticking rate $\lambda$). Such a decomposition is called a uniformization or randomization and was first proposed by Jensen~\cite{Jensen:1953}. The series~\eref{eq:Intro_MasterEqSolution_Uniform} can be evaluated via numerically stable algorithms and truncation errors can be bounded~\cite{Hellander:2008,Didier:2009}. Nevertheless, the uniformization method requires the computation of the powers of a matrix having as many rows and columns as the system has states. Consequently, a numerical implementation of the method is only feasible for sufficiently small state spaces. Further information on the method and on its improvements can be found in~\cite{Jensen:1953,Gross:1984,Reibman:1988,vanMoorsel:1994,Hellander:2008,Didier:2009}.

    The mathematical study of the existence and uniqueness of solutions of the forward and backward master equations was pioneered by Feller and Doob in the 1940s~\cite{Feller:1940,Doob:1945}. Feller derived an integral recurrence formula~\cite{Feller:1940,Anderson:1991}, which essentially constitutes a single step of the ``path summation representation'' that we derive further below. In the following, we assume that the forward and the backward master equations have the same unique solution and we restrict our attention to processes performing only a finite number of jumps during any finite time interval. These conditions, and the conservation of total probability, are, for example, violated by processes that ``explode'' after a finite time.\footnote{Just as a population whose growth is described by the deterministic equation $\partial_\tVar n =  n^2$ explodes after a finite time, so does a population whose growth is described by the master equation~\eref{eq:Intro_Mescoscopic_MasterEqGainLoss} with transition rate $\transitionRate(n, m) = \delta_{n,m+1} m(m-1)$~\cite{Anderson:1991}. This transition rate models the elementary reaction $2\, A \to 3\, A$ as explained in \sref{subsec:Intro_PathInts}. An explosion also occurs for the rate $\transitionRate(n, m) = \delta_{n,m+1} m^2$.} More information on such processes is provided in~\cite{Anderson:1991, Norris:1998}. 
      
    In the following, we complement the above representations of the master equations's solution with a ``path summation representation''. This representation can be derived by examining the steps of the stochastic simulation algorithm (SSA) of Gillespie (its ``direct'' version)~\cite{Gillespie:1976,Gillespie:1977,Gillespie:2007} or by performing a Laplace transformation of the forward master equation. Here we follow the former, qualitative, approach. A formal derivation of the representation via the master equation's Laplace transform is provided in \aref{sec:A_PathSummationProof}. Although the basic elements of the SSA had already been known before Gillespie's work~\cite{Feller:1940,Doob:1942,Doob:1945,Kendall:1949,Kendall:1950,Bartlett:1953,Bortz:1975}, its popularity largely increased after Gillespie applied it to the study of chemical reactions. As the SSA is restricted to time-independent transition rates, so is the following derivation. 
         
    \begin{figure}[tb] 
      \centering
      \includegraphics{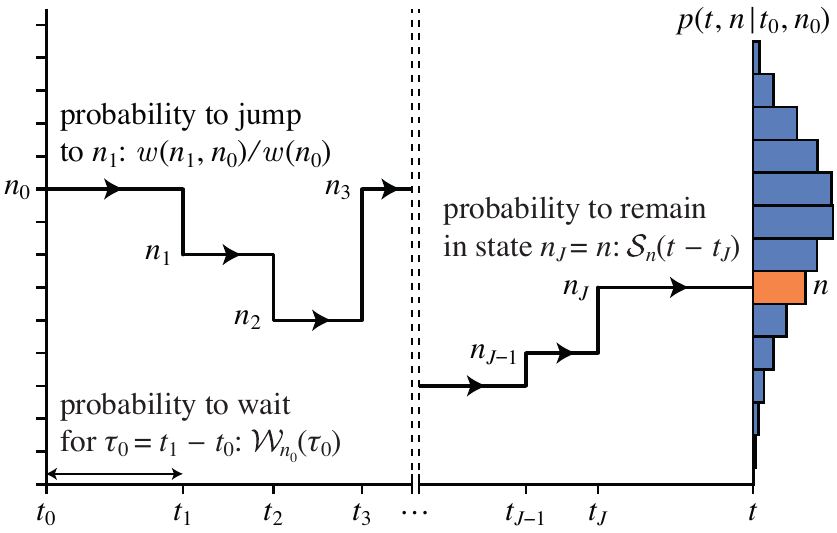}
       \caption{\label{fig:Gillespie}
        Illustration of the stochastic simulation algorithm (SSA) and of the path summation representation of the probability distribution $\prob(t, n| t_0, n_0)$. In a numerical implementation of the SSA, the system is prepared in state $n_0$ at time $t_0$. After a waiting time $\tVar_0$ that is drawn from the exponential distribution $\waitingTimeDist_{n_0}(\tVar_0) = \exitRate(n_0) \ee^{-\exitRate(n_0) \tVar_0}\HeavisideStep(\tVar_0)$, the system transitions into a new state. An arrival state $n_1$ is chosen with probability $\transitionRate(n_1, n_0)/\exitRate(n_0)$. The procedure is repeated until after $J$ steps, the current time $t_J \leq t$ plus an additional waiting time exceeds~$t$. The sample path has thus resided in state $n_J$ at time~$t$. This information is recorded in a histogram approximation of $\prob(t, n| t_0, n_0)$. The path summation representation of $\prob(t, n| t_0, n_0)$ requires $n_J$ to coincide with $n$. The probability that the system has remained in state $n$ over the last time interval $[t_J,t]$ is given by the survival probability $\survivalProb_n(t-t_J) = \ee^{-\exitRate(n) (t-t_J)}\HeavisideStep(t-t_J)$. The total probability of the generated path, integrated over all possible waiting times, is represented by $\prob_\tVar(\stochPath_J)$ in~\eref{eq:Intro_Mescoscopic_Solution_Probability}. A summation of this probability over all possible sample paths results in the path summation representation~\eref{eq:Intro_Mescoscopic_Solution}.
      }
    \end{figure}    
    
    To derive the path summation representation, we prepare a system in state $n_0$ at time $t_0$ as illustrated in \fref{fig:Gillespie}. Since the process is homogeneous in time, we may choose $t_0=0$. The total rate of leaving state $n_0$ is given by the exit rate $\exitRate(n_0) = \sum_{n_1} \transitionRate(n_1, n_0)$. In order to determine how long the system actually stays in state $n_0$, one may draw a random waiting time $\tVar_0$ from the exponential distribution $\waitingTimeDist_{n_0}(\tVar_0) \coloneqq \exitRate(n_0) \ee^{-\exitRate(n_0) \tVar_0}\HeavisideStep(\tVar_0)$. The Heaviside step function $\HeavisideStep$ prevents the sampling of negative waiting times and is here defined as $\HeavisideStep(\tVar) = 1$ for $\tVar \geq 0$ and $\HeavisideStep(\tVar) = 0$ for $\tVar < 0$. Thus far, we only know that the system leaves $n_0$ but not where it ends up. It could end up in any state $n_1$ for which the transition rate $\transitionRate(n_1, n_0)$ is positive. The probability that a particular state $n_1$ is realized is given by $\transitionRate(n_1, n_0)/\exitRate(n_0)$. In a numerical implementation of the SSA, the state $n_1$ is determined by drawing a second (uniformly-distributed) random number. Our goal is to derive an analytic representation of the probability $\prob(t, n | t_0, n_0)$ of finding the system in state $n$ at time $t$. Thus, after taking $J-1$ further steps, the sample path $\stochPath_J \coloneqq \{n_J \leftarrow \cdots \leftarrow n_1 \leftarrow n_0\}$ should eventually visit state $n_J=n$ at some time $t_J \leq t$. The total time $\tVar_{J-1} + \cdots + \tVar_0$ until the jump to state $n$ occurs is distributed by the convolutions of the individual waiting time distributions, i.e.\ by $\mathop{\star}\limits^{J-1}_{j=0} \waitingTimeDist_{n_j}$. For example, $\tVar\coloneqq \tVar_1+\tVar_0$ is distributed by
    \begin{eqnarray}
      (\waitingTimeDist_{n_1}&\star\waitingTimeDist_{n_0})(\tVar)  
      = \int_{\reals} \diff{\tVar_0}\,
      \waitingTimeDist_{n_1}(\tVar-\tVar_0) \waitingTimeDist_{n_0}(\tVar_0)\\
      &= \frac{\exitRate(n_1)\exitRate(n_0) \bigl(\ee^{-\exitRate(n_0) \tVar} - \ee^{-\exitRate(n_1) \tVar}\bigr)}{\exitRate(n_1)-\exitRate(n_0)}   \HeavisideStep(\tVar)  \,.\label{eq:Intro_Mescoscopic_Convolution}
    \end{eqnarray}
    The probability that the system still resides in state $n_J=n$ at time $t$ is determined by the ``survival probability'' $\survivalProb_n(t-t_J) = \ee^{-\exitRate(n) (t-t_J)}\HeavisideStep(t-t_J)$. After putting all of these pieces together, we arrive at the following path summation representation of the conditional probability distribution:
    \begin{eqnarray}
      \prob(t, n | t_0, n_0 )    \label{eq:Intro_Mescoscopic_Solution}
      = \sum_{J=0}^\infty \sum_{\{\stochPath_J\}} \prob_{t-t_0}(\stochPath_J)   \mathtext{ with} \\
      \prob_\tVar(\stochPath_J)     \label{eq:Intro_Mescoscopic_Solution_Probability}
      = \biggl(\survivalProb_{n} \star \Bigl(\mathop{\star}\limits^{J-1}_{j=0} \frac{\transitionRate(n_{j+1},n_j)}{\exitRate(n_j)} \waitingTimeDist_{n_j}  \Bigr)\biggr)(\tVar)   \,. 
    \end{eqnarray}  
    Here, $\sum_{\{\stochPath_J\}}\coloneqq \sum_{n_1} \cdots \sum_{n_{J-1}}$ generates every path with $J$ jumps between $n_0$ and $n_J=n$. The probability of such a path, integrated over all possible waiting times, is represented by $\prob_\tVar(\stochPath_J)$. By an appropriate choice of integration variables, the probability $\prob_\tVar(\stochPath_J)$ can also be written as
    \begin{eqnarray}
      \prob_\tVar(\stochPath_J) 
      = &\Bigl(\prod_{j=0}^{J-1} \int_{0}^\tVar \diff{\tVar_j}\, \frac{\transitionRate(n_{j+1},n_j)}{\exitRate(n_j)} \waitingTimeDist_{n_j}(\tVar_j)  \Bigr)    \\
        &\cdot \survivalProb_{n}\bigl(\tVar - (\tVar_{J-1} + \dots + \tVar_0)\bigr)  \nonumber\,.
    \end{eqnarray}  
    The survival probability $\survivalProb_{n}$ is included in these integrations. Without the integrations, the expression would represent the probability of a path with $J$ jumps and fixed waiting times. That probability is, for example, used in the master equation formulation of stochastic thermodynamics in associating an entropy to individual paths~\cite{Seifert:2012}. In \aref{sec:A_PathSummationProof}, we formally derive the path summation representation~\eref{eq:Intro_Mescoscopic_Solution} from the Laplace transform of the forward master equation~\eref{eq:Intro_Mescoscopic_MasterEqGainLoss}.
        
    The path summation representation~\eref{eq:Intro_Mescoscopic_Solution} does not only form the basis of the SSA but also of some alternative algorithms~\cite{Empacher:1992,Helbing:1994,Helbing:1996,Sun:2006,Harland:2007,Jackson:2009}. These algorithms either infer the path probability $\prob_\tVar(\stochPath_J)$ numerically from its Laplace transform or evaluate the convolutions in~\eref{eq:Intro_Mescoscopic_Solution_Probability} analytically. The analytic expressions that arise are, however, rather cumbersome generalizations of the convolution in~\eref{eq:Intro_Mescoscopic_Convolution}~\cite{Jasiulewicz:2003,Akkouchi:2008}. They simplify only for the most basic processes (e.g.\ for a random walk or for the Poisson process). Moreover, care has to be taken when the analytic expressions are evaluated numerically because they involve pairwise differences of the exit rate $\exitRate(n)$ (cf.~\eref{eq:Intro_Mescoscopic_Convolution}). When these exit rates differ only slightly along a path, a substantial loss of numerical significance may occur due to finite precision arithmetic. Future studies could explore how the convolutions of exponential distributions in~\eref{eq:Intro_Mescoscopic_Solution_Probability} can be approximated efficiently (for example, in terms of a Gamma distribution or by analytically determining the Laplace transform of~\eref{eq:Intro_Mescoscopic_Solution_Probability}, followed by a saddle-point approximation~\cite{Daniels:1954} of the corresponding inverse Laplace transformation). In general, both the SSA as well as its competitors suffer from the enormous number of states of non-trivial systems, as well as from the even larger number of paths connecting different states. In~\cite{Helbing:1996}, these paths were generated using a deterministic depth-first search algorithm, combined with a filter to select the paths that arrive at the right place at the right time. In~\cite{Harland:2007}, a single path was first generated using the SSA and then gradually changed into new paths through a Metropolis Monte Carlo scheme. Thus far, the two methods have only been applied to relatively simple systems and their prevalence is low compared to the prevalence of the SSA. Further research is needed to explore how relevant paths can be sampled more efficiently.
    
    The true power of the SSA lies in its generation of sample paths with the correct probability of occurrence. Thus, just a few sample paths generated with the SSA are often sufficient to infer the ``typical'' dynamics of a process. A look at individual paths may, for example, reveal that the dynamics of a system are dominated by some spatial pattern, e.g.\ by spirals~\cite{Reichenbach:2007a}. Efficient variations of the above ``direct'' version of the SSA are, for example, described in~\cite{Gibson:2000,Gillespie:2007,Slepoy:2008,Gillespie:2013}. Algorithms for the fast simulation of biochemical networks or processes with spatial degrees of freedom are implemented in the simulation packages~\cite{Tomita:1999,Adalsteinsson:2004,Ander:2004,Hattne:2005,Sanft:2011,Hepburn:2012,Drawert:2012,StochSS:2015}. 
    
    The evaluation of the average $\langle \observable \rangle = \sum_{n} \observable(n) \prob(t, n | \cdot )$ of an observable $A$ typically requires the computation of a larger number of sample paths. However, since the occurrence probability of sample paths generated with the SSA is statistically correct, such an average typically converges comparatively fast. Furthermore, each path can be sampled independently of every other path. Therefore, the computation of paths can be distributed to individual processing units, saving real time, albeit no computation time. A distributed computation of the sample paths is most often required, but possibly not even sufficient, if one wishes to compute the full probability distribution $\prob(t, n | t_0, n_0)$. Vastly more sample paths are required for this purpose, especially if ``rare event probabilities'' in the distribution's tails are sought for. In particular, if the probability of finding a system in state $n$ at time $t$ is only $10^{-10}$, an average of $10^{10}$ sample paths are needed to observe that event just once. Moreover, the probability of observing any particular state decreases with the size of a system's state space. Thus, the sampling of full distributions becomes less and less feasible as systems become larger. Various other challenges remain open as well; for example, the efficient simulation of processes evolving on multiple time scales. These processes are typically simulated using approximative techniques such as $\tVar$-leaping~\cite{Gillespie:2001,Gillespie:2003,Rathinam:2003,Tian:2004,Chatterjee:2005,Cao:2005,Cao:2006,Auger:2006,Gillespie:2007,Cao:2007,Anderson:2008,Gillespie:2013}. Another challenge is posed by the evaluation of processes with time-dependent transition rates~\cite{Lu:2004,Anderson:2007,Roberts:2015}. 
    
    For completeness, let us mention yet another numerical approach to the (forward) master equation. Since the master equation~\eref{eq:Intro_Mescoscopic_MasterEq} constitutes a set of coupled linear first-order ODEs, it can of course be treated as such and be integrated numerically. The integration is, however, only feasible if the state space is sufficiently small (or appropriately truncated) and if all transitions occur on comparable time scales (otherwise, the master equation is quite probably stiff~\cite{Press:2007}). Nevertheless, a numerical integration of the master equation may be preferable over the use of the SSA if the full probability distribution is to be computed.
    
    Neither the matrix exponential representation $\prob(t| t_0) = \ee^{\qMatrix(t-t_0)}\unitMatrix$ of the conditional probability distribution, nor its path summation representation~\eref{eq:Intro_Mescoscopic_Solution} is universally applicable. Moreover, even if the requirements of these solutions are met, the size of the state space or the complexity of the transition matrix may make it infeasible to evaluate them. In the next sections, we formulate conditions under which the conditional probability distribution can be represented in terms of the ``forward'' path integral
    \begin{equation}
      \prob(t, n | t_0, n_0)    \label{eq:Intro_Mesoscopic_FwdPI}
      =  \braA{n}_{t} 
         \pathintegral{[t_0}{t)} 
        \ee^{-\action}    
        \, \ketA{n_0}_{t_0}        
    \end{equation}
    and in terms of the ``backward'' path integral
    \begin{equation}
      \prob(t, n | t_0, n_0)    \label{eq:Intro_Mesoscopic_BwdPI}
      = \braA{n_0}_{t_0} 
        \pathintegral{(t_0}{t]}
        \, \ee^{-\actionAd} 
        \, \ketA{n}_{t} \,.
    \end{equation} 
    The meaning of the integral signs and of the bras $\braA{n}$ and kets $\ketA{n}$ will become clear over the course of this review. Let us only note that the integrals do not proceed along paths of the discrete variable $n$, but over the paths of two continuous auxiliary variables that are introduced for this purpose. The relevance of each path is weighed by the exponential factors inside the integrals. 
    
    Besides these exact representations of the conditional probability distribution solving the master equations, there exist powerful ways of approximating this distribution and the values of averaged observables. These methods include the Kramers-Moyal~\cite{Kramers:1940,Moyal:1949} and the system-size expansion~\cite{vanKampen:1961,vanKampen:2007}, as well as the derivation of moment equations. Information on these methods can be found in classic text books~\cite{vanKampen:2007,Gardiner:2009} and in a recent review~\cite{Schnoerr:2016}. Although moment equations encode the complete information about a stochastic process, they typically constitute an infinite hierarchy whose evaluation requires a truncation by some closure scheme~\cite{Goodman:1953,Whittle:1957,Keeling:2000,Nasell:2003,Lee:2009,Smadbeck:2013,Schnoerr:2014,Lakatos:2015,Schnoerr:2015}. On the other hand, the Kramers-Moyal and the system-size expansion approximate the master equation in terms of a Fokker-Planck equation. Both expansions work best if the system under consideration is ``large'' (more precisely, they work best if the dynamics are centred around a stable or meta-stable state at a distance $N \gg 1$ from a potentially absorbing state; the standard deviation of its surrounding distribution is then of order $\sqrt{N}$). An extension of the system-size expansion to absorbing boundaries has recently been proposed in~\cite{DiPatti:2011}. In sections~\ref{subsec:PathInts_BackwardKramersMoyal} and~\ref{subsec:PathInts_ForwardKramersMoyal}, we show how Kramers-Moyal expansions of the backward and forward master equations can be used to derive path integral representations of processes with continuous state spaces. When the expansion of the backward master equation stops (or is truncated) at the level of a diffusion approximation, one recovers a classic path integral representation of the (backward) Fokkers-Planck equation~\cite{Martin:1973,deDominicis:1976,Janssen:1976,Bausch:1976}.

  \subsection{History of stochastic path integrals}\label{subsec:Intro_PathInts}  
  
    The oldest path integral, both in the theory of stochastic processes and beyond~\cite{Dirac:1933,Feynman:2005,Feynman:1948,Feynman:2010}, is presumably Wiener's integral for Brownian motion~\cite{Wiener:1921a,Wiener:1921b}. The path integrals we consider here were devised somewhat later, namely in the 1970s and 80s: first for the Fokker-Planck (or Langevin) equation~\cite{Martin:1973,deDominicis:1976,Janssen:1976,Bausch:1976} and soon after for the master equation. For the master equation, the theoretical basis of these ``stochastic'' path integrals was developed by Doi~\cite{Doi:1976a,Doi:1976b}. He first expressed the creation and annihilation of molecules in a chemical reaction by the corresponding operators for the quantum harmonic oscillator~\cite{Dirac:1958} (modulo a normalization factor), introducing the concept of the Fock space~\cite{Fock:1932} into non-equilibrium theory. Furthermore, he employed coherent states~\cite{Schroedinger:1926b,Sudarshan:1963,Glauber:1963b} to express averaged observables. Similar formalisms as the one of Doi were  independently developed by Zel'dovich and Ovchinnikov~\cite{Zeldovich:1978}, as well as by Grassberger and Scheunert~\cite{Grassberger:1980}. Introductions to the Fock space approach, which are in part complementary to our review, are, for example, provided in~\cite{Mattis:1998,Cardy:2008,Altland:2010,Taeuber:2014,Baez:2015}. The review of Mattis and Glasser~\cite{Mattis:1998} provides a chronological list of contributions to the field. These contributions include Rose's renormalized kinetic theory~\cite{Rose:1979}, Mikhailov's path integral~\cite{Mikhailov:1981a,Mikhailov:1981b,Mikhailov:1985}, which is based on Zel'dovich's and Ovchinnikov's work, and Goldenfeld's extension of the Fock space algebra to polymer crystal growth~\cite{Goldenfeld:1984}. Furthermore, Peliti reviewed the Fock space approach and provided  derivations of path integral representations of averaged observables and of the probability generating function~\cite{Peliti:1985}. Peliti also expressed the hope that future ``rediscoveries'' of the path integral formalism would be unnecessary in the future. However, we believe that the probabilistic structures behind path integral representations of stochastic processes have not yet been clearly exposed. As illustrated  in \fref{fig:Outline}, we show that the forward and backward master equations admit not only one but two path integral representations: the forward representation~\eref{eq:Intro_Mesoscopic_FwdPI} and the novel backward representation~\eref{eq:Intro_Mesoscopic_BwdPI}. Although the two path integrals resemble each other, they differ conceptually. While the forward path integral representation provides a probability generating function in an intermediate step, the backward representation provides a distribution that is marginalized over an initial state. Both path integrals can be used to represent averaged observables as shown in \sref{sec:PathInts_Observables}. The backward path integral, however, will turn out to be more convenient for this purpose (upon choosing a Poisson distribution as the basis function, i.e.\ $\ketA{n} \coloneqq \frac{\xAd^{n} \ee^{-\xAd}}{n!}$, the representation is obtained by summing~\eref{eq:Intro_Mesoscopic_BwdPI} over an observable $A(n)$). Let us note that even though we adopt some of the notation of quantum theory, our review is guided by the notion that quantum (field) theory is ``totally unnecessary'' for the theory of stochastic path integrals. Michael E. Fisher once stated the same about the relevance of quantum field theory for renormalization group theory~\cite{Fisher:1998} (while acknowledging that in order to do certain types of calculations, familiarity with quantum field theory can be very useful; the same applies to the theory of stochastic path integrals).
    
    Thus far, path integral representations of the master equation have primarily been applied to processes whose transition rates can be decomposed additively into the rates of simple chemical reactions. Simple means that the transition rate of such a reaction is determined by combinatoric counting. Consider, for example, a reaction of the form $k\, A \to l\, A$ in which $k\in\naturals_0$ molecules of type $A$ are replaced by $l\in\naturals_0$ molecules of the same type. Assuming that the $k$ reactants are drawn from an urn with a total number of $m$ molecules, the total rate of the reaction should be proportional to the falling factorial $(m)_k \coloneqq m (m-1)\cdots (m-k+1)$. The global time scale of reaction is set by the rate coefficient $\rateCoeffGeneric_\tVar$, which we allow to depend on time. Thus, the rate at which the chemical reaction $k\, A \to l\, A$ induces a transition from state $m$ to state $n$ is $\transitionRate_\tVar(n, m) = \rateCoeffGeneric_\tVar  (m)_k \delta_{n, m-k+l}$. The Kronecker delta inside the transition rate ensures that $k$ molecules are indeed replaced by $l$ new ones. Note that the number of particles in the system can never become negative, provided that the initial number of particles was non-negative. Hence, the state space of $n$ is $\naturals_0$. Insertion of the above transition rate into the forward master equation~\eref{eq:Intro_Mescoscopic_MasterEqGainLoss} results in the ``chemical'' master equation
    \begin{eqnarray} 
      \partial_\tVar \prob(\tVar, n | t_0, n_0)      \label{eq:Intro_Mescoscopic_ChemicalMasterEq} \\
      =   \rateCoeffGeneric_\tVar \bigl[ (n-l+k)_k \prob(\tVar, n-l+k | \cdot)      
         - (n)_k \prob(\tVar, n | \cdot) \bigr]        \nonumber     \,.
    \end{eqnarray} 
    Microphysical arguments for its applicability to chemical reactions can be found in~\cite{Gillespie:1992}. According to the chemical master equation, the mean particle number $\langle n \rangle = \sum_{n} n\, \prob(\tVar, n | \cdot)$ obeys the equation $\partial_\tVar \langle n \rangle = \rateCoeffGeneric_\tVar (l-k) \langle (n)_k \rangle$. For a system with a large number of particles ($n\gg k$), fluctuations can often be neglected in a first approximation, leading to the deterministic rate equation
    \begin{equation}
      \partial_\tVar \bar{n} = \rateCoeffGeneric_\tVar (l-k) \bar{n}^k    \,,
    \end{equation}
    obeyed by a continuous particle number $\bar{n}(\tVar) \in \reals$.
    
    Path integral representations of the chemical master equation~\eref{eq:Intro_Mescoscopic_ChemicalMasterEq} are sometimes said to be ``bosonic''. First, because an arbitrarily large number of molecules may in principle be present in the system (both globally and, upon extending the system to a spatial domain, locally). Second, because the path integral representations are typically derived with the help of ``creation and annihilation operators'' fulfilling a ``bosonic'' commutation relation (see \sref{subsec:GenFct_ForwardBases_ChemReactions}). ``Fermionic'' path integrals, on the other hand, have been developed for systems in which the particles exclude one another. Thus, the number of particles in these systems is locally restricted to the values $0$ and $1$. For systems with excluding particles, the master equation's solution may be represented in terms of a path integral whose underlying creation and annihilation operators fulfil an anti-commutation relation~\cite{Sandow:1993,Patzlaff:1994,Bares:1999,Brunel:2000,Schulz:2005,Silva:2008}. However, van Wijland recently showed that the use of operators fulfilling the bosonic commutation relation is also possible~\cite{Wijland:2001}. These approaches are considered in \sref{subsec:GenFct_ForwardBases_Exclusion}.
    
    We do not intend to delve further into the historic development and applications of stochastic path integrals at this point. Doing so would require a proper introduction into renormalization group theory, which is of pivotal importance for the evaluation of the path integrals. Readers can find information on the application of renormalization group techniques in the review of T\"auber, Howard, and Vollmayr-Lee~\cite{Taeuber:2005} and in the book of T\"auber~\cite{Taeuber:2014}. Introductory texts are also provided by Cardy's (lecture) notes~\cite{Cardy:1997,Cardy:2008}. Roughly speaking, path integral representations of the chemical master equation~\eref{eq:Intro_Mescoscopic_ChemicalMasterEq} have been used to assess how a macroscopic law of mass action changes due to fluctuations, both below~\cite{Peliti:1986,Lee:1994b,Howard:1995,Howard:1996b,Oerding:1996,Cardy:1996,Rey:1997,Sasamoto:1997,Cardy:1998,Wijland:1998,Taeuber:1998,Hinrichsen:1999,Goldschmidt:1999,Hnatich:2000,Vernon:2003} and above the (upper) critical dimension~\cite{Howard:1995,Winkler:2012,Homrighausen:2013}, using either perturbative~\cite{Mikhailov:1985,Peliti:1986,Droz:1993,Lee:1994b,Lee:1994c,Howard:1995,Lee:1995,Howard:1996a,Howard:1996b,Oerding:1996,Cardy:1996,Rey:1997,Sasamoto:1997,Wijland:1998,Taeuber:1998,Konkoli:1999,Goldschmidt:1999,Hnatich:2000,Dickman:2002,Vernon:2003,Hilhorst:2004,Janssen:2004,Janssen:2005,Whitelam:2005,Hnatic:2013,Taeuber:2014} or non-perturbative~\cite{Canet:2004a,Canet:2004b,Canet:2005,Winkler:2012,Homrighausen:2013} techniques. All of these articles focus on stochastic processes with spatial degrees of freedom for which alternative analytical approaches are scarce. Path integral representations of these processes, combined with renormalization group techniques, have been pivotal in understanding non-equilibrium phase transitions and they contributed significantly to the classification of these transitions in terms of universality classes~\cite{Odor:2004,Elgart:2006,Taeuber:2014}. Moreover, path integral representations of master equations have recently been employed in such diverse contexts as the study of neural networks~\cite{Buice:2007,Buice:2009,Buice:2013}, of ecological populations~\cite{Bettelheim:2001,Butler:2009a,Butler:2009b,Taeuber:2011,Taeuber:2012,Shih:2014}, and of the differentiation of stem-cells~\cite{Zhang:2014}.
    
  \subsection{R\'esum\'e}\label{subsec:Intro_Resume}  
  
    Continuous-time Markov processes with discontinuous sample paths describe a broad range of phenomena, ranging from traffic jams on highways~\cite{Nagel:1992} and on cytoskeletal filaments~\cite{Lipowsky:2001,Parmeggiani:2003,Parmeggiani:2004,Chou:2011} to novel forms of condensation in bosonic systems~\cite{Vorberg:2013,Knebel:2015,Vorberg:2015}. In the introduction, we laid out the mathematical theory of these processes and derived the fundamental equations governing their evolution: the forward and the backward master equations. Whereas the forward master equation~\eref{eq:Intro_Mescoscopic_MasterEq} evolves a conditional probability distribution forward in time, the backward master equation~\eref{eq:Intro_Mescoscopic_BackwardMasterEq} evolves the distribution backward in time. In the following main part of this review, we represent the conditional probability distribution solving the master equations in terms of path integrals. The framework upon which these path integrals are based unifies a broad range of approaches to the master equations, including, for example, the spectral method of Walzcak, Mugler, and Wiggins~\cite{Walczak:2009} and the Poisson representation of Gardiner and Chaturvedi~\cite{Gardiner:1977,Chaturvedi:1978}.

\section{The probability generating function}\label{sec:GenFct}  
  
    The following two sections~\ref{sec:GenFct} and~\ref{sec:GenFctnl} are devoted to mapping the forward and backward master equations \eref{eq:Intro_Mescoscopic_MasterEq} and \eref{eq:Intro_Mescoscopic_BackwardMasterEq} to linear partial differential equations (PDEs). For brevity, we refer to such linear PDEs as ``flow equations''. In sections~\ref{sec:PathInts_Backward} and~\ref{sec:PathInts_Forward}, the derived flow equations are solved in terms of path integrals.
    
    It has been known since at least the 1940s that the (forward) master equation can be cast into a flow equation obeyed by the ordinary probability generating function
    \begin{equation}
      \gen(\tVar; \q | t_0, n_0) \label{eq:GenFct_GenFct_Ordinary}
      \coloneqq \sum_{n\in\naturals_0} \q^n \prob(\tVar, n | t_0, n_0)  \,,
    \end{equation}
    at least when the corresponding transition rate describes a simple chemical reaction~\cite{Feller:1949,Singer:1953,Bartholomay:1958,Krieger:1960,McQuarrie:1963,McQuarrie:1964,Ishida:1964,McQuarrie:1967}. The generating function effectively replaces the discrete variable $n$ by the continuous variable $\q$. The absolute convergence of the sum in~\eref{eq:GenFct_GenFct_Ordinary} is ensured (at least) for $|\q|\leq 1$. The generating function ``generates'' probabilities in the sense that
    \begin{equation}
      \prob(\tVar, n | t_0, n_0) 
      = \frac{1}{n!} \partial_{\q}^n \gen(\tVar; \q | t_0, n_0) \Big|_{\q=0}  \,.
    \end{equation}
    This inverse transformation from $\gen$ to $\prob$ involves the application of the (real) linear functional $L_n[f] \coloneqq \frac{1}{n!} \partial_{\q}^n f(\q) \big|_{\q=0}$, which maps a (real-valued) function $f$ to a (real) number. Moreover, it fulfils $L_n[f+\alpha g] = L_n[f]+\alpha L_n[g]$ for two functions $f$ and $g$, and $\alpha\in\reals$. A more convenient notation for linear functionals is introduced shortly. In the following, we generalize the probability generating function~\eref{eq:GenFct_GenFct_Ordinary} and formulate conditions under which the generalized function obeys a linear PDE, i.e.\ a \textit{flow equation}. But before proceeding to the general case, let us exemplify the use of a generalized probability generating function for a specific process (for brevity, we often drop the terms ``probability'' and ``generalized'' in referring to this function).
    
    As the example, we consider the bi-directional reaction $\emptyset \rightleftharpoons A$ in which molecules of type $A$ form at rate $\rateCoeffSpontGrowth \geq 0$ and degrade at per capita rate $\rateCoeffLinDecay > 0$. According to the chemical master equation~\eref{eq:Intro_Mescoscopic_ChemicalMasterEq}, the probability of observing $n\in\naturals_0$ such molecules obeys the equation
    \begin{eqnarray} 
      \partial_\tVar \prob(\tVar, &n | t_0, n_0) 
      = \rateCoeffSpontGrowth \bigl[ \prob(\tVar, n-1 | \cdot)   -  \prob(\tVar, n | \cdot) \bigr]    \label{eq:GenFct_MastertEq}  \\
        &+  \rateCoeffLinDecay \bigl[ (n+1) \prob(\tVar, n+1 | \cdot)   - n \prob(\tVar, n | \cdot) \bigr]          \nonumber     \,,
    \end{eqnarray} 
    with initial condition $\prob(t_0, n | t_0, n_0 ) = \delta_{n,n_0}$. This master equation respects the fact that the number of molecules cannot become negative through the reaction $A \to \emptyset$. By differentiating the probability generating function $\gen(\tVar; \q | t_0, n_0)$ in~\eref{eq:GenFct_GenFct_Ordinary} with respect to the current time $\tVar$, one finds that it obeys the flow equation
    \begin{equation}
      \partial_\tVar \gen  \label{eq:GenFct_FlowEq_0}
      = (\q-1)(\rateCoeffSpontGrowth - \rateCoeffLinDecay  \partial_\q) \gen  \,.
    \end{equation}
    Its time evolution starts out from $\gen(t_0 ; \q | t_0, n_0) = \q^{n_0}$. Instead of solving the flow equation right away, let us first simplify it by changing the basis function $\q^n$ of the generating function~\eref{eq:GenFct_GenFct_Ordinary}. As a first step, we change it to $(\q+1)^n$, turning the corresponding flow equation into $\partial_\tVar \gen = \q(\rateCoeffSpontGrowth - \rateCoeffLinDecay  \partial_\q) \gen$. As a second step, we multiply the new basis function by $\ee^{-\frac{\rateCoeffSpontGrowth}{\rateCoeffLinDecay} (\q + 1)}$ and arrive at the simplified flow equation
    \begin{equation}
      \partial_\tVar \gen   \label{eq:GenFct_FlowEq_1}
      = - \rateCoeffLinDecay \q \partial_\q \gen    \,.
    \end{equation}
    The generating function obeying this equation reads
    \begin{equation}
      \gen(\tVar; \q | \cdot)   \label{eq:GenFct_GenFct_Orig}
      = \sum_{n\in\naturals_0}(\q + 1)^n \ee^{-\frac{\rateCoeffSpontGrowth}{\rateCoeffLinDecay} (\q + 1)}   \, p(\tVar, n | \cdot)  \,.
    \end{equation}
    As before, the dots inside the functions' arguments abbreviate the initial parameters $t_0$ and $n_0$. 
    
    The simplified flow equation~\eref{eq:GenFct_FlowEq_1} is now readily solved by separation of variables. But before doing so, let us introduce some new notation. From now on, we write the basis function as
    \begin{equation}
      \ketA{n}_\q       \label{eq:GenFct_BaseFct}
      \coloneqq (\q + 1)^n \ee^{-\frac{\rateCoeffSpontGrowth}{\rateCoeffLinDecay} (\q + 1)}    
    \end{equation}
    and the corresponding generalized probability generating function~\eref{eq:GenFct_GenFct_Orig} as
    \begin{equation}
      \ketA{\gen(\tVar | \cdot)}_{\q}
      \coloneqq \sum_{n\in\naturals_0} \ketA{n}_{\q}\, p(\tVar, n | \cdot)  \,.    \label{eq:GenFct_GenFct}  
    \end{equation}
    In quantum mechanics, an object written as $\ketA{\cdot}$ is called a ``ket'', a notation that was originally introduced by Dirac~\cite{Dirac:1937}. In the above two expressions, the kets simply represent ordinary functions. For brevity, we write the arguments of the kets as subscripts and occasionally drop these subscripts altogether. In principle, the basis function could also depend on time (i.e.\ $\ketA{n}_{\tVar,\q}$). 
    
    Later, in \sref{subsec:GenFct_ForwardBases}, we introduce various basis functions for the study of different stochastic processes, including the Fourier basis function $\ketA{n}_\q \coloneqq \ee^{\ii n \q}$ for the solution of a random walk (with $n \in \integers$). Moreover, we consider a ``linear algebra'' approach in which $\ketA{n} \coloneqq \hat{\vect{e}}_n$ represents the unit column vector in direction $n\in\naturals_0$  (this vector equals one at position $n+1$ and is zero everywhere else; cf.~\sref{subsec:GenFct_ForwardBases_Algebraic}). The generating function~\eref{eq:GenFct_GenFct} corresponding to this ``unit vector basis'' coincides with the column vector $\vect{\prob}(\tVar | t_0, n_0)$ of the probabilities $\prob(\tVar, n | t_0, n_0)$. The unit vector basis will prove useful later on in recovering a path integral representation of averaged observables  in \sref{subsec:PathInts_Observables_Algebraic}. 
    
    In our present example and in most of this review, however, the generating function~\eref{eq:GenFct_GenFct} represents an ordinary function and obeys a linear PDE. For the basis function~\eref{eq:GenFct_BaseFct}, this PDE reads
    \begin{equation}
      \partial_\tVar \ketA{\gen}    \label{eq:GenFct_FlowEq_2}
      = - \rateCoeffLinDecay \q \partial_\q \ketA{\gen}     \,.
    \end{equation}
    Its time evolution starts out from $\ketA{\gen(t_0 | t_0, n_0)} = \ketA{n_0}$.
    
    In order to recover the conditional probability distribution from the generating function~\eref{eq:GenFct_GenFct}, we now complement the ``kets'' with ``bras''. Such a bra is written as $\braA{\cdot}$ and represents a linear functional in our present example. In particular, we define a bra $\braA{m}$ for every $m\in\naturals_0$ by its following action on a test function~$f$:
    \begin{equation}
      \braA{m} f   \label{eq:GenFct_BasisFunctional}
      \coloneqq \frac{1}{m!} \Bigl(\partial_\q + \frac{\rateCoeffSpontGrowth}{\rateCoeffLinDecay}\Bigr)^m f(\q) |_{\q = -1}  \,.
    \end{equation}
    The evaluation at $\q=-1$ could also be written in integral form as $\int_{\reals} \diff{\q} \, \delta(\q+1) (\cdots)$. The functional $\braA{m}$ is obviously linear and it maps the basis function~\eref{eq:GenFct_BaseFct} to
    \begin{equation} 
      \braketA{m}{n} = \delta_{m,n}    \label{eq:GenFct_Orthogonal}  \,.
    \end{equation} 
    Thus, the ``basis functionals'' in $\{\braA{m}\}_{m\in\naturals_0}$ are orthogonal to the basis functions in $\{\ketA{n}\}_{n\in\naturals_0}$. The orthogonality condition can be used to recover the conditional probability distribution from~\eref{eq:GenFct_GenFct} via
    \begin{equation}
      \prob(\tVar, n | \cdot) = \braketA{n}{\gen(\tVar | \cdot)}    \label{eq:GenFct_Ordinary_Inverse}  \,.
    \end{equation}
    Besides being orthogonal to one another, the kets~\eref{eq:GenFct_BaseFct} and bras~\eref{eq:GenFct_BasisFunctional} fulfil the completeness relation $\sum_{n} \ketA{n}_{\q} \braA{n} f = f(\q)$ with respect to analytic functions\footnote{As before, sums whose range is not specified cover the whole state space.} (note that $\braA{n} f$ as defined in~\eref{eq:GenFct_BasisFunctional} is just a real number and does not depend on $\q$). Above, we mentioned that we later introduce alternative basis kets, including the Fourier basis function $\ketA{n}_\q = \ee^{\ii n \q}$ for the solution of a random walk (with $n\in\integers$), and the unit column vector $\ketA{n} = \hat{\vect{e}}_n$ with $n\in\naturals_0$. These kets can also be complemented to obtain orthogonal and complete bases, namely by complementing the Fourier basis function with the basis functional $\braA{m} f \coloneqq \int_{-\pi}^\pi \frac{\diff{\q}}{2\pi} \ee^{-\ii m \q} f(\q)$ ($m\in\integers$), and by complementing the unit column vector with the unit row vector $\braA{m} \coloneqq \hat{\vect{e}}_m^\transpose$ ($m\in\naturals_0$). For the unit vector basis $\{\ketA{n},\braA{m}\}_{n,m\in\naturals_0}$, the completeness condition $\sum_{n} \ketA{n} \braA{n} = \unitMatrix$ involves the (infinitely large) unit matrix $\unitMatrix$.
    
    Thanks to the new basis function~\eref{eq:GenFct_BaseFct}, the simplified flow equation~\eref{eq:GenFct_FlowEq_2} can be easily solved by separation of variables. Making the ansatz $\ketA{\gen} = f(\tVar)h(\q)$, one obtains an equation whose two sides depend either on $f$ or on $h$ but not on both. The equation is solved by $f(\tVar) = \ee^{-k \rateCoeffLinDecay (\tVar - t_0)}$ and $h(\q) = \q^k$, with $k$ being a non-negative parameter. The non-negativity of $k$ ensures the finiteness of the initial condition $\ketA{\gen(t_0|t_0, n_0)}_{\q} = \ketA{n_0}_{\q}$ in the limit $\q\to 0$. By the completeness of the polynomial basis, the values of $k$ can be restricted to $\naturals_0$. It proves convenient to represent also the standard polynomial basis in terms of bras and kets, namely by defining $\braB{k} f \coloneqq \frac{1}{k!} \partial_\q^k f(\q) |_{\q=0}$ and $\ketB{k}_\q \coloneqq \q^k$. These bras and kets are again orthogonal to one another in the sense of $\braketB{k}{l} = \delta_{k,l}$ and they also fulfil a completeness relation ($\sum_{k} \ketB{k}  \braB{k}$ represents a Taylor expansion around $\q=0$ and thus acts as an identity on analytic functions). Using the auxiliary bras and kets, the solution of the flow equation~\eref{eq:GenFct_FlowEq_2} can be written as
    \begin{equation}
      \ketA{\gen(\tVar | \cdot)}
      = \sum_{k\in\naturals_0} \ketB{k} \,\ee^{-k \rateCoeffLinDecay (\tVar - t_0)}  \braketBA{k}{n_0} \,.  \label{eq:GenFct_GenFct_Solution}
    \end{equation} 
    We wrote the expansion coefficient in this solution as $\braketBA{k}{n_0}$ to respect the initial condition $\ketA{\gen(t_0 | \cdot)} = \ketA{n_0}$.
    
    The conditional probability distribution can be recovered from the generating function~\eref{eq:GenFct_GenFct_Solution} via the inverse transformation~\eref{eq:GenFct_Ordinary_Inverse} as
    \begin{equation}
      \prob(\tVar, n | t_0, n_0)
      = \sum_{k\in\naturals_0} \braketAB{n}{k} \, \ee^{-k \rateCoeffLinDecay (\tVar - t_0)}  \braketBA{k}{n_0} \,.\label{eq:GenFct_GenFct_SolutionDist}
    \end{equation} 
    The coefficients $\braketAB{n}{k}$ and $\braketBA{k}{n_0}$ can be computed recursively as explained in~\cite{Walczak:2012}. Here we are interested in the asymptotic limit $\tVar \to \infty$ of the distribution~\eref{eq:GenFct_GenFct_SolutionDist} for which only the $k=0$ ``mode'' survives. Therefore, the distribution converges to the stationary Poisson distribution
    \begin{equation}
      \prob(\infty, n | t_0, n_0) 
      = \braketAB{n}{0}  \braketBA{0}{n_0} = \frac{(\rateCoeffSpontGrowth/\rateCoeffLinDecay)^n \ee^{-\rateCoeffSpontGrowth/\rateCoeffLinDecay}}{n!}    \,.
    \end{equation}

    The above example illustrates how the master equation can be transformed into a linear PDE obeyed by a generalized probability generating function and how this PDE simplifies for the right basis function. The explicit choice of the basis function depends on the problem at hand. Moreover, the above example introduced the bra-ket notation used in this review. In \sref{subsec:PathInts_SimpleGrowthLinearDecay}, the reaction $\emptyset \rightleftharpoons A$ will be reconsidered using a path integral representation of the probability distribution. We will then see that this process is not only solved by a Poisson distribution in the stationary limit, but actually for all times (at least, if the number of molecules in the system was initially Poisson distributed).
    
    In the remainder of this section, as well as in \sref{sec:GenFctnl}, we generalize the above approach and derive flow equations for the following four series expansions (with dynamic time variable $\tVar \in [t_0,t]$):\\
    \noindent\begin{minipage}{.45\linewidth}
      \begin{eqnarray}
          \sum_{n}\ketA{n}  \prob(\tVar, n | t_0, n_0)            \label{eq:GenFct_GenFct_mul}  \,,\\
         \sum_{n_0} \prob(t, n |\tVar,n_0) \ketA{n_0}      \label{eq:GenFct_Marginalized}    \,,
      \end{eqnarray}
    \end{minipage}%
    \noindent\begin{minipage}{.1\linewidth}
    \hfill
    \end{minipage}%
    \begin{minipage}{.45\linewidth}
      \begin{eqnarray}
          \sum_{n} \braA{n} \prob(\tVar, n | t_0, n_0)           \label{eq:GenFct_PoissonRep}  \,,  \\
          \sum_{n_0} \prob(t, n |\tVar,n_0) \braA{n_0}     \label{eq:GenFct_XXX}    \,. 
      \end{eqnarray}
    \end{minipage}\ \\
    Apparently, the series~\eref{eq:GenFct_GenFct_mul} coincides with the generalized probability generating function~\eref{eq:GenFct_GenFct}. In the next \sref{subsec:GenFct_Flow}, we formulate general conditions under which this function obeys a linear PDE. The remaining series~\eref{eq:GenFct_Marginalized}--\eref{eq:GenFct_XXX} may not be as familiar. We call the series~\eref{eq:GenFct_Marginalized} a ``marginalized distribution''. It will be shown in \sref{sec:GenFctnl} that this series does not only solve the (forward) master equation, but that it also obeys a backward-time PDE under certain conditions. The marginalized distribution proves useful in the computation of mean extinction times as we demonstrate in \sref{subsec:GenFctnl_ExtTimes}. In \sref{subsec:GenFctnl_Flow}, we consider the ``probability generating functional''~\eref{eq:GenFct_PoissonRep}. For a ``Poisson basis function'', the inverse transformation, which maps this functional to the conditional probability distribution, coincides with the Poisson representation of Gardiner and Chaturvedi~\cite{Gardiner:1977,Chaturvedi:1978}. The potential use of the series~\eref{eq:GenFct_XXX} remains to be explored. 
    
    The goal of the subsequent sections~\ref{sec:PathInts_Backward} and~\ref{sec:PathInts_Forward} lies in the solution of the derived flow equations by path integrals. In \sref{sec:PathInts_Backward}, we first solve the flow equations obeyed by the marginalized distribution~\eref{eq:GenFct_Marginalized} and by the generating functional~\eref{eq:GenFct_PoissonRep} in terms of a ``backward'' path integral. Afterwards, in \sref{sec:PathInts_Forward}, the flow equations obeyed by the generating function~\eref{eq:GenFct_GenFct_mul} and by the series expansion~\eref{eq:GenFct_XXX} are solved in terms of a ``forward'' path integral. Inverse transformations, such as~\eref{eq:GenFct_Ordinary_Inverse}, will then provide distinct path integral representations of the forward and backward master equations.
        
  \subsection{Flow of the generating function}\label{subsec:GenFct_Flow}  

    We now formulate general conditions under which the forward master equation~\eref{eq:Intro_Mescoscopic_MasterEq} can be cast into a linear PDE obeyed by the generalized probability generating function
    \begin{equation}
      \ketA{\gen(\tVar | t_0, n_0)}_{\q}  \label{eq:GenFct_Flow_GenFct}
      \coloneqq \sum_{n}\ketA{n}_{\tVar,\q} \, \prob(\tVar, n | t_0, n_0)        \,.
    \end{equation} 
    The basis function $\ketA{n}_{\tVar,\q}$ is a function of the variable $\q$ and possibly of the time variable $\tVar$. But unless one of these variables is of direct relevance, its corresponding subscript will be dropped. The explicit form of the basis function depends on the problem at hand and is chosen so that the four conditions~\eref{eq:GenFct_Flow_OrthogonalityCondition}, \eref{eq:GenFct_Flow_CompletenessCondition}, \eref{eq:GenFct_Flow_TCondition}, and~\eref{eq:GenFct_Flow_FCondition} below are satisfied (the conditions~\eref{eq:GenFct_Flow_OrthogonalityCondition} and~\eref{eq:GenFct_Flow_CompletenessCondition} concern the orthogonality and completeness of the basis, which we already required in the introductory example). The variable $n$ again represents some state from a countable state space. For example, $n$ could describe the position of a molecular motor along a cytoskeletal filament ($n\in\integers$), the copy number of a molecule ($n\in\naturals_0$), the local copy numbers of the molecule on a lattice ($n \equiv \{n_i\in \naturals_0\}_{i\in\integers}$), or the copy numbers of multiple kinds of molecules ($n \equiv (n^{A}, n^{B}, n^{C}) \in \naturals_0^3$). For the multivariate configurations, the basis function is typically decomposed into a product $\ketA{n_1}\ketA{n_2}\cdots$ of individual basis functions $\ketA{n_i}$, each depending on its own variable $\q_i$. A process with spatial degrees of freedom is considered in \sref{subsubsec:PathInts_BackwardSolutions_DiffusionNetwork}. Besides, we also consider a system of excluding particles in \sref{subsec:GenFct_ForwardBases_Exclusion}. There, the (local) number of particles $n$ is restricted to the values $0$ and $1$.  Note that the sum in~\eref{eq:GenFct_Flow_GenFct} extends over the whole state space. 
    
    The definition of the generating function~\eref{eq:GenFct_Flow_GenFct} assumes the existence of a set $\{ \ketA{n}_\tVar\}$ of basis functions for every time $\tVar \in [t_0, t]$. In addition, we assume that there exists a set $\{\braA{m}_\tVar\}$ of linear basis functionals for every time $\tVar \in [t_0, t]$. These bras shall be orthogonal to the kets in the sense that at each time point $\tVar$, they act on the kets as
    \renewcommand{\theequation}{O}%
    \begin{equation} 
      \braA{m}_\tVar \ketA{n}_\tVar = \delta_{m,n}     \label{eq:GenFct_Flow_OrthogonalityCondition}  
    \end{equation} 
    \renewcommand{\theequation}{\arabic{equation}}%
    (for all $m$ and $n$). Here we note the possible time-dependence of the basis because the \eref{eq:GenFct_Flow_OrthogonalityCondition}rthogonality condition will only be required for equal times of the bras and kets. In addition to orthogonality, the basis shall fulfil the \eref{eq:GenFct_Flow_CompletenessCondition}ompleteness condition
    \renewcommand{\theequation}{C}%
    \begin{equation} 
      \sum_{n}\ketA{n}_\tVar \braA{n}_\tVar f = f  \,,
      \label{eq:GenFct_Flow_CompletenessCondition}  
    \end{equation} 
    \renewcommand{\theequation}{\arabic{equation}}%
    where $f$ represents an appropriate test function. The completeness condition implies that the function $f$ can be decomposed in the basis functions $\ketA{n}$ with expansion coefficients $\braA{n} f$. In the introduction to this section, we introduced various bases fulfilling both the orthogonality and the completeness condition. As in the introductory example, the orthogonality condition allows one to recover the conditional probability distribution via the inverse transformation
    \begin{equation}
      \prob(\tVar, n | t_0, n_0) = \braketA{n}{\gen(\tVar | t_0, n_0)}  \,.        \label{eq:GenFct_Flow_InverseTransformation} 
    \end{equation} 

    Before deriving the flow equation obeyed by the generating function, let us note that the \eref{eq:GenFct_Flow_OrthogonalityCondition}rthogonality condition differs slightly from the corresponding conditions used in most other texts on stochastic path integrals (see, for example, \cite{Goldenfeld:1984,Peliti:1985} or the ``exclusive scalar product'' in~\cite{Grassberger:1980}). Typically, the orthogonality condition includes an additional factorial $n!$ on its right hand side. The inclusion of this factorial is advantageous in that it accentuates a symmetry between the bases that we consider in sections~\ref{subsec:GenFct_ForwardBases_ChemReactions} and~\ref{subsec:GenFctnl_BackwardBases_ChemReactions} for the study of chemical reactions. Its inclusion would be rather unusual, however, for the Fourier basis introduced in sections~\ref{subsec:GenFct_ForwardBases_RandomWalk} and~\ref{subsec:GenFctnl_BackwardBases_RandomWalk}. The Fourier basis will be used to solve a simple random walk. Moreover, the factorial obscures a connection between the probability generating functional introduced in \sref{subsec:GenFctnl_Flow} and the Poisson representation of Gardiner and Chaturvedi~\cite{Gardiner:1977,Chaturvedi:1978}. We discuss this connection in \sref{subsec:GenFctnl_PoissonRep}.
    
    To derive the flow equation obeyed by the generating function $\ketA{\gen}$, we differentiate its definition~\eref{eq:GenFct_Flow_GenFct} with respect to the time variable $\tVar$. The resulting time derivative of the conditional probability distribution $\prob(\tVar, n | t_0, n_0)$ can be replaced by the right-hand side of the forward master equation~\eref{eq:Intro_Mescoscopic_MasterEq}. In matrix notation, this equation reads $\partial_\tVar \prob(\tVar | t_0) = \qMatrix_\tVar \prob(\tVar | t_0)$. Eventually, one finds that
    \begin{equation}
      \partial_\tVar \ketA{\gen}    
      = \sum_{n} 
          \Bigl(\partial_\tVar \ketA{n}
          +
          \sum_{m} 
            \ketA{m}
            \qMatrix_\tVar(m, n) 
          \Bigr)
          \prob(\tVar, n | \cdot )      \label{eq:GenFct_Flow_FlowEq_Derivation}  \,.
    \end{equation} 
    Our goal is to turn this expression into a partial differential equation for $\ketA{\gen}$. For this purpose, we require two differential operators. First, we require a differential operator $\evolutionOp_\tVar(\q,\partial_\q)$ encoding the time evolution of the basis function. In particular, this operator should fulfil, for all values of $n$,
    \renewcommand{\theequation}{E}%
    \begin{equation}   
      \evolutionOp_\tVar \ketA{n} 
      =  \partial_\tVar \ketA{n}  \,.
      \label{eq:GenFct_Flow_TCondition}  
    \end{equation} 
    \renewcommand{\theequation}{\arabic{equation}}%
    By the \eref{eq:GenFct_Flow_OrthogonalityCondition}rthogonality condition, one could also define this operator in a ``constructive'' way as
    \begin{equation}   
      \evolutionOp_\tVar 
      \coloneqq \sum_{n} \bigl(\partial_\tVar \ketA{n}  \bigr) \braA{n}    \,.
    \end{equation} 
    We call $\evolutionOp_\tVar$ the basis \eref{eq:GenFct_Flow_TCondition}volution operator. In order to arrive at a proper PDE for $\ketA{\gen}$, $\evolutionOp_\tVar(\q,\partial_\q)$ should be polynomial in $\partial_\q$ (later, in \sref{subsec:GenFct_ForwardBases_Exclusion} we also encounter a case in which it constitutes a power series with infinitely high powers of $\partial_\q$). For now, the pre-factors of $1$, $\partial_\q$, $\partial_\q^2$,\hldots\ may be arbitrary functions of $\q$. Later, in our derivation of a path integral in \sref{sec:PathInts_Forward}, we will also require that the pre-factors can be expanded in powers of $\q$. Note that for a multivariate configuration $n\equiv {(n^{A}, n^{B}, \hldots)}$, the derivative $\partial_\q$ represents individual derivatives with respect to $\q \equiv {(\q^{A}, \q^{B}, \hldots)}$. 
    
    The actual dynamics of a jump process are encoded by its transition rate matrix $\qMatrix_\tVar$ (see \sref{subsec:Intro_Mesoscopic}). The off-diagonal elements of this matrix are the transition rates $\transitionRate_\tVar(n, m)$ from a state $m$ to a state $n$, and its diagonal elements are the negatives of the exit rates $\exitRate_{\tVar}(m) = \sum_{n} \transitionRate_\tVar(n, m)$ from a state $m$ (with $\transitionRate_\tVar(m, m)=0$; see~\eref{eq:Intro_Mescoscopic_TransitionRate}). We encode the information stored in $\qMatrix_\tVar$ by a second differential operator called $\transitionOp_\tVar(\q,\partial_\q)$. This operator should fulfil, for all values of $n$,
    \renewcommand{\theequation}{\qMatrix}%
    \begin{equation} 
      \transitionOp_\tVar \ketA{n} 
      = \sum_{m} \ketA{m} \qMatrix_\tVar(m, n)   \,.  \label{eq:GenFct_Flow_FCondition}  
    \end{equation} 
    \renewcommand{\theequation}{\arabic{equation}}%
    In analogy with the transition (rate) matrix $\qMatrix_\tVar$, we call $\transitionOp_\tVar$ the transition (rate) operator (note that we only speak of ``operators'' with respect to differential operators, but not with respect to matrices). Just like the basis \eref{eq:GenFct_Flow_TCondition}volution operator, $\transitionOp_\tVar(\q,\partial_\q)$ should be polynomial in $\partial_\q$. In \sref{sec:PathInts_Forward}, it will be assumed that $\transitionOp_\tVar(\q,\partial_\q)$ can be expanded in powers of both $\q$ and $\partial_\q$. In a constructive approach, one could also define the operator as
    \begin{equation} 
      \transitionOp_\tVar
      \coloneqq \sum_{m,n} \ketA{m} \qMatrix_\tVar(m, n) \braA{n}   \label{eq:GenFct_Flow_FCondition_Construct}  \,.  
    \end{equation} 
    This constructive definition does not guarantee, however, that $\transitionOp_\tVar(\q,\partial_\q)$ has the form of a differential operator. This property is, for example, not immediately clear for the Fourier basis function $\ketA{n}_\q = \ee^{\ii n \q}$ (with $n\in\integers$), which we complemented with the functional $\braA{n} f = \int_{-\pi}^\pi \frac{\diff{\q}}{2\pi} \ee^{-\ii n \q} f(\q)$ in the introduction to this section. Most of the processes that we solve in later sections have polynomial transition rates. Suitable bases and operators for these processes are provided in the next \sref{subsec:GenFct_ForwardBases}. It remains an open problem for the field to find such bases and operators for processes whose transition rates have different functional forms. That is, for example, the case for transition rates that saturate with the number of particles and have the form of a Hill function. 
    
    Provided that one has found a transition operator $\transitionOp_\tVar$ and a basis \eref{eq:GenFct_Flow_TCondition}volution operator $\evolutionOp_\tVar$ for a \eref{eq:GenFct_Flow_CompletenessCondition}omplete and \eref{eq:GenFct_Flow_OrthogonalityCondition}rthogonal basis, it follows from~\eref{eq:GenFct_Flow_FlowEq_Derivation} that the generalized generating function $\ketA{\gen(\tVar | t_0, n_0)}$  obeys the flow equation\footnote{Later, in our derivation of the forward path integral in~\sref{subsec:PathInts_Forward_Derivation}, we employ the finite difference approximation
      \begin{eqnarray*}
        \lim_{\Delta t\to 0} \bigl(\ketA{\gen(\tVar+\Delta t | \cdot )} - \ketA{\gen(\tVar | \cdot)}\bigr)/\Delta t
        = \lim_{\Delta t\to 0} \transEvoOp_{\tVar,\Delta t}\ketA{\gen(\tVar | \cdot)}  \,.
      \end{eqnarray*} 
      This scheme conforms with the derivation of the forward master equation~\eref{eq:Intro_Mescoscopic_MasterEq}.}
    \begin{equation}
      \partial_\tVar \ketA{\gen}    
      = (\evolutionOp_\tVar + \transitionOp_\tVar) \ketA{\gen}   
      \eqqcolon \transEvoOp_\tVar \ketA{\gen}       \,.      
      \label{eq:GenFct_Flow_FlowEq}
    \end{equation} 
    Its initial condition reads $\ketA{\gen(t_0 | t_0, n_0)} = \ketA{n_0}$, with the possibly time-dependent basis function $\ketA{n_0}$ being evaluated at time $t_0$. Although time-dependent bases prove useful in \sref{sec:StationaryPaths}, we mostly work with time-independent bases in the following. The operators $\transEvoOp_\tVar$ and $\transitionOp_\tVar$ then agree because $\evolutionOp_\tVar$ is zero. Therefore, we refer to both $\transEvoOp_\tVar$ and $\transitionOp_\tVar$ as transition (rate) operators.
    
    In our above derivation, we assumed that the ket $\ketA{n}$ represents an ordinary function and that the bra $\braA{n}$ represents a linear functional. To understand why we chose similar letters for the $\qMatrix_\tVar$-matrix and the $\transitionOp_\tVar$-operator, it is insightful to consider the unit column vectors $\ketA{n} \coloneqq \hat{\vect{e}}_n$ and the unit row vectors $\braA{n} \coloneqq \hat{\vect{e}}_n^\transpose$ (with $m,n\in\naturals_0$). For these vectors, the right hand side of the transition operator~\eref{eq:GenFct_Flow_FCondition_Construct} simply constitutes a representation of the $\qMatrix_\tVar$-matrix. Hence, $\transitionOp_\tVar$ is not a differential operator in this case but coincides with the $\qMatrix_\tVar$-matrix. This observation does not come as a surprise because we already noted that the generating function~\eref{eq:GenFct_Flow_GenFct} represents the vector $\vect{\prob}(\tVar| t_0, n_0)$ of all probabilities in this case. Moreover, the corresponding flow equation~\eref{eq:GenFct_Flow_FlowEq} does not constitute a linear PDE but a vector representation of the forward master equation~\eref{eq:Intro_Mescoscopic_MasterEq}. 

    Following Doi, the transition operator $\transEvoOp_\tVar$ could be called a ``time evolution'' or ``Liouville'' operator~\cite{Doi:1976a,Doi:1976b}, or, following Zel'dovich and Ovchinnikov, a ``Hamiltonian'' ~\cite{Zeldovich:1978}. The latter name is due to the formal resemblance of the flow equation~\eref{eq:GenFct_Flow_FlowEq} to the Schr\"odinger equation in quantum mechanics~\cite{Schroedinger:1926c} (this resemblance holds for any linear PDE that is first order in $\partial_\tVar$). The name ``Hamiltonian'' has gained in popularity throughout the recent years, possibly because the path integrals derived later in this review share many formal similarities with the path integrals employed in quantum mechanics~\cite{Dirac:1933,Feynman:2005,Feynman:1948,Feynman:2010}. Nevertheless, let us point out that $\transEvoOp_\tVar$ is generally not Hermitian and that the generating function $\ketA{\gen}$ does not represent a wave function and also not a probability (unless one chooses the unit vectors $\ketA{n} = \hat{\vect{e}}_n$ and $\braA{n} = \hat{\vect{e}}_n^\transpose$ as basis). In this review, we stick to the name transition (rate) operator for $\transEvoOp_\tVar$ (and $\transitionOp_\tVar$) to emphasize its connection to the transition (rate) matrix $\qMatrix_\tVar$.

  \subsection{Bases for particular stochastic processes}\label{subsec:GenFct_ForwardBases}  
    
    In the previous section, we formulated four conditions under which the master equation~\eref{eq:Intro_Mescoscopic_MasterEq} can be cast into the linear PDE~\eref{eq:GenFct_Flow_FlowEq} obeyed by the generalized generating function~\eref{eq:GenFct_Flow_GenFct}. In the following, we illustrate how these conditions can be met for various stochastic processes.
    
    \subsubsection{Random walks.}\label{subsec:GenFct_ForwardBases_RandomWalk} 
        
      A simple process that can be solved by the method from the previous section is the one-dimensional random walk. We model this process in terms of a particle sitting at position $n$ of the one-dimensional lattice $\lattice \coloneqq \integers$. The particle may jump to the nearest lattice site on its right with the (possibly time-dependent) rate $r_\tVar>0$, and to the site on its left with the rate $l_\tVar>0$. Given that the particle was at position $n_0$ at time $t_0$, the probability of finding it at position $n$ at time $\tVar$ obeys the master equation
      \begin{eqnarray}
        \partial_\tVar \prob(\tVar, n | t_0, n_0 )    \label{eq:GenFct_ForwardBases_RandomWalk_Master}
         & =  r_\tVar \bigl[\prob(\tVar, n-1 | \cdot ) - \prob(\tVar, n | \cdot )\bigr]  \\
           &\hspace{0.95mm} +  l_\tVar \bigl[\prob(\tVar, n+1 | \cdot ) - \prob(\tVar, n | \cdot )\bigr]    \nonumber\,,  
      \end{eqnarray} 
      with initial condition $\prob(t_0, n | t_0, n_0 ) = \delta_{n,n_0}$. One can solve this equation by solving the associated flow equation~\eref{eq:GenFct_Flow_FlowEq}. But for this purpose, we first require an \eref{eq:GenFct_Flow_OrthogonalityCondition}rthogonal and \eref{eq:GenFct_Flow_CompletenessCondition}omplete basis, as well as a basis \eref{eq:GenFct_Flow_TCondition}volution operator $\evolutionOp_\tVar$ and a transition operator $\transitionOp_\tVar$ (condition~\eref{eq:GenFct_Flow_FCondition}).
      
      An appropriate choice of the orthogonal and complete basis proves to be the time-independent Fourier basis
      \begin{equation}
        \ketA{n}_\q \coloneqq \ee^{\ii n \q}
        \mathtext{ and }
        \braA{n}f \coloneqq \int_{-\pi}^\pi \frac{\diff{\q}}{2\pi} \ee^{-\ii n \q} f(\q)    
      \end{equation}
      with $n\in\integers$ and test function $f$. For this basis, the generalized generating function $\ketA{\gen}  = \sum_{n}\ee^{\ii n \q}  \prob(\tVar, n | \cdot) = \langle \ee^{\ii n \q} \rangle$ coincides with the characteristic function. Moreover, the corresponding orthogonality condition $\braketA{m}{n} = \delta_{m,n}$ agrees with a common integral representation of the Kronecker delta. The completeness of the basis can be shown with the help of a Fourier series representation of the ``Dirac comb''
      \begin{equation}
        \sha(\q)   \label{eq:GenFct_ForwardBases_RandomWalk_DiracComb}
        = \sum_{n\in\integers} \delta(\q - 2\pi n)
        = \frac{1}{2\pi} \sum_{n\in\integers} \ee^{\ii n \q}  \,.
      \end{equation}
      It thereby follows that
      \begin{equation}
        \sum_{n\in\integers} \ketA{n}_{\q} \braA{n} f 
        =\int_{-\pi}^\pi \diff{\q^\prime}\, \sha(\q-\q^\prime) f(\q^\prime)
        =f(\q)  \,.
      \end{equation}
      Since the Fourier basis function is time-independent, the condition~\eref{eq:GenFct_Flow_TCondition} is trivially fulfilled for the evolution operator $\evolutionOp_\tVar \coloneqq 0$. The only piece still missing is the transition operator $\transitionOp_\tVar$. Its defining condition~\eref{eq:GenFct_Flow_FCondition} requires knowledge of the transition matrix $\qMatrix_\tVar$ whose elements
      \begin{eqnarray} 
        \qMatrix_\tVar(m, n)       \label{eq:GenFct_ForwardBases_RandomWalk_TransitionMatrix}\\
        = r_\tVar (\delta_{m,n+1} - \delta_{m,n}) + l_\tVar (\delta_{m,n-1} - \delta_{m,n})    \nonumber
      \end{eqnarray} 
      can be inferred by comparing the master equation~\eref{eq:GenFct_ForwardBases_RandomWalk_Master} with its general form~\eref{eq:Intro_Mescoscopic_MasterEq}. The condition~\eref{eq:GenFct_Flow_FCondition} therefore reads
      \begin{equation} 
        \transitionOp_\tVar \ketA{n}  
        =   r_\tVar \bigl(\ketA{n+1} - \ketA{n}\bigr) + l_\tVar \bigl(\ketA{n-1} - \ketA{n}\bigr)  \,. \label{eq:GenFct_ForwardBases_RandomWalk_FCondition}
      \end{equation} 
      One can construct a transition operator with this property with the help of the functions $\cre(\q) \coloneqq \ee^{\ii\q}$ and $\ann(\q) \coloneqq \ee^{-\ii\q}$. The function $\cre$ shifts the basis function $\ketA{n}_\q = \ee^{\ii n \q}$ to the right via $\cre\ketA{n} = \ketA{n+1}$, and the function $\ann$ shifts it to the left via $\ann\ketA{n} = \ketA{n-1}$.  Thus, an operator with the property~\eref{eq:GenFct_ForwardBases_RandomWalk_FCondition} can be defined as
      \begin{equation}
        \transitionOp_\tVar(\q,\partial_\q)   \label{eq:GenFct_ForwardBases_RandomWalk_FOperator}
        \coloneqq r_\tVar \bigl[\cre(\q) - 1\bigr] + l_\tVar \bigl[\ann(\q) - 1\bigr]    \,.
      \end{equation}
      This operator can also be inferred from its constructive definition~\eref{eq:GenFct_Flow_FCondition_Construct} by making use of the Dirac comb.

      After putting the above pieces together, one finds that the generating function $\ketA{\gen(\tVar | t_0, n_0)}_\q$ obeys the flow equation
      \begin{equation}
        \partial_\tVar \ketA{\gen}        \label{eq:GenFct_ForwardBases_RandomWalk_FlowEq}
        = \bigl[ r_\tVar (\ee^{\ii\q} - 1) + l_\tVar (\ee^{-\ii\q} - 1)\bigr] \ketA{\gen}    
      \end{equation}
      for $ t_0 \leq \tVar \leq t$ with the initial value $\ketA{\gen(t_0|t_0,n_0)} = \ketA{n_0} = \ee^{\ii n_0\q}$. The flow equation is readily solved for $\ketA{\gen}$ by
      \begin{equation*}
        %\ketA{\gen} = 
        \exp\Bigl( (\ee^{\ii\q} - 1) \int_{t_0}^{\tVar}\diff{s}\, r_s+   (\ee^{-\ii\q} - 1) \int_{t_0}^{\tVar}\diff{s}\,l_s  \Bigr)\ketA{n_0}    \,.
      \end{equation*}
      The conditional probability distribution can be recovered from this generating function by performing the inverse Fourier transformation~\eref{eq:GenFct_Flow_InverseTransformation}, i.e.\ by evaluating $\prob(\tVar, n | t_0, n_0) = \braketA{n}{\gen}$. A series expansion of all of the involved exponentials and some laborious rearrangement of sums eventually result in a Skellam distribution as the solution of the process~\cite{Skellam:1946}. We outline the derivation of this distribution in \aref{sec:A_RandomWalk}. The distribution's mean is $\mu = n_0 + \int_{t_0}^{\tVar} \diff{s}\, (r_s - l_s)$ and its variance $\sigma^2 = \int_{t_0}^{\tVar} \diff{s}\, (r_s + l_s)$. These moments also follow from the fact that the Skellam distribution describes the difference of two Poisson random variables; one for jumps to the right, the other for jumps to the left. The two moments can, also be obtained more easily by deriving equations for their time evolution from the master equation. Those equations do not couple for the simple random walk. Let us also note that if the two jump rates $r_\tVar$ and $l_\tVar$ agree, the Skellam distribution tends to a Gaussian for large times. 

    \subsubsection{Chemical reactions.}\label{subsec:GenFct_ForwardBases_ChemReactions} 
      
      As a second example, we turn to processes whose transition rates can be decomposed additively into the transition rates of simple chemical reactions. In such a reaction, $k_1$, $k_2$,\,\dots\ molecules of types $A_1$, $A_2$,\,\dots\ come together to be replaced by $l_1$, $l_2$,\,\dots\ molecules of the same types (with $k_j, l_j \in\naturals_0$). Besides reacting with each other, the molecules could also diffuse in space, which can be modelled in terms of hopping processes on a regular, $d$-dimensional lattice such as $\lattice \coloneqq \integers^d$. Upon labelling particles on different lattice sites by their positions, the hopping of a molecule of type $A_1$ from lattice site $i\in\lattice$ to lattice site $j\in \lattice$ could be regarded as the chemical reaction $A_1^{(i)} \to A_1^{(j)}$. We consider the ``chemical'' master equation associated to such hopping processes in \sref{subsubsec:PathInts_BackwardSolutions_DiffusionNetwork}. 
      
      To demonstrate the generating function approach from \sref{subsec:GenFct_Flow}, we focus on a system with only a single type of molecule $A$ engaging in the ``well-mixed'' reaction $k\, A \to l\, A$ ($k, l \in\naturals_0$). Since the forward master equation~\eref{eq:Intro_Mescoscopic_MasterEq} is linear in the transition rate $\qMatrix_\tVar(n, m)$, the following considerations readily extend to networks of multiple reactions, multiple types of molecules, and processes with spatial degrees of freedom. The basis functions and functionals introduced below can, for example, be used to study branching and annihilating random walks, which are commonly modelled in terms of diffusing particles that engage in the binary annihilation reaction $2\, A \to \emptyset$ and the linear growth process $A \to (1+m) A$~\cite{Takayasu:1992,Jensen:1993,Cardy:1996,Cardy:1998,Canet:2004a}. For an odd number of offspring, these walks exhibit an absorbing state phase transition falling into the universality class of directed percolation (according to perturbative calculations in one and two spatial dimensions~\cite{Cardy:1996,Cardy:1998}, according to non-perturbative calculations in up to six dimensions~\cite{Canet:2004b}). The decomposition of a process into elementary reactions of the form $k\, A \to l\, A$ is also possible in the contexts of growing polymer crystals~\cite{Goldenfeld:1984}, aggregation phenomena~\cite{Peliti:1986}, and predator-prey ecosystems~\cite{Shih:2014}.

      As explained in \sref{subsec:Intro_PathInts}, the transition rate $\transitionRate_\tVar(m, n) = \rateCoeffGeneric_\tVar  \delta_{m, n-k+l} (n)_k$ of the reaction $k\, A \to l\, A$ is determined by combinatorial counting. Its proportionality to the falling factorial $(n)_k = n (n-1)\cdots (n-k+1)$ derives from picking $k$ molecules out of a population of $n$ molecules. The overall time scale of the reaction is set by the rate coefficient $\rateCoeffGeneric_\tVar$ (this coefficient may also absorb a factorial $k!$ accounting for the indistinguishability of molecules). The above transition rate guarantees that the number of molecules (or ``particles'') in the system never drops below zero, provided that it was non-negative initially. Thus, the state space of $n$ is $\naturals_0$ and we require basis functions $\ketA{n}$ and basis functionals $\braA{n}$ only for such values. 
      
      Before specifying the \eref{eq:GenFct_Flow_CompletenessCondition}omplete and \eref{eq:GenFct_Flow_OrthogonalityCondition}rthogonal basis as well as the basis \eref{eq:GenFct_Flow_TCondition}volution operator $\evolutionOp_\tVar$, let us first specify an appropriate transition operator $\transitionOp_\tVar$. Its corresponding condition~\eref{eq:GenFct_Flow_FCondition} depends on the transition matrix $\qMatrix_\tVar$, whose elements $\qMatrix_\tVar(m, n) = \rateCoeffGeneric_\tVar  (n)_k (\delta_{m, n-k+l} - \delta_{m, n})$ can be inferred from the rate $\transitionRate_\tVar(m, n)$ with the help of the relation~\eref{eq:Intro_Mescoscopic_TransitionRate}. Consequently, the condition~\eref{eq:GenFct_Flow_FCondition} reads
      \begin{equation} 
        \transitionOp_\tVar \ketA{n}  
        =  \rateCoeffGeneric_\tVar  (n)_k \bigl(    \ketA{n-k+l} - \ketA{n}\bigr)  \,. \label{eq:GenFct_Flow_Poisson_FCondition}
      \end{equation} 
      This condition is met by the transition operator
      \begin{equation}
        \transitionOp_\tVar(\cre, \ann)    \label{eq:GenFct_Flow_Poisson_HamiltonianF}
        \coloneqq \rateCoeffGeneric_\tVar (\cre^l - \cre^k) \ann^k \,,
      \end{equation}
      provided that there exist a ``creation'' operator $\cre$ and an ``annihilation'' operator $\ann$ acting as
      \begin{eqnarray}
        \cre \ketA{n} = \ketA{n+1}   \label{eq:GenFct_Flow_Poisson_CreationAnnihilation_CreAction} \mathtext{ and} \\
        \ann \ketA{n} = n \ketA{n-1}   \label{eq:GenFct_Flow_Poisson_CreationAnnihilation_AnnAction}   \,.
      \end{eqnarray}
      The second of these relations ensures that basis functions with $n<0$ do not appear because $\ann \ketA{0}$ vanishes. Both $\cre$ and $\ann$ may depend on time, just as the basis functionals and functions $\braA{n}$ and $\ketA{n}$ may do. It follows from~\eref{eq:GenFct_Flow_Poisson_CreationAnnihilation_CreAction} and~\eref{eq:GenFct_Flow_Poisson_CreationAnnihilation_AnnAction} that the two operators fulfil the commutation relation
      \begin{equation}
        [\ann_\tVar, \cre_\tVar] \coloneqq \ann_\tVar \cre_\tVar - \cre_\tVar \ann_\tVar = 1    \label{eq:GenFct_Flow_Poisson_Commutator}
      \end{equation} 
      at a fixed time $\tVar$. The commutation relation is meant with respect to functions that can be expanded in the basis functions yet to be defined. For brevity, we commonly drop the subscript $\tVar$ of the creation and annihilation operators. Together with the orthogonality condition, the commutation relation implies that the operators act on the basis functionals as
      \begin{eqnarray}
        \braA{n} \cre = {\braA{n-1}} \mathtext{ and}\label{eq:GenFct_Flow_Poisson_CreationAnnihilation_CreActionFctnl} \\
        \braA{n} \ann = {(n+1)} {\braA{n+1}}   \label{eq:GenFct_Flow_Poisson_CreationAnnihilation_AnnActionFctnl}  \,.
      \end{eqnarray}

      In quantum mechanics, creation and annihilation operators prove useful in solving the equation of motion of a quantum particle in a quadratic potential (i.e.\ in solving the Schr\"odinger equation of the quantum harmonic oscillator)~\cite{Schroedinger:1926b,Ballentine:1998}. There, the ket $\ketA{n}$ is interpreted as carrying $n$ quanta of energy $\hbar\omega$ in addition to the ground state energy $\hbar\omega/2$ of the oscillator's ``vacuum state'' $\ketA{0}$ (with Dirac constant $\hbar$ and angular frequency $\omega$). The creation operator then adds a quantum of energy to state $\ketA{n}$ via the relation $\cre \ketA{n} = \sqrt{n+1}\ketA{n+1}$, and the annihilation operator removes an energy quantum via $\ann \ketA{n} = \sqrt{n} \ketA{n-1}$ ($a$ also destroys the vacuum state). These relations differ from the ones in~\eref{eq:GenFct_Flow_Poisson_CreationAnnihilation_CreAction} and~\eref{eq:GenFct_Flow_Poisson_CreationAnnihilation_AnnAction} because in quantum mechanics, the creation and annihilation operators are defined in terms of self-adjoint position and momentum operators, forcing the former operators to be hermitian adjoints of each other (i.e.\ $\cre=\ann^\dagger$ and $\ann=\cre^\dagger$). Consequently, the relations corresponding to~\eref{eq:GenFct_Flow_Poisson_CreationAnnihilation_CreActionFctnl} and~\eref{eq:GenFct_Flow_Poisson_CreationAnnihilation_AnnActionFctnl} read $\braA{n} \ann^\dagger = \sqrt{n}\braA{n-1}$ and $\braA{n} \ann = \sqrt{n+1}\braA{n+1}$. Nevertheless, the commutation relation~\eref{eq:GenFct_Flow_Poisson_Commutator} also holds in the quantum world in which its validity ultimately derives from the non-commutativity of the position operator $Q$ and the momentum operator $P$ (i.e.\ from $[Q,P]=\ii\hbar$, which follows from $P$ being the generator of spatial displacements in state space~\cite{Ballentine:1998}). Besides, let us note that in quantum field theory, the creation operator is interpreted as adding a bosonic particle to an energy state, and the annihilation operator as removing one~\cite{Peskin:1995}. 
      
      One may wonder whether the above interpretations can be transferred to the theory of stochastic processes in which the creation and annihilation operators need not be each other's adjoints. For example, the creation operator in~\eref{eq:GenFct_Flow_Poisson_CreationAnnihilation_CreAction} could be interpreted as adding a molecule to a system, and the corresponding annihilation operator~\eref{eq:GenFct_Flow_Poisson_CreationAnnihilation_AnnAction} as removing a molecule. The commutation relation~\eref{eq:GenFct_Flow_Poisson_Commutator} may then be interpreted as that the addition of a particle to the system (one way to do it) and the removal of a particle (many ways to do it) do not commute. Let us, however, note that these interpretations only apply to the processes discussed in the present section, which can be decomposed into reactions of the form $k\, A \to l\, A$ (with possibly multiple types of particles and spatial degrees of freedom). The interpretations do not apply, for example, to the random walk of a single particle with state space $\integers$, which we discussed in the previous section (there, the ``shift operators'' with actions $\cre\ketA{n} = \ketA{n+1}$ and $\ann\ketA{n} = \ketA{n-1}$ commute).
      
      Whether creation and annihilation operators with the properties~\eref{eq:GenFct_Flow_Poisson_CreationAnnihilation_CreAction} and~\eref{eq:GenFct_Flow_Poisson_CreationAnnihilation_AnnAction} exist depends on the choice of the basis functions. For the study of chemical reactions, a useful choice proves to be the basis function
      \begin{equation}
        \ketA{n}_{\q}   \label{eq:GenFct_Flow_Poisson_BaseFunction}
        \coloneqq (\basisPrefactor\q + \tq)^n \ee^{-\tx (\basisPrefactor\q + \tq)}  \,.
      \end{equation}
      Here, $\tq(\tVar)$ and $\tx(\tVar)$ are arbitrary, possibly time-dependent functions and $\basisPrefactor\neq 0$ is a free parameter. This parameter only becomes relevant in \sref{sec:PathInts_Observables} in recovering a path integral representation of averaged observables. There, its value is set to $\basisPrefactor \coloneqq \ii$ but typically we choose $\basisPrefactor \coloneqq 1$. For the latter choice, the basis function~\eref{eq:GenFct_Flow_Poisson_BaseFunction} simplifies to
      \begin{equation}
        \ketA{n}_{\q}   \label{eq:GenFct_Flow_Poisson_BaseFunction_1}
        = (\q + \tq)^n \ee^{-\tx (\q + \tq)}  \,.
      \end{equation}
      Alternatively, the parameter $\basisPrefactor$ could be used to rescale the variable $\q$ by a system size parameter $N$ if such a parameter is available. 
      
      The two functions $\tq$ and $\tx$ may prove helpful in simplifying the flow equation obeyed by the generating function. For example, we chose $\tq \coloneqq 1$ and $\tx \coloneqq \rateCoeffSpontGrowth/\rateCoeffLinDecay$ in the introduction to \sref{sec:GenFct} to simplify the flow equation of the reaction $\emptyset \rightleftharpoons A$ (with growth rate coefficient $\rateCoeffSpontGrowth$ and decay rate coefficient $\rateCoeffLinDecay$). Later, in \sref{sec:StationaryPaths}, $\tq$ and $\tx$ will act as the stationary paths of a path integral with $\q$ and an auxiliary variable $\x$ being deviations from them. If both $\tq$ and $\tx$ are chosen as zero, the basis function~\eref{eq:GenFct_Flow_Poisson_BaseFunction_1} simplifies to the basis function
      \begin{equation}
        \ketA{n}_{\q} = \q^n \label{eq:GenFct_Flow_Poisson_BaseFunction_2}
      \end{equation}
       of the ordinary probability generating function. 
      
      For the general basis function~\eref{eq:GenFct_Flow_Poisson_BaseFunction}, creation and annihilation operators with the properties~\eref{eq:GenFct_Flow_Poisson_CreationAnnihilation_CreAction} and~\eref{eq:GenFct_Flow_Poisson_CreationAnnihilation_AnnAction} can be defined as
      \begin{eqnarray}
        \cre(\q, \partial_\q)   \label{eq:GenFct_Flow_Poisson_Creation}  
        \coloneqq \basisPrefactor\q + \tq 
        \mathtext{ and}    \\
        \ann(\q, \partial_\q) 
        \coloneqq \partial_{\basisPrefactor\q} + \tx \,.   \label{eq:GenFct_Flow_Poisson_Annihilation}  
      \end{eqnarray}
      Apparently, these operators are not each other's adjoints. For the basis function $\ketA{n}_{\q} = \q^n$ of the ordinary probability generating function, the operators simplify to $\cre=\q$ and $\ann=\partial_\q$. The corresponding  transition operator~\eref{eq:GenFct_Flow_Poisson_HamiltonianF} reads $\transitionOp_\tVar(\q, \partial_\q)  = \rateCoeffGeneric_\tVar (\q^l - \q^k) \partial_\q^k$, resulting in the flow equation
      \begin{equation}
        \partial_\tVar \ketA{\gen}
        = \rateCoeffGeneric_\tVar (\q^l - \q^k) \partial_{\q}^k \ketA{\gen} \,.
      \end{equation}

      Thus far, we have not specified the basis functionals. These functionals can be defined using the annihilation operator~\eref{eq:GenFct_Flow_Poisson_Annihilation}. In particular, we complement the basis function~\eref{eq:GenFct_Flow_Poisson_BaseFunction} with the functionals
      \begin{equation}
        \braA{n} f     \label{eq:GenFct_Flow_Poisson_BaseFunctional}
        \coloneqq \frac{\ann^n}{n!} f(\q) \big|_{\q = -\tq/\basisPrefactor}     \,,
      \end{equation}
      where $f$ represents a test function. The \eref{eq:GenFct_Flow_OrthogonalityCondition}rthogonality of the basis $\{\braA{m},\ketA{n}\}_{m,n\in\naturals_0}$ follows directly from the action of the annihilation operator on the basis function $\ketA{n}$, and from the vanishing of this function at $\q=-\tq/\basisPrefactor$, except for $n=0$. The \eref{eq:GenFct_Flow_CompletenessCondition}ompleteness of the basis is also readily established.\footnote{Use $(\partial_{\q} + \basisPrefactor \tx)^n f(\q) |_{\q = -\tq/\basisPrefactor}  = \partial_{\q}^n ( \ee^{\tx (\basisPrefactor\q + \tq)} f(\q) ) |_{\q = -\tq/\basisPrefactor} $ and the analyticity of $\ee^{\tx (\basisPrefactor\q + \tq)} f(\q)$ for an analytic test function $f$.}
      
      The only piece still missing is the basis \eref{eq:GenFct_Flow_TCondition}volution operator $\evolutionOp_\tVar$ to encode the time-dependence of the basis function~\eref{eq:GenFct_Flow_Poisson_BaseFunction}. Differentiation of this function with respect to time and use of the creation and annihilation operators~\eref{eq:GenFct_Flow_Poisson_Creation} and~\eref{eq:GenFct_Flow_Poisson_Annihilation} show that this operator is given by
      \begin{equation}
        \evolutionOp_\tVar(\cre, \ann)     \label{eq:GenFct_Flow_Poisson_HamiltonianT}
        \coloneqq   (\partial_\tVar \tq) (\ann - \tx)
                    - (\partial_\tVar \tx) \cre  \,.     
      \end{equation}
            
      Let us illustrate the use of the above basis for a process whose transition operator can be reduced to a mere constant by an appropriate choice of $\tq$ and $\tx$. This simplification is, however, bought by making the basis functions time-dependent. In particular, we consider the simple growth process $\emptyset \to A$ for a time-independent growth rate coefficient $\rateCoeffSpontGrowth$. By our previous discussions, one can readily verify that the ordinary probability generating function with basis function $\ketA{n}_{\q} = \q^n$ obeys the flow equation $\partial_\tVar \ketA{\gen} = \rateCoeffSpontGrowth (\q - 1) \ketA{\gen}$ for this process. This flow equation can be simplified by redefining the basis function of the generating function as $\ketA{n}_{\tVar,\q} \coloneqq (\q + 1)^n \ee^{-\tx (\q + 1)}$, with $\tx$ solving the rate equation $\partial_\tVar \tx = \rateCoeffSpontGrowth$ of the process. Hence, the basis function depends explicitly on time through its dependence on $\tx(\tVar) = \tx(t_0) + \rateCoeffSpontGrowth (\tVar - t_0)$. 
      Upon combining the transition operator~\eref{eq:GenFct_Flow_Poisson_HamiltonianF} with the basis evolution operator~\eref{eq:GenFct_Flow_Poisson_HamiltonianT}, one finds that the generating function now obeys the flow equation $\partial_\tVar \ketA{\gen} = -\rateCoeffSpontGrowth \ketA{\gen}$. The equation is readily solved by $\ketA{\gen(\tVar | t_0, n_0)}  = \ee^{-\rateCoeffSpontGrowth (\tVar - t_0)} \ketA{n_0}_{t_0}$. The conditional probability distribution can be recovered via the inverse transformation~\eref{eq:GenFct_Flow_InverseTransformation} as
      \begin{equation}
        \prob(\tVar, n | t_0, n_0)   \label{eq:GenFct_ForwardBases_ChemReactions_Example1}
        = \ee^{-\rateCoeffSpontGrowth (\tVar - t_0)} \braA{n}_\tVar \ketA{n_0}_{t_0}  \,. 
      \end{equation}
      The coefficient $\braA{n}_\tVar \ketA{n_0}_{t_0}$ can be evaluated by determining how the functional $\braA{n}_\tVar$ acts at time $t_0$. Using $\ann_{\tVar} - \tx(\tVar) = \ann_{t_0} - \tx(t_0)$ and the binomial theorem, one can rewrite the action of this functional as
      \begin{equation}
        \braA{n}_{\tVar} f   
        = \sum_{m=0}^{n} \frac{\bigl(\rateCoeffSpontGrowth(\tVar-t_0)\bigr)^{n-m}}{(n-m)!} \braA{m}_{t_0} f      \,.
      \end{equation}
      By the \eref{eq:GenFct_Flow_OrthogonalityCondition}rthogonality condition, the conditional probability distribution~\eref{eq:GenFct_ForwardBases_ChemReactions_Example1} therefore evaluates to the shifted Poisson distribution
      \begin{equation}
        \prob(\tVar , n| t_0, n_0)    \label{eq:GenFct_ForwardBases_ChemReactions_ShiftedPoisson}
        = \frac{\ee^{-\rateCoeffSpontGrowth (\tVar - t_0)} \bigl(\rateCoeffSpontGrowth(\tVar-t_0)\bigr)^{n-n_0}}{(n-n_0)!} \HeavisideStepDiscrete_{n-n_0}  \,,
      \end{equation}
      where $\HeavisideStepDiscrete_{n} \coloneqq 1$ for $n \geq 0$ and $\HeavisideStepDiscrete_{n} \coloneqq 0$ otherwise. The validity of this solution can be verified by solving the corresponding flow equation of the ordinary generating function.

      In our above approach, we first established the form of the transition operator because the basis function~\eref{eq:GenFct_Flow_Poisson_BaseFunction} is not the only possible choice. Ohkubo recently proposed the use of orthogonal polynomials for this purpose~\cite{Ohkubo:2012}. In analogy to the eigenfunctions of the quantum harmonic oscillator~\cite{Schroedinger:1926b,Ballentine:1998}, one can, for example, choose the basis function as the Hermite polynomial
      \begin{equation}
        \ketA{n}_\q \coloneqq \hermite_n(\q)   \label{eq:GenFct_ForwardBases_ChemReactions_Hermite}
        = (-1)^n \ee^{\q^2/2} \partial_{\q}^n \, \ee^{-\q^2/2}  \,,
      \end{equation}
      with $n\in\naturals_0$. Hermite polynomials constitute an Appell sequence, i.e.\ they fulfil $\partial_\q\hermite_{n}(\q) = n \hermite_{n-1}(\q)$ (18.9.27 in~\cite{NIST:2010}). With $\ann \coloneqq \partial_{\q}$, this property coincides with the defining relation $\ann \ketA{n} = n\ketA{n-1}$ of the annihilation operator in~\eref{eq:GenFct_Flow_Poisson_CreationAnnihilation_AnnAction}. Furthermore, Hermite polynomials obey the recurrence relation $\hermite_{n+1}(\q) = \q \hermite_{n}(\q) - n \hermite_{n-1}(\q)$ (18.9.1 in~\cite{NIST:2010}). Combined with the Appell property, $\cre \coloneqq  \q - \partial_{\q}$ therefore fulfils the defining relation $\cre \ketA{n}  = \ketA{n+1}$ of a creation operator in~\eref{eq:GenFct_Flow_Poisson_CreationAnnihilation_CreAction}. After complementing the basis function~\eref{eq:GenFct_ForwardBases_ChemReactions_Hermite} with the functional
      \begin{equation}
        \braA{m} f 
        \coloneqq \frac{1}{m!}\int_{-\infty}^\infty  \diff{\q}\, \frac{\ee^{-\q^2/2}}{\sqrt{2\pi}}  \hermite_m(\q) f(\q)  \,,
      \end{equation}
      the \eref{eq:GenFct_Flow_OrthogonalityCondition}rthogonality and \eref{eq:GenFct_Flow_CompletenessCondition}ompleteness of the basis are also established (18.3 and 18.18.6 in~\cite{NIST:2010}). Thus, the chemical master equation~\eref{eq:Intro_Mescoscopic_ChemicalMasterEq} can be transformed into a flow equation obeyed by a generating function based on Hermite polynomials. To our knowledge, however, no stochastic process has thus far been solved or been approximated along these lines. Besides Hermite polynomials, Ohkubo also proposed the use of Charlier polynomials and mentioned their relation with certain birth-death processes~\cite{Ohkubo:2012}.
      
    \subsubsection{Intermezzo: The unit vector basis.}\label{subsec:GenFct_ForwardBases_Algebraic} 
    
      One may wonder why we actually bother with explicit representations of the bras $\braA{n}$ and kets $\ketA{n}$. Often, these objects are introduced only formally as the basis of a ``bosonic Fock space''~\cite{Doi:1976a,Doi:1976b,Mattis:1998,Cardy:2008,Wiese:2015}, leaving the impression that the particles under consideration are in fact bosonic quantum particles. Although this impression takes the analogy with quantum theory too far, the analogy has helped in developing new methods for solving the master equations. In the previous sections, we showed how the forward master equation can be cast into a linear PDE obeyed by a probability generating function. Later, in \sref{sec:PathInts_Forward}, this PDE is solved in terms of a path integral by which we then recover a path integral representation of averaged observables in \sref{sec:PathInts_Observables}. This ``analytic'' derivation of the path integral is, however, not the only possible way. In the following, we sketch the mathematical basis of an alternative approach, which employs the unit vectors $\hat{\vect{e}}_n$ as the basis. The approach ultimately results in the same path integral representation of averaged observables as we show in \sref{subsec:PathInts_Observables_Algebraic}. The duality between the two approaches resembles the duality between the matrix mechanics formulation of quantum mechanics by Heisenberg, Born, and Jordan~\cite{Heisenberg:1925,Born:1925,Born:1926} and its analytic formulation in terms of the Schr\"odinger equation~\cite{Schroedinger:1926c}.  

      The alternative derivation of the path integral also starts out from the forward master equation $\partial_\tVar \prob(\tVar | t_0) = \qMatrix \prob(\tVar | t_0)$. As before, $\prob(\tVar | t_0)$ represents the matrix of the conditional probabilities $\prob(\tVar , n| t_0, n_0)$, and $\qMatrix$ is the transition rate matrix. For simplicity, we assume the transition rate matrix to be independent of time. Moreover, we focus on processes that can be decomposed additively into chemical reactions of the form $k\, A \to l\, A$ with time-independent rate coefficients. The elements of the transition matrix associated to this reaction read $\qMatrix(m, n) = \rateCoeffGeneric  (n)_k (\delta_{m, n-k+l} - \delta_{m, n})$. Since the particle numbers $n$ and $n_0$ may assume any values in $\naturals_0$, the probability matrix $\prob(\tVar | t_0)$ and the transition matrix $\qMatrix$ have infinitely many rows and columns. 
      
      In the introductory \sref{subsec:Intro_MasterEqSolution}, we formulated conditions under which the forward master equation is solved by the matrix exponential $\prob(\tVar | t_0) = \ee^{\qMatrix(\tVar-t_0)} \unitMatrix$ (in the sense of the expansion~\eref{eq:Intro_MasterEqSolution_Uniform}, the state space must countable and the exit rates bounded). For the above chemical reaction, those conditions are not necessarily met, and thus the matrix exponential may not exist. Nevertheless, we regard $\prob(\tVar | t_0) = \ee^{\qMatrix(\tVar-t_0)} \unitMatrix$ as a ``formal'' solution in the following. With the help of certain mathematical ``tricks'', this solution will be cast into a path integral in \sref{sec:PathInts_Observables}. But before, let us rewrite the transition matrix in the exponential in terms of creation and annihilation \textit{matrices}. For this purpose, we employ the (infinitely large) unit column vectors $\ketA{n} \coloneqq \hat{\vect{e}}_n$ and their orthogonal unit row vectors $\braA{n} \coloneqq \hat{\vect{e}}_n^\transpose$ as basis (with $n\in\naturals_0$). Individual probabilities can therefore be inferred from the above solution as
      \begin{equation}
        \prob(\tVar, n | t_0, n_0)   \label{eq:GenFct_ForwardBases_Prob}
        = \braA{n}\ee^{\qMatrix(\tVar-t_0)}\ketA{n_0}  \,.
      \end{equation}
      For multivariate or spatial processes, the basis vectors can be generalized to tensors (i.e.\ $\ketA{\vect{n}} = \ketA{n_1}\otimes\ketA{n_2}\otimes\hldots$).
      
      The orthogonality of the basis allows us to rewrite the elements of the transition rate matrix of the reaction $k\, A \to l\, A$ as
      \begin{equation}
        \qMatrix(m, n) \label{eq:GenFct_ForwardBases_QMatrix_Elts}
        = \braA{m} \Bigl(\rateCoeffGeneric\, (n)_k \bigl(\ketA{n-k+l} - \ketA{n}\bigr)  \Bigr)    \,.
      \end{equation} 
      A matrix with these elements can be written as
      \begin{equation}
        \qMatrix          \label{eq:GenFct_ForwardBases_QMatrix}
        = \rateCoeffGeneric (\cre^l - \cre^k) \ann^k   \,,
      \end{equation}
      with $\cre$ acting as a ``creation matrix'' and $\ann$ acting as an ``annihilation matrix''. In particular, we define $\cre$ in terms of its sub-diagonal $(1,1,1,\hldots)$ and $\ann$ in terms of its super-diagonal $(1,2,3,\hldots)$. All of their other matrix elements are set to zero. It is readily shown that these matrices fulfil $\cre \ketA{n} = \ketA{n+1}$ and $\ann \ketA{n} = n \ketA{n-1}$ with respect to the basis vectors. These relations coincide with our previous relations~\eref{eq:GenFct_Flow_Poisson_CreationAnnihilation_CreAction} and~\eref{eq:GenFct_Flow_Poisson_CreationAnnihilation_AnnAction}. Besides, the matrices also fulfil $\braA{n} \cre = {\braA{n-1}}$ and $\braA{n} \ann = {(n+1)} {\braA{n+1}}$, as well as the commutation relation $[\ann,\cre] = \unitMatrix$. Further properties of the matrices are addressed in \sref{subsec:PathInts_Observables_Algebraic}. By our previous comments in \sref{subsec:GenFct_Flow}, it is not a surprise that the transition matrix~\eref{eq:GenFct_ForwardBases_QMatrix} has the same form as the transition operator~\eref{eq:GenFct_Flow_Poisson_HamiltonianF}. Both of them are normal-ordered polynomials in $\cre$ and $\ann$. This property will help us in \sref{subsec:PathInts_Observables_Algebraic} in recovering a path integral representation of averaged observables.

    \subsubsection{Processes with locally excluding particles.}\label{subsec:GenFct_ForwardBases_Exclusion} 
  
      As noted in the introductory \sref{subsec:Intro_Mesoscopic}, one can model the movement of molecular motors along a cytoskeletal filament in terms of a master equation~\cite{Lipowsky:2001,Parmeggiani:2003,Parmeggiani:2004}. In its simplest form, such a model describes the movement of mutually excluding motors as a hopping process on a one-dimensional lattice $\lattice \subset \integers$, with attachment and detachment of motors at certain boundaries. To respect the mutual exclusion of motors, their local number $n_i$ on a lattice site $i\in\lattice$ is restricted to $0$ and $1$. Various other processes can be modelled in similar ways, for example, aggregation processes~\cite{Sandow:1993}, adsorption processes~\cite{Stinchcombe:1993,Grynberg:1994}, and directed percolation~\cite{Brunel:2000}. In the following, we show how the generating function approach from \sref{subsec:GenFct_Flow} and its complementary approach from \sref{subsec:GenFct_ForwardBases_Algebraic} can be applied to such processes. To illustrate the mathematics behind these approaches while not burdening ourselves with too many indices, we demonstrate the approaches for the simple, non-spatial telegraph process~\cite{Goldstein:1951,Kac:1974}. This process describes a system that randomly switches between two states that are called the ``on'' and ``off'' states, or, for brevity, the ``$1$'' and ``$0$'' states. The rate at which the system switches from state $1$ to state $0$ is denoted as $\rateCoeffLinDecay$, and the rate of the reverse transition as $\rateCoeffSpontGrowth$. For simplicity, we assume these rate coefficients to be time-independent. Consequently, the master equation of the telegraph process reads
      \begin{equation}
        \partial_\tVar   
        \vect{\prob}
        =
        \biggl({\begin{array}{cc} 
          -\rateCoeffLinDecay & \rateCoeffSpontGrowth \\ 
          \rateCoeffLinDecay & -\rateCoeffSpontGrowth
        \end{array}}\biggr)
        \vect{\prob}
        \,,  \label{eq:GenFct_ForwardBases_Exclusion_Master}
      \end{equation}
      with $\vect{\prob}(\tVar | \cdot) = (\prob(\tVar,1|\cdot), \prob(\tVar,0|\cdot))^\transpose$ being the probability vector. Alternatively, the master equation can be written as $\partial_\tVar \prob(\tVar, n | \cdot) = \sum_{m \in \{1,0\}} \qMatrix(n, m) \prob(\tVar, m | \cdot)$ with transition matrix elements
      \begin{equation}
        \qMatrix(n, m) 
        = \rateCoeffLinDecay (\delta_{n,0} - \delta_{n,1})\delta_{m,1} + \rateCoeffSpontGrowth (\delta_{n,1} - \delta_{n,0}) \delta_{m,0}  \,. \,   
      \end{equation}
      One can solve the above master equation in various ways, for example, by evaluating the matrix exponential in $\prob(\tVar| \tVar_0) = \ee^{\qMatrix(\tVar-\tVar_0)}\unitMatrix$ after diagonalizing the transition matrix. In the following, the telegraph process serves as the simplest representative of processes with excluding particles and it helps in explaining how the methods from the previous sections are applied to such processes. The inclusion of spatial degrees of freedom into the procedure is straightforward.
      
      To apply the generating function technique from \sref{subsec:GenFct_Flow}, we require orthogonal and complete basis functions $\{\ketA{0}, \ketA{1}\}$ and basis functionals $\{\braA{0}, \braA{1}\}$, as well as a transition operator $\transitionOp(\q,\partial_\q)$ fulfilling condition~\eref{eq:GenFct_Flow_FCondition}, i.e., for $n\in\{0,1\}$,
      \begin{equation} 
        \transitionOp \ketA{n}  \label{eq:GenFct_ForwardBases_Exclusion_FOperator}
        =   \rateCoeffLinDecay \bigl( \ketA{0} - \ketA{1}) \delta_{n,1} + \rateCoeffSpontGrowth \bigl( \ketA{1} - \ketA{0})\delta_{n,0} 
          \,.        
      \end{equation} 
      An operator with this property can be constructed in (at least) two ways: using creation and annihilation operators fulfilling the ``bosonic'' commutation relation $[\ann, \cre] = \ann \cre - \cre \ann = 1$, or using operators fulfilling the ``fermionic'' anti-commutation relation $\{\ann, \cre\} \coloneqq \ann \cre + \cre \ann = 1$.
      
      Let us first outline an approach based on the commutation relation. The approach was recently proposed by van Wijland~\cite{Wijland:2001}. To illustrate the approach, we take the ``analytic'' point of view with $\cre(\q,\partial_\q)$ and $\ann(\q,\partial_\q)$ being differential operators. However, the following considerations also apply to the approach from the previous section where $\cre$ and $\ann$ represented creation and annihilation \textit{matrices}. To cast a master equation such as~\eref{eq:GenFct_ForwardBases_Exclusion_Master} into a path integral, van Wijland employed creation and annihilation operators with the same actions as in \sref{subsec:GenFct_ForwardBases_ChemReactions}, i.e.\ $\cre \ketA{n} = \ketA{n+1}$ and $\ann \ketA{n} = n \ketA{n-1}$. Besides complying with the commutation relation $[\ann, \cre] = 1$, these relations imply that the ``number operator'' $\numberOperator \coloneqq \cre\ann$ fulfils $\numberOperator\ketA{n} = n \ketA{n}$. This operator may be used to define the ``Kronecker operator''
      \begin{equation}
        \delta_{\numberOperator, m}   \label{eq:GenFct_ForwardBases_Exclusion_KroneckerOp}
        \coloneqq \int_{-\pi}^\pi \frac{\diff{u}}{2\pi} \ee^{\ii u (\numberOperator-m)}    
      \end{equation}
      in terms of a series expansion of its exponential. The Kronecker operator acts on the basis functions as $\delta_{\numberOperator, m} \ketA{n} = \delta_{n, m}\ketA{n}$, and thus it can be used to replace the Kronecker deltas in~\eref{eq:GenFct_ForwardBases_Exclusion_FOperator}. One thereby arrives at the flow equation
      \begin{equation}
        \partial_\tVar \ketA{\gen}   \label{eq:GenFct_ForwardBases_Exclusion_FlowEqGen}
        = \transitionOp  \ketA{\gen}  
        = \bigl[\rateCoeffLinDecay (\ann-1) \delta_{\numberOperator, 1} + \rateCoeffSpontGrowth (\cre-1) \delta_{\numberOperator, 0}\bigr] \ketA{\gen}    \,.
      \end{equation}
      On the downside, the flow equation now involves power series with arbitrarily high derivatives with respect to $\q$ (upon choosing the operators as $\cre=\q$ and $\ann=\partial_\q$). In \sref{sec:PathInts_Forward}, we show how the solution of a flow equation such as~\eref{eq:GenFct_ForwardBases_Exclusion_FlowEqGen} can be expressed in terms of a path integral. The derivation of the path integral requires that the transition operator $\transitionOp(\q,\partial_\q)$ is normal-ordered with respect to $\q$ and $\partial_\q$, i.e.\ that all the $\q$ are to the left of all the $\partial_\q$ in every summand. This order can be achieved by the repeated use of $[\partial_\q,\q]f(\q)=f(\q)$. More information on the procedure can be found in~\cite{Wijland:2001}. Van Wijland applied the resulting path integral to the asymmetric diffusion of excluding particles on a one-dimensional lattice. Moreover, Mobilia, Georgiev, and T\"auber have employed the method in studying the stochastic Lotka-Volterra model on $d$-dimensional lattices~\cite{Mobilia:2006}.
      
      An alternative to the above approach lies in the use of operators fulfilling the anti-commutation relation $\{\ann, \cre\} = \ann \cre + \cre \ann = 1$. This relation is clearly \textit{not} fulfilled by $\cre\coloneqq\q$ and $\ann\coloneqq \partial_\q$, at least not if $\q$ represents an ordinary real variable. However, $\partial_\q \q + \q \partial_\q = 1$ holds true if $\q$ represents a Grassmann variable (see~\cite{Berezin:1966,Combescure:2014} for details). Grassmann variables commute with real and complex numbers (i.e.\ $[\q,\alpha]=0$ for $\alpha\in\complex$), but they anti-commute with themselves and with other Grassmann variables (i.e.\ $\{\q,\tilde{\q}\}=0$). Grassmann variables have, for example, proven useful in calculating the correlation functions of kinetic Ising models~\cite{Aliev:2000}, even when the system is driven far from equilibrium~\cite{Mobilia:2004}.
            
      Instead of using Grassmann variables for the basis, let us consider a representation of the basis defined by the unit row vectors $\braA{0} = (0,1)$ and $\braA{1} = (1,0)$, and by the unit column vectors $\ketA{0} = (0,1)^\transpose$ and $\ketA{1} = (1,0)^\transpose$. Obviously, these vectors fulfil the orthogonality condition $\braketA{m}{n}=\delta_{m,n}$, and $\sum_{n} \ketA{n} \braA{n} = \unitMatrix$ is the $2$-by-$2$ unit matrix. The anti-commutation relation $\{\ann, \cre\} = 1$ is met by the creation and annihilation \textit{matrices}
      \begin{equation}
        \cre \coloneqq \sigma^+ = 
        \biggl({\begin{array}{cc} 
          0 & 1 \\ 
          0 & 0
        \end{array}}\biggr)
        \mathtext{ and }
        \ann \coloneqq \sigma^- = 
        \biggl({\begin{array}{cc} 
          0 & 0 \\ 
          1 & 0
        \end{array}}\biggr) \,.
      \end{equation}
      The creation matrix acts as $\cre \ketA{0}=\ketA{1}$ and $\cre \ketA{1}=\vect{0}$, and the annihilation matrix as $\ann\ketA{1} = \ketA{0}$ and $\ann\ketA{0} = \vect{0}$ (zero vector). Consequently, the matrix product $\cre\ann$ fulfils $\cre\ann\ketA{0}=\vect{0}$ and $\cre\ann\ketA{1}=\ketA{1}$, and thus it serves the same purpose as the Kronecker delta $\delta_{n,1}$ in~\eref{eq:GenFct_ForwardBases_Exclusion_FOperator}. Analogously, the operator $\ann \cre$ serves the same purpose as $\delta_{n,0}$. The master equation~\eref{eq:GenFct_ForwardBases_Exclusion_Master} can therefore be written in a form resembling the flow equation~\eref{eq:GenFct_ForwardBases_Exclusion_FlowEqGen}, namely as
      \begin{equation}
        \partial_\tVar \vect{\prob}      
        =   \bigl[\rateCoeffLinDecay (\ann-\unitMatrix)\cre\ann + \rateCoeffSpontGrowth(\cre-\unitMatrix) \ann\cre \bigr] \vect{\prob}    \,.
      \end{equation}
      In this form, the stochastic process mimics a spin-$1/2$ problem (especially, if the creation and annihilation matrices are written in terms of the Pauli spin matrices $\sigma_x$, $\sigma_y$, and $\sigma_z$). More information on spin-representations of master equations is provided in~\cite{Stinchcombe:1993,Alcaraz:1994,Schutz:1994,Grynberg:1994,Schutz:1995,Henkel:1997,Hinrichsen:2000}. A spin-representation has, for example, been employed in the analysis of reaction-diffusion master equations via the density matrix renormalization group~\cite{Carlon:1999,Carlon:2001}. In their study of directed percolation of excluding particles on the one-dimensional lattice $\integers$, Brunel, Oerding, and van Wijland performed a Jordan-Wigner transformation of the spatially extended ``spin'' matrices $\cre_i\coloneqq\sigma^+_i$ and $\ann_i\coloneqq\sigma^-_i$ ($i\in\integers$)~\cite{Brunel:2000}. The transformed operators fulfill anti-commutation relations not only locally but also non-locally, and thus represent the stochastic process in terms of a ``fermionic'' (field) theory. While the Jordan-Wigner transformation provides an exact reformulation of the stochastic process, its applicability is largely limited to systems with one spatial dimension. Moreover, the transformation requires that the stochastic process is first rewritten in terms of a spin-$1/2$ chain (typically possible only for processes with a single species). Based on the (coherent) eigenstates of the resulting fermionic creation and annihilation operators, Brunel et al.\ then derived a path integral representation of averaged observables. The paths of these integrals proceed along the values of Grassmann variables. Further information on the Jordan-Wigner transformation, Grassmann path integrals, and alternative approaches can be found in~\cite{Sandow:1993,Patzlaff:1994,Bares:1999,Park:2000,Schulz:2005,Park:2005,Silva:2008,Tailleur:2008b}.

  \subsection{Methods for the analysis of the generating function's flow equation}\label{subsec:GenFct_Spectral} 
  
    In the following, we outline various approaches that have recently been proposed for the analysis of the generating function's flow equation.  
    
    \subsubsection{A spectral method for the computation of stationary distributions.}\label{subsec:GenFct_Spectral_Walczak} 
      
      In their study of a linear transcriptional regulatory cascade of genes and proteins, Walczak, Mugler, and Wiggins developed a spectral method for the computation of stationary probability distributions~\cite{Walczak:2009}. They described the regulatory cascade on a coarse-grained level in terms of the copy numbers of certain chemical ``species'' at its individual steps. By imposing a Markov approximation, the dynamics of the cascade was reduced to a succession of two-species master equations. The solution of each master equation served as input for the next equation downstream. Every of the reduced master equations allowed for the following processes: First, each of the master equation's two species $i\in\{1,2\}$ is produced in a one-step process whose rate $\rateCoeffSpontGrowth_i(n_1)$ depends only on the copy number $n_1$ of the species coming earlier along the cascade. Second, each of the two species degrades at a constant per-capita rate $\rateCoeffLinDecay_i$. With $\vect{n}\coloneqq (n_1,n_2)^\transpose\in\naturals_0^2$, the corresponding master equation reads
      \begin{eqnarray}    
        \partial_\tVar \prob(\tVar, \vect{n} | \cdot)        
        &= \rateCoeffSpontGrowth_1(n_1-1) \prob(\tVar, \vect{n} - \hat{\vect{e}}_1 |\cdot) - \rateCoeffSpontGrowth_1(n_1) \prob(\tVar, \vect{n} |\cdot)    \nonumber\\
        &+ \rateCoeffSpontGrowth_2(n_1) \bigl[ \prob(\tVar, \vect{n} - \hat{\vect{e}}_2 |\cdot) - \prob(\tVar, \vect{n} |\cdot)\bigr]   \label{eq:GenFct_Spectral_MasterEq}  \\
        &+ \rateCoeffLinDecay_1 \bigl[ (n_1+1) \prob(\tVar, \vect{n}  + \hat{\vect{e}}_1 | \cdot) - n_1 \prob(\tVar, \vect{n} |\cdot) \bigr]    \nonumber\\
        &+ \rateCoeffLinDecay_2 \bigl[(n_2+1) \prob(\tVar, \vect{n} + \hat{\vect{e}}_2 |\cdot) - n_2 \prob(\tVar, \vect{n} |\cdot) \bigr]  \nonumber  \,.
      \end{eqnarray} 
      Here, the unit vector $\hat{\vect{e}}_i$ points in the direction of the $i$-th species. The master equation can be cast into a flow equation for the generating function $\ketA{\gen(\tVar |\cdot)} = \sum_{\vect{n}} \prob(\tVar,\vect{n}|\cdot) \ketA{\vect{n}} $ by following the  steps in  \sref{subsec:GenFct_Flow}. For this purpose, we choose the basis function as a multivariate extension of the basis function from the introduction to \sref{sec:GenFct}, i.e.\ as $\ketA{\vect{n}}_{\vect{\q}}  \coloneqq \ketA{n_1}_{\q_1}\ketA{n_2}_{\q_2}$ with 
      \begin{equation}
        \ketA{n_i}_{\q_i}  \coloneqq (\q_i + 1)^{n_i} \ee^{-\frac{\bar{\rateCoeffSpontGrowth}_i}{\rateCoeffLinDecay_i} (\q_i + 1)}  \,.
      \end{equation}
      The values of the auxiliary parameters $\bar{\rateCoeffSpontGrowth}_1$ and $\bar{\rateCoeffSpontGrowth}_2$ are only specified in a numerical implementation of the method and affect its stability. Information on how their values are chosen is provided in~\cite{Walczak:2009}. For our present purposes, the values of the parameters remain unspecified. By differentiating the generating function with respect to time, one finds that the generating function obeys the flow equation $\partial_\tVar \ketA{\gen}   = (\transitionOp_0 + \transitionOp_1) \ketA{\gen}$ with the transition operators
      \begin{eqnarray}
        \transitionOp_0  \coloneqq  - \!\!\sum_{i\in\{1,2\}} \rateCoeffLinDecay_i \q_i \partial_{\q_i}   \mathtext{ and }   \\
        \transitionOp_1  \coloneqq -\!\! \sum_{i\in\{1,2\}} \q_i(\bar{\rateCoeffSpontGrowth}_i - \hat{\rateCoeffSpontGrowth}_i)     \,.
      \end{eqnarray} 
      Here, the two new operators $\hat{\rateCoeffSpontGrowth}_1$ and $\hat{\rateCoeffSpontGrowth}_2$ are defined in terms of their actions $\hat{\rateCoeffSpontGrowth}_i \ketA{n_1} = \rateCoeffSpontGrowth_i(n_1) \ketA{n_1}$ on the basis functions. Surprisingly, the explicit form of these operators is not needed. Thus, the spectral method even allows for non-polynomial growth rates $\rateCoeffSpontGrowth_i(n_1)$. 
      
      The operator $\transitionOp_0$ has the same form as the transition operator of the bi-directional reaction $\emptyset \rightleftharpoons A$ from the introductory example. Without the perturbation $\transitionOp_1$, the above flow equation could thus be solved by extending the previous ansatz~\eref{eq:GenFct_GenFct_Solution} to two species. To accommodate $\transitionOp_1$ as well, it proves useful to generalize that ansatz to
      \begin{equation}
        \ketA{\gen(\tVar | \cdot)} = \sum_{\vect{k} \in\naturals_0^2} \ketB{\vect{k}} G_{\vect{k}}(\tVar | \cdot) \,,
      \end{equation} 
      with yet to be determined expansion coefficients $G_{\vect{k}}(\tVar | \cdot)$. The auxiliary ket is defined as $\ketB{\vect{k}}_{\vect{\q}} \coloneqq \ketB{k_1}_{\q_1} \ketB{k_2}_{\q_2}$ with $\ketB{k_i}_{\q_i} = \q_i^{k_i}$ and is orthogonal to the bra $\braB{\vect{k}} \coloneqq \braB{k_1}\braB{k_2}$ with $\braB{k_i} f \coloneqq \frac{1}{k_i!} \partial_{\q_i}^{k_i} f(\q_i) |_{\q_i=0}$. These bras can be used to extract the expansion coefficient via $G_{\vect{k}}(\tVar | \cdot) = \braketBA{\vect{k}}{\gen(\tVar | \cdot)} $. Differentiation of this coefficient with respect to time and imposing stationarity eventually results in a recurrence relation for $G_{\vect{k}}$ that can be solved iteratively. The steady-state probability distribution of the stochastic process is then recovered from the generating function via the inverse transformation~\eref{eq:GenFct_Flow_InverseTransformation}.
      
      In~\cite{Walczak:2009}, Walczak et al.\ computed the steady-state distribution of the transcriptional regulatory cascade by solving the recurrence relation for $G_{\vect{k}}^s$ numerically. The resulting distribution was compared to distributions acquired via an iterative method and via the stochastic simulation algorithm (SSA) of Gillespie. The spectral method was found to be about $10^8$ times faster than the SSA in achieving the same accuracy. But as we already mentioned in the introductory \sref{subsec:Intro_MasterEqSolution}, the SSA generally performs poorly in the estimation of full distributions, especially in the estimation of their tails. As the spectral method has only been applied to cascades with steady-state copy numbers below $n\approx 30$ thus far, it may be challenged by a direct integration of the two-species master equation (after introducing a reasonable cut-off in the copy numbers). The integration of $\sim\! 1000$ coupled ODEs does not pose a problem for modern integrators and the integration provides the full temporal dynamics of the process (see~\cite{Press:2007} for efficient algorithms). In its current formulation, the spectral method is limited to the evaluation of steady-state distributions and to simple one-step birth-death dynamics. It would be interesting to advance the method for the application to more complex processes (possibly with spatial degrees of freedom), and to extend its scope to the temporal evolution of distributions.

    \subsubsection{WKB approximations and related approaches.} \label{subsubsec:GenFct_Spectral_WKB}
    
      The probability distribution describing the transcriptional regulatory cascade from the previous section approaches a non-trivial stationary shape in the asymptotic time limit $\tVar\to\infty$ (cf.\ figure~2 in~\cite{Walczak:2009}). Often, however, a non-trivial shape of the probability distribution persists only transiently and is said to be quasi-stationary or metastable. The lifetime and shape of such a distribution can often be approximated in terms of a WKB approximation~\cite{Bender:1999}. A WKB approximation of a jump process starts out from an exponential (eikonal) ansatz for the shape of the metastable probability distribution (``real-space'' approach) or for the generating function discussed in the previous sections~ (``momentum-space'' approach). Information on the real-space approach can be found in~\cite{Kubo:1973,Gang:1987,Dykman:1994,Kessler:2007,Meerson:2008,Escudero:2009,Assaf:2010,Mobilia:2010b,Assaf:2010c,Ovaskainen:2010,Meerson:2011,Black:2011,Black:2012b,Bressloff:2014,Bressloff:2014b,Smith:2016}. The recent review of Assaf and Meerson provides an in-depth discussion of the applicability of the real- and momentum-space approaches~\cite{Assaf:2016}.
      
      \begin{figure}[tb] 
        \centering
        \includegraphics{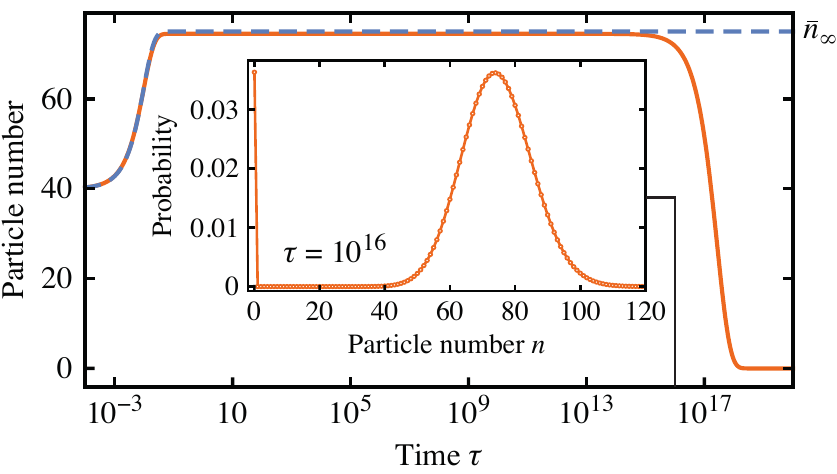}
        \caption{\label{fig:WKBparticleNumber}
          Comparison of the deterministic particle number $\bar{n}(\tVar)$ (blue dashed line) and the mean particle number $\langle n \rangle(\tVar)$ of the stochastic model (orange line) of the combined processes $2\, A \to \emptyset$ and $A \to 2\, A$. The annihilation rate coefficient was set to $\rateCoeffAnnihilation=1$, the growth rate coefficient to $\rateCoeffLinGrowth=150$. The deterministic trajectory started out from $\bar{n}(0)=40$, the numerical integration of the master equation from $\prob(0,n|0,40)=\delta_{n,40}$. The deterministic trajectory converged to $\bar{n}_\infty=75$ for very large times, whereas the mean particle number approached a quasi-stationary value close to $\bar{n}_\infty$ before converging to the absorbing state $n=0$. The inset shows the conditional probability distribution at time $\tVar=10^{16}$. At this time, a significant share of particles was already in the absorbing state and the distribution was bimodal.
        }
      \end{figure} 
    
      In the following, we outline the momentum-space WKB approximation for a system in which particles of type $A$ annihilate in the binary reaction $2\, A \to \emptyset$ with rate coefficient $\rateCoeffAnnihilation$, and are replenished in the linear reaction $A \to 2\, A$ with rate coefficient $\rateCoeffLinGrowth \gg \rateCoeffAnnihilation$. According to a deterministic model of the combined processes with rate equation $\partial_\tVar \bar{n} = \rateCoeffLinGrowth \bar{n} - 2 \rateCoeffAnnihilation \bar{n}^2$, the particle number $\bar{n}$ converges to an asymptotic value $\bar{n}_\infty = \rateCoeffLinGrowth / (2\rateCoeffAnnihilation) \gg 1$. However, a numerical integration of the (truncated) master equation of the stochastic process shows that the mean particle number stays close to $\bar{n}_\infty$ only for a long but finite time (cf.~\fref{fig:WKBparticleNumber}). Asymptotically, all particles become trapped in the ``absorbing'' state $n=0$. Consequently, the conditional probability distribution converges to $\prob(\infty, n | t_0, n_0) = \delta_{n,0}$. Up to a pre-exponential factor, the time after which the absorbing state is reached can be readily estimated using a momentum-space WKB approximation as shown below~\cite{Elgart:2004}. The value of the pre-exponential factor was determined by Turner and Malek-Mansour using a recurrence relation~\cite{Turner:1978}, by Kessler and Shnerb using a real-space WKB approximation~\cite{Kessler:2007}, and by Assaf and Meerson upon combining the generating function technique with Sturm-Liouville theory~\cite{Assaf:2007} (the latter method was developed in~\cite{Assaf:2006a,Assaf:2006b}; see also~\cite{Assaf:2010b}). Using this method, Assaf and Meerson also succeeded in computing the shape of the metastable distribution. Moreover, Assaf et al.\ showed how a momentum-space WKB approximation can be used to determine mean extinction times for processes with time-modulated rate coefficients~\cite{Assaf:2008}.
     
\begin{figure}[tb] 
        \centering
        \includegraphics{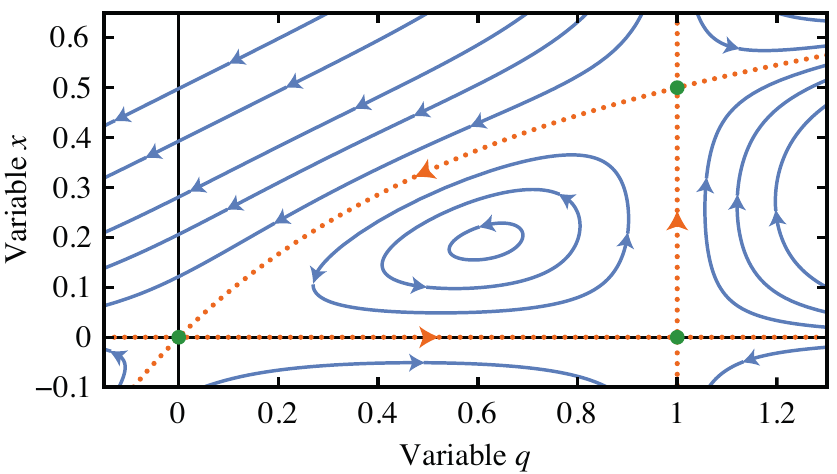}
        \caption{\label{fig:WKBFlow}
          Phase portrait of Hamilton's equations~\eref{eq:GenFct_Spectral_WKB_ansatz_H1} and~\eref{eq:GenFct_Spectral_WKB_ansatz_H2}. The Hamiltonian $H(\q, \x)$ in~\eref{eq:GenFct_Spectral_WKB_Hamiltonian} vanishes along ``zero-energy'' lines (orange dotted lines). These lines connect fixed points of Hamilton's equations (green disks). Hamilton's equations reduce to the (rescaled) rate equation $\partial_\sVar \x =  \x - 2 \x^2$ along the line connecting the fixed points $(1,0)$ and $(1,1/2)$. The zero-energy line connecting the fixed point $(1,1/2)$ to $(0,0)$ denotes the ``optimal path to extinction'' and can be used to approximate the mean extinction time of the process.
        }
      \end{figure}

     Upon rescaling time as $\rateCoeffLinGrowth\,\tVar \to \tVar$, the chemical master equation~\eref{eq:Intro_Mescoscopic_ChemicalMasterEq} of the combined processes $A \to 2\, A$ and $2\, A \to \emptyset$ translates into the flow equation
      \begin{equation}
        \partial_\tVar \ketA{\gen}  \label{eq:GenFct_Spectral_WKB_flow}
        = \Bigl[  (\q^2 - \q) \partial_{\q} +\frac{1}{2\bar{n}_\infty} (1 - \q^2) \partial_{\q}^2 \Bigr]\ketA{\gen} 
      \end{equation}
      of the ordinary generating function $\gen(\tVar; \q | \cdot) = \sum_{n} \q^n \prob(\tVar, n | \cdot)$ (cf.\ sections~\ref{subsec:GenFct_Flow} and~\ref{subsec:GenFct_ForwardBases_ChemReactions}). A WKB approximation of the flow equation can be performed by inserting the ansatz
      \begin{equation}  
        \gen(\tVar; \q | \cdot) = \ee^{S(\tVar,\q)}     \label{eq:GenFct_Spectral_WKB_ansatz}
      \end{equation}
      with the ``action''
      \begin{equation}
        S(\tVar,\q) = 2\, \bar{n}_\infty \sum_{k=0}^\infty \frac{S_k(\tVar,\q)}{\bar{n}_\infty^k}     \label{eq:GenFct_Spectral_WKB_ansatz_series}
      \end{equation}
      into the equation, followed by a successive analysis of terms that are of the same order with respect to the power of the small parameter $1/\bar{n}_\infty$. Note that the exponential ansatz~\eref{eq:GenFct_Spectral_WKB_ansatz} often includes a minus sign in front of the action, which we neglect for convenience. The pre-factor ``$2$'' of the action~\eref{eq:GenFct_Spectral_WKB_ansatz_series} is also included just for convenience. Upon inserting the ansatz~\eref{eq:GenFct_Spectral_WKB_ansatz} into the flow equation~\eref{eq:GenFct_Spectral_WKB_flow}, one obtains the equality
      \begin{equation}
        \partial_\tVar S_0 + \bigO(1/\bar{n}_\infty) 
        = H(\q,\partial_\q S_0) + \bigO(1/\bar{n}_\infty)
      \end{equation}
      with the ``Hamiltonian'' 
      \begin{equation}
        H(\q,\x) \coloneqq (\q^2 - \q) \x +  (1 - \q^2) \x^2  \,.    \label{eq:GenFct_Spectral_WKB_Hamiltonian}
      \end{equation}
      Thus, at leading order of $1/\bar{n}_\infty$, we have obtained a closed equation for $S_0$, which has the form of a Hamilton-Jacobi equation~\cite{Evans:2010}. A Hamilton-Jacobi equation can be solved by the method of characteristics~\cite{Evans:2010}, with the characteristic curves $\q(\sVar)$ and $\x(\sVar)$ obeying Hamilton's equations
      \begin{eqnarray}
        \partial_\sVar \x          \label{eq:GenFct_Spectral_WKB_ansatz_H1}
         = \frac{\partial H(\q, \x)}{\partial \q}           
         =   (2\q - 1) \x - 2\q \x^2
          \quad\mathtext{and} \\
        \partial_{-\sVar} \q       \label{eq:GenFct_Spectral_WKB_ansatz_H2}
        =  \frac{\partial H(\q, \x)}{\partial \x} 
        = (\q^2 - \q) +  2(1 - \q^2) \x   \,.
      \end{eqnarray}
      These equations are, for example, solved by $\q(\sVar)=1$ and $\x(\sVar)$ being a solution of $\partial_\sVar \x =  \x - 2 \x^2$. Note that this equation corresponds to a rescaled rate equation. The Hamiltonian vanishes along the characteristic curve because $H(1,\x)=0$. Further ``zero-energy'' lines of the Hamiltonian are given by $\x=0$ and by $\x(\q)=\frac{\q}{1+\q}$. These lines partition the phase portrait of Hamilton's equations into separate regions as shown in \fref{fig:WKBFlow}. The path $(\q,\frac{\q}{1+\q})$ from the ``active state'' $(\q,\x)=(1,\frac{1}{2})$ to the ``passive state'' $(0,0)$ constitutes the ``optimal path to extinction'' (see below)~\cite{Elgart:2004,Assaf:2008,Schwartz:2011}.
      
      In addition to Hamilton's equations, the method of characteristics implies that the action $S_0$ obeys
      \begin{equation}
        \frac{\diff}{\diff{\sVar}} S_0 = - \bigl( \x\, \partial_{-\sVar} \q - H(\q,\x)\bigr) 
      \end{equation}
      along characteristic curves (the minus signs are only included to emphasize a similarity with an action encountered later in \sref{subsec:StationaryPaths_GenFct}). According to Elgart and Kamenev, the (negative) value of the action~\eref{eq:GenFct_Spectral_WKB_ansatz_series} along the optimal path to extinction determines the logarithm of the mean extinction time $\bar{\tVar}$ from the metastable state in leading order of $\bar{n}_\infty$~\cite{Elgart:2004}. As this path proceeds along a zero-energy line, the action~\eref{eq:GenFct_Spectral_WKB_ansatz_series} evaluates to
      \begin{equation}
        S \approx 2 \bar{n}_\infty S_0 = 2 \bar{n}_\infty \int_{1}^0 \frac{\q \diff{\q}}{1+\q}  
        = -2 \bar{n}_\infty (1-\ln{2})  \,.
      \end{equation}
      Upon returning to the original time scale via $\tVar\to\rateCoeffLinGrowth\,\tVar$, the mean extinction time follows as
      \begin{equation}
      \bar{\tVar} = A\, \rateCoeffLinGrowth^{-1} \ee^{2 \bar{n}_\infty (1-\ln{2} )}  
      \end{equation}
      with pre-exponential factor $A$. The value of this factor has been determined in~\cite{Turner:1978,Kessler:2007,Assaf:2007} and reads, in terms of our rate coefficients, $A=2\sqrt{\pi/\bar{n}_\infty}$. More information on the above procedure can be found in~\cite{Elgart:2004,Assaf:2008}. The method has also been applied to classic epidemiological models~\cite{Schwartz:2011} and to a variant of the Verhulst logistic model~\cite{Assaf:2009} (exact results on this model with added immigration have been obtained in~\cite{Meerson:2013} using the generating function technique).

    \subsubsection{A variational method.}
      
      The generating function's flow equation can also be analysed using a variational method as proposed by Sasai and Wolynes~\cite{Sasai:2003,Lan:2006}. Their approach is based on a method that Eyink had previously developed (primarily) for Fokker-Planck equations~\cite{Eyink:1996}. Instead of dealing with flow equation $\partial_\tVar \ketA{\gen} = \transEvoOp_\tVar \ketA{\gen}$ of the generating function in its differential form (or in terms of its spin matrix representation~\sref{subsec:GenFct_ForwardBases_Exclusion}), the variational method involves a functional variation of the ``effective action'' $\Gamma = \int_{t_0}^t\diff{\tVar}\,\braA{\psi^L} (\partial_\tVar - \transEvoOp_\tVar)\ketA{\psi^R}$ with $\ketA{\psi^R} = \ketA{\gen}$. To perform this variation, one requires an ansatz for the objects $\braA{\psi^L}$ and $\ketA{\psi^R}$, being parametrized by the variables $\{\alpha_i^L\}_{i=1,\hldots,S}$ and $\{\alpha_i^R\}_{i=1,\hldots,S}$. The values of these parameters are determined by requiring that $\braA{\psi^L}$ is an extremum of the action. Besides its application to networks of genetic switches~\cite{Sasai:2003}, the variational method has been applied to signalling in enzymatic cascades in~\cite{Lan:2006}. An extension of the method to multivariate processes is described in~\cite{Ohkubo:2008}. Whether or not the variational method provides useful information mainly depends on making the right ansatz for $\braA{\psi^L}$ and $\ketA{\psi^R}$. The method itself does not suggest their choice. It remains to be seen whether the method can be applied to processes for which only little is known about the generating function.

  \subsection{R\'esum\'e}\label{subsec:GenFct_Resume}        

    In the present section, we formulated general conditions under which the forward master equation~\eref{eq:Intro_Mescoscopic_MasterEq} can be transformed into a linear partial differential equation, a ``flow equation'', obeyed by the probability generating function
    \begin{equation}
      \ketA{\gen(\tVar | t_0, n_0)} 
      = \sum_{n}\ketA{n} \, \prob(\tVar, n | t_0, n_0)  \,.      \label{eq:GenFct_Resume_GenFct}
    \end{equation} 
    First, the conditions~\eref{eq:GenFct_Flow_CompletenessCondition} and~\eref{eq:GenFct_Flow_OrthogonalityCondition} require that there exists a complete and orthogonal basis comprising a set of basis functions $\{\ketA{n}\}$ and a set of basis functionals $\{\braA{n}\}$. The basis functionals recover the conditional probability distribution via $\prob(\tVar, n | t_0, n_0) = \braketA{n}{\gen(\tVar | t_0, n_0)}$ and the right choice of the basis functions may help to obtain a simplified flow equation. We introduced different bases for the study of random walks, of chemical reactions, and of processes with locally excluding particles in \sref{subsec:GenFct_ForwardBases}. Moreover, the conditions~\eref{eq:GenFct_Flow_TCondition} and~\eref{eq:GenFct_Flow_FCondition} require that there exist two differential operators: a basis evolution operator $\evolutionOp_\tVar$ encoding the possible time-dependence of the basis, and a transition operator $\transitionOp_\tVar$ encoding the actual dynamics of the process. The generating function~\eref{eq:GenFct_Resume_GenFct} then obeys the \emph{flow equation}
    \begin{equation}
      \partial_\tVar \ketA{\gen}    \label{eq:GenFct_Resume_Flow}
      = (\evolutionOp_\tVar + \transitionOp_\tVar) \ketA{\gen}  = \transEvoOp_\tVar \ketA{\gen}   \,.
    \end{equation} 
    Various methods have recently been proposed for the study of such a flow equation and were outlined in \sref{subsec:GenFct_Spectral}. These methods include the variational approach of Eyink~\cite{Eyink:1996} and of Sasai and Wolynes~\cite{Sasai:2003}, WKB approximations and spectral formulations of Elgart and Kamenev and of Assaf and Meerson~\cite{Elgart:2004,Assaf:2006a,Assaf:2006b}, and the spectral method of Walczak, Mugler, and Wiggins~\cite{Walczak:2009}. Thus far, most of these methods have only been applied to systems without spatial degrees of freedom and with only one or a few types of particles. Future research is needed to overcome these limitations. Later, in \sref{sec:PathInts_Forward}, we show how the solution of the flow equation~\eref{eq:GenFct_Resume_Flow} can be represented by a path integral. The evaluation of the path integral is demonstrated in computing the generating function of general linear processes. Moreover, we explain in \sref{sec:StationaryPaths} how this path integral connects to a recent method of Elgart and Kamenev for the computation of rare event probabilities~\cite{Elgart:2004}. One can also use the path integral to derive a path integral representation of averaged observables. But a simpler route to this representation starts out from a different flow equation, a flow equation obeyed by the ``marginalized distribution''.
  
\section{The marginalized distribution and the probability generating functional}\label{sec:GenFctnl}  

    In the following, we discuss two further ways of casting the forward and backward master equations into linear PDEs. The first of these flow equations is obeyed by a ``marginalized distribution'' and is easily derived  from the backward master equation~\eref{eq:Intro_Mescoscopic_BackwardMasterEq}. The equation proves useful in the computation of mean extinction times. Moreover, it provides a most direct route to path integral representations of the conditional probability distribution and of averaged observables. To our knowledge, the flow equation of the marginalized distribution has not been considered thus far. In \sref{subsec:GenFctnl_Flow}, we then introduce a probability generating ``functional'' whose flow equation is derived from the forward master equation. The transformation mapping the functional to the conditional probability distribution is shown to generalize the Poisson representation of Gardiner and Chaturvedi~\cite{Gardiner:1977,Chaturvedi:1978}.

  \subsection{Flow of the marginalized distribution}\label{subsec:GenFctnl_MarginalizedDist}  
  
    The generalized probability generating function~\eref{eq:GenFct_Flow_GenFct} was defined by summing the conditional probability distribution $\prob(\tVar, n | t_0, n_0)$ over a set of basis functions, the kets $\{\ketA{n}\}$. In the following, we consider the distribution $\prob(t, n | \tVar, n_0)$ instead, i.e.\ we fix the final time $t$ while keeping the initial time $\tVar$ variable. By summing this distribution over a set of basis functions $\{\ketA{n_0}_\tVar\}$, one can define the series
    \begin{equation} 
      \ketA{\prob(t, n | \tVar)} 
      \coloneqq \sum_{n_0} \prob(t, n|\tVar,n_0) \ketA{n_0}   \,.      \label{eq:GenFctnl_MarginalizedDist_Expansion} 
    \end{equation} 
    As before, the subscript denoting the time-dependence of the basis function is typically dropped. The variables $n$ and $n_0$ again represent states from some countable state space. Assuming that the basis functions $\{\ketA{n}\}$ and appropriately chosen basis functionals $\{\braA{n}\}$ form a \eref{eq:GenFct_Flow_CompletenessCondition}omplete and \eref{eq:GenFct_Flow_OrthogonalityCondition}rthogonal basis, the conditional probability distribution can be recovered from~\eref{eq:GenFctnl_MarginalizedDist_Expansion} via
    \begin{equation} 
      \prob(t, n | \tVar, n_0) =  \braketA{n_0} {\prob(t,n|\tVar)}\,.    \label{eq:GenFctnl_MarginalizedDist_Inverse}
    \end{equation} 

    We call the function $\ketA{\prob(t,n|\tVar)}$ a ``marginalized distribution'' because it proves most useful when the summation in~\eref{eq:GenFctnl_MarginalizedDist_Expansion} constitutes a marginalization of the conditional probability distribution $\prob(t, n|\tVar,n_0)$ over a probability distribution $\ketA{n_0}$. The marginalized distribution then represents a ``single-time distribution'' with respect to the random variable $n$ in the sense of \sref{subsec:Intro_Mesoscopic}. A basis function to which these considerations apply is the ``Poisson basis function'' $\ketA{n_0}_{\xAd} \coloneqq \frac{\xAd^{n_0} \ee^{-\xAd}}{n_0!}$. We make heavy use of this basis function in the study of chemical reactions in \sref{subsec:GenFctnl_BackwardBases_ChemReactions}. Since the definition of the marginalized distribution in~\eref{eq:GenFctnl_MarginalizedDist_Expansion} does not affect the random variable $n$, the marginalized distribution of course solves the forward master equation $\partial_t \ketA{\prob(t, n | \tVar)} = \sum_m \qMatrix_t(n,m) \ketA{\prob(t, m | \tVar)}$ with the initial condition $\ketA{\prob(\tVar, n | \tVar)} = \ketA{n_0}_\tVar$. In the following, we formulate conditions under which $\ketA{\prob}$ also obeys a linear PDE evolving backward in time.
    
    Before proceeding, let us briefly note that if the basis function of the marginalized distribution is not chosen as a probability distribution, the name marginalized ``distribution'' is somewhat of a misnomer; that is, for example, the case for the Fourier basis $\ketA{n_0}_{\xAd} \coloneqq \ee^{\ii n_0 \xAd}$, which we consider in \sref{subsec:GenFctnl_BackwardBases_RandomWalk}.
  
    The derivation of the linear PDE obeyed by the marginalized distribution proceeds analogously to the derivation in \sref{subsec:GenFct_Flow}. But instead of employing the forward master equation, we now employ the backward master equation $\partial_{-\tVar} \prob(t | \tVar) = \prob(t | \tVar) \qMatrix_\tVar$ for this purpose (recall that $\prob(t | \tVar)$ is the matrix with elements $\prob(t, n | \tVar, n_0)$). Upon differentiating the marginalized distribution~\eref{eq:GenFctnl_MarginalizedDist_Expansion} with respect to the time parameter $\tVar$, one obtains the equation
    \begin{eqnarray}
      \partial_{-\tVar} \ketA{\prob}      \label{eq:GenFctnl_MarginalizedDist_Derivation}    \\
      = \sum_{n_0} 
          \prob(\cdot | \tVar, n_0 )
          \Bigl(
          -\partial_\tVar \ketA{n_0}
          +
          \sum_{m} 
            \ketA{m}
            \qMatrix^\transpose_\tVar(m, n_0) 
          \Bigr)    \,.      \nonumber
    \end{eqnarray} 
    The rate $\qMatrix_{\tVar}^\transpose(m,n_0) = \qMatrix_{\tVar}(n_0,m)$ represents an element of the transposed transition matrix $\qMatrix_{\tVar}^\transpose$.
    
    As in \sref{subsec:GenFct_Flow}, two differential operators are required to turn the above expression into a linear PDE. First, we require a basis evolution operator $\evolutionOpAd_\tVar(\xAd,\partial_\xAd)$ fulfilling $\evolutionOp_\tVar \ketA{n_0} =  \partial_\tVar \ketA{n_0}$ for all values of $n_0$. The previous evolution operator~\eref{eq:GenFct_Flow_TCondition} serves this purpose. 
    A second differential operator $\transitionOpAd_\tVar(\xAd,\partial_\xAd)$ is required to encode the information stored in the transition matrix. This operator should be a power series in $\partial_\xAd$ and fulfil, for all $n_0$,
    \renewcommand{\theequation}{\qMatrix$^\transpose$}%
    \begin{equation} 
      \transitionOpAd_\tVar  \ketA{n_0}
      = \sum_{m} \ketA{m} \qMatrix^\transpose_\tVar(m, n_0)  
      \,.
      \label{eq:GenFctnl_Flow_BCondition}
    \end{equation}  
    \renewcommand{\theequation}{\arabic{equation}}%
    By the \eref{eq:GenFct_Flow_CompletenessCondition}ompleteness of the basis, one could also define this operator constructively as
    \begin{equation} 
      \transitionOpAd_\tVar  
      \coloneqq \sum_{m,n_0} \ketA{m} \qMatrix^\transpose_\tVar(m, n_0) \braA{n_0}   \label{eq:GenFctnl_Flow_BCondition_Construct}  \,.  
    \end{equation} 
    We wrote these expressions in terms of the transposed transition matrix because for the unit column vectors $\ketA{m} = \hat{\vect{e}}_m$ and the unit row vectors $\braA{n} = \hat{\vect{e}}_n^\transpose$, $\transitionOpAd_\tVar$ and $\qMatrix^\transpose_\tVar$ coincide (with $m,n\in\naturals_0$). As we mostly consider bases whose kets and bras represent functions and functionals, we call $\transitionOpAd_\tVar$ the ``adjoint'' transition operator in the following. Often, an operator $O^\dagger(\xAd,\partial_\xAd)$ is said to be the adjoint of an operator $O(\xAd,\partial_\xAd)$ if the following relation holds with respect to two test functions $f$ and $g$:
    \begin{eqnarray}
      \int \diff{\xAd} \, \bigl[O^\dagger(\xAd,\partial_\xAd) f(\xAd)\bigr] g(\xAd)       \label{eq:GenFctnl_MarginalizedDist_Adjoint}  \\
      =\int \diff{\xAd} \, f(\xAd) \bigl[O(\xAd,\partial_\xAd) g(\xAd)\bigr]   \nonumber  \,.    
    \end{eqnarray}
    However, whether an operator $\transitionOp_\tVar$ complementing the above $\transitionOpAd_\tVar$ actually exists will not be important in the following (except in our discussion of the Poisson representation in \sref{subsec:GenFctnl_PoissonRep}).
    
    Provided that both a basis evolution operator $\evolutionOpAd_\tVar$ and an adjoint transition operator $\transitionOpAd_\tVar$ are found for a particular process, it follows from the backward-time equation~\eref{eq:GenFctnl_MarginalizedDist_Derivation} that the marginalized distribution $\ketA{\prob(t, n | \tVar)}$ obeys the flow equation\footnote{
      In the derivation of the backward path integral in~\sref{subsec:PathInts_Backward_Derivation}, we use the finite difference approximation
      \begin{eqnarray*}
        \lim_{\Delta t\to 0} (\ketA{\prob(\cdot | \tVar-\Delta t)} - \ketA{\prob(\cdot | \tVar)})/\Delta t
        = \lim_{\Delta t\to 0} \transEvoOpAd_{\tVar-\Delta t,\Delta t}\ketA{\prob(\cdot | \tVar)}  \,.
      \end{eqnarray*} 
      The discretization scheme conforms with the derivation of the backward master equation~\eref{eq:Intro_Mescoscopic_BackwardMasterEq}.%
      \label{ftn:Bwd_Discretization}}
    \begin{equation}
      \partial_{-\tVar} \ketA{\prob}         \label{eq:GenFctnl_MarginalizedDist_Flow}
      = (-\evolutionOpAd_\tVar + \transitionOpAd_\tVar)\ketA{\prob}     
      \eqqcolon \transEvoOpAd_\tVar\ketA{\prob}     \,.
    \end{equation} 
    The evolution of this equation proceeds backward in time, starting out from the final condition $\ketA{\prob(t, n | t)} = \ketA{n}$. 
    
    In section~\ref{sec:PathInts_Backward}, we show how the flow equation~\eref{eq:GenFctnl_MarginalizedDist_Flow} can be solved in terms of a ``backward'' path integral. Provided that the basis function $\ketA{n}$ is chosen as a probability distribution, this path integral represents a true probability distribution: the marginalized distribution. The fact that the backward path integral represents a probability distribution distinguishes it from the ``forward'' path integral in section~\ref{sec:PathInts_Forward}. The forward path integral represents the probability generating function~\eref{eq:GenFct_Flow_GenFct}. Both the forward path integral and the backward path integral can be used to derive a path integral representation of averaged observables, as we show in section~\ref{sec:PathInts_Observables}. The derivation of this representation from the backward path integral, however, is significantly easier. In fact, the path integral representation of the average of an observable $A$ will follow directly by summing the backward path integral representation of the marginalized distribution over $A(n)$, i.e.\ via $\langle A \rangle = \sum_{n} A(n) \ketA{\prob(t, n | \tVar)}$. Note that this average also obeys the flow equation~\eref{eq:GenFctnl_MarginalizedDist_Flow}. In section~\ref{subsec:GenFctnl_ExtTimes}, we demonstrate how the flow equation can be used to compute mean extinction times.
    
    Before introducing bases for the analysis of different stochastic processes, let us briefly note that if a process is homogeneous in time, its marginalized distribution depends only on the difference $t-\tVar$. The above flow equation can then be rewritten so that it evolves $\ketA{\prob(\tVar, n | 0)}$ forward in time $\tVar$, starting out from the initial condition $\ketA{\prob(0, n | 0)} = \ketA{n}$. In \sref{subsec:GenFctnl_ExtTimes}, we make use of this property to compute mean extinction times.

  \subsection{Bases for particular stochastic processes}\label{subsec:GenFctnl_BackwardBases}  
  
    To demonstrate the application of the marginalized distribution~\eref{eq:GenFctnl_MarginalizedDist_Expansion}, let us reconsider the random walk from \sref{subsec:GenFct_ForwardBases_RandomWalk} and the chemical reaction from \sref{subsec:GenFct_ForwardBases_ChemReactions}. The ``Poisson basis function'' introduced in the latter section will be employed in the computation of mean extinction times in \sref{subsec:GenFctnl_ExtTimes}. Moreover, it will allow us to recover the Poisson representation of  Gardiner and Chaturvedi~\cite{Gardiner:1977,Chaturvedi:1978} in \sref{subsec:GenFctnl_PoissonRep}.
    
    \subsubsection{Random walks.}\label{subsec:GenFctnl_BackwardBases_RandomWalk} 
  
      The following solution of the random walk largely parallels the previous derivation in \sref{subsec:GenFct_ForwardBases_RandomWalk}. In particular, we again use the orthogonal and complete Fourier basis $\ketA{n}_\xAd = \ee^{\ii n \xAd}$ and $\braA{n} f = \int_{-\pi}^\pi \frac{\diff{\xAd}}{2\pi} \ee^{-\ii n \xAd} f(\xAd)$ with $n\in\integers$. Due to the time-independence of the basis, the \eref{eq:GenFct_Flow_TCondition}volution operator $\evolutionOpAd_\tVar$ is zero. The condition~\eref{eq:GenFctnl_Flow_BCondition} on the adjoint transition operator $\transitionOpAd_\tVar$ is specified by the transition matrix~\eref{eq:GenFct_ForwardBases_RandomWalk_TransitionMatrix} of the process and reads
      \begin{equation}
        \transitionOpAd_\tVar \ketA{n_0}  
        =   r_\tVar (\ketA{n_0-1} - \ketA{n_0}) + l_\tVar (\ketA{n_0+1} - \ketA{n_0})  \,. \label{eq:GenFctnl_BackwardBases_RandomWalk_Adjoint}
      \end{equation} 
      As before, we employ the operators $\cre(\xAd) = \ee^{\ii\xAd}$ and $\ann(\xAd) = \ee^{-\ii\xAd}$, which act on the basis functions as $\cre\ketA{n} = \ketA{n+1}$ and $\ann \ketA{n} = \ketA{n-1}$, respectively. Therefore, the adjoint transition operator with the above property can be defined as
      \begin{equation}
        \transitionOpAd_\tVar(\xAd,\partial_\xAd)   
        \coloneqq r_\tVar \bigl(\ann(\xAd) - 1) + l_\tVar (\cre(\xAd) - 1\bigr)    \,.
      \end{equation}
      As this operator does not contain any derivatives, it is self-adjoint in the sense of~\eref{eq:GenFctnl_MarginalizedDist_Adjoint}. Due to a mismatch in signs, it is, however, not the adjoint of the previous operator $\transitionOp_\tVar$ in~\eref{eq:GenFct_ForwardBases_RandomWalk_FOperator}. This mismatch could be corrected by redefining the above basis function as $\ketA{n}_\xAd \coloneqq \ee^{-\ii n \xAd}$. Ignoring this circumstance, the flow equation of the marginalized distribution follows as
      \begin{equation}
        \partial_{-\tVar} \ketA{\prob}    \label{eq:GenFctnl_BackwardBases_RandomWalk_FlowEq}
        = \bigl[r_\tVar (\ee^{-\ii\xAd} - 1) + l_\tVar (\ee^{\ii\xAd} - 1)\bigr] \ketA{\prob}     \,,
      \end{equation}
      and is solved by
      \begin{equation}
        \ketA{\prob} 
        = \exp\Bigl( (\ee^{-\ii\xAd} - 1) \int_{\tVar}^{t}\diff{\sVar}\, r_\sVar+   (\ee^{\ii\xAd} - 1) \int_{\tVar}^{t}\diff{\sVar}\,l_\sVar  \Bigr)   \ketA{n}    \,.
      \end{equation}
      The conditional probability distribution is recovered via the inverse Fourier transformation $\prob(t, n | \tVar, n_0) =  \braketA{n_0}{\prob}$. Upon inserting the explicit representation of the basis, the derivation proceeds as in \aref{sec:A_RandomWalk} (with the substitutions $\xAd \to -\q$, $\tVar\to t_0$ and $t\to\tVar$). Eventually, one recovers a Skellam distribution as the solution of the process.

    \subsubsection{Chemical reactions.}\label{subsec:GenFctnl_BackwardBases_ChemReactions} 

      To prepare the computation of mean extinction times in the next section as well as the derivation of the Poisson representation in \sref{subsec:GenFctnl_PoissonRep}, we now reconsider processes that can be decomposed additively into chemical reactions of the form $k\, A \to l\, A$. Our later derivation of a path integral representation of averaged observables is also restricted to such processes. The state space of the number of molecules is again $\naturals_0$. Analogous to \sref{subsec:GenFct_ForwardBases_ChemReactions}, we require an \eref{eq:GenFct_Flow_OrthogonalityCondition}rthogonal and \eref{eq:GenFct_Flow_CompletenessCondition}omplete basis, as well as an \eref{eq:GenFct_Flow_TCondition}volution operator $\evolutionOpAd_\tVar$ and an adjoint transition operator $\transitionOpAd_\tVar$ (condition~\eref{eq:GenFctnl_Flow_BCondition}) to specify the flow equation of the marginalized distribution.
      
      As discussed in \sref{subsec:GenFct_ForwardBases_ChemReactions}, the elements of the transition rate matrix of the reaction $k\, A \to l\, A$ with rate coefficient $\rateCoeffGeneric_\tVar$ are given by $\qMatrix_\tVar(m, n) = \rateCoeffGeneric_\tVar  (n)_k (\delta_{m, n-k+l} - \delta_{m, n})$. The condition~\eref{eq:GenFctnl_Flow_BCondition} on the adjoint transition operator therefore reads
      \begin{equation}
        \transitionOpAd_\tVar\ketA{n_0} 
        =  \rateCoeffGeneric_\tVar  \bigl((n_0-l+k)_k \ketA{n_0-l+k} - (n_0)_k \ketA{n_0}\bigr)    \,. \label{eq:GenFctnl_Flow_Poisson_BCondition}
      \end{equation}
      This condition is met by
      \begin{equation}
         \transitionOpAd_\tVar(\cre, \ann)   \label{eq:GenFctnl_Flow_Poisson_HamiltonianB}
         \coloneqq \rateCoeffGeneric_\tVar \cre^k (\ann^l - \ann^k) \,,
      \end{equation}
      provided that there exist operators $\cre$ and $\ann$ fulfilling
      \begin{eqnarray}
        \cre \ketA{n}  = (n+1) \ketA{n+1} \mathtext{ and}    \label{eq:GenFctnl_Flow_Poisson_CreationAnnihilation_CreAction} \\
        \ann \ketA{n} = \ketA{n-1}  \label{eq:GenFctnl_Flow_Poisson_CreationAnnihilation_AnnAction}   \,,
      \end{eqnarray}
      respectively. We again call $\cre$ the creation and $\ann$ the annihilation operator, even though the pre-factors in the above relations differ from the ones in~\eref{eq:GenFct_Flow_Poisson_CreationAnnihilation_CreAction} and~\eref{eq:GenFct_Flow_Poisson_CreationAnnihilation_AnnAction}. Similar relations also hold with respect to the basis functionals, namely $\braA{n} \cre = n \braA{n-1}$ and $\braA{n} \ann  = \braA{n+1}$ (assuming the orthogonality of the basis). The operators also fulfil the commutation relation $[\ann, \cre ] = 1$.
            
      The actions of the creation and annihilation operators on the basis functions hint at how the basis functions and functionals from \sref{subsec:GenFct_ForwardBases_ChemReactions} can be adapted to meet the present requirements. In particular, the relations~\eref{eq:GenFctnl_Flow_Poisson_CreationAnnihilation_CreAction} and~\eref{eq:GenFctnl_Flow_Poisson_CreationAnnihilation_AnnAction} can be fulfilled by moving the factorial from the basis functional~\eref{eq:GenFct_Flow_Poisson_BaseFunctional} to the basis function~\eref{eq:GenFct_Flow_Poisson_BaseFunction} so that
      \begin{eqnarray}
        \ketA{n}_{\xAd}       \label{eq:GenFctnl_Flow_Poisson_BaseFunction}
          \coloneqq \frac{(\basisPrefactor\xAd + \txAd)^n \ee^{-\tqAd (\basisPrefactor\xAd + \txAd)}}{n!}   \mathtext{ and}\\
        \braA{n} f     \label{eq:GenFctnl_Flow_Poisson_BaseFunctional}
          \coloneqq \ann^n f(\xAd) \big|_{\xAd = -\txAd/\basisPrefactor}    \,.
      \end{eqnarray}
      Let us briefly note that we changed the argument of the basis function from $\q$ to $\xAd$ as compared to \sref{subsec:GenFct_ForwardBases_ChemReactions} because both the generating function approach and the marginalized distribution approach thereby result in the same path integral representation of averaged observables (cf. \sref{sec:PathInts_Observables}). Apart from this notational change, the basis evolution operator%
      \begin{equation}
        \evolutionOpAd_\tVar(\cre, \ann)   \label{eq:GenFctnl_Flow_Poisson_HamiltonianT}
          \coloneqq  (\partial_\tVar \txAd) (\ann - \tqAd)   - (\partial_\tVar \tqAd) \cre    
      \end{equation}
      keeps the form it had in~\eref{eq:GenFct_Flow_Poisson_HamiltonianT}. The creation and annihilation operators also keep their previous forms in~\eref{eq:GenFct_Flow_Poisson_Creation} and~\eref{eq:GenFct_Flow_Poisson_Annihilation}, i.e.
      \begin{eqnarray}
        \cre(\xAd, \partial_\xAd)  \label{eq:GenFctnl_Flow_Poisson_Creation}  
          \coloneqq \basisPrefactor\xAd + \txAd \,\mathtext{ and }\\
        \ann(\xAd, \partial_\xAd)  \label{eq:GenFctnl_Flow_Poisson_Annihilation}  
          \coloneqq \partial_{\basisPrefactor\xAd} + \tqAd \,.  
      \end{eqnarray}
      Despite their similar appearance, the operator $\transitionOpAd_\tVar$ in~\eref{eq:GenFctnl_Flow_Poisson_HamiltonianB} is not the adjoint of the operator $\transitionOp_\tVar$ in~\eref{eq:GenFct_Flow_Poisson_HamiltonianF}. Nevertheless, the two operators fulfil $\transitionOp_\tVar(\q, \x) = \transitionOpAd_\tVar(\x, \q)$ for scalar arguments. This relation is essentially the reason why we interchanged the letters $\xAd$ and $\qAd$ as compared to \sref{subsec:GenFct_ForwardBases_ChemReactions}. Both the generating function approach and the marginalized distribution approach thereby lead to the same path integral representation of averaged observables, as will be shown in \sref{subsec:PathInts_Observables_Derivation}.
      
      The choice of the parameters $\basisPrefactor\neq 0$, $\txAd(\tVar)$, and $\tqAd(\tVar)$ in the basis function~\eref{eq:GenFctnl_Flow_Poisson_BaseFunction} depends on the problem at hand. In \sref{sec:StationaryPaths}, $\txAd$ and $\tqAd$ will act as ``stationary'' or ``extremal'' paths, with $\xAd$ and an auxiliary variable $\qAd$ being deviations from them. For $\basisPrefactor\coloneqq\tqAd\coloneqq1$ and $\txAd\coloneqq 0$, the basis function instead simplifies to the Poisson distribution 
      \begin{equation}
        \ketA{n}_{\xAd}       \label{eq:GenFctnl_Flow_Poisson_BaseFunction_Poisson}
          = \frac{\xAd^n \ee^{-\xAd}}{n!}   \,.
      \end{equation}
      In the following, we make heavy use of this ``Poisson basis function''. It will play a crucial role in the formulation of a path integral representation of averaged observables in \sref{sec:PathInts_Observables} and in recovering the Poisson representation in \sref{subsec:GenFctnl_PoissonRep}.

      Let us demonstrate the use of the Poisson basis function for the linear decay process $A \to \emptyset$ with rate coefficient $\rateCoeffLinDecay_\tVar$. For the above choice of $\basisPrefactor$, $\txAd$, and $\tqAd$, the creation operator~\eref{eq:GenFctnl_Flow_Poisson_Creation} reads $\cre = \xAd$ and the annihilation operator~\eref{eq:GenFctnl_Flow_Poisson_Annihilation} reads $\ann = \partial_{\xAd} + 1$. With the transition operator~\eref{eq:GenFctnl_Flow_Poisson_HamiltonianB}, the flow equation~\eref{eq:GenFctnl_MarginalizedDist_Flow} obeyed by the marginalized distribution $\ketA{\prob(t, n | \tVar)}_\xAd$ follows as
      \begin{equation}
        \partial_{-\tVar} \ketA{\prob}   \label{eq:GenFctnl_MarginalizedDist_Flow_Decay}
        = -\rateCoeffLinDecay_\tVar \xAd\partial_{\xAd} \ketA{\prob}     \,.
      \end{equation} 
      This equation is solved by the Poisson distribution
      \begin{equation}
        \ketA{\prob(t, n|\tVar)}_{\xAd}   \label{eq:GenFctnl_MarginalizedDist_Solution}
        = \frac{(\alpha_{t,\tVar}\xAd)^n \ee^{-\alpha_{t,\tVar}\xAd}}{n!}
      \end{equation} 
      whose mean $\alpha_{t,\tVar}\, \xAd$ decays proportionally to $\alpha_{t,\tVar} \coloneqq \ee^{-\int_{\tVar}^t \diff{\sVar}\, \rateCoeffLinDecay_\sVar}$. To interpret this solution, let us recall that the sum in the definition of the marginalized distribution~\eref{eq:GenFctnl_MarginalizedDist_Expansion} does not affect the particle number $n$. Therefore, upon relabelling the time parameters, the above solution~\eref{eq:GenFctnl_MarginalizedDist_Solution} also solves the forward master equation of the process, namely
      \begin{eqnarray}
        \partial_\tVar \ketA{\prob(\tVar, n | t_0)}_{\xAd} \\
        = \rateCoeffLinDecay_\tVar \bigl((n+1)\ketA{\prob(\tVar, n+1 | t_0)}_{\xAd} - n\ketA{\prob(\tVar, n | t_0)}_{\xAd} \bigr)    \nonumber    \,.
      \end{eqnarray}
      Unlike the conditional distribution $\prob(\tVar, n | t_0, n_0)$, however, the marginalized distribution $\ketA{\prob(\tVar, n | t_0)}_\x$ describes the dynamics of a population whose particle number at time $t_0$ is Poisson distributed with mean $\x$. The conditional distribution is recovered from it via the inverse transformation $\prob(\tVar, n | t_0, n_0) =  \braketA{n_0}{\prob(\tVar,n|t_0)}$ with $\braA{n_0} f = (\partial_{\xAd} + 1)^{n_0} f(\xAd)|_{\xAd = 0}$. This transformation results in the Binomial distribution
      \begin{equation}
        \prob(\tVar, n | t_0, n_0)          \label{eq:PathInts_SimpleGrowthLinearDecay_BinomialDist}
          =  \binomCoeff{n_0}{n} \bigl(\alpha_{\tVar,t_0}\bigr)^n \bigl(1- \alpha_{\tVar,t_0} \bigr)^{n_0-n}      \,.  
      \end{equation} 
      Both the mean value $\ee^{-\int_{t_0}^{\tVar} \diff{\sVar}\, \rateCoeffLinDecay_\sVar} n_0$ and the variance $(1-\ee^{-\int_{t_0}^{\tVar} \diff{\sVar}\, \rateCoeffLinDecay_\sVar})\, \ee^{-\int_{t_0}^{\tVar} \diff{\sVar}\, \rateCoeffLinDecay_\sVar} n_0$ of this distribution decay exponentially for large times, provided that $\rateCoeffLinDecay_{\sVar}>0$.

      If at most two reactants and two products are involved in a reaction, the flow equation~\eref{eq:GenFctnl_MarginalizedDist_Flow} of the marginalized distribution has the mathematical form of a backward Fokker-Planck equation. That is, for example, the case for the coagulation reaction $2\, A \to A$. For the Poisson basis function, the marginalized distribution $\ketA{\prob(t, n | \tVar)}_{\xAd}$ of this process obeys the flow equation
      \begin{equation}
        \partial_{-\tVar} \ketA{\prob} 
        = \alpha_\tVar(\xAd) \partial_{\xAd} \ketA{\prob}+ \frac{1}{2}\beta_\tVar(\xAd)\partial_{\xAd}^2 \ketA{\prob}    \,,     \label{eq:GenFctnl_MarginalizedDist_Fokker}
      \end{equation} 
      with the drift coefficient $\alpha_\tVar(\xAd) \coloneqq -\rateCoeffLinDecay_\tVar \xAd^2$ and the diffusion coefficient $\beta_\tVar(\xAd) \coloneqq -2\rateCoeffLinDecay_\tVar \xAd^2$. Unlike the diffusion coefficient of the ``true'' backward Fokker-Planck equation~\eref{eq:Intro_Mescoscopic_FP_BW}, this diffusion coefficient may be negative (e.g.\ for $\xAd\in\reals{\setminus}\{0\}$). The final condition $\ketA{\prob(t, n | t)}_\xAd = \frac{\xAd^n \ee^{-\xAd}}{n!}$ and the Feynman-Kac formula~\eref{eq:A_FeynmanKacProof_FK} imply that the above flow equation is solved by
      \begin{equation}
        \ketA{\prob(t, n | \tVar)}_\xAd  \label{eq:GenFctnl_BackwardBases_ChemReactions_WienerAverage}
        = \BigLLangle \frac{\xAd(t)^n \ee^{-\xAd(t)}}{n!}  \BigRRangle_\wienerProcess    \,.
      \end{equation}
      Here, $\xAd(\sVar)$ obeys the It\^{o} SDE
      \begin{equation}
        \diff{\xAd(\sVar)}   
        = \alpha_\sVar(\xAd(\sVar)) \diff{\sVar}  + \sqrt{\beta_\sVar(\xAd(\sVar))}\diff{\wienerProcess}(\sVar)      \,,
      \end{equation} 
      which evolves $\xAd(\sVar)$ from $\xAd(\tVar)=\xAd$ to $\xAd(t)$. The symbol $\LLangle \cdot \RRangle_{\wienerProcess}$ represents an average over realizations of the Wiener process $\wienerProcess$. Since the diffusion coefficient $\beta_\tVar(\xAd) = -2\rateCoeffLinDecay_\tVar \xAd^2$ of the coagulation process can take on negative values, the sample paths of the SDE may acquire imaginary components. This circumstance does not prevent the use of~\eref{eq:GenFctnl_BackwardBases_ChemReactions_WienerAverage} for the calculation of $\ketA{\prob(t, n | \tVar)}$, although it may complicate numerical evaluations. Recently, Wiese attempted the evaluation of the average in~\eref{eq:GenFctnl_BackwardBases_ChemReactions_WienerAverage} via the generation of sample paths~\cite{Wiese:2015}. Over a short time interval $[\tVar, t]$, he found a good agreement between the resulting distribution and a distribution effectively acquired via the stochastic simulation algorithm (SSA) in \sref{subsec:Intro_Mesoscopic}. Over larger time intervals, however, the integration of the SDE encountered problems regarding its numerical convergence. Future research is needed to overcome this limitation. Moreover, it remains an open challenge to specify the boundary conditions of the PDE~\eref{eq:GenFctnl_MarginalizedDist_Fokker} to enable its direct numerical integration (an analogous problem is encountered for the flow equation of the generating function, cf.~\cite{Assaf:2007}).

  \subsection{Mean extinction times}\label{subsec:GenFctnl_ExtTimes}  
    
    One often wishes to know the mean time at which a process first hits some target in state space. Such a target could, for example, be a state in which no more particles are left in the system. If the particles only replenish through auto-catalysis, the process will then come to a halt. The mean time after which that happens is called the mean extinction time. For Markov processes with continuous sample paths, mean extinction times and, more generally, first-passage times, are commonly calculated with the help of the backward Fokker-Planck equation~\eref{eq:Intro_Mescoscopic_FP_BW}. For jump processes, one can use the backward master equation for this purpose. The calculation, however, is typically feasible only for one-step processes and involves the solution of a recurrence relation~\cite{Doering:2005,Gardiner:2009}. In~\cite{Drummond:2010}, Drummond et al.\ recently showed how the calculation can be simplified using the Poisson representation of Gardiner and Chaturvedi~\cite{Gardiner:1977,Chaturvedi:1978}. We introduce this representation in \sref{subsec:GenFctnl_PoissonRep}. In the following, we outline how mean extinction times can instead be inferred in an analogous way using the marginalized distribution from the previous sections. For the Poisson basis function $\ketA{n_0}_{\xAd} = \frac{\xAd^{n_0} \ee^{-\xAd}}{n_0!}$, the marginalized distribution is a true (single-time) probability distribution and is more easily interpreted than the integral kernel of the Poisson representation.
    
    We again consider a process that can be decomposed additively into chemical reactions of the form $k\, A \to l\, A$. In addition, the transition matrix $\transitionOpAd_\tVar=\transitionOpAd$ of the process shall now be time-independent and the particles in the system shall be Poisson distributed with mean $\xAd$ at time $\tVar=0$. Consequently, the probability of finding $n$ particles in the system at time $\tVar \geq 0$ is described by the marginalized distribution $\ketA{\prob(\tVar, n | 0)}_\xAd$, provided that its basis function is chosen as $\ketA{n_0}_{\xAd} \coloneqq \frac{\xAd^{n_0} \ee^{-\xAd}}{n_0!}$ (cf.\ the definition~\eref{eq:GenFctnl_MarginalizedDist_Expansion} of the marginalized distribution). Since the Poisson basis function is independent of time and the process under consideration homogeneous in time, the marginalized distribution $\ketA{\prob(\tVar, n | 0)}_\xAd$ obeys the forward-time flow equation (cf.~\eref{eq:GenFctnl_MarginalizedDist_Flow})
    \begin{equation}
      \partial_{\tVar} \ketA{\prob}       \label{eq:GenFctnl_ExtTimes_Flow}
      = \transitionOpAd \ketA{\prob}       \,.
    \end{equation} 

    We define the probability of finding the system in an ``active'' state with $n>0$ particles at time $\tVar \geq 0$ as
    \begin{equation}
      \alpha(\tVar; \xAd) \coloneqq \sum_{n=1}^\infty \ketA{\prob(\tVar, n | 0)}_\xAd 
      = 1 -  \ketA{\prob(\tVar, 0 | 0)}_\xAd   
      \,.  \label{eqc:GenFctnl_ExtTimes_ExtTime_Dev}
    \end{equation}
     Over time, this probability flows into the ``absorbing'' state $n=0$ at rate $f(\tVar,\xAd) \coloneqq - \partial_\tVar \alpha(\tVar; \xAd)$. Since $f(\tVar,\xAd) \diff{\tVar}$ is the probability of becoming absorbed during the time interval $[\tVar, \tVar + \Delta t]$, one can define the mean extinction time as $\langle \tVar \rangle_\xAd \coloneqq \int_{0}^\infty \diff{\tVar}\, \tVar f(\tVar,\xAd)$. An integration by parts transforms this average into
    \begin{equation}
      \langle \tVar \rangle_\xAd     \label{eqc:GenFctnl_ExtTimes_ExtTime1}
      = - \int_{0}^\infty \diff{\tVar}\, \tVar \partial_\tVar \alpha(\tVar;\xAd)
      = \int_0^\infty \diff{\tVar}\, \alpha(\tVar; \xAd)   \,.
    \end{equation}
    The boundary terms of the integration by parts vanished because we assume that all particles are eventually absorbed (in particular, we assume $\lim_{\tVar\to\infty} \tVar \alpha(\tVar; \xAd) = 0$). The definition of the probability $\alpha(\tVar; \xAd)$ in~\eref{eqc:GenFctnl_ExtTimes_ExtTime_Dev} implies that it fulfils the same forward-time flow equation as the marginalized distribution $\ketA{\prob(\tVar, n | 0)}_\xAd$. Since $\lim_{\tVar\to\infty} \alpha(\tVar; \xAd) = 0$, the mean extinction time~\eref{eqc:GenFctnl_ExtTimes_ExtTime1} therefore obeys
    \begin{equation}
      - \transitionOpAd(\xAd, \partial_\xAd) \langle \tVar \rangle_\xAd   \label{eqc:GenFctnl_ExtTimes_ExtTime2}
      = \alpha(0;\xAd) = 1 - \ee^{-\xAd}  \,.
    \end{equation}
    The last equality follows from the fact that a certain fraction of all particles, namely $\ee^{-\xAd}$, has already been in the absorbing state initially. The above derivation readily extends to higher moments of the mean extinction time. 
    
    \begin{figure}[tb] 
      \centering
      \includegraphics{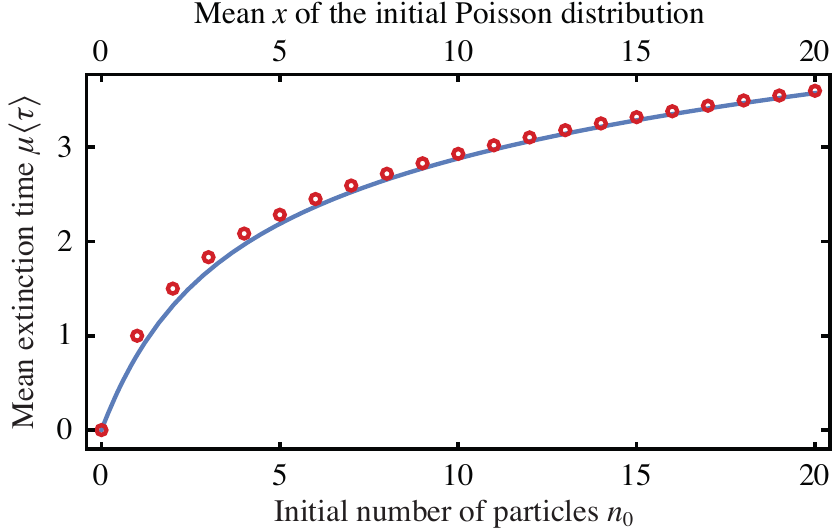}
      \caption{\label{fig:MeanExtinctionTime}
        Mean extinction time $\rateCoeffLinDecay \langle \tVar \rangle$ of the linear decay process $A \to \emptyset$ with decay rate coefficient $\rateCoeffLinDecay$. The blue line represents the mean extinction time if the number of particles is initially Poisson distributed with mean $\xAd\in\reals_{\geq 0}$ ($\rateCoeffLinDecay \langle \tVar \rangle_\xAd = \ln{\xAd} + \gamma - \mathtext{Ei}(-\xAd)$). The red circles represent the mean extinction time if the initial number of particles is set to $n_0\in\naturals_0$ ($\rateCoeffLinDecay \langle \tVar \rangle_{n_0} = H_{n_0}$).
      }
    \end{figure}    
    
    For the linear decay process $A \to \emptyset$ with decay rate coefficient $\rateCoeffLinDecay$, the equation~\eref{eqc:GenFctnl_ExtTimes_ExtTime2} for the mean extinction time reads
    \begin{equation}
      \rateCoeffLinDecay \partial_\xAd \langle \tVar \rangle_\xAd = (1 - \ee^{-\xAd})/\xAd  \,.
    \end{equation}
    This equation implies that the mean extinction time $\rateCoeffLinDecay \langle \tVar \rangle_\xAd$ increases logarithmically with the particles' mean initial distance $\xAd$ from the absorbing state (see \fref{fig:MeanExtinctionTime}). The explicit solution of the equation is $\rateCoeffLinDecay \langle \tVar \rangle_\xAd = \ln{\xAd} + \gamma - \mathtext{Ei}(-\xAd)$ with Euler's constant $\gamma$ and the exponential integral $\mathtext{Ei}$ (6.2.6 in~\cite{NIST:2010}; the exponential integral ensures that $\langle \tVar \rangle_0 = 0$). The application of the functional $\braA{n_0} f = (\partial_\xAd+1)^{n_0} f(\xAd) \big|_{\xAd =0}$ to the solution returns the mean extinction time of particles whose initial number is not Poisson distributed but that is fixed to some value $n_0 \geq 0$. The corresponding mean extinction time $\rateCoeffLinDecay \langle \tVar \rangle_{n_0}$ is given by the harmonic number $H_{n_0} \coloneqq \sum_{i=1}^{n_0} \frac{1}{i}$. One can also infer this result directly from the backward master equation using the methods discussed in~\cite{Gardiner:2009,Doering:2005}.
    
    Drummond et al.\ have extended the above computation to chemical reactions with at most two reactants and two products~\cite{Drummond:2010}. For these reactions, the flow equation~\eref{eq:GenFctnl_ExtTimes_Flow} has the mathematical form of a (forward) Fokker-Planck equation.

  \subsection{Flow of the generating functional}\label{subsec:GenFctnl_Flow}  
    
    Both the probability generating function~\eref{eq:GenFct_Flow_GenFct} and the marginalized distribution~\eref{eq:GenFctnl_MarginalizedDist_Expansion} were defined by summing the conditional probability distribution over a set of basis functions, either at the final or at the initial time. In addition, one can define the ``probability generating functional''
    \begin{equation}
      \braA{\gen(\tVar | t_0, n_0)}  \label{eq:GenFctnl_Flow_GenFctnl}
      \coloneqq \sum_{n} \braA{n} \prob(\tVar, n | t_0, n_0)     
    \end{equation} 
    by summing the conditional distribution over the set $\{\braA{n}\}$ of basis functionals. The definition of the generating functional is meant with respect to test functions that can be expanded in the basis functions $\{\ketA{n}\}$. As before, we assume that the basis functions and functionals constitute a \eref{eq:GenFct_Flow_CompletenessCondition}omplete and \eref{eq:GenFct_Flow_OrthogonalityCondition}rthogonal basis. The probability generating functional ``generates'' probabilities in the sense that
    \begin{equation} 
      \prob(\tVar, n | t_0, n_0) =  \braketA{\gen(\tVar|t_0,n_0)}{n}                   \label{eq:GenFctnl_Flow_InverseTransformation} \,.
    \end{equation} 
    In the next section, we show how this inverse transformation reduces to the Poisson representation of Gardiner and Chaturvedi~\cite{Gardiner:1977,Chaturvedi:1978} upon choosing the basis function $\ketA{n}$ as a Poisson distribution.
    
    The derivation of the (functional) flow equation of $\braA{\gen}$ proceeds analogously to the derivations in sections~\ref{subsec:GenFct_Flow} and~\ref{subsec:GenFctnl_MarginalizedDist}. Differentiation of its definition~\eref{eq:GenFctnl_Flow_GenFctnl} with respect to $\tVar$ and use of the forward master equation $\partial_\tVar \prob(\tVar | t_0) = \qMatrix_\tVar \prob(\tVar | t_0)$ result in
    \begin{equation}
      \partial_\tVar \braA{\gen}  \label{eq:GenFctnl_Flow_GenFctnl_Derivation}
      =  \sum_{n} 
          \Bigl(\partial_\tVar \braA{n} + \sum_{m} \qMatrix_\tVar^\transpose(n,m) \braA{m}\Bigr) 
          \prob(\tVar, n | \cdot )  \,.
    \end{equation} 
    This equation can be cast into a linear PDE by using the basis \eref{eq:GenFct_Flow_TCondition}volution operator $\evolutionOpAd_\tVar$ and the adjoint transition operator $\transitionOpAd_\tVar$ (condition~\eref{eq:GenFctnl_Flow_BCondition}). In particular, since the Kronecker delta $\braketA{n}{m} = \delta_{n,m}$ is independent of time, it holds that $\braA{n} \partial_\tVar \ketA{m} = (-\partial_{\tVar} \braA{n})\ketA{m}$. Consequently, the basis evolution operator in condition~\eref{eq:GenFct_Flow_TCondition} fulfils
    \begin{equation}
      \braA{n} \evolutionOpAd_\tVar = -\partial_{\tVar} \braA{n}
    \end{equation}
    with respect to functions that can be expanded in the basis functions. 
    Moreover, the \eref{eq:GenFct_Flow_OrthogonalityCondition}rthogonality and \eref{eq:GenFct_Flow_CompletenessCondition}ompleteness of the basis imply that the adjoint transition operator in condition~\eref{eq:GenFctnl_Flow_BCondition} acts on basis functionals as
    \begin{equation} 
      \braA{n} \transitionOpAd_\tVar  
      = \sum_{m} \qMatrix^\transpose_\tVar(n, m)  \braA{m} \,.
    \end{equation}  
    Both of the above expressions hold for all values of $n$. If the two differential operators $\evolutionOpAd_\tVar$ and $\transitionOpAd_\tVar$ exist, the generating functional $\braA{\gen(\tVar | t_0, n_0)}$ obeys the functional flow equation
    \begin{equation}   
      \partial_\tVar \braA{\gen}
      = \braA{\gen} (- \evolutionOpAd_\tVar + \transitionOpAd_\tVar )   
      = \braA{\gen} \transEvoOpAd_\tVar        \label{eq:GenFctnl_Flow_FlowEq}   \,,
    \end{equation} 
    with initial condition $\braA{\gen(t_0|t_0,n_0)} = \braA{n_0}$. Note that this flow equation employs the same operator as the flow equation~\eref{eq:GenFctnl_MarginalizedDist_Flow} of the marginalized distribution. Both equations can be used to derive the ``backward'' path integral representation considered in \sref{sec:PathInts_Backward}.
  
    As a side note, let us remark that the flow equation $\partial_\tVar \ketA{\gen}  = (\evolutionOp_\tVar + \transitionOp_\tVar) \ketA{\gen}$ of the generating function in~\eref{eq:GenFct_Flow_FlowEq} also admits a functional counterpart. In particular, the series
    \begin{equation} 
      \braA{\prob(t, n | \tVar)}   \label{eq:GenFctnl_Flow_XXX}
      \coloneqq \sum_{n_0} \prob(t, n|\tVar,n_0) \braA{n_0}    
    \end{equation}
    obeys the flow equation
    \begin{equation}
      \partial_{-\tVar} \braA{\prob}     \label{eq:GenFctnl_Flow_XXX_Flow}
      = \braA{\prob} (\evolutionOp_\tVar + \transitionOp_\tVar)   
      = \transEvoOp_\tVar \braA{\prob}     \,,
    \end{equation} 
    with final value $\braA{\prob(t, n | t)} = \braA{n}$. The corresponding inverse transformation reads $\prob(t, n|\tVar,n_0) =  \braketA{\prob(t, n | \tVar)}{n_0}$.
    Both the flow equation obeyed by the generating function and the above flow equation can be used to derive the ``forward'' path integral representation in \sref{sec:PathInts_Forward}. Further uses of the series~\eref{eq:GenFctnl_Flow_XXX} remain to be explored.

  \subsection{The Poisson representation}\label{subsec:GenFctnl_PoissonRep}  

    Assuming that the action of the generating functional $\braA{\gen}$ on a function $f$ can be expressed in terms of an integral kernel (also called $\gen$) as
    \begin{equation} 
      \braA{\gen(\tVar | t_0, n_0)} f    \label{eq:GenFctnl_PoissonRep_Def}
      =  \int_{-\infty}^\infty \diff{\xAd}\,  
        \gen(\tVar; \xAd | t_0, n_0)  f(\xAd)   
      \,,
    \end{equation} 
    the insertion of the Poisson basis function $\ketA{n}_{\xAd} = \frac{\xAd^n \ee^{-\xAd}}{n!}$ into the inverse transformation~\eref{eq:GenFctnl_Flow_InverseTransformation} results in
    \begin{equation} 
      \prob(\tVar, n | t_0, n_0) 
      =  \int_{-\infty}^\infty \diff{\xAd}\,  
        \gen(\tVar; \xAd | t_0, n_0)  
        \frac{\xAd^n \ee^{-\xAd}}{n!}  \,.               \label{eq:GenFctnl_PoissonRep_Representation}
    \end{equation} 
    A representation of the probability distribution of this form is called a ``Poisson representation'' and was first proposed by Gardiner and Chaturvedi~\cite{Gardiner:1977,Chaturvedi:1978}. Since the integration in~\eref{eq:GenFctnl_PoissonRep_Representation} proceeds along the real line, the above representation is referred to as a ``real'' Poisson representation~\cite{Drummond:1981}. Although the use of a real variable may seem convenient, its use typically results in the kernel being a ``generalized function'', i.e.\ a distribution. For example, the initial condition $\prob(t_0, n | t_0, n_0) = \delta_{n,n_0}$ is recovered for the integral kernel $\gen(t_0; \xAd | t_0, n_0) = \delta(\xAd) (\partial_\xAd+1)^{n_0}$. By the definition
    \begin{equation}
      \int_{-\infty}^\infty \diff{\xAd}\, \bigl(\partial_\xAd^j \delta(\xAd) \bigr)f(\xAd) 
      \coloneqq \int_{-\infty}^\infty \diff{\xAd}\, \delta(\xAd) (-\partial_\xAd)^j f(\xAd)    \nonumber
    \end{equation}
    of distributional derivatives (1.16.12 in \cite{NIST:2010}, $j\in\naturals_0$), this kernel can also be written as $[(1-\partial_\xAd)^{n_0}\delta(\xAd)]$. The integral kernel $\delta(\xAd - \xAd_0)$ instead results for a Poisson distribution with mean $\xAd_0$. In~\cite{Gardiner:1977}, Gardiner and Chaturvedi used the real Poisson representation to calculate steady-state probability distributions of various elementary reactions. Furthermore, the real Poisson representation was employed by Elderfield~\cite{Elderfield:1985} to derive a stochastic path integral representation. Droz and McKane~\cite{Droz:1994} later argued that this representation is equivalent to a path integral representation based on Doi's Fock space algebra. Our discussion of the backward and forward path integral representations in sections~\ref{sec:PathInts_Backward} and~\ref{sec:PathInts_Forward} clarifies the similarities and differences between these two approaches.
    
    To circumvent the use of generalized functions, various alternatives to the real Poisson representation have been proposed: the ``complex'' and ``positive'' Poisson representations~\cite{Drummond:1981,Gardiner:2009} as well as the ``gauge'' Poisson representation~\cite{Drummond:2004}. The former two representations as well as the real representation are discussed in the book of Gardiner~\cite{Gardiner:2009}. The complex Poisson representation is obtained by continuing the Poisson basis function in~\eref{eq:GenFctnl_PoissonRep_Representation} into the complex domain and performing the integration around a closed path $\complexPath$ around $0$ (once in counter-clockwise direction). Upon using Cauchy's differentiation formula to redefine the basis functional as
    \begin{equation}
      \braA{n} f 
      \coloneqq \partial_\xAd^{n} \ee^\xAd f(\xAd) |_{\xAd=0}
      = \oint_\complexPath\diff{\xAd}  \frac{n!}{2\pi\ii} \frac{\ee^\xAd}{\xAd^{n+1}} f(\xAd)        \label{eq:GenFctnl_PoissonRep_ComplexBaseFctnl}\,,
    \end{equation}
     it becomes apparent that the initial condition $\prob(t_0, n | t_0, n_0) = \delta_{n,n_0}$ is then recovered for the kernel $\gen(t_0; \xAd | t_0, n_0) = \frac{n_0!}{2\pi\ii} \frac{\ee^\xAd}{\xAd^{n_0+1}}$. As the interpretation of this kernel is also not straightforward, we refrain from calling it a ``quasi-probability distribution''~\cite{Gardiner:1977}. 
     
    Let us exemplify the flow equation obeyed by the integral kernel for the reaction $k\, A \to l\, A$ with rate coefficient $\rateCoeffGeneric_\tVar$. The flow equation can be inferred from the corresponding flow equation~\eref{eq:GenFctnl_Flow_FlowEq} of the generating functional and the kernel's definition in~\eref{eq:GenFctnl_PoissonRep_Def}. Since we employ the Poisson basis function, the results from \sref{subsec:GenFctnl_BackwardBases_ChemReactions} imply that the adjoint transition operator~\eref{eq:GenFctnl_Flow_Poisson_HamiltonianB} reads
    \begin{equation}
      \transitionOpAd_\tVar
       = \rateCoeffGeneric_\tVar  \xAd^k \bigl((\partial_\xAd+1)^l - (\partial_\xAd+1)^k\bigr) \,. 
    \end{equation}
    Consequently, one finds that the integral kernel $\gen(\tVar; \xAd | \cdot)$ obeys the flow equation
    \begin{eqnarray}  \label{eq:GenFctnl_PoissonRep_Flow}
      \partial_\tVar \gen
      &= \transitionOp_\tVar(\xAd, \partial_\xAd) \gen \\
      &=  \rateCoeffGeneric_\tVar  \bigl((1-\partial_\xAd)^l - (1-\partial_\xAd)^k\bigr) \xAd^k \gen \,.
    \end{eqnarray} 
    To arrive at this equation, we performed repeated integrations by parts while ignoring any potential boundary terms (cf.\ the definition of the adjoint operator in~\eref{eq:GenFctnl_MarginalizedDist_Adjoint}). The importance of boundary terms is discussed in~\cite{Drummond:2004}.
    
    Thus far, most studies employing the Poisson representation have focused on networks of bimolecular reactions $\sum_j k_j\, A_j \to \sum_j  l_j\, A_j$ with $\sum_j k_j \leq 2$ and $\sum_j l_j \leq 2$. For these networks, the flow equation~\eref{eq:GenFctnl_PoissonRep_Flow} assumes the mathematical form of a forward Fokker-Planck equation with the derivatives being of at most second order (the corresponding diffusion coefficient may be negative). It has been attempted to map the resulting equation to an It\^{o} SDE~\cite{Gardiner:1977}, but it should be explored whether this procedure is supported by the Feynman-Kac formula in \aref{sec:A_FeynmanKacProof}. The numerical integration of an SDE with potentially negative diffusion coefficient was attempted for the bi-directional reaction $2\, A \leftrightharpoons \emptyset$ in~\cite{Deloubriere:2002}. The value of the integration has remained inconclusive. Recently, an exponential ansatz for the integral kernel $\gen(\tVar; \xAd | \cdot)$ has been considered to approximate its flow equation in the limit of weak noise~\cite{Petrosyan:2014} (cf.\ \sref{subsubsec:GenFct_Spectral_WKB}). Moreover, a gauge Poisson representation was recently employed in a study of the coagulation reaction $2\, A \to A$~\cite{Burnett:2015}.
    
  \subsection{R\'esum\'e}\label{subsec:GenFctnl_Resume}      
  
    In the previous section, we formulated general conditions under which the master equation can be transformed into a partial differential equation obeyed by the probability generating function~\eref{eq:GenFct_Flow_GenFct}. In the present section, we complemented this flow equation by a backward-time flow equation obeyed by the marginalized distribution~\eref{eq:GenFctnl_MarginalizedDist_Expansion} and by a functional flow equation obeyed by the probability generating functional~\eref{eq:GenFctnl_Flow_GenFctnl}. Whereas the marginalized distribution
    \begin{equation} 
      \ketA{\prob(t, n | \tVar)} 
      = \sum_{n_0} \prob(t, n|\tVar,n_0) \ketA{n_0}
    \end{equation} 
    was defined as the sum of the conditional probability distribution over a set of basis functions, the generating functional
    \begin{equation}
      \braA{\gen(\tVar | t_0, n_0)} 
      = \sum_{n} \braA{n} \prob(\tVar, n | t_0, n_0)     
    \end{equation} 
    was defined as the sum of the distribution over a set of basis functionals. In \sref{subsec:GenFctnl_BackwardBases}, we introduced \eref{eq:GenFct_Flow_OrthogonalityCondition}rthogonal and \eref{eq:GenFct_Flow_CompletenessCondition}omplete basis functions and functionals for the study of different stochastic processes, including the Poisson basis for the study of chemical reactions ($\ketA{n}_{\xAd} = \frac{\xAd^n \ee^{-\xAd}}{n!}$ and $\braA{m}f = (\partial_{\xAd} + 1)^m f(\xAd) \big|_{\xAd = 0}$). Provided that there also exist a basis \eref{eq:GenFct_Flow_TCondition}volution operator~$\evolutionOp_\tVar$ and an adjoint transition operator $\transitionOpAd_\tVar$ (condition~\eref{eq:GenFctnl_Flow_BCondition}), the marginalized distribution obeys the backward-time flow equation
    \begin{equation}
      \partial_{-\tVar} \ketA{\prob}       \label{eq:GenFctnl_Resume_FlowEq}
      = (-\evolutionOpAd_\tVar + \transitionOpAd_\tVar)\ketA{\prob}     
      = \transEvoOpAd_\tVar\ketA{\prob}   
    \end{equation} 
    with final condition $\ketA{\prob(t, n | t)} = \ketA{n}$. Moreover, the probability generating functional~\eref{eq:GenFctnl_Flow_GenFctnl} then obeys the functional flow equation~\eref{eq:GenFctnl_Flow_FlowEq}. The inverse transformation $\prob(\tVar, n | t_0, n_0) =  \braketA{\gen(\tVar|t_0,n_0)}{n}$ of this functional generalizes the Poisson representation of Gardiner and Chaturvedi~\cite{Gardiner:1977,Chaturvedi:1978} as shown in \sref{subsec:GenFctnl_PoissonRep}. In \sref{subsec:GenFctnl_ExtTimes}, we showed how the flow equation~\eref{eq:GenFctnl_Resume_FlowEq} obeyed by the marginalized distribution can be used to compute mean extinction times. Furthermore, the equation will prove useful in the derivation of path integral representations of the master equation and of averaged observables  in sections~\ref{sec:PathInts_Backward} and~\ref{sec:PathInts_Observables}. Future studies could explore whether the flow equation obeyed by the marginalized distribution can be evaluated in terms of WKB approximations or spectral methods.

\section{The backward path integral representation}\label{sec:PathInts_Backward}  
    
    In the previous two sections, we showed how the forward and backward master equations can be cast into four linear PDEs for the series expansions~\eref{eq:GenFct_GenFct_mul}--\eref{eq:GenFct_XXX}. In this section, as well as in \sref{sec:PathInts_Forward}, the solutions of these four equations are expressed in terms of two path integrals. Upon applying inverse transformations, the path integrals provide distinct representations of the conditional probability distribution solving the master equations. The flow equations obeyed by the generating function~\eref{eq:GenFct_GenFct_mul} and by the series~\eref{eq:GenFct_XXX} will lead us to the ``forward'' path integral representation
    \begin{equation}
      \prob(t, n | t_0, n_0)    
      =  \braA{n}_{t} 
         \pathintegral{[t_0}{t)} 
        \ee^{-\action}    
        \, \ketA{n_0}_{t_0}         \,,  \label{eq:PathInts_FwdFlow}
    \end{equation}
    and the flow equations obeyed by the marginalized distribution~\eref{eq:GenFct_Marginalized} and by the generating functional~\eref{eq:GenFct_PoissonRep} to the ``backward'' path integral representation
    \begin{equation}
      \prob(t, n | t_0, n_0)  
      = \braA{n_0}_{t_0} 
        \pathintegral{(t_0}{t]}
        \, \ee^{-\actionAd} 
        \, \ketA{n}_{t}   \,.  \label{eq:PathInts_BwdFlow}
    \end{equation} 
    The meanings of the integral signs $\pathintegral{[t_0}{t)}$ and $\pathintegral{(t_0}{t]}$, as well as of the exponential weights are explained below. We choose the above terms for the two representations because the forward path integral representation propagates the basis function $\ketA{n_0}_{t_0}$ to time $t$, where it is then acted upon by the functional $\braA{n}_{t}$. Counter-intuitively, however, this procedure requires us to solve a stochastic differential equation proceeding backward in time (cf.~sections~\ref{subsec:PathInts_LinearProcesses} and~\ref{subsec:PathInts_ForwardKramersMoyal}). Analogously, the backward path integral representation propagates the basis function $\ketA{n}_{t}$ backward in time to $t_0$, where one then applies the functional $\braA{n_0}_{t_0}$. This procedure requires us to solve an ordinary, forward-time It\^{o} SDE (cf.~\sref{subsec:PathInts_BackwardAlongPaths}). The name ``adjoint'' path integral representation may also be used for the backward representation. As we will see below, its ``action'' $\actionAd$ involves the adjoint transition operator $\transEvoOpAd_\tVar$.
    
    To make the following derivations as explicit as possible, we now assume the discrete variables $n$ and $n_0$ to be one-dimensional. Likewise, the variables $\xAd$ and $\q$ of the associated flow equations are one-dimensional real variables. The derivation below employs an exponential representation of the Dirac delta function. Although such a representation also exists for Grassmann variables~\cite{Swanson:1992}, we do not consider that case. The extension of the following derivations to processes with multiple types of particles is straightforward and proceeds analogously to the inclusion of spatial degrees of freedom. A process with spatial degrees of freedom is considered in sections~\ref{subsubsec:PathInts_BackwardSolutions_DiffusionNetwork} to \ref{subsubsec:PathInts_BackwardSolutions_DiffusionAndDecay}.
    
    After deriving the backward path integral representation in \sref{subsec:PathInts_Backward_Derivation}, we exemplify how this representation can be used to solve the bi-directional reaction $\emptyset \rightleftharpoons A$ (\sref{subsec:PathInts_SimpleGrowthLinearDecay}), the pair generating process $\emptyset \to 2\, A$ (\sref{subsubsec:PathInts_BackwardSolutions_PairGeneration}), and a process with diffusion and linear decay (sections~\ref{subsubsec:PathInts_BackwardSolutions_DiffusionNetwork} to \ref{subsubsec:PathInts_BackwardSolutions_DiffusionAndDecay}). The forward path integral representation is derived in \sref{sec:PathInts_Forward} and is exemplified in deriving the generating function of linear processes $A \to l\, A$ with $l \geq 0$. For the linear growth process $A \to 2\, A$, we recover a negative Binomial distribution as the solution of the master equation. Observables of the particle number are considered in \sref{sec:PathInts_Observables}.
     
     As an intermezzo, we show in sections~\ref{subsec:PathInts_BackwardKramersMoyal} and~\ref{subsec:PathInts_ForwardKramersMoyal} how one can derive path integral representations for jump processes with continuous state spaces (or processes whose transition rates can be extended to such spaces). The corresponding derivations are based on Kramers-Moyal expansions of the backward and forward master equations. Since the backward and forward Fokker-Planck equations constitute special cases of these expansions, we recover a classic path integral representation whose development goes back to works of Martin, Siggia, and Rose~\cite{Martin:1973}, de Dominicis~\cite{deDominicis:1976}, Janssen~\cite{Janssen:1976,Bausch:1976}, and Bausch, Janssen, and Wagner~\cite{Bausch:1976}. Moreover, we recover the Feynman-Kac formula from \aref{sec:A_FeynmanKacProof}, an Onsager-Machlup representation~\cite{Onsager:1953}, and Wiener's path integral for Brownian motion~\cite{Wiener:1921a,Wiener:1921b}.
     
     Before starting out with the derivations of the two path integral representations~\eref{eq:PathInts_FwdFlow} and~\eref{eq:PathInts_BwdFlow}, let us note that these representations only apply if their underlying transition operators $\transEvoOp_\tVar$ or $\transEvoOpAd_\tVar$ can be written as power series in their arguments ($\xAd$ and $\partial_\xAd$ for the backward path integral and $\q$ and $\partial_\q$ for the forward path integral; the names of the variables differ because both path integrals then lead to the same path integral representation of averaged observables in \sref{sec:PathInts_Observables} without requiring a change of variable names). In addition, the power series need to be ``normal-ordered'', meaning that in every summand, all the $\xAd$ are to the left of all the $\partial_\xAd$ (all the $\q$ to the left of all the $\partial_\q$). This order can be established by the repeated use of the commutation relation $[\partial_\xAd,\xAd]f(\xAd)=f(\xAd)$ (or, more directly, by invoking $(\partial_\xAd\, \xAd)^n f(\xAd)= \sum_{m=0}^n {n+1 \brace m+1} \xAd^m \partial_\xAd^m f(\xAd)$ or $(\xAd\partial_\xAd)^n f(\xAd)= \sum_{m=0}^n {n \brace m}   \xAd^m \partial_\xAd^m$, with the curly braces representing Stirling numbers of the second kind; cf.\ section~26.8 in~\cite{NIST:2010}). The transition operator $\transitionOp_\tVar  = \rateCoeffGeneric_\tVar (\cre^l - \cre^k) \ann^k$ in~\eref{eq:GenFct_Flow_Poisson_HamiltonianF} and the adjoint transition operator $\transitionOpAd_\tVar = \rateCoeffGeneric_\tVar \cre^k (\ann^l - \ann^k)$ in~\eref{eq:GenFctnl_Flow_Poisson_HamiltonianB} of the chemical reaction $k\, A \to l\, A$ are already in their normal-ordered forms (with the creation and annihilation operators in~\eref{eq:GenFct_Flow_Poisson_Creation} and \eref{eq:GenFct_Flow_Poisson_Annihilation}, or~\eref{eq:GenFctnl_Flow_Poisson_Creation} and \eref{eq:GenFctnl_Flow_Poisson_Annihilation}, respectively).

     \begin{figure*}[tb] 
      \includegraphics{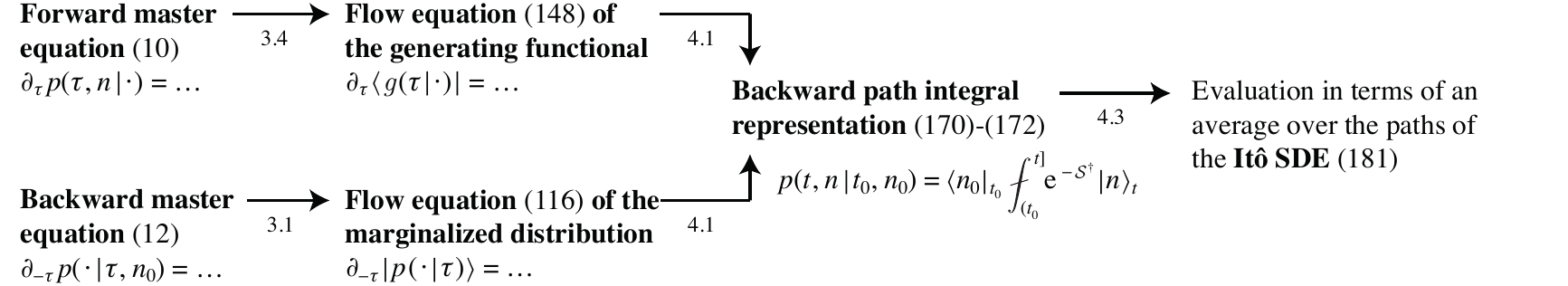}
      \caption{\label{fig:BackwardPathIntegral}
        Outline of the derivation of the backward path integral representation and of its evaluation in terms of an average over the paths of an It\^{o} stochastic differential equation.
      }
    \end{figure*}    
    
  \subsection{Derivation}\label{subsec:PathInts_Backward_Derivation}  
  
    We first derive the backward path integral representation~\eref{eq:PathInts_BwdFlow} because its application to actual processes is more intuitive than the application of the forward path integral representation. For this purpose, we consider the flow equation $\partial_{-\tVar} \ketA{\prob(\cdot | \tVar)}   = \transEvoOpAd_\tVar\ketA{\prob(\cdot | \tVar)}$ in~\eref{eq:GenFctnl_MarginalizedDist_Flow} obeyed by the marginalized distribution. Its final condition reads $\ketA{\prob(t, n | t)} = \ketA{n}$. The conditional probability distribution is recovered from the marginalized distribution via $\prob(t, n | t_0, n_0) = \braketA{n_0}{\prob(t, n | t_0)}$. As the first step, we split the time interval $[t_0,t]$ into $N$ pieces $t_0 \leq t_1 \leq \hldots \leq t_N  \coloneqq t$ of length $\Delta t \coloneqq (t-t_0)/N\ll 1$. Over the time interval $[t_0, t_1]$, the flow equation is then solved by\footnote{This discretization of time conforms with the It\^{o} prescription for forward-time SDEs and with the derivation of the backward master equation~\eref{eq:Intro_Mescoscopic_BackwardMasterEq} (see \sref{subsec:PathInts_BackwardAlongPaths} and also the footnote on page~\pageref{ftn:Bwd_Discretization}).}
    \begin{equation} 
      \ketA{\prob(t, n | t_0)}_{\xAd_0}  
      = \generatorAd_{t_0}(\xAd_0,\partial_{\xAd_0}) \ketA{\prob(t, n | t_1)}_{\xAd_0}   \label{eq:PathInts_Backward_Step1}
    \end{equation} 
    with the generator $\generatorAd_\tVar \coloneqq 1 + \transEvoOpAd_\tVar \Delta t + O\bigl((\Delta t)^2\bigr)$. Alternatively, the following derivation could be performed by solving the flow equation~\eref{eq:GenFctnl_Flow_FlowEq} obeyed by the generating functional over the time interval $[t_{N-1},t]$ as (cf.~\fref{fig:BackwardPathIntegral})
    \begin{equation}   
      \braA{\gen(t | t_0, n_0)}
      = \braA{\gen(t_{N-1} | t_0, n_0)} \generatorAd_{t_{N-1}}(\xAd_{N-1},\partial_{\xAd_{N-1}})         \,.\,
    \end{equation} 
    Here, the generating functional $\braA{\gen}$ acts on functions $f(\xAd_{N-1})$, meets the initial condition $\braA{\gen(t_0 | t_0, n_0)} = \braA{n_0}$, and is transformed back into the conditional probability distribution via $\prob(t, n | t_0, n_0) = \braketA{\gen(t_0 | t_0, n_0)}{n}$. Both of the above approaches result in the same path integral representation. In the following, we use the equation~\eref{eq:PathInts_Backward_Step1} for this purpose.
    
    As the next step of the derivation, we insert the integral representation of a Dirac delta between the generator $\generatorAd$ and the marginalized distribution $\ketA{\prob}$ in~\eref{eq:PathInts_Backward_Step1}, turning the right-hand side of this expression into
    \begin{equation}
      \generatorAd_{t_0}(\xAd_0, \partial_{\xAd_0})  
      \int_{\reals^2}\! \frac{\diff{\xAd_{1}}\diff{\qAd_1}}{2\pi}
      \,\ee^{-\ii\qAd_1(\xAd_1 - \xAd_0)} 
      \ketA{\prob(t, n | t_1)}_{\xAd_1}
        \label{eq:PathInts_Backward_Step2}   \,.
    \end{equation}       
    The integrations over $\xAd_{1}$ and $\qAd_1$ are both performed along the real line from $-\infty$ to $+\infty$. If the adjoint transition operator $\transEvoOpAd_\tVar$, and thus also $\generatorAd_\tVar$, are normal-ordered power series, we can replace $\partial_{\xAd_0}$ by $\ii\qAd_{1}$ in the above expression and pull $\generatorAd_{t_{0}}$ to the right of the exponential. Making use of the final condition $\ketA{\prob(t, n | t)}_{\xAd_N} = \ketA{n}_{t,\xAd_N}$, the above steps can be repeated until~\eref{eq:PathInts_Backward_Step2} reads
    \begin{equation}
       \pathintegral{1}{N}
        \,\Bigl( 
          \prod_{j=1}^N
          \ee^{-\ii\qAd_j(\xAd_j - \xAd_{j-1})} 
          \generatorAd_{t_{j-1}}(\xAd_{j-1}, \ii\qAd_j)  
        \Bigr) 
        \ketA{n}_{t, \xAd_N}     \label{eq:PathInts_Backward_Step3} \,,
    \end{equation} 
    with the abbreviation
    \begin{equation}
       \pathintegral{k}{l} \coloneqq \prod_{j=k}^{l}  \int_{\reals^2}\!\frac{\diff{\xAd_j}\diff{\qAd_{j}}}{2\pi}    \,.\label{eq:PathInts_Backward_DiscreteTime_Measure}    
    \end{equation} 

    To proceed, we now replace $\generatorAd_\tVar = 1 + \transEvoOpAd_\tVar \Delta t + \bigO\bigl((\Delta t)^2\bigr)$ by the exponential $\ee^{\transEvoOpAd_\tVar \Delta t}$ (for brevity, we drop the correction term in the following). Note that the exponential $\ee^{\transEvoOpAd_\tVar \Delta t}$ does not involve differential operators because those were all replaced by the new variables $\ii\qAd_j$. Whether the exponentiation can be made mathematically rigorous is an open question and may possibly be answered positively only for a restricted class of stochastic processes. Upon performing the exponentiation, one obtains the following discrete-time path integral representation of the marginalized distribution:
    \begin{eqnarray} 
      \ketA{\prob(t, n | t_0)}_{\xAd_0}
      =   \pathintegral{1}{N}  
                \,\ee^{-\actionAd_N}    
                \,\ketA{n}_{t, \xAd_N}          \mathtext{ with }     \label{eq:PathInts_Backward_DiscreteTime_Solution_MarginalizedDistribution}     \\
      \actionAd_N
      \coloneqq  \sum_{j=1}^N \Delta t  
                  \Bigl(
                    \ii\qAd_j \frac{\xAd_j - \xAd_{j-1}}{\Delta t} 
                    - \transEvoOpAd_{t_{j-1}}(\xAd_{j-1}, \ii\qAd_j) 
                  \Bigr)      \,.        \label{eq:PathInts_Backward_DiscreteTime_Solution_Action}   
    \end{eqnarray} 
    The final condition $\ketA{\prob(t, n | t)}_\xAd = \ketA{n}_{t,\xAd}$ is trivially fulfilled because only $N = 0$ time slices fit between $t$ and $t$. The object $\actionAd_N$ is called an ``action''. The exponential factor $\ee^{-\actionAd_N}$ weighs the contribution of every path $(\xAd_1,\qAd_1) \to (\xAd_2,\qAd_2) \to \hldots \to (\xAd_N,\qAd_N)$.
    
    As the final step of the derivation, we take the continuous-time limit $N \to \infty$, or $\Delta t\to 0$, at least formally. The following continuous-time expressions effectively serve as abbreviations of the above discretization scheme with the identifications $\xAd(t_0 + j \Delta t) \coloneqq \xAd_j$ and $\qAd(t_0 + j \Delta t) \coloneqq \qAd_j$. Further comments on the discretization scheme are provided in \sref{subsec:PathInts_BackwardKramersMoyal}. Combined with the inverse transformation~\eref{eq:GenFctnl_MarginalizedDist_Inverse}, we thus arrive at the following backward path integral representation of the conditional probability distribution:
    \begin{eqnarray}
      \prob(t, n | t_0, n_0)  = \braA{n_0}_{t_0, \xAd(t_0)} \ketA{\prob(t, n | t_0)}_{\xAd(t_0)}          \label{eq:PathInts_Backward_ContinuousTime_Solution}  \\ 
       \mathtext{with }   
      \ketA{\prob(t, n | t_0)}_{\xAd(t_0)}
      =  \pathintegral{(t_0}{t]}
        \, \ee^{-\actionAd} 
        \, \ketA{n}_{t, \xAd(t)}            \label{eq:PathInts_Backward_ContinuousTime_Solution_MarginalizedDistribution}        \\
      \mathtext{and }  
      \actionAd
      \coloneqq  \int_{t_0}^{t} \diff{\tVar}\, 
                  \bigl[
                    \ii\qAd \partial_\tVar \xAd 
                    - \transEvoOpAd_\tVar(\xAd, \ii\qAd)  
                  \bigr]\,.                               \label{eq:PathInts_Backward_ContinuousTime_Action} 
    \end{eqnarray} 
    Here we included $\xAd(t_0)$ as an argument of the functional $\braA{n_0}_{t_0, \xAd(t_0)}$ to express that this functional acts on $\xAd(\tVar)$ only for $\tVar = t_0$. Moreover, we defined $\pathintegral{(t_0}{t]} \coloneqq \lim_{N \to \infty} \pathintegral{1}{N}$ to indicate that the path integral involves integrations over $\xAd(t)$ and $\qAd(t)$, but not over $\xAd(t_0)$ and $\qAd(t_0)$.
    
    The evaluation of the above path integral representation for an explicit process involves two steps. First, the path integral~\eref{eq:PathInts_Backward_ContinuousTime_Solution_MarginalizedDistribution} provides the marginalized distribution $\ketA{\prob(t, n | t_0)}$. Second, this marginalized distribution is mapped to the conditional probability distribution $\prob(t, n | t_0, n_0)$ by the action of the functional $\braA{n_0}$. 

  \subsection{Simple growth and linear decay}\label{subsec:PathInts_SimpleGrowthLinearDecay}  
    
    Let us demonstrate the above two steps for the bi-directional reaction $\emptyset \rightleftharpoons A$. The stationary distribution of this process was already derived in the introduction to \sref{sec:GenFct}. Both the growth rate coefficient $\rateCoeffSpontGrowth$ and the decay rate coefficient $\rateCoeffLinDecay$ shall be time-independent, but this assumption is easily relaxed. As the first step, the backward path integral~\eref{eq:PathInts_Backward_ContinuousTime_Solution} requires us to choose an appropriate basis. The time-independent Poisson basis with the basis function $\ketA{n}_{\xAd} = \frac{\xAd^n \ee^{-\xAd}}{n!} $ proves to be convenient for this purpose (cf.\ \sref{subsec:GenFctnl_BackwardBases_ChemReactions}). The adjoint transition operator follows from~\eref{eq:GenFctnl_Flow_Poisson_HamiltonianB}, \eref{eq:GenFctnl_Flow_Poisson_Creation}, and~\eref{eq:GenFctnl_Flow_Poisson_Annihilation} as $\transEvoOpAd(\xAd, \partial_\xAd) = (\rateCoeffSpontGrowth - \rateCoeffLinDecay \xAd)\partial_\xAd$. Insertion of this operator into the action~\eref{eq:PathInts_Backward_ContinuousTime_Action} results in
    \begin{equation}
      \actionAd    
      =  \int_{t_0}^{t} \diff{\tVar}\, 
            \ii\qAd \, \bigl[
               \partial_\tVar \xAd 
              - (\rateCoeffSpontGrowth - \rateCoeffLinDecay \xAd )  
            \bigr]    
    \end{equation}
    The integration over the path $\qAd(\tVar)$ from $\tVar=t_0$ to $\tVar=t$ is performed most easily in the discrete-time approximation. The marginalized distribution $\ketA{\prob(t, n | t_0)}_{\xAd(t_0)}$ with $\xAd(t_0) \equiv \xAd_0$ thereby follows as the following product of Dirac deltas:
    \begin{equation*}
      \Bigl(
        \prod_{j=1}^N \int_{\reals} \diff{\xAd_j} \,
        \delta\bigl(\xAd_j - \xAd_{j-1} - (\rateCoeffSpontGrowth - \rateCoeffLinDecay\, \xAd_{j-1})\Delta t\bigr)
      \Bigr)  
      \ketA{n}_{t,\xAd_N}\,.    
    \end{equation*}
    The function $\ketA{n}_{t,\xAd_N}$ is also integrated over. Upon taking the continuous-time $\Delta t \to 0$ and performing the integration over the path $\xAd(\tVar)$, one finds that the marginalized distribution is given by the Poisson distribution
    \begin{equation}
      \ketA{\prob(t, n | t_0)}_{\xAd(t_0)} \label{eq:PathInts_SimpleGrowthLinearDecay_PoissonDist}
      =   \ketA{n}_{t, \xAd(t)}    
      = \frac{\xAd(t)^n \ee^{-\xAd(t)}}{n!}  \,,
    \end{equation}
    with $\xAd(\tVar)$ solving $\partial_\tVar \xAd = \rateCoeffSpontGrowth - \rateCoeffLinDecay \xAd$. This equation coincides with the rate equation of the process $\emptyset \rightleftharpoons A$. The solution $\xAd(\tVar) = \ee^{-\rateCoeffLinDecay(\tVar-t_0)}(\xAd(t_0) - \frac{\rateCoeffSpontGrowth}{\rateCoeffLinDecay}) + \frac{\rateCoeffSpontGrowth}{\rateCoeffLinDecay}$ of the rate equation specifies the mean and the variance of the above Poisson distribution. For $\rateCoeffLinDecay>0$, both of them converge to the asymptotic value $\rateCoeffSpontGrowth/\rateCoeffLinDecay$. In the next section and in \sref{subsec:PathInts_BackwardSolutions}, we generalize these results to more general processes.
    
    The marginalized distribution~\eref{eq:PathInts_SimpleGrowthLinearDecay_PoissonDist} does not only solve the master equation of the reaction $\emptyset \rightleftharpoons A$, but it also establishes the link between the path integral variable $\xAd$ and the moments of the particle number $n$. In particular, the mean particle number evaluates to $\langle n \rangle(t) = \xAd(t)$, while higher order moments can determined via $\langle n^{k} \rangle(t) = \sum_{l=0}^{k} {k \brace l} \xAd(t)^l$. The curly braces denote a Stirling number of the second kind (cf.\ section~26.8 in~\cite{NIST:2010}).
    
    The conditional probability distribution can be calculated from the marginalized distribution by applying the functional $\braA{n_0} f = (\partial_{\xAd(t_0)} + 1)^{n_0} f(\xAd(t_0)) |_{\xAd(t_0) = 0}$ to the latter (cf.~\eref{eq:GenFctnl_Flow_Poisson_BaseFunctional} and~\eref{eq:GenFctnl_Flow_Poisson_Annihilation}). In the limit of a vanishing decay rate ($\rateCoeffLinDecay\to 0$), for which $\xAd(\tVar) =\xAd(t_0) +\rateCoeffSpontGrowth (\tVar-t_0)$, we thereby recover the shifted Poisson distribution~\eref{eq:GenFct_ForwardBases_ChemReactions_ShiftedPoisson}, i.e.
    \begin{equation}
      \prob(t, n | t_0, n_0)
      =  \frac{\ee^{-\rateCoeffSpontGrowth (t-t_0)}(\rateCoeffSpontGrowth (t-t_0))^{n-n_0} }{(n-n_0)!}     \HeavisideStepDiscrete_{n-n_0}     \,.
    \end{equation} 
    Thus, the mean and variance of the conditional probability distribution grow linearly with time. If the particles decay but are not replenished ($\rateCoeffLinDecay > 0$ but $\rateCoeffSpontGrowth= 0$), we instead recover the Binomial distribution~\eref{eq:PathInts_SimpleGrowthLinearDecay_BinomialDist}. 
        
  \subsection{Feynman-Kac formula for jump processes}\label{subsec:PathInts_BackwardAlongPaths}  
  
    We now show how the backward path integral representation~\eref{eq:PathInts_Backward_ContinuousTime_Solution_MarginalizedDistribution} of the marginalized distribution can be expressed in terms of an average over the paths of an It\^{o} stochastic differential equation (SDE). The resulting expression bears similarities with the Feynman-Kac formula~\eref{eq:Intro_Mescoscopic_FP_BW_Solution}, especially when the adjoint transition operator $\transEvoOpAd_\tVar(\xAd, \partial_\xAd)$ of the stochastic process under consideration is quadratic in $\partial_\xAd$. In the general case, however, functional derivatives act on the average over paths. These derivatives can, for example, be evaluated in terms of perturbation expansions as we demonstrate in \sref{subsubsec:PathInts_BackwardSolutions_DiffusionAndDecay}. The procedure outlined below serves as a general starting point for the exact or approximate evaluation of the backward path integral~\eref{eq:PathInts_Backward_ContinuousTime_Solution_MarginalizedDistribution}. 
    
    In the following, we consider a stochastic process whose adjoint transition operator has the generic form
    \begin{equation}
       \transEvoOpAd_\tVar(\xAd, \partial_\xAd)    
        = \alpha_\tVar(\xAd) \partial_\xAd + \frac{1}{2}\beta_\tVar(\xAd) \partial_\xAd^2  + \perturbationOpAd_\tVar(\xAd, \partial_\xAd)    \,.\label{eq:PathInts_BackwardAlongPaths_Hamiltonian}  
    \end{equation} 
    As before, we call $\alpha_\tVar$ a drift and $\beta_\tVar$ a diffusion coefficient. The object $\perturbationOpAd_\tVar$ is referred to as the (adjoint) perturbation operator, or simply as the perturbation. The perturbation operator absorbs all the terms of higher order in $\partial_\xAd$ and possibly also terms of lower order. Thus, the above form of $\transEvoOpAd_\tVar$ is not unique and $\alpha_\tVar$, $\beta_\tVar$, and $\perturbationOpAd_\tVar$ should be chosen so that the evaluation of the expressions below becomes as simple as possible. If the perturbation operator $\perturbationOpAd_\tVar$ is zero, those expressions simplify considerably. 
    
    As the first step, let us rewrite the backward path integral representation~\eref{eq:PathInts_Backward_ContinuousTime_Solution_MarginalizedDistribution} of the marginalized distribution as
    \begin{equation}
      \ketA{\prob(t, n | t_0)}_{\xAd(t_0)} 
      =  \int_{\reals^{2}}\!\frac{\diff{\xAd_N}\diff{\qAd_N}}{2\pi} 
        \ee^{  - \ii\qAd_N  \xAd_N} \fieldGenFctAd_{0, 0}
        \ketA{n}_{t, \xAd_N}  \,,    \label{eq:PathInts_BackwardAlongPaths_MarginalizedDistribution}  
    \end{equation}
    where $\fieldGenFctAd_{\QAd=0, \XAd=0}$ represents a value of the $(\QAd, \XAd)$-generating functional
    \begin{equation}
        \fieldGenFctAd_{\QAd, \XAd}  \coloneqq    
        \pathintegral{(t_0}{t)}
        \ee^{  \ii\qAd_N \xAd_N  - \actionAd + \int_{t_0}^t \diff{\tVar}\, [ \QAd(\tVar)  \xAd(\tVar) +\XAd(\tVar)  \ii \qAd(\tVar) ]  }    \,.
    \end{equation} 
    Hence, we have singled out the integrations over $\xAd_N$ and $\qAd_N$ before performing the continuous-time limit (cf.~\eref{eq:PathInts_Backward_DiscreteTime_Solution_MarginalizedDistribution} and~\eref{eq:PathInts_Backward_ContinuousTime_Solution_MarginalizedDistribution}).
    We call $\fieldGenFctAd_{\QAd, \XAd}$ a generating functional because a functional differentiation of $\fieldGenFctAd_{\QAd, \XAd}$ with respect to $\QAd(\tVar)$ generates a factor $\xAd(\tVar)$, and a functional differentiation of $\fieldGenFctAd_{\QAd, \XAd}$ with respect to $\XAd(\tVar)$ generates a factor $\ii\qAd(\tVar)$. 
    
    Given the action~\eref{eq:PathInts_Backward_ContinuousTime_Action} with the adjoint transition operator~\eref{eq:PathInts_BackwardAlongPaths_Hamiltonian}, the above properties of the $(\QAd, \XAd)$-generating functional can now be used to rewrite this function as (cf.~\aref{sec:A_BackwardAlongPaths})
    \begin{eqnarray}
      \fieldGenFctAd_{\QAd, \XAd}          
      = \ee^{
              \ii\qAd_N   \frac{\delta}{\delta \QAd(t)}
              + \int_{t_0}^t \diff{\tVar}\, \perturbationOpAd_\tVar(\frac{\delta}{\delta \QAd(\tVar)}, \frac{\delta}{\delta \XAd(\tVar)})
            }
      \fieldGenFctAd^0_{\QAd,\XAd}      \label{eq:PathInts_BackwardAlongPaths_PhiXGeneratingFunction}\\
      \mathtext{with } \fieldGenFctAd^0_{\QAd, \XAd}      
      \coloneqq \bigLLangle
                \ee^{  \int_{t_0}^t\! \diff{\tVar}  \QAd(\tVar) \xAd(\tVar)   }  
              \bigRRangle_{\wienerProcess}        \label{eq:PathInts_BackwardAlongPaths_PhiXGeneratingFunction_0}    \,.
    \end{eqnarray}
     Here, $\LLangle \cdot \RRangle_{\wienerProcess}$ represents the average over realizations of a Wiener process $\wienerProcess(\tVar)$. This Wiener process influences the evolution of the path $\xAd(\tVar)$ through the It\^{o} SDE
    \begin{equation}
      \diff{\xAd}(\tVar) 
      = \bigl[\alpha_\tVar(\xAd) + \XAd(\tVar)\bigr]\diff{\tVar}  + \sqrt{\beta_\tVar(\xAd)}\diff{\wienerProcess}(\tVar)   \,.     \label{eq:PathInts_BackwardAlongPaths_SDE}
    \end{equation} 
    The temporal evolution of $\xAd(\tVar)$ starts out from $\xAd(t_0)$, {which is determined by the argument of the marginalized distribution~\eref{eq:PathInts_BackwardAlongPaths_MarginalizedDistribution}}. In \sref{subsubsec:PathInts_BackwardSolutions_DiffusionAndDecay}, we demonstrate how the marginalized distribution can be evaluated in terms of a perturbation expansion of the $(\QAd, \XAd)$-generating functional~\eref{eq:PathInts_BackwardAlongPaths_PhiXGeneratingFunction} for a process of diffusing particles that are also decaying. In order to perform the perturbation expansion, let us already note that the value $\xAd(\tVar)$ of the path at time $\tVar$ depends on the ``source'' $X(\tVar^\prime)$ only for times $\tVar^\prime < \tVar$, and that $\frac{\delta}{\delta \QAd(\tVar)} \int_{t_0}^t \diff{\tVar^\prime} \QAd(\tVar^\prime) f(\tVar^\prime) = f(\tVar)$ holds for all $\tVar\in (t_0,t]$. Moreover, let us note that the $(\QAd, \XAd)$-generating functional depends on $\qAd_N$ and $\xAd(t_0)$ but not on $\xAd_N$. These properties follow from the derivation of the representation~\eref{eq:PathInts_BackwardAlongPaths_PhiXGeneratingFunction}, which we outline in \aref{sec:A_BackwardAlongPaths}.
    
    An important special case of the above representation is constituted by processes whose perturbation operator $\perturbationOpAd_\tVar$ is zero. For $\QAd=\XAd=0$, the generating functional~\eref{eq:PathInts_BackwardAlongPaths_PhiXGeneratingFunction} then simplifies to $\fieldGenFctAd_{0,0} = \ee^{\ii\qAd_N \xAd(t)}$ and the marginalized distribution~\eref{eq:PathInts_BackwardAlongPaths_MarginalizedDistribution} follows in terms of the Feynman-Kac like fomula
    \begin{equation}
      \ketA{\prob(t, n | t_0)}_{\xAd(t_0)}  \label{eq:PathInts_BackwardAlongPaths_Marg2}
      =   \bigLLangle \ketA{n}_{t, \xAd(t)}  \bigRRangle_\wienerProcess    \,.
    \end{equation}
    Here, $\xAd(\tVar)$ solves the It\^{o} SDE
    \begin{equation}
      \diff{\xAd}(\tVar)  \label{eq:PathInts_BackwardAlongPaths_SDE_Special}
      = \alpha_\tVar(\xAd) \diff{\tVar}  + \sqrt{\beta_\tVar(\xAd)}\diff{\wienerProcess}(\tVar)     
    \end{equation} 
    with the initial value $\xAd(t_0)$. For the Poisson basis function $\ketA{n}_{\xAd} = \frac{\xAd^{n} \ee^{-\xAd}}{n!}$, the above representation of the marginalized distribution coincides with our earlier result~\eref{eq:GenFctnl_BackwardBases_ChemReactions_WienerAverage}, which we encountered in the discussion of the coagulation reaction $2\, A \to A$. For the bi-directional reaction $\emptyset \rightleftharpoons A$ from the previous section with adjoint transition operator $\transEvoOpAd(\xAd, \partial_\xAd) = (\rateCoeffSpontGrowth - \rateCoeffLinDecay \xAd)\partial_\xAd$, the It\^{o} SDE~\eref{eq:PathInts_BackwardAlongPaths_SDE_Special} simplifies to the deterministic rate equation $\partial_\tVar \xAd = \rateCoeffSpontGrowth - \rateCoeffLinDecay \xAd$. Thus, no averaging is required to evaluate the marginalized distribution~\eref{eq:PathInts_BackwardAlongPaths_Marg2} and one recovers our earlier solution~\eref{eq:PathInts_SimpleGrowthLinearDecay_PoissonDist}.
    
    Our above discussion only applies to processes in well-mixed environments with a single type of particles. A generalization of the results to processes with multiple types of particles or with spatial degrees of freedom is straightforward. A spatial process will be discussed in \sref{subsec:PathInts_BackwardSolutions}. In a multivariate generalization of the above procedure, the adjoint transition operator~\eref{eq:PathInts_BackwardAlongPaths_Hamiltonian} includes a vector-valued drift coefficient $\vect{\alpha}_\tVar(\vect{\xAd})$ and a diffusion matrix $\beta_\tVar(\vect{\xAd})$. Moreover, the derivation of the corresponding generating functional~\eref{eq:PathInts_BackwardAlongPaths_PhiXGeneratingFunction} requires that there exists a matrix $\sqrt{\beta_\tVar} \coloneqq \gamma_\tVar$ fulfilling $\gamma_\tVar \gamma_\tVar^\transpose = \beta_\tVar$. If the diffusion matrix $\beta_\tVar$ is positive-semidefinite, its positive-semidefinite and symmetric square root $\sqrt{\beta_\tVar}$ can be determined via diagonalization~\cite{Harville:1997}.
    
    In \sref{subsec:PathInts_BackwardSolutions}, we exemplify the use of our above results for various well-mixed and spatial processes. But before, let us show how the results from the previous sections can be used to derive a path integral representation for processes with continuous state spaces.

  \subsection{Intermezzo: The backward Kramers-Moyal expansion}\label{subsec:PathInts_BackwardKramersMoyal}  
    
    The transition rate $\transitionRate_\tVar(m,n)$ denotes the rate at which probability flows from a state $n$ to a state $m$, or, considering an individual sample path, the rate at which particles jump from state $n$ to state $m$. Thus, $\transitionRateJumpSize_\tVar(\Delta n,n) \coloneqq \transitionRate_\tVar(n+\Delta n,n)$ denotes the rate at which the state $n$ is left via jumps of size $\Delta n$ (with $n\in\naturals_0$ and $\Delta n\in\integers$). Thus far, we only considered jumps between the states of a discrete state space. But in the following, we derive a path integral representation for processes having a continuous state space. 
    
    \subsubsection{Processes with continuous state spaces.}\label{subsubsec:PathInts_BackwardKramersMoyal_ContSpace}  
  
      We consider a process whose state is characterized by a continuous variable $\xAd\in\reals_{\geq 0}$. The change from the letter $n$ to the letter $\xAd$ is purely notational and emphasizes that the state space is now continuous (the change also highlights a formal similarity between the linear PDEs discussed so far and the ones derived below). The conditional probability distribution $\prob(\tVar, \xAd | t_0, \xAd_0)$ describing the system shall be normalized as $ \int_\reals \diff{\x}\, \prob(\tVar, \x | \cdot ) = 1$ and obey the master equation
      \begin{eqnarray} 
        \partial_\tVar \prob(\tVar, \x | \cdot)        \label{eq:PathInts_BackwardKramersMoyal_ForwardMaster}
        = \int_\reals \diff{\Delta \x}\, \bigl[
            &\transitionRateJumpSize_\tVar(\Delta \x, \x-\Delta \x) \prob(\tVar, \x-\Delta \x | \cdot )   \nonumber\\
             &- \transitionRateJumpSize_\tVar(\Delta \x, \x) \prob(\tVar, \x | \cdot )   
          \bigr]    
      \end{eqnarray} 
      with the initial condition $\prob(t_0, \x | t_0, \x_0 ) = \delta(\x-\x_0)$. The structure of this master equation is equivalent to the structure of the master equation~\eref{eq:Intro_Mescoscopic_MasterEqGainLoss}. Provided that the product $\transitionRateJumpSize_\tVar(\Delta \x, \x) \prob(\tVar, \x | \cdot )$ is analytic in $\x$, one can perform a Taylor expansion of the above master equation to obtain the (forward) Kramers-Moyal expansion~\cite{Einstein:1905,Kramers:1940,Moyal:1949}
      \begin{equation}
        \partial_\tVar \prob(\tVar, \x | \cdot )  \label{eq:PathInts_BackwardKramersMoyal_ForwardKramersMoyal}
        = \sum_{m=1}^\infty \frac{(-1)^m}{m!} \partial_{\x}^m 
          \bigl[\jumpMoment{m}_\tVar(\x) \prob(\tVar, \x | \cdot )\bigr]  \,.
      \end{equation}
      The ``jump moments'' $\jumpMoment{m}_\tVar$ are defined as
      \begin{equation}
        \jumpMoment{m}_\tVar(\x)    \label{eq:PathInts_BackwardKramersMoyal_JumpMoments}
        \coloneqq \int_{\reals} \diff{\Delta \x}\,
        (\Delta \x)^m \transitionRateJumpSize_\tVar(\Delta \x, \x)  \,.
      \end{equation}
      By Pawula's theorem~\cite{Pawula:1967a,Pawula:1967b,Risken:1996}, the positivity of the conditional probability distribution requires that the Kramers-Moyal expansion either stops at its first or second summand, or that it does not stop at all. If the expansion stops at its second summand, it assumes the form of a (forward) Fokker-Planck equation. The drift coefficient of this Fokker-Planck equation is given by $\jumpMoment{1}_\tVar(\x)$ and its non-negative diffusion coefficient by $\jumpMoment{2}_\tVar(\x)$. The sample paths of the Fokker-Planck equation are continuous, however, contradicting our earlier assumption of the process making discontinuous jumps in state space. For a jump process, the Kramers-Moyal expansion cannot stop. Nevertheless, a truncation of the Kramers-Moyal expansion at the level of a Fokker-Planck equation often provides a decent approximation of a process, provided that fluctuations cause only small relative changes of its state $\xAd$.
      
      The backward analogue of the master equation~\eref{eq:PathInts_BackwardKramersMoyal_ForwardMaster} reads
      \begin{eqnarray} 
         \partial_{-\tVar} \prob(t, \xAd | \tVar, \xAd_0 )   \label{eq:PathInts_BackwardKramersMoyal_BackwardMaster}\\
         = \int_\reals \diff{\Delta \xAd}\,
           \bigl[
            \prob(\cdot | \tVar, \xAd_0 +\Delta \xAd) 
            - \prob(\cdot | \tVar, \xAd_0)      
          \bigr] \transitionRateJumpSize_\tVar(\Delta \xAd, \xAd_0)    \nonumber    \,,
      \end{eqnarray} 
      with the final condition $\prob(t, \x | t, \x_0 ) = \delta(\x-\x_0)$. Given the analyticity of $\prob(\cdot | \tVar, \xAd_0)$ in $\xAd_0$, the backward master equation can be rewritten in terms of the backward Kramers-Moyal expansion
      \begin{equation} 
         \partial_{-\tVar} \prob(\cdot | \tVar, \xAd_0 )   \label{eq:PathInts_BackwardKramersMoyal_BackwardKramersMoyal}
         = \transEvoOpAd_\tVar(\xAd_0, \partial_{\xAd_0})  \prob(\cdot | \tVar, \xAd_0 )
      \end{equation} 
      with the adjoint transition operator
      \begin{equation}
        \transEvoOpAd_\tVar(\xAd_0, \partial_{\xAd_0})  \label{eq:PathInts_BackwardKramersMoyal_BackwardKramersMoyal_TransOpAd}
        \coloneqq \sum_{m=1}^\infty \frac{1}{m!} \jumpMoment{m}_\tVar(\xAd_0) \partial_{\xAd_0}^m   \,.
      \end{equation}
      This operator is the adjoint of the operator in the forward Kramers-Moyal expansion~\eref{eq:PathInts_BackwardKramersMoyal_ForwardKramersMoyal} in the sense that $\int \diff{\xAd} \, \bigl[\transEvoOpAd_\tVar(\xAd,\partial_\xAd) f(\xAd)\bigr] g(\xAd) = \int \diff{\xAd} f(\xAd) \bigl[\transEvoOp_\tVar(\xAd,\partial_\xAd) g(\xAd)\bigr]$ (provided that all boundary terms in the integrations by parts vanish).
      
    \subsubsection{Path integral representation of the backward Kramers-Moyal expansion.}\label{subsubsec:PathInts_BackwardKramersMoyal_PathInt}  
  
      The adjoint transition operator $\transEvoOpAd_\tVar(\xAd_0, \partial_{\xAd_0})$ of the backward Kramers-Moyal expansion~\eref{eq:PathInts_BackwardKramersMoyal_BackwardKramersMoyal_TransOpAd} is normal-ordered with respect to $\xAd_0$ and $\partial_{\xAd_0}$. Moreover, the backward Kramers-Moyal expansion~\eref{eq:PathInts_BackwardKramersMoyal_BackwardKramersMoyal} has the same form as the flow equation~\eref{eq:GenFctnl_MarginalizedDist_Flow} obeyed by the marginalized distribution. One can therefore follow the steps in \sref{subsec:PathInts_Backward_Derivation} to represent the backward Kramers-Moyal expansion by the path integral
      \begin{eqnarray}
        \prob(t, \xAd | t_0, \xAd_0 )    \label{eq:PathInts_BackwardKramersMoyal_PathIntegral}
        =  \pathintegral{(t_0}{t]}
          \, \ee^{-\actionAd} 
          \, \delta(\xAd - \xAd(t))  \big|_{\xAd(t_0)=\xAd_0}                    
      \end{eqnarray} 
      with the action
      \begin{eqnarray}
        \actionAd      \label{eq:PathInts_BackwardKramersMoyal_PathIntegral_Action}
        \coloneqq  \int_{t_0}^{t} \diff{\tVar}\, 
                    \bigl[
                      \ii\qAd(\tVar) \partial_\tVar \xAd(\tVar) 
                      - \transEvoOpAd_\tVar(\xAd(\tVar), \ii\qAd(\tVar))  
                    \bigr]\,.                       
      \end{eqnarray} 
      The integral sign in~\eref{eq:PathInts_BackwardKramersMoyal_PathIntegral} is again defined as the continuous-time limit of~\eref{eq:PathInts_Backward_DiscreteTime_Measure} and traces out all paths of $\xAd(\tVar)$ and $\qAd(\tVar)$ for $\tVar\in(t_0,t]$ (note that $\xAd(\tVar)$ differs from $\xAd$ and $\xAd_0$, which are fixed parameters; for brevity, however, $\xAd(\tVar)$ is occasionally abbreviated as $\xAd$ below). A diagrammatic computation of multi-time correlation functions based on a path integral representation equivalent to~\eref{eq:PathInts_BackwardKramersMoyal_PathIntegral} has recently been considered in~\cite{Thomas:2014}.
      
      Let us specify the path integral representation~\eref{eq:PathInts_BackwardKramersMoyal_PathIntegral} for a model of the chemical reaction $k\, A \to l\, A$ with rate coefficient $\rateCoeffGeneric_\tVar$ (and $k,l\in\naturals_0$). As we assume the particle ``number'' $\xAd$ to be continuous, the model is only reasonable in an approximate sense for large values of $\xAd$. In defining the transition rate of the reaction, one has to ensure the non-negativity of $\xAd$ and the conservation of probability. A possible choice of the transition rate is
      \begin{equation}
        \transitionRateJumpSize_\tVar(\Delta \xAd,\xAd)  \label{eq:PathInts_BackwardKramersMoyal_PathIntegral_Transition}
        \coloneqq \rateCoeffGeneric_\tVar \xAd^k \HeavisideStep(\xAd - k) \delta\bigl(\Delta \xAd - (l-k)\bigr)   \,.
      \end{equation}
      Here, the Heaviside step function $\HeavisideStep(\xAd - k)$ ensures that a sufficient number of particles are present to engage in a reaction (and it prevents the loss of probability to negative values of $\xAd$).
      
      Assuming a smooth approximation of the Heaviside step function that vanishes for $\xAd < 0$, the jump moments~\eref{eq:PathInts_BackwardKramersMoyal_JumpMoments} follow as
      \begin{equation}
        \jumpMoment{m}_\tVar(\x) = \rateCoeffGeneric_\tVar (l-k)^m \xAd^k \HeavisideStep(\xAd - k) \,.
      \end{equation}
      The corresponding adjoint transition operator evaluates to
      \begin{equation}
        \transEvoOpAd_\tVar(\xAd, \ii\qAd) 
        = \rateCoeffGeneric_\tVar (\xAd \ee^{-\ii\qAd})^k \bigl[(\ee^{\ii\qAd})^l - (\ee^{\ii\qAd})^k\bigr] \HeavisideStep(\xAd - k) \,.\label{eq:PathInts_BackwardKramersMoyal_PathIntegral_Operator}
      \end{equation}
      Path integrals with such an operator have been noted in~\cite{Andreanov:2006,Lefevre:2007,Itakura:2010}, but their potential use remains to be fully explored. Curiously, upon ignoring the step function, the operator~\eref{eq:PathInts_BackwardKramersMoyal_PathIntegral_Operator} has the same structure as the transition operator~\eref{eq:GenFctnl_Flow_Poisson_HamiltonianB} upon identifying $\xAd \ee^{-\ii\qAd}$ with the creation operator $\cre$, and $\ee^{\ii\qAd}$ with the annihilation operator $\ann$ (Cole-Hopf transformation~\cite{Janssen:1999,Frey:1999,Andreanov:2006,Lefevre:2007,Itakura:2010}). Note, however, that the stochastic processes associated to the two transition operators differ from each other (discrete vs.\ continuous state space; different transition rates). The connection between the two associated path integrals~\eref{eq:PathInts_Backward_ContinuousTime_Solution_MarginalizedDistribution} and~\eref{eq:PathInts_BackwardKramersMoyal_PathIntegral} should be further explored.

    \subsubsection{Path integral representation of the backward Fokker-Planck equation.}\label{subsubsec:PathInts_BackwardKramersMoyal_BwFP}
    
      With the drift coefficient $\alpha_\tVar(\xAd) \coloneqq \jumpMoment{1}_\tVar(\xAd)$ and the non-negative diffusion coefficient $\beta_\tVar(\xAd) \coloneqq \jumpMoment{2}_\tVar(\xAd)$, a truncation of the adjoint transition operator~\eref{eq:PathInts_BackwardKramersMoyal_BackwardKramersMoyal_TransOpAd} at its second summand reads
      \begin{equation}
        \transEvoOpAd_\tVar(\xAd_0, \partial_{\xAd_0})      \label{eq:PathInts_BackwardKramersMoyal_FPAdjointOperator}
          = \alpha_\tVar(\xAd_0) \partial_{\xAd_0} + \frac{1}{2}\beta_\tVar(\xAd_0) \partial_{\xAd_0}^2     \,.
      \end{equation} 
      Thus, the corresponding Kramers-Moyal expansion~\eref{eq:PathInts_BackwardKramersMoyal_BackwardKramersMoyal} recovers the backward Fokker-Planck equation~\eref{eq:Intro_Mescoscopic_FP_BW}. Since the action~\eref{eq:PathInts_BackwardKramersMoyal_PathIntegral_Action} evaluates to
      \begin{eqnarray}
        \actionAd      
        \coloneqq  \int_{t_0}^{t} \diff{\tVar}\, 
                    \Bigl(
                      \ii\qAd \big[\partial_\tVar \xAd - \alpha_\tVar(\xAd)\bigr]
                      + \frac{1}{2}\beta_\tVar(\xAd) \qAd^2  
                    \Bigr)      \,,                       
      \end{eqnarray} 
      the corresponding path integral~\eref{eq:PathInts_BackwardKramersMoyal_PathIntegral} coincides with a classic path integral representation of the (backward) Fokker-Planck equation. The original development of this representation goes back to works of Martin, Siggia, and Rose~\cite{Martin:1973}, de Dominicis~\cite{deDominicis:1976}, Janssen~\cite{Janssen:1976,Bausch:1976}, and Bausch, Janssen, and Wagner~\cite{Bausch:1976}. The application of this path integral to stochastic processes is, for example, discussed in the book of T\"auber~\cite{Taeuber:2014}. 
      
      The  transition operator~\eref{eq:PathInts_BackwardKramersMoyal_FPAdjointOperator} of the backward Fokker-Planck equation has the same form as the transition operator~\eref{eq:PathInts_BackwardAlongPaths_Hamiltonian} in \sref{subsec:PathInts_BackwardAlongPaths}, but it does not involve a perturbation operator $\perturbationOpAd_\tVar$. Therefore, one can follow the steps in that section to evaluate the path integral~\eref{eq:PathInts_BackwardKramersMoyal_PathIntegral} in terms of an average over the paths of an It\^o SDE. This procedure shows that the backward Fokker-Planck equation is solved by
      \begin{equation}
        \prob(t, \xAd | t_0, \xAd_0)  
        = \bigLLangle \delta (\xAd - \xAd(t)) \bigRRangle_\wienerProcess  \,,
      \end{equation}
      with $\xAd(\tVar)$ solving the It\^{o} SDE
      \begin{equation}
        \diff{\xAd}(\tVar)     
        = \alpha_\tVar(\xAd(\tVar)) \diff{\tVar}  + \sqrt{\beta_\tVar(\xAd(\tVar))}\diff{W}(\tVar)   
      \end{equation} 
      with initial value $\xAd(t_0) = \xAd_0$. Hence, we have recovered the Feynman-Kac formula~\eref{eq:Intro_Mescoscopic_FP_BW_Solution} (apart from a notational change in the time parameters). 
      
    \subsubsection{The Onsager-Machlup function.}\label{subsubsec:PathInts_BackwardKramersMoyal_OnsagerMachlup}  
      
      The connection between the above path integral representation of the (backward) Fokker-Planck equation and the work of Onsager and Machlup~\cite{Onsager:1953} becomes apparent upon the completion of a square (as in a Hubbard-Stratonovich transformation~\cite{Stratonovich:1957,Hubbard:1959}). For this purpose, the diffusion coefficient $\beta_\tVar$ in the transition operator~\eref{eq:PathInts_BackwardKramersMoyal_FPAdjointOperator} must not only be non-negative but positive(-definite). One can then complete the square in the variable $\qAd$ and perform the path integration over this variable. Returning to the discrete-time approximation~\eref{eq:PathInts_Backward_DiscreteTime_Solution_Action} of the action, one obtains the following representation of the conditional probability distribution $\prob(t, \xAd | t_0, \xAd_0 )$ in terms of convolutions of Gaussian distributions:
      \begin{equation}
        \lim_{N\to\infty} \Bigl(\prod_{j=0}^{N-1} \int_\reals \diff{\xAd_{j+1}}\,    \label{eq:PathInts_BackwardKramersMoyal_OnsagerMachlupConv}
          \gaussianDistribution_{\mu_{j}, \sigma_{j}^2}(\xAd_{j+1} - \xAd_{j}) \Bigr)\delta(\xAd - \xAd_N)  \,.
      \end{equation} 
      The Dirac delta function is included in the integrations. Moreover, $\mu_j \coloneqq \alpha_{t_{j}}(\xAd_{j})\Delta t$ acts as the mean and $\sigma_j^2 \coloneqq \beta_{t_{j}}(\xAd_{j})\Delta t$ as the variance of the Gaussian distribution
      \begin{equation}
        \gaussianDistribution_{\mu, \sigma^2}(\xAd)   \label{eq:PathInts_BackwardKramersMoyal_Gaussian}
        \coloneqq \frac{\ee^{-(\xAd-\mu)^2/(2\sigma^2)}}{\sqrt{2\pi\sigma^2}}    \,.
      \end{equation}
      As before, $\Delta t = (t-t_0)/N$ denotes the time intervals between $t_0 \leq t_1 \leq \hldots \leq t_N  = t$. 
      
      Upon identifying $\xAd(t_0 + j \Delta t)$ with $\xAd_j$, the representation~\eref{eq:PathInts_BackwardKramersMoyal_OnsagerMachlupConv} can be rewritten as
      \begin{equation}
        \prob(t, \xAd | t_0, \xAd_0 )    \label{eq:PathInts_BackwardKramersMoyal_OnsagerMachlup}
        =  \pathintegral{(t_0}{t]}
          \ee^{-\actionAd} 
          \, \delta(\xAd - \xAd(t))  \big|_{\xAd(t_0)=\xAd_0}    \,.
      \end{equation}
      This integral proceeds only over paths $\xAd(\tVar)$ with $\tVar \in (t_0,t]$ because the $\qAd$-variables have already been integrated over. Paths are weighed by the exponential factor $\ee^{-\actionAd}$ with the action
      \begin{equation} 
        \actionAd    \label{eq:PathInts_BackwardKramersMoyal_OnsagerMachlupAction}
        \coloneqq \lim_{N\to\infty}
           \sum_{j=0}^{N-1}  \frac{\bigl[\xAd_{j+1} - \xAd_{j} - \alpha_{t_{j}}(\xAd_{j})\Delta t\bigr]^2}{2 \beta_{t_{j}}(\xAd_{j})\Delta t}      \,. 
      \end{equation} 
      One may abbreviate the continuous-time limit of this action by the integral
      \begin{equation}
        \actionAd 
        =  \int_{t_0}^t \diff{\tVar} \, 
          \frac{\bigl[\partial_\tVar \xAd - \alpha_\tVar(\xAd)\bigr]^2}{2\beta_\tVar(\xAd)}
        \,.
      \end{equation} 
      The integrand of this action is called an Onsager-Machlup function~\cite{Onsager:1953,Horsthemke:1975} and the representation~\eref{eq:PathInts_BackwardKramersMoyal_OnsagerMachlup} may, consequently, be called an Onsager-Machlup representation (or ``functional'' in the sense of functional integration). Note that the Onsager-Machlup representation involves the limit
      \begin{equation}
        \pathintegral{(t_0}{t]} 
        \coloneqq \lim_{N\to\infty} 
          \prod_{j=0}^{N-1} \int_\reals \frac{\diff{\xAd_{j+1}}}{\sqrt{2\pi\beta_{t_{j}}(\xAd_{j})\Delta t}}    \,.
      \end{equation}
      Mathematically rigorous formulations of the Onsager-Machlup representation have been attempted in~\cite{Durr:1978,Ito:1978,Takahashi:1981}. For the other path integrals discussed in this review, such attempts have not yet been made to our knowledge. 
      
    \subsubsection{Alternative discretization schemes.}\label{subsubsec:PathInts_BackwardKramersMoyal_Discretization}  
    
      Up to this point, we have discretized time in such a way that the evaluation of the path integrals eventually proceeds via the solution of an It\^{o} stochastic differential equation (cf.~\sref{subsec:PathInts_BackwardAlongPaths}). Alternative discretization schemes have been proposed as well and are, for example, employed in stochastic thermodynamics~\cite{Seifert:2012}. To illustrate these schemes, let us, for simplicity, assume that the drift coefficient $\alpha(\xAd)$ does not depend on time and that the diffusion coefficient $\rateCoeffDiffCont \coloneqq \beta_\tVar$ is constant. A general discretization scheme --- called the $\alpha$-scheme but here we use a $\kappa$ --- consists of shifting the argument $\xAd_j$ of the drift coefficient in the action~\eref{eq:PathInts_BackwardKramersMoyal_OnsagerMachlupAction} to $\bar{\xAd}_j \coloneqq \kappa\, \xAd_{j+1} + (1-\kappa) \xAd_j$ by writing
      \begin{equation}
        \xAd_j = \bar{\xAd}_j - \kappa(\xAd_{j+1}-\xAd_j) 
      \end{equation}
      (with $\kappa \in [0,1]$). Afterwards, the drift coefficient is expanded in powers of $\xAd_{j+1}-\xAd_j$, which is of order $\sqrt{\Delta t}$ along relevant paths. Following~\cite{ZinnJustin:2002}, it suffices to keep only the first two terms of the expansion of $\alpha(\xAd_j)$ so that
      \begin{eqnarray}
        \xAd_{j+1} - \xAd_{j} - \alpha(\xAd_{j})\Delta t    \\
        \approx \bigl[1 + \kappa\, \alpha^\prime(\bar{\xAd}_j) \Delta t\bigr] (\xAd_{j+1} - \xAd_{j}) - \alpha(\bar{\xAd}_j)\Delta t  \,.  \nonumber
      \end{eqnarray}
      As the next step, the integration variables $\xAd_1$, \dots, $\xAd_N$ are transformed according to
      \begin{equation}
        \bigl[1 + \kappa\, \alpha^\prime(\bar{\xAd}_j) \Delta t\bigr] (\xAd_{j+1} - \xAd_{j})
        \mapsto
        \xAd_{j+1} - \xAd_{j}  \,.
      \end{equation}
      The Jacobian of this transformation vanishes everywhere except on its diagonal and sub-diagonal. Its determinant therefore contributes an additional factor to the path integral and requires a redefinition of the action~\eref{eq:PathInts_BackwardKramersMoyal_OnsagerMachlupAction} as
      \begin{equation*} 
        \actionAd  
        \coloneqq \lim_{N\to\infty}
           \sum_{j=0}^{N-1}
           \Bigl(
             \frac{\bigl[\xAd_{j+1} - \xAd_{j} - \alpha(\bar{\xAd}_j)\Delta t\bigr]^2}{2 D \Delta t}  
             + \kappa \, \alpha^\prime(\bar{\xAd}_j) \Delta t
           \Bigr)    \,.      
      \end{equation*} 
      One may abbreviate this limit by
      \begin{equation}
        \actionAd 
        =  \int_{t_0}^t \diff{\tVar} \, 
          \Bigl(
            \frac{\bigl[\partial_\tVar \xAd - \alpha(\xAd)\bigr]^2}{2D}
            + \kappa \, \alpha^\prime(\xAd)
          \Bigr)
        \,.      
      \end{equation} 
      The It\^{o} version of this action with $\kappa = 0$ has proved to be convenient in perturbation expansions of path integrals because so called ``closed response loops'' can be omitted right from the start (see section 4.5 in~\cite{Taeuber:2014}). Moreover, this prescription does not require a change of variables and makes the connection to the Feynman-Kac formula in \aref{sec:A_FeynmanKacProof} most apparent. The Stratonovich version of the action with $\kappa = \frac{1}{2}$ is, for example, employed in Seifert's review on stochastic thermodynamics~\cite{Seifert:2012}. For a recent, more thorough discussion of the above discretization schemes, as well as of conflicting approaches, we refer the reader to~\cite{Tang:2014} (the above action is discussed in the appendices).

    \subsubsection{Wiener's path integral.}\label{subsubsec:PathInts_BackwardKramersMoyal_Wiener}  

      Before returning to processes with a discrete state space, let us briefly note how the Onsager-Machlup representation~\eref{eq:PathInts_BackwardKramersMoyal_OnsagerMachlup} relates to Wiener's path integral of Brownian motion~\cite{Wiener:1921a,Wiener:1921b}, as it is discussed in~\cite{Chaichian:2001}. It turns out that the Onsager-Machlup representation in fact coincides with that path integral for one-dimension Brownian motion, for which the drift coefficient vanishes and the diffusion coefficient $D=\beta_\tVar$ is constant. Since the convolution of two Gaussian distributions is again a Gaussian distribution, with means and variances being summed, the solution of the process follows as $\prob(t, \xAd | t_0, \xAd_0 )= \gaussianDistribution_{0, \rateCoeffDiffCont (t-t_0)}(\xAd-\xAd_0)$. 
            
  \subsection{Further exact solutions and perturbation expansions}\label{subsec:PathInts_BackwardSolutions}  
    
      After this detour to processes with continuous state spaces, let us return to the evaluation of the backward path integral representation~\eref{eq:PathInts_Backward_ContinuousTime_Solution_MarginalizedDistribution}. In the following, we show how the method introduced in \sref{subsec:PathInts_BackwardAlongPaths} can be applied to several elementary jump processes. We already applied a simplified version of the method in \sref{subsec:PathInts_SimpleGrowthLinearDecay} to solve the bi-directional reaction $\emptyset \rightleftharpoons A$. We now consider a generic reaction $k\, A \to l\, A$ with rate coefficient $\rateCoeffGeneric_\tVar$. Using the Poisson basis function $\ketA{n}_{\xAd} = \frac{\xAd^n \ee^{-\xAd}}{n!}$, the adjoint transition operator of this reaction can be written as  (cf.\ \sref{subsec:GenFctnl_BackwardBases_ChemReactions})
      \begin{eqnarray}
        &\transEvoOpAd_\tVar(\xAd, \partial_\xAd)  
        = \transitionOpAd(\cre, \ann) 
        = \rateCoeffGeneric_\tVar \cre^k (\ann^l - \ann^k)  \\
        &=  \bigl[\rateCoeffGeneric_\tVar (l-k) \xAd^k\bigr]  \partial_\xAd
            + \bigl[\rateCoeffGeneric_\tVar\bigl(l(l-1) - k(k-1)\bigr) \xAd^k\bigr]  \frac{\partial_\xAd^2}{2}    \nonumber\\
            &
            \phantom{{}={}}+ \perturbationOpAd_\tVar(\xAd, \partial_\xAd)        \label{eq:PathInts_BackwardSolutions_Chemical_Hamiltonian} \,.
      \end{eqnarray} 
      Here we employed the creation operator $\cre = \xAd$ and the annihilation operator $\ann = \partial_{\xAd} + 1$, and we performed a series expansion with respect to $\partial_\xAd$. Terms of third and higher order in $\partial_\xAd$ were shoved into the perturbation operator $\perturbationOpAd_\tVar(\xAd, \partial_\xAd)$. In the following, we approximate the marginalized distribution of the reaction $k\, A \to l\, A$ by first dropping both the diffusion coefficient and the perturbation operator from~\eref{eq:PathInts_BackwardSolutions_Chemical_Hamiltonian}. Afterwards, we reintroduce the diffusion coefficient and show how the marginalized distribution follows as the average of a Poisson distribution over the paths of an It\^{o} SDE. For the pair generation process $\emptyset \to 2\, A$, this representation is exact because the perturbation operator associated to this process is zero (cf.\ \sref{subsubsec:PathInts_BackwardSolutions_PairGeneration}). Later, in \sref{subsubsec:PathInts_BackwardSolutions_DiffusionAndDecay}, we solve a process with a non-vanishing perturbation operator $\perturbationOpAd_\tVar$.
      
      As a first approximation of the reaction $k\, A \to l\, A$, let us drop all the terms of the adjoint transition operator~\eref{eq:PathInts_BackwardSolutions_Chemical_Hamiltonian} except for the drift coefficient $\alpha_\tVar(\xAd) = \rateCoeffGeneric_\tVar (l-k) \xAd^k$. The SDE~\eref{eq:PathInts_BackwardAlongPaths_SDE} then simplifies to the deterministic rate equation $\partial_\tVar \xAd = \rateCoeffGeneric_\tVar (l-k) \xAd^k$ of the reaction. Its solution $\xAd(\tVar)$ acts as the mean of the marginalized distribution, which, according to~\eref{eq:PathInts_BackwardAlongPaths_MarginalizedDistribution}--\eref{eq:PathInts_BackwardAlongPaths_PhiXGeneratingFunction_0}, is again given by the Poisson distribution $\ketA{\prob(t, n | t_0)}_{\xAd(t_0)} =   \ketA{n}_{t, \xAd(t)}$ in~\eref{eq:PathInts_SimpleGrowthLinearDecay_PoissonDist}. 
      
      Going one step further, one may keep the diffusion coefficient in~\eref{eq:PathInts_BackwardSolutions_Chemical_Hamiltonian}. The marginalized distribution then reads
      \begin{equation}
        \ketA{\prob(t, n|t_0)}_{\xAd(t_0)}     \label{eq:PathInts_BackwardSolutions_Chemical_PoissonDist}
        = \BigLLangle 
            \frac{\xAd(t)^n \ee^{-\xAd(t)}}{n!}
          \BigRRangle_{\wienerProcess} 
      \end{equation} 
      with $\xAd(\tVar)$ solving the It\^{o} SDE
      \begin{eqnarray}
        \diff{\xAd}       \label{eq:PathInts_BackwardSolutions_Chemical_SDE}
        =& \rateCoeffGeneric_\tVar (l-k) \xAd^k \diff{\tVar}\\  
          &  + \sqrt{\rateCoeffGeneric_\tVar \bigl[l(l-1) - k(k-1)\bigr] \xAd^k}\,\diff{\wienerProcess(\tVar)}    \nonumber \,.           
      \end{eqnarray} 
      Hence, the Poisson distribution in~\eref{eq:PathInts_BackwardSolutions_Chemical_PoissonDist} is averaged over all possible sample paths of the SDE~\eref{eq:PathInts_BackwardSolutions_Chemical_SDE}, whose evolution starts out from the initial value $\xAd(t_0)$ (cf.~\eref{eq:GenFctnl_BackwardBases_ChemReactions_WienerAverage}). The above expression has also recently been derived by Wiese~\cite{Wiese:2015}, although based on the forward path integral that we will discuss in \sref{sec:PathInts_Forward}. Apparently, the expression under the square root in~\eref{eq:PathInts_BackwardSolutions_Chemical_SDE} is non-negative only if  $k \in \{0,1\}$ or if $l \geq k\geq 2$. If that is not the case, $\xAd(\tVar)$ strays off into the complex domain. SDEs with imaginary noise (or the corresponding Langevin equations) have been studied in several recent articles, most notably for the binary annihilation reaction $2\, A \to \emptyset$ and for the coagulation process $2\, A \to A$~\cite{HowardMJ:1997,Hochberg:2006,Wiese:2015}. The numerical evaluation of such SDEs, however, often encountered severe convergence problems~\cite{Deloubriere:2002,Wiese:2015}. The appearance of imaginary noise has been linked to the anti-correlation of particles in spatial systems~\cite{HowardMJ:1997}, but as~\eref{eq:PathInts_BackwardSolutions_Chemical_SDE} shows, no spatial degrees of freedoms are actually required for its emergence. As Wiese pointed out, imaginary noise generally appears when, over time, the marginalized distribution~\eref{eq:PathInts_BackwardSolutions_Chemical_PoissonDist} becomes narrower than a Poisson distribution~\cite{Wiese:2015}.

    \subsubsection{Pair generation.}\label{subsubsec:PathInts_BackwardSolutions_PairGeneration} 
    
\begin{figure}[tb] 
        \centering
        \includegraphics{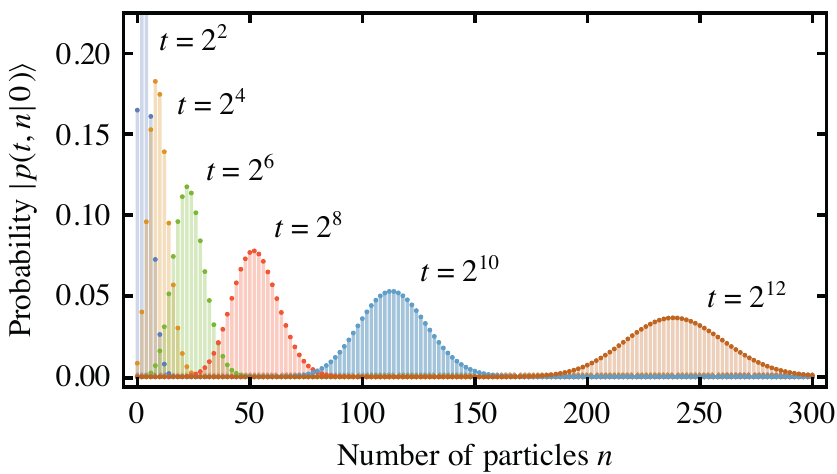}
        \caption{\label{fig:PairGeneration}
          Evolution of the marginalized distribution $\ketA{\prob(t,n|t_0)}$ in~\eref{eq:PathInts_BackwardSolutions_PairGeneration_Solution}, which solves the pair generation process $\emptyset \to 2\, A$ (with $t_0\coloneqq 0$). The initial mean of the distribution was chosen as $\xAd(t_0) = 0$ and the rate coefficient of the reaction as $\rateCoeffSpontGrowth_\tVar = 1/(1+\sqrt{\tVar})$.
        }
      \end{figure}

      For the pair generation process $\emptyset \to 2\, A$ with growth rate coefficient $\rateCoeffSpontGrowth_\tVar$, the drift and the (squared) diffusion coefficients agree: $\alpha_\tVar=\beta_\tVar=2 \rateCoeffSpontGrowth_\tVar$. As neither of them depends on $\xAd(\tVar)$, the It\^{o} SDE is readily solved by $\xAd(t) = \xAd(t_0) + \int_{t_0}^t\diff{\tVar} 2\rateCoeffSpontGrowth_\tVar + \int_{t_0}^t\diff{\wienerProcess(\tVar)}\sqrt{2\rateCoeffSpontGrowth_\tVar}$. After introducing the (rather daunting) parameter
      \begin{eqnarray}
        \eta_k &\coloneqq     \label{eq:PathInts_BackwardSolutions_PairGeneration_Coeff}
            \BigLLangle 
          \ee^{-(\int_{t_0}^t\diff{\tVar}\, \rateCoeffSpontGrowth_\tVar + \int_{t_0}^t\diff{\wienerProcess(\tVar)}\sqrt{2\rateCoeffSpontGrowth_\tVar})}   \\
          &\phantom{{}\coloneqq}
          \cdot \biggl(
            \Bigl(\int_{t_0}^t\diff{\tVar}\, 2 \rateCoeffSpontGrowth_\tVar\Bigr)^{1/2} + \frac{ \int_{t_0}^t\diff{\wienerProcess(\tVar)}\sqrt{2\rateCoeffSpontGrowth_\tVar}}{\Bigl(\int_{t_0}^t\diff{\tVar}\, 2 \rateCoeffSpontGrowth_\tVar\Bigr)^{1/2}}
          \biggr)^k
          \BigRRangle_{\wienerProcess}  \nonumber  \,,
      \end{eqnarray} 
      one can employ the binomial theorem to rewrite the right-hand side of the marginalized distribution~\eref{eq:PathInts_BackwardSolutions_Chemical_PoissonDist} as
      \begin{equation*}
        \frac{\ee^{-(\xAd(t_0) + \int_{t_0}^t\diff{\tVar}\, \rateCoeffSpontGrowth_\tVar)} }{n!} 
        \sum_{k=0}^n \binomCoeff{n}{k}
        \Bigl(\int_{t_0}^t\diff{\tVar}\, 2 \rateCoeffSpontGrowth_\tVar\Bigr)^{k/2}
        \eta_k
        \, x(t_0)^{n-k}       \,.    
      \end{equation*} 
      If the rate coefficient $\rateCoeffSpontGrowth$ is independent of time, one can show that the parameter $\eta_k$ coincides with the $k$-th moment of a Gaussian distribution with zero mean and unit variance (i.e.\ $\eta_k = 0$ for odd $k$ and $\eta_k = (k-1)!!$ for even $k$; note that the sum of two Gaussian random variables is again a Gaussian random variable, with its mean and variance following additively). The marginalized distribution $\ketA{\prob(t,n|t_0)}_{\xAd(t_0)}$ therefore follows as
      \begin{equation}
          \ee^{-(\xAd(t_0) + \int_{t_0}^t\diff{\tVar}\, \rateCoeffSpontGrowth_\tVar)}  \label{eq:PathInts_BackwardSolutions_PairGeneration_Solution}
            \sum_{k=0}^{\lfloor n/2 \rfloor}
            \frac{\bigl(\int_{t_0}^t\diff{\tVar}\, \rateCoeffSpontGrowth_\tVar\bigr)^{k}}{k!}
            \, \frac{(\xAd(t_0))^{n-2k}}{(n-2k)!}     \,.
      \end{equation} 
      Here, $\lfloor n/2 \rfloor$ represents the integral part of $n/2$. The marginalized distribution~\eref{eq:PathInts_BackwardSolutions_PairGeneration_Solution} solves the master equation of the pair generation process and is initially of Poisson shape. For $\xAd(t_0) = 0$, the distribution effectively keeps that shape, although only on the set of all even numbers (the reaction $\emptyset \to 2\, A$ cannot create an odd number of particles when starting out from zero particles). Using Mathematica by Wolfram Research, we verified that the distribution also applies when $\rateCoeffSpontGrowth_\tVar$ depends on time. The evolution of the distribution is shown in \fref{fig:PairGeneration} for the rate coefficient $\rateCoeffSpontGrowth_\tVar = 1/(1+\sqrt{\tVar})$. We refrain from computing the conditional distribution from the marginalized distribution because the computation is unwieldy and does not shed any more light on the path integral approach.
    
    \subsubsection{Diffusion on networks.}\label{subsubsec:PathInts_BackwardSolutions_DiffusionNetwork} 
    
      The backward path integral can also be used to solve spatial processes. In fact, stochastic path integrals have been most useful in the study of such processes. If particles engaging in a chemical reaction can also diffuse in space, their density may evolve in ways that are not expected from the well-mixed, non-spatial limit. That is, for example, the case for particles that annihilate one another in the reaction $2\, A \to \emptyset$ while diffusing along a one-dimensional line. From the well-mixed limit, one would expect that the particle density decays asymptotically as $t^{-1}$ with time (see \sref{sec:StationaryPaths}), but instead it decays as $t^{-1/2}$~\cite{Torney:1983a,Peliti:1986,Lee:1994b}. The reason behind this surprising decay law is the rapid condensation of the system into a state in which isolated particles are separated by large voids. From that point on, further decay of the particle density requires that two particles first find each other through diffusion. Therefore, the process is ``diffusion-limited'' and exhibits a slower decay of the particles (see e.g.~\cite{Redner:2001} on the related ``first-passage'' problem). Path integral representations of the master equation, combined with renormalization group techniques, have been successfully applied to the computation of decay laws in spatially extended systems, both regarding the (universal) exponents and the pre-factors of these laws~\cite{Peliti:1986,Jensen:1994,Lee:1994b,Lee:1995,Cardy:1996,Winkler:2012}. For a broader discussion of systems exhibiting a transition into an absorbing state see~\cite{Henkel:2008}. Before turning to the particle density and, more generally, to observables of the particle number, we now show how the backward path integral helps in computing the full probability distribution of a spatial process.
      
      As a first step, we consider a pure diffusion process with particles hopping between neighbouring nodes of a network~$\lattice$. For now, the topology of the network may be arbitrary, being either random or regular. In the next section, the network topology will be chosen as a regular lattice. In the limit of a vanishing lattice spacing (and an infinite number of lattice sites), a ``field theory'' will be obtained. In \sref{subsubsec:PathInts_BackwardSolutions_DiffusionAndDecay}, the particles will also be allowed to decay.
      
      The configuration of particles on the network may be represented by the vector $\vect{n} \in \naturals_0^{|\lattice|}$, with $|\lattice|$ being the total number of network nodes. The configuration changes whenever a particle hops from some node $i\in\lattice$ to a neighbouring node $j \in \neighbors_i \subset \lattice$. The probability $\prob(\tVar, \vect{n} | t_0, \vect{n}_0)$ of finding the system in configuration $\vect{n}$ then obeys the master equation
      \begin{eqnarray}
        \partial_\tVar \prob(\tVar, \vect{n} | \cdot)           
        = \sum_{i\in\lattice} \rateCoeffDiffDisc_{\tVar, i} \sum_{j\in\neighbors_i} &\bigl[ (n_i+1)  \prob(\tVar, \vect{n} + \hat{\vect{e}}_i - \hat{\vect{e}}_j | \cdot)     \nonumber  \\
        & - n_i  \prob(\tVar, \vect{n} | \cdot)  \bigr]        \label{eq:PathInts_BackwardSolutions_DiffusionNetwork_MasterEq} \,.
      \end{eqnarray} 
      Here, $\hat{\vect{e}}_i$ represents a unit vector that points in direction $i\in\lattice$. Moreover, $\rateCoeffDiffDisc_{\tVar, i}$ acts as a hopping rate and may depend both on time and on the node from which a particle departs. Apparently, the above master equation has the same structure as the chemical master equation~\eref{eq:Intro_Mescoscopic_ChemicalMasterEq}. It may therefore be cast into a linear PDE by extending the operators and basis functions from \sref{subsec:GenFctnl_BackwardBases_ChemReactions} to multiple variables. In particular, we extend the Poisson basis function to
      \begin{equation} 
        \ketA{\vect{n}}_{\vect{\xAd}} \coloneqq \prod_{i\in\lattice} \frac{\xAd_i^{n_i} \ee^{-\xAd_i}}{n_i!}   
      \end{equation}
      and employ the creation and annihilation operators
      \begin{equation} 
        \vect{\cre} \coloneqq \vect{\xAd} \mathtext{ and }  
        \vect{\ann} \coloneqq \nabla + \vect{1}     \,. 
      \end{equation}
      According to the flow equation~\eref{eq:GenFctnl_MarginalizedDist_Flow}, the marginalized distribution evolves via $\partial_{-\tVar} \ketA{\prob(t, n | \tVar)}  = \transEvoOpAd_\tVar\ketA{\prob(t, n | \tVar)}$ with the adjoint transition operator
      \begin{equation}
        \transEvoOpAd_\tVar(\vect{\xAd}, \nabla)    \label{eq:PathInts_BackwardSolutions_DiffusionNetwork_Hamiltonian}
         = \sum_{i\in\lattice}  (\discreteLaplaceOp \rateCoeffDiffDisc_{\tVar,i} \xAd_i) \partial_{\xAd_i} \,.
      \end{equation}
      Here we introduced the discrete Laplace operator $\discreteLaplaceOp f_i \coloneqq \sum_{j\in\neighbors_i} (f_j - f_i)$, which acts both on $\rateCoeffDiffDisc_{\tVar,i}$ and on $\xAd_i$.
      
      By following the steps in \sref{subsec:PathInts_BackwardAlongPaths}, one finds that the marginalized distribution is given by the multivariate Poisson distribution
      \begin{equation}
        \ketA{\prob(t, \vect{n} | t_0)}_{\vect{\xAd}(t_0)}      \label{eq:PathInts_BackwardSolutions_DiffusionNetwork_PoissonDist}
        = \prod_{i\in\lattice}\frac{\xAd_i(t)^{n_i} \ee^{-\xAd_i(t)}}{n_i!}   \,.
      \end{equation}
      Its mean $\vect{\xAd}(t)$ solves the discrete diffusion equation
      \begin{equation}
        \partial_\tVar \xAd_i     \label{eq:PathInts_BackwardSolutions_Diffusion_ODEs}
        = \sum_{j\in\neighbors_i} \bigl( \rateCoeffDiffDisc_{\tVar, j} \xAd_j - \rateCoeffDiffDisc_{\tVar, i} \xAd_i\bigr)
        = \discreteLaplaceOp \rateCoeffDiffDisc_{\tVar, i} \xAd_i    \,,
      \end{equation} 
      with the initial condition $\vect{\xAd}(t_0)$. Let us note that the marginalized distribution solves the master equation~\eref{eq:PathInts_BackwardSolutions_DiffusionNetwork_MasterEq}, but with the initial number of particles being Poisson distributed locally. 
      
      The conditional distribution $\prob(t, \vect{n} | t_0, \vect{n}_0)$ follows from the marginalized distribution~\eref{eq:PathInts_BackwardSolutions_DiffusionNetwork_PoissonDist} by applying the functional
      \begin{equation}
        \braA{\vect{n_0}}_{\vect{\xAd}(t_0)} f
        = \Bigl[\prod_{i\in\lattice} \bigl(\partial_{\xAd_i(t_0)} + 1\bigr)^{n_{0,i}} \Bigr] f(\vect{\xAd}(t_0)) \Big|_{\vect{\xAd}(t_0)=\vect{0}}   
      \end{equation}
       to it (cf.~\eref{eq:GenFctnl_Flow_Poisson_BaseFunctional} with~\eref{eq:GenFctnl_Flow_Poisson_Annihilation}). The evaluation requires a prior solution of the discrete diffusion equation~\eref{eq:PathInts_BackwardSolutions_Diffusion_ODEs}. This equation can be written in matrix form as $\partial_\tVar \vect{\xAd} = M_\tVar \vect{\xAd}$ with $M_{\tVar, ij} = (A_{ij} - |\neighbors_i|\delta_{ij}) \rateCoeffDiffDisc_{\tVar, j}$. Here, $A$ represents the symmetric adjacency matrix of the network and $|\neighbors_i|$ represents the number of neighbours of node $i\in\lattice$. The matrix equation can in principle be solved through a Magnus expansion. The solution has the generic form $\vect{\xAd}(\tVar) = \propagator(\tVar | t_0) \vect{\xAd}(t_0)$ with the ``propagator'' $\propagator$ solving $\partial_\tVar \propagator(\tVar | t_0) = M_\tVar \propagator(\tVar | t_0)$. We write the elements of the propagator as $\propagator(\tVar, i | t_0, j)$. Its flow starts out from $\propagator(t_0 | t_0) = \unitMatrix$. For later purposes, let us note the time-reversal property in terms of the matrix inverse $\propagator(t | \tVar)^{-1} = \propagator(\tVar | t)$ and also the conservation law $\vect{1}^\transpose \propagator(\tVar | t_0) = \vect{1}^\transpose$.

      It proves insightful to consider the master equation~\eref{eq:PathInts_BackwardSolutions_DiffusionNetwork_MasterEq} of the multi-particle hopping process for the random walk of a single particle on the one-dimensional lattice $\lattice=\integers$, which we already considered in sections~\ref{subsec:GenFct_ForwardBases_RandomWalk} and~\ref{subsec:GenFctnl_BackwardBases_RandomWalk} (with symmetric hopping rates $\rateCoeffDiffDisc_{\tVar, i} = l_\tVar = r_\tVar$). The presence of only a single particle can be enforced by choosing $\vect{n}_0$ as $n_{0,k}=1$ for one $k\in\integers$ and $n_{0,j}=0$ for all $j\neq k$. By following the above steps, one eventually finds that the master equation~\eref{eq:PathInts_BackwardSolutions_DiffusionNetwork_MasterEq} is solved by the conditional probability distribution
      \begin{equation}
        \prob(t, \vect{n} | t_0, \vect{n}_0)
        = \sum_{i\in\integers} \bigl(\prod_{j\neq i} \delta_{n_j,0}\bigr) \delta_{n_i,1} \propagator(t,i | t_0,k)   \,,
      \end{equation}
      with the propagator $\propagator$ solving the master equation~\eref{eq:GenFct_ForwardBases_RandomWalk_Master} of the simple random walk. Hence, the propagator evaluates to a Skellam distribution.
      
    \subsubsection{Diffusion in continuous space.}\label{subsubsec:PathInts_BackwardSolutions_DiffusionLattice} 
    
      To make the transition to a field theory, one may specify the network as the the $d$-dimensional lattice $\lattice = (l\integers)^d$ with the lattice spacing $l>0$ going to zero. In order to take this limit, we define the variable $\xAd(\tVar, \vect{r})  \coloneqq \xAd_{\vect{r}}(\tVar)$, the rescaled Laplace operator $\laplaceOp \coloneqq \discreteLaplaceOp/l^2$, and the rescaled diffusion coefficient $\rateCoeffDiffCont_\tVar(\vect{r}) \coloneqq l^2 \rateCoeffDiffDisc_{\tVar, \vect{r}}$ for $\vect{r}\in\lattice$. Assuming that these definitions can be continued to $\vect{r}\in\reals^d$, the discrete diffusion equation~\eref{eq:PathInts_BackwardSolutions_Diffusion_ODEs} becomes a PDE for the ``field'' $\xAd(\tVar,\vect{r})$ in the limit $l \to 0$, namely $\partial_\tVar \xAd(\tVar,\vect{r}) = \laplaceOp (\rateCoeffDiffCont_\tVar(\vect{r}) \xAd(\tVar,\vect{r}))$. Here, $\laplaceOp$ represents the ordinary Laplace operator.\footnote{With $\neighbors_{\vect{r}}$ denoting the neighbouring lattice site of $\vect{r} \in (l\integers)^d$, the ordinary Laplace operator follows as the limit of a finite-difference approximation, i.e. as
      \begin{equation}
        \sum_{\vect{r}^\prime \in \neighbors_{\vect{r}}} \frac{f_{\vect{r}^\prime}-f_{\vect{r}}}{l^2} 
        \to \sum_{i=1}^d \partial_{r_i}^2 f(\vect{r})  = \laplaceOp f(\vect{r}) \,.
      \end{equation}
      } The solution of the PDE acts as the mean of the multivariate Poisson distribution~\eref{eq:PathInts_BackwardSolutions_DiffusionNetwork_PoissonDist} whose extension to $\vect{r}\in\reals$ is, however, not quite obvious. If the diffusion coefficient is homogeneous in space, the PDE is solved by
      \begin{equation}
        \xAd(\tVar, \vect{r}) 
        = \int_{\reals^d} \diff{\vect{r}_0}\, \propagator(\tVar, \vect{r} | t_0, \vect{r}_0) \xAd(t_0, \vect{r}_0)   \,,
      \end{equation} 
      with the Gaussian kernel
      \begin{equation}
        \propagator(\tVar, \vect{r} | \tVar^\prime, \vect{r}^\prime)
        = \frac{\ee^{-(\vect{r}-\vect{r}^\prime)^2/4\int_{\tVar^\prime}^\tVar \diff{s}\, \rateCoeffDiffCont_s} }{\sqrt{4\pi \int_{\tVar^\prime}^\tVar \diff{s}\, \rateCoeffDiffCont_s}} \,.
      \end{equation}
      The action~\eref{eq:PathInts_Backward_ContinuousTime_Action} can also be extended into continuous space. For that purpose, one may rescale the second path integral variable as $\qAd(\tVar,\vect{r})  \coloneqq l^{-d} \qAd_{\vect{r}}(\tVar)$ so that in the limit $l \to 0$,
      \begin{eqnarray}
          \actionAd  
            &=  \int_{t_0}^{t} \diff{\tVar}\, \sum_{\vect{r}\in\lattice} 
              \Bigl(
                \ii\qAd_{\vect{r}}  \bigl( \partial_\tVar \xAd_{\vect{r}}  - \discreteLaplaceOp \rateCoeffDiffDisc_{\tVar,\vect{r}} \xAd_{\vect{r}}\bigr)   
              \Bigr)      \\      
        &\to \int_{t_0}^{t} \diff{\tVar}\, \int_{\reals^d} \diff{\vect{r}}\,
          \ii\qAd \bigl(
             \partial_\tVar \xAd 
            - \laplaceOp \rateCoeffDiffCont_\tVar  \xAd
          \bigr)\,.                             
      \end{eqnarray} 

      The above rescaling of $\xAd$ and $\qAd$ is not unique and depends on the problem at hand. Instead of dividing $\qAd_{\vect{r}}$ by the volume factor $l^{d}$, this factor is sometimes employed to cast $\xAd_{\vect{r}}$ and the particle number $n_{\vect{r}}$ into densities~\cite{Lee:1994b}. In the study of branching and annihilating random walks with an odd number of offspring, yet another kind of rescaling may bring the action into a Reggeon field theory~\cite{Moshe:1978,Frey:1994a,Frey:1994b,Taeuber:2014} like form~\cite{Cardy:1998}. The critical behaviour of these random walks falls into the universality class of directed percolation (DP)~\cite{Cardy:1996,Cardy:1998}. Information on this universality class can be found in the book~\cite{Henkel:2008}, as well as in the original articles of Janssen and Grassberger, which established the extensive scope of the DP class~\cite{Janssen:1981,Grassberger:1982}.
      
    \subsubsection{Diffusion and decay.}\label{subsubsec:PathInts_BackwardSolutions_DiffusionAndDecay} 
    
      Let us exemplify how one can accommodate a non-vanishing perturbation operator $\perturbationOpAd_\tVar$ in the adjoint transition operator~\eref{eq:PathInts_BackwardAlongPaths_Hamiltonian}. As in the section before the last, we consider a system of particles that are hopping between the nodes of an arbitrary network $\lattice$. The corresponding diffusive transition operator~\eref{eq:PathInts_BackwardSolutions_DiffusionNetwork_Hamiltonian} shall specify the drift coefficient $\alpha_{\tVar,i}(\vect{\xAd}) = \discreteLaplaceOp \rateCoeffDiffDisc_{\tVar,i} \xAd_i$ in the multivariate extension of~\eref{eq:PathInts_BackwardAlongPaths_Hamiltonian}. The coefficient $\beta_{\tVar,ij}(\vect{\xAd})$ is zero. Besides allowing for diffusion, we now allow the particles to decay in the linear reaction $A \to \emptyset$. For the sake of brevity, we assume that the corresponding decay rate coefficient $\rateCoeffLinDecay_{\tVar}$ is spatially homogeneous. This assumption can be relaxed but the equations that result cannot be written in matrix form and involve many indices. We treat the adjoint transition operator of the decay process as the perturbation, which reads, according to the flow equation~\eref{eq:GenFctnl_MarginalizedDist_Flow_Decay},
      \begin{equation}
        \perturbationOpAd_\tVar(\vect{\xAd}, \nabla)    \label{eq:PathInts_BackwardSolutions_DiffusionAndDecay_Perturbation}
         = -\rateCoeffLinDecay_{\tVar} \sum_{i\in\lattice} \xAd_i \partial_{\xAd_i}  \,.
      \end{equation}  
      The combined process could also be solved directly by treating the decay process as part of the drift coefficient $\alpha_{\tVar,i}(\vect{\xAd})$. For pedagogic reasons, however, we wish to outline a perturbative solution using Feynman diagrams. 
      
      The first step of the derivation is to solve the differential equation~\eref{eq:PathInts_BackwardAlongPaths_SDE} for $\vect{\xAd}$, which now reads $\partial_\tVar \xAd_i  = \discreteLaplaceOp \rateCoeffDiffDisc_{\tVar,i} \xAd_i + \XAd_i(\tVar)$. The homogeneous solution of this equation is given by $\vect{\xAd}_h(\tVar) \coloneqq \propagator(\tVar | t_0) \vect{\xAd}(t_0)$, with $\propagator$ being the propagator from the end of \sref{subsubsec:PathInts_BackwardSolutions_DiffusionNetwork}. The solution of the full equation can be written as
      \begin{equation}
        \vect{\xAd}(\tVar)     \label{eq:PathInts_BackwardSolutions_DiffusionAndDecay_DiffEqSol}
        = \vect{\xAd}_h(\tVar) + \int_{t_0}^\tVar \diff{\tVar^\prime} \propagator(\tVar | \tVar^\prime) \vect{\XAd}(\tVar^\prime)  \,.
      \end{equation}
      As the next step, the evaluation of the marginalized distribution~\eref{eq:PathInts_BackwardAlongPaths_MarginalizedDistribution} requires us to compute the $(\vect{\QAd}, \vect{\XAd})$-generating functional~\eref{eq:PathInts_BackwardAlongPaths_PhiXGeneratingFunction} for $\vect{\QAd} = \vect{\XAd} = \vect{0}$. By performing a series expansion of its leading exponential, this function can be written as
      \begin{eqnarray}
        \fieldGenFctAd_{\vect{0},\vect{0}}        \label{eq:PathInts_BackwardSolutions_DiffusionAndDecay_Expansion}
        =&   \sum_{k=0}^\infty   \sum_{l=0}^k  
            \frac{1}{l!} \Bigl( \ii\vect{\qAd}_N \cdot  \frac{\delta}{\delta \vect{\QAd}(t)} \Bigr)^l  
            \\&
            \cdot \int_{t_0}^t \diff{\tVar_{k-l}}\, \perturbationOpAd_{\tVar_{k-l}}
            \cdots  
            \int_{t_0}^{\tVar_2} \diff{\tVar_{1}}\, \perturbationOpAd_{\tVar_1}
            \, \fieldGenFctAd^0_{\vect{\QAd},\vect{\XAd}}  
            \big|_{\vect{\QAd}=\vect{\XAd}=\vect{0}}
            \nonumber      \,,
      \end{eqnarray}
      with $\perturbationOpAd_\tVar=  \frac{\delta}{\delta \vect{\XAd}(\tVar)} \cdot (-\rateCoeffLinDecay_{\tVar})\frac{\delta}{\delta \vect{\QAd}(\tVar)}$. To evaluate $\fieldGenFctAd_{\vect{0},\vect{0}}$, it helps to note that $\ln \fieldGenFctAd^0_{\vect{\QAd},\vect{\XAd}}  =  \int_{t_0}^t\! \diff{\tVar} \, \vect{\QAd}\cdot \vect{\xAd}$, from which it follows that for $t_0 \leq \tVar \leq t$:  
      \begin{eqnarray}
        \frac{\delta \ln \fieldGenFctAd^0_{\vect{\QAd},\vect{\XAd}}}{\delta \vect{\QAd}(\tVar)} 
        \bigg|_{\vect{\QAd}=\vect{\XAd}=\vect{0}}
        = \vect{\xAd}_h(\tVar)
        \label{eq:PathInts_BackwardSolutions_DiffusionAndDecay_Diff1}
        \mathtext{ and }\\
        \frac{\delta^2 \ln \fieldGenFctAd^0_{\vect{\QAd},\vect{\XAd}}}{\delta \QAd_i(\tVar) \delta \XAd_j(\tVar^\prime)}
         \bigg|_{\vect{\QAd}=\vect{\XAd}=\vect{0}}
        = \propagator(\tVar,i | \tVar^\prime,j) \HeavisideStep(\tVar - \tVar^\prime)   \,.
        \label{eq:PathInts_BackwardSolutions_DiffusionAndDecay_Diff2}
      \end{eqnarray}
      The Heaviside step function is used with $\HeavisideStep(0) \coloneqq 0$ to take into account that $\vect{\xAd}_h(\tVar)$ depends on $\vect{\XAd}(\tVar^\prime)$ only for $\tVar^\prime < \tVar$ (see \aref{sec:A_BackwardAlongPaths}). All other derivatives of $\ln \fieldGenFctAd^0_{\vect{\QAd},\vect{\XAd}}$, as well as itself, vanish for $\vect{\QAd}=\vect{\XAd}=\vect{0}$.  
  
      Upon inserting the perturbation~\eref{eq:PathInts_BackwardSolutions_DiffusionAndDecay_Perturbation} into~\eref{eq:PathInts_BackwardSolutions_DiffusionAndDecay_Expansion}, one observes that every summand of the expansion can be written in terms of a combination of the derivatives~\eref{eq:PathInts_BackwardSolutions_DiffusionAndDecay_Diff1} and~\eref{eq:PathInts_BackwardSolutions_DiffusionAndDecay_Diff2}, and a terminal factor $\ii\vect{\qAd}_N$.
      For example, the summand with $k=2$ and $l=1$ reads
      \begin{equation}
            \ii\vect{\qAd}_N \cdot
            \frac{\delta}{\delta \vect{\QAd}(t)} 
            \int_{t_0}^t \diff{\tVar_{1}}\, 
            \frac{\delta}{\delta \vect{\XAd}(\tVar_1)} \cdot (-\rateCoeffLinDecay_{\tVar_1})\frac{\delta}{\delta \vect{\QAd}(\tVar_1)}
            \, \fieldGenFctAd^0_{\vect{\QAd},\vect{\XAd}}  
            \,,  
      \end{equation}
      with $\vect{\QAd}=\vect{\XAd}=\vect{0}$ being taken after the evaluation of the functional derivatives. The evaluation of these derivatives results in
      \begin{equation}
          \ii\vect{\qAd}_N \cdot  \int_{t_0}^t \diff{\tVar_1}\, \propagator(t | \tVar_1) (-\rateCoeffLinDecay_{\tVar_1})  \vect{\xAd}_h(\tVar_1) \,.
            \label{eq:PathInts_BackwardSolutions_DiffusionAndDecay_Feynman}
      \end{equation}
      One can represent this expression graphically by a Feynman diagram according to the following rules. First, every diagram ends in a sink 
      \mbox{
        \begin{tikzpicture}[baseline=-0.55ex, node distance=0.6cm]
          \coordinate[vertex] (sink);
          \coordinate[coordinate, right=of sink] (source);
          \draw[lineWithArrowInline] (source) -- (sink);
        \end{tikzpicture}
        ,
      }
      which contributes the factor $\ii\vect{\qAd}_N$. The incoming line, or ``leg'', of the sink represents the derivative $\frac{\delta}{\delta \vect{\QAd}(t)}$. This leg may either be left dangling, resulting in a factor $\vect{\xAd}_h(t)$ according to~\eref{eq:PathInts_BackwardSolutions_DiffusionAndDecay_Diff1}, or it may be ``contracted'' with the outgoing line of a vertex 
      \mbox{
        \begin{tikzpicture}[baseline=-0.55ex, node distance=0.6cm]
          \coordinate[vertex] (dualVertex);
          \coordinate[coordinate, left=of dualVertex] (sink);
          \coordinate[coordinate, right=of dualVertex] (source);
          \draw[lineWithArrowInline] (dualVertex) -- (sink);
          \draw[lineWithArrowInline] (source) -- (dualVertex);
        \end{tikzpicture}
        .
      }
      The two legs of this vertex reflect the two derivatives in the perturbation $\perturbationOpAd_\tVar=  \frac{\delta}{\delta \vect{\XAd}(\tVar)} \cdot (-\rateCoeffLinDecay_{\tVar})\frac{\delta}{\delta \vect{\QAd}(\tVar)}$. According to~\eref{eq:PathInts_BackwardSolutions_DiffusionAndDecay_Diff2}, the contraction results in a propagator $\propagator(t,i | \tVar,j)$ and the vertex itself contributes a factor $-\rateCoeffLinDecay_{\tVar}$. The incoming leg of the vertex may again be left dangling or it may be connected to another vertex. Therefore, each Feynman diagram is a straight line for the linear decay process. The expression~\eref{eq:PathInts_BackwardSolutions_DiffusionAndDecay_Feynman} can therefore be represented graphically by
      \begin{center}
        \begin{equation}
          \begin{tikzpicture}[node distance=2.5cm,baseline=(current  bounding  box.center)]
            \coordinate[vertex, label=below: $i \vect{\qAd}_{N} \cdot$] (sink);
            \coordinate[vertex, right=of sink, label=below: $-\rateCoeffLinDecay_{\tVar_1}$] (dualVertex);
            \coordinate[right=of dualVertex] (source);
            \draw[lineWithArrow2] (dualVertex) -- (sink) node[midway,above=0.1cm] {$\int_{t_0}^t \diff{\tVar_1} \propagator(t|\tVar_1)$} ;
            \draw[lineWithArrow2] (source) -- (dualVertex) node[midway,above=0.17cm] {$\vect{\xAd}_{h}(\tVar_1)$} ;
            \coordinate[right=of source, xshift=-2.25cm,label=right:.] (period);
          \end{tikzpicture}
        \end{equation}
      \end{center}
      Note that dangling outgoing lines are not permitted because the derivative $\delta \ln \fieldGenFctAd^0_{\vect{\QAd},\vect{\XAd}}/\delta \vect{\XAd}(\tVar)$ vanishes for $\vect{\QAd}=\vect{\XAd}=\vect{0}$.
      
      For more complex, non-linear processes, the Feynman diagrams may contain multiple kinds of vertices, each representing an individual summand of $\perturbationOpAd_\tVar$. If there exist vertices with more than two legs, the diagrams may exhibit internal loops (see \sref{subsec:PathInts_Observables_Binary}). It is then usually impossible to evaluate the full perturbation series and it needs to be truncated at a certain order in the number of loops. Further information about these techniques, and about how renormalization group theory comes into play, is provided, for example, by the book of T\"auber~\cite{Taeuber:2014}. Information on a non-perturbative renormalization group technique can be found in~\cite{Canet:2011}.
      
      For the simple diffusion process with decay, all the summands of~\eref{eq:PathInts_BackwardSolutions_DiffusionAndDecay_Expansion} are readily cast into Feynman diagrams. However, it turns out that individual summands of the expansion may be associated to multiple diagrams that are not connected to one another. Furthermore, diagrams that represent summands of lower order keep reappearing as unconnected components of summands of higher order. Hence, there appears to be redundant information involved. This redundancy is removed by a classical theorem from diagrammatic analysis. This theorem states that the logarithm $\ln \fieldGenFctAd_{\vect{0},\vect{0}}$ is given by the sum of only the connected diagrams~\cite{ZinnJustin:2007}, i.e.\ by
      \begin{center}
        \begin{equation} 
          \begin{tikzpicture}[node distance=0.707cm and 1.cm,baseline=(current  bounding  box.center)]
            \coordinate[vertex,label={[xshift=-0.92cm,yshift=-0.36cm]$\ln \fieldGenFctAd_{\vect{0},\vect{0}} =$}] (sink1);
            \coordinate[right=of sink1,label={[xshift=0.28cm,yshift=-0.24cm]$+$}] (source1);
            \draw[lineWithArrow1] (source1) -- (sink1);
            \coordinate[vertex,right=of source1,xshift=-0.4cm] (sink2);
            \coordinate[vertex,right=of sink2] (dualVertex2);
            \coordinate[right=of dualVertex2] (source2);
            \draw[lineWithArrow1] (dualVertex2) -- (sink2);
            \draw[lineWithArrow1] (source2) -- (dualVertex2);
            \coordinate[vertex,below=of sink1,xshift=0.43cm,label={[xshift=-0.42cm,yshift=-0.31cm]$+$}] (sink3);
            \coordinate[vertex,right=of sink3] (dualVertex13);
            \coordinate[vertex,right=of dualVertex13] (dualVertex23);
            \coordinate[right=of dualVertex23,label={[xshift=0.8cm,yshift=-0.24cm]$+\hspace{2mm}\hldots \,\,.$}] (source3);
            \draw[lineWithArrow1] (dualVertex13) -- (sink3);
            \draw[lineWithArrow1] (dualVertex23) -- (dualVertex13);
            \draw[lineWithArrow1] (source3) -- (dualVertex23);
          \end{tikzpicture}
          \end{equation}
      \end{center}
      This sum can be evaluated with the help of $\propagator(t | \tVar) \propagator(\tVar | t_0) = \propagator(t | t_0)$ as
      \begin{eqnarray}
        \ln \fieldGenFctAd_{\vect{0},\vect{0}} 
        & = \sum_{k=0}^\infty \ii\vect{\qAd}_N\cdot  \int_{t_0}^t \diff{\tVar_k}\, \propagator(t | \tVar_k) (-\rateCoeffLinDecay_{\tVar_k}) \\
        &\hspace{10mm} \cdots \int_{t_0}^{\tVar_2} \diff{\tVar_1}\, \propagator(\tVar_2 | \tVar_1) (-\rateCoeffLinDecay_{\tVar_1})  \vect{\xAd}_h(\tVar_1)    \nonumber\\
        &=  \ii\vect{\qAd}_N\cdot \vect{\xAd}_h(t) \ee^{-\int_{t_0}^t \diff{\tVar} \, \rateCoeffLinDecay_\tVar}    \,.
      \end{eqnarray}
      After inserting this expression into the marginalized distribution~\eref{eq:PathInts_BackwardAlongPaths_MarginalizedDistribution}, one recovers the multivariate Poisson distribution~\eref{eq:PathInts_BackwardSolutions_DiffusionNetwork_PoissonDist}. Its mean, however, has now acquired the pre-factor $\ee^{-\int_{t_0}^t \diff{\tVar} \, \rateCoeffLinDecay_\tVar} $, reflecting the decay of the particles.
              
  \subsection{R\'esum\'e}\label{subsec:PathInts_Backward_Resume}  
  
    In the present section, we introduced the novel backward path integral representation
    \begin{equation}
      \ketA{\prob(t, n | t_0)}    \label{eq:PathInts_Backward_Resume_PathInt}
      =  \pathintegral{(t_0}{t]}
        \, \ee^{-\actionAd} 
        \, \ketA{n}_{t}    
    \end{equation} 
    of the marginalized distribution  (cf.\ \sref{sec:GenFctnl})
    \begin{equation}
      \ketA{\prob(t, n | t_0)} = \sum_{n_0} \prob(t, n|t_0,n_0) \ketA{n_0}_{t_0}    \,.
    \end{equation}
    When the basis function $\ketA{n}_\xAd$ is chosen as a probability distribution (e.g.\ $\ketA{n}_\xAd = \frac{\xAd^{n} \ee^{-\xAd}}{n!}$ for all $n\in\naturals_0$), the backward path integral represents a true probability distribution: the marginalized distribution $\ketA{\prob(t, n | t_0)}$. This distribution solves the forward master equation~\eref{eq:Intro_Mescoscopic_MasterEq} for the initial condition $\ketA{\prob(t_0, n | t_0)} = \ketA{n}$ and it transforms into the conditional probability distribution as $\prob(t, n | \tVar, n_0) =  \braketA{n_0}{\prob(t,n|\tVar)}$. In \sref{subsec:PathInts_BackwardAlongPaths}, we showed how the backward path integral~\eref{eq:PathInts_Backward_Resume_PathInt} can be expressed in terms of an average over the paths of an It\^{o} stochastic differential equation. This method provided the exact solutions of various elementary stochastic processes, including the simple growth, the linear decay, and the pair generation processes. Moreover, we showed how the path integral can be evaluated perturbatively using Feynman diagrams for a process with diffusion and linear particle decay. We hope that the backward path integral~\eref{eq:PathInts_Backward_Resume_PathInt} will prove useful in the study of reaction-diffusion processes. Thus far, the critical behaviour of such processes could only be approached via path integral representations of averaged observables or of the generating function~\cite{Dickman:2003,Taeuber:2005}. In \sref{sec:PathInts_Observables}, we show how the former representation readily follows from the backward path integral~\eref{eq:PathInts_Backward_Resume_PathInt} upon summing the marginalized distribution over an observables $A(n)$. The corresponding representation is commonly applied in the study of diffusion-limited reactions (e.g.\ \cite{Lee:1994b,Lee:1995,Cardy:1996,Cardy:1998,Taeuber:2005}, but we here show how it can be freed of some of its quantum mechanical ballast (such as ``second-quantized'' or ``normal-ordered'' observables and coherent states). Besides, we showed in \sref{subsec:PathInts_BackwardKramersMoyal} how one can derive a path integral representation of processes with continuous state spaces whose (backward) master equations admit a Kramers-Moyal expansion. Provided that this expansion stops at the level of a diffusion approximation, one recovers a classic path integral representation of the (backward) Fokker-Planck equation and also the Feynman-Kac formula~\eref{eq:Intro_Mescoscopic_FP_BW_Solution}. Moreover, the representation can be rewritten in terms of an Onsager-Machlup function and, for diffusive Brownian motion, it simplifies to the path integral of Wiener.
    
\section{The forward path integral representation}\label{sec:PathInts_Forward}  

  Thus far, we have focused on the backward path integral~\eref{eq:PathInts_BwdFlow}. This integral will be used again in \sref{sec:PathInts_Observables} to derive a path integral representation of averaged observables. Moreover, we use it in \sref{subsec:StationaryPaths_GenFctnl_BinaryAnnihilation} to approximate the binary annihilation reaction $2\, A \to \emptyset$. In the present section, however, we shift our focus to the forward path integral~\eref{eq:PathInts_FwdFlow}. Its derivation proceeds analogously to the derivation of the backward solution, so we keep it brief.  
  
   \begin{figure*}[tb] 
    \includegraphics{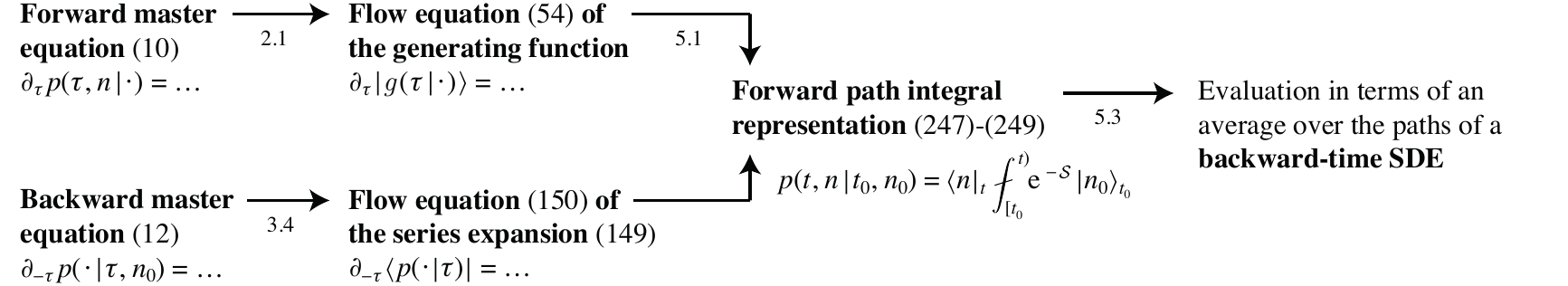}
    \caption{\label{fig:ForwardPathIntegral}
      Outline of the derivation of the forward path integral representation and of its evaluation in terms of an average over the paths of a backward-time stochastic differential equation.
    }
  \end{figure*}
  
  \subsection{Derivation}\label{subsec:PathInts_Forward_Derivation} 

    The forward path integral can be derived both from the flow equation~\eref{eq:GenFct_Flow_FlowEq} obeyed by the generating function or from the flow equation~\eref{eq:GenFctnl_Flow_XXX_Flow} obeyed by the series~\eref{eq:GenFctnl_Flow_XXX} (cf.\ \fref{fig:ForwardPathIntegral}). As it is more convenient to work with functions than with functionals, we use the flow equation of the generating function for this purpose. The derivation parallels a derivation of Elgart and Kamenev~\cite{Elgart:2004}. As in \sref{subsec:PathInts_Backward_Derivation}, we first split the time interval $[t_0,t]$ into $N$ pieces $t_0 \leq t_1 \leq \hldots \leq t_N  \coloneqq t$ of length $\Delta t$. Over a sufficiently small interval $\Delta t$, the flow equation~\eref{eq:GenFct_Flow_FlowEq} is then solved by
    \begin{equation} 
      \ketA{\gen(t | t_0, n_0)}_{\q_N}  
      =\generator_{t_{N-1}}(\q_N,\partial_{\q_N}) \ketA{\gen(t_{N-1} | t_0, n_0)}_{\q_N}     \,,
    \end{equation} 
    with the generator $\generator_\tVar \coloneqq 1 + \transEvoOp_\tVar \Delta t + O\bigl((\Delta t)^2\bigr)$. After inserting the integral form of a Dirac delta between $\generator$ and $\ketA{\gen}$, the right-hand side of the equation reads
    \begin{eqnarray}
      &\generator_{t_{N-1}}(\q_N,\partial_{\q_N})   
      \int_{\reals^2}\! \frac{\diff{\q_{N-1}}\diff{\x_{N-1}}}{2\pi} 
      \ee^{-\ii\x_{N-1}(\q_{N-1}-\q_N)}  
      \nonumber\\&\hspace{45mm}\cdot 
      \ketA{\gen(t_{N-1} | \cdot)}_{\q_{N-1}}  \,.        
    \end{eqnarray} 
    Assuming that the transition operator $\transEvoOp_\tVar$, and therefore also $\generator_\tVar$, are normal-ordered with respect to $\q$ and $\partial_\q$, we may replace $\partial_{\q_N}$ by $\ii\x_{N-1}$ and interchange $\generator_{t_{N-1}}$ with the exponential. This procedure is repeated $N$ times before invoking the exponentiation $\generator_\tVar = \exp{(\transEvoOp_\tVar \Delta t})$. Using the inverse transformation~\eref{eq:GenFct_Flow_InverseTransformation} and the initial condition $\ketA{\gen(t_0 | t_0, n_0)} = \ketA{n_0}$, one obtains the discrete-time path integral representation
    \begin{eqnarray}
      &\prob(t, n | t_0, n_0)  =  
        \braA{n}_{t, \q_N} \ketA{\gen(t | t_0, n_0)}_{\q_N}       \label{eq:PathInts_Forward_DiscreteTime_Solution}     \\
      &\mathtext{with }
      \ketA{\gen(t | t_0, n_0)}_{\q_N} =
        \pathintegral{0}{N-1}
        \, \ee^{-\action_N}
        \, \ketA{n_0}_{t_0, \q_0}      \mathtext{ and}  \label{eq:PathInts_Forward_DiscreteTime_Solution_GeneratingFunction}  \\
      &
      \action_N  \coloneqq  
        \sum_{j=0}^{N-1} \Delta t \Bigl(
        \ii\x_j \frac{\q_j - \q_{j+1}}{\Delta t} - \transEvoOp_{t_j}(\q_{j+1}, \ii\x_j)      
        \Bigr)     \,.                  \label{eq:PathInts_Forward_DiscreteTime_Solution_Action}
    \end{eqnarray}
    Here we again employed the abbreviation~\eref{eq:PathInts_Backward_DiscreteTime_Measure}, i.e.\
    \begin{equation}
      \pathintegral{k}{l}
      = \prod_{j=k}^{l}  \int_{\reals^2}\!\frac{\diff{\q_j}\diff{\x_{j}}}{2\pi}    \label{eq:PathInts_Forward_DiscreteTime_Measure} \,.  
    \end{equation} 
    Moreover, the initial condition $\prob(t_0, n | t_0, n_0) = \delta_{n,n_0}$ is again trivially fulfilled for $N=0$. Upon taking the continuous-time limit $N \to \infty$, so that $\Delta t\to 0$, the forward path integral representation of the master equation follows as
    \begin{eqnarray}
      &  \prob(t, n | t_0, n_0)    =   \braA{n}_{t, \q(t)} \ketA{\gen(t | t_0, n_0)}_{\q(t)}         \label{eq:PathInts_Forward_ContinuousTime_Solution}   \\
      &  \mathtext{with }
        \ketA{\gen(t | t_0, n_0)}_{\q(t)}
          = \pathintegral{[t_0}{t)} 
          \ee^{-\action}    
          \, \ketA{n_0}_{t_0, \q(t_0)}        \label{eq:PathInts_Forward_ContinuousTime_Solution_GeneratingFunction} \\
      &\mathtext{and }  
      \action 
      \coloneqq  \int_{t_0}^{t}\!\diff{\tVar}\, \bigl[
                  \ii\x \partial_{-\tVar} \q - \transEvoOp_\tVar(\q, \ii\x) 
                  \bigr] \,.                  \label{eq:PathInts_Forward_ContinuousTime_Solution_Action} 
    \end{eqnarray} 
    The limit $\pathintegral{[t_0}{t)} \coloneqq \lim_{N \to \infty} \pathintegral{0}{N-1}$ now involves integrations over $\x(t_0)$ and $\q(t_0)$, but not over  $\x(t)$ and $\q(t)$. 
  
  \subsection{Linear processes}\label{subsec:PathInts_LinearProcesses}  
  
    The forward path integral~\eref{eq:PathInts_Forward_ContinuousTime_Solution_GeneratingFunction} can, for example, be used to derive the generating function of the generic linear process $A \to l\, A$ with rate coefficient $\rateCoeffLinDecay_\tVar$ (and $l\in\naturals_0$). For that purpose, we choose the basis function as $\ketA{n}_{\tVar,\q} = \q^n$ so that $\ketA{\gen}$ coincides with the ordinary generating function. Since the basis function does not depend on time, \eref{eq:GenFct_Flow_Poisson_HamiltonianF} alone specifies the transition operator
    \begin{equation}
      \transEvoOp_\tVar(\q, \partial_\q) 
      = \rateCoeffLinDecay_\tVar (\q^l - \q) \partial_\q     
    \end{equation} 
    of the flow equation $\partial_\tVar \ketA{\gen(\tVar | \cdot)}   = \transEvoOp_\tVar(\q, \partial_\q) \ketA{\gen(\tVar | \cdot)}$. Consequently, the action
    \begin{equation}
      \action 
      =  \int_{t_0}^{t}\!\diff{\tVar}\, \ii\x \bigl[
                    \partial_{-\tVar} \q - \rateCoeffLinDecay_\tVar (\q^l - \q)
                  \bigr]
    \end{equation} 
    is linear in $\ii\x$. To evaluate the path integral~\eref{eq:PathInts_Forward_ContinuousTime_Solution_GeneratingFunction}, it helps to reconsider the discrete-time approximation
    \begin{equation}
      \action_N  =  
        \sum_{j=0}^{N-1}  \ii\x_j \Bigl(
           \q_j - (\q_{j+1} +\rateCoeffLinDecay_{t_j} (\q_{j+1}^l - \q_{j+1}) \Delta t )
        \Bigr)     \,.                  
    \end{equation}
    of the action. Upon integrating over all the $\q_j$-variables and taking the limit $\Delta t\to 0$, one obtains the generating function
    \begin{equation}
      \ketA{\gen(t|t_0, n_0)}_{\q(t)} 
      =  \q(t_0)^{n_0}   \
    \end{equation} 
    with $\q(\tVar)$ solving $\partial_{-\tVar} \q = \rateCoeffLinDecay_\tVar(\q^l - \q)$. The unique real solution of this equation with final value $\q(t)$ reads
    \begin{equation}
      \q(\tVar)
      =\frac{\q(t)}{\bigl[\q(t)^{l-1} + \ee^{(l-1)\int_{\tVar}^t \diff{\sVar}\, \rateCoeffLinDecay_\sVar}(1-\q(t)^{l-1}) \bigr]^{1/(l-1)}}     \,.\,
    \end{equation} 

    For the linear growth, or Yule-Furry~\cite{Yule:1925,Furry:1937}, process $A \to 2\, A$ ($l=2$), the inverse transformation~\eref{eq:PathInts_Forward_ContinuousTime_Solution} casts the generating function into
    \begin{eqnarray}
      &\prob(t,n|t_0,n_0)     
      = \Bigl(\,\frac{1}{n!}  \partial_{\q(t)}^n \ketA{\gen(t|t_0,n_0)}_{\q(t)}\Bigr) \Big|_{\q(t)=0}  \label{eq:PathInts_LinearProcesses_NegBinom_Deriv} \\
      &= \binomCoeff{n-1}{n-n_0} \bigl(\ee^{-\int_{t_0}^t \diff{\tVar}\, \rateCoeffLinDecay_\tVar}\bigr)^{n_0}  \bigl(1-\ee^{-\int_{t_0}^t \diff{\tVar}\, \rateCoeffLinDecay_\tVar}\bigr)^{n-n_0}    \label{eq:PathInts_LinearProcesses_NegBinom}
    \end{eqnarray} 
    for $n > 0$ and into $\delta_{0,n_0}$ for $n=0$. The above distribution has the form of a negative Binomial distribution with probability of success $\ee^{-\int_{t_0}^t \diff{\tVar}\, \rateCoeffLinDecay_\tVar}$, number of failures $n-n_0$, and number of successes $n_0$~\cite{DeGroot:2012}. The mean value $\ee^{\int_{t_0}^t \diff{\tVar}\, \rateCoeffLinDecay_\tVar} n_0$ of the marginalized distribution~\eref{eq:PathInts_LinearProcesses_NegBinom} grows exponentially with time as long as $\rateCoeffLinDecay_\tVar>0$, and so does its variance $(\ee^{\int_{t_0}^t \diff{\tVar}\, \rateCoeffLinDecay_\tVar}-1)\, \ee^{\int_{t_0}^t \diff{\tVar}\, \rateCoeffLinDecay_\tVar} n_0$.
    
    For the linear decay process $A \to \emptyset$ ($l=0$), the inverse transformation in~\eref{eq:PathInts_LinearProcesses_NegBinom_Deriv} instead recovers the Binomial distribution~\eref{eq:PathInts_SimpleGrowthLinearDecay_BinomialDist}.

  \subsection{Intermezzo: The forward Kramers-Moyal expansion}\label{subsec:PathInts_ForwardKramersMoyal}  
    
    The above procedure can be generalized to processes whose transition operator is of the form
    \begin{equation}
       \transEvoOp_\tVar(\q, \partial_\q)      \label{eq:PathInts_ForwardKramersMoyal_TransitionOp}
        = \alpha_\tVar(\q) \partial_\q   + \frac{1}{2} \beta_\tVar(\q) \partial_\q^2 + \perturbationOp_\tVar(\q, \partial_\q)        \,.
    \end{equation}  
    The derivation proceeds analogously to \sref{subsec:PathInts_BackwardAlongPaths} and \aref{sec:A_BackwardAlongPaths}. The probability distribution is thereby expressed as an average over the paths of a stochastic differential equation proceeding backward in time. The merit of such a representation remains to be explored. In the following, we briefly outline how the procedure is applied to processes with continuous sample paths. 
    
    In \sref{subsec:PathInts_BackwardKramersMoyal}, we explained how the forward master equation~\eref{eq:PathInts_BackwardKramersMoyal_ForwardMaster} of a process with a continuous state space can be written in terms of the (forward) Kramers-Moyal expansion
    \begin{equation}
      \partial_\tVar \prob(\tVar, \q | \cdot)  
      = \sum_{m=1}^\infty \frac{(-1)^m}{m!} \partial_{\q}^m 
        \bigl[\jumpMoment{m}_\tVar(\q)   \prob(\tVar, \q | \cdot )  \bigr]
    \end{equation}
    with initial condition $\prob(\tVar, \q | t_0, \q_0) = \delta(\q - \q_0)$. Here we changed the letter from $\x$ to $\q$ to keep in line with the notation used in sections~\ref{subsec:PathInts_Forward_Derivation} and~\ref{subsec:PathInts_LinearProcesses}. If the Kramers-Moyal expansion stops with its second summand, it coincides with the (forward) Fokker-Planck equation
    \begin{equation}
      \partial_\tVar \prob(\tVar, \q | \cdot )  
      = -\partial_{\q} \bigl[\jumpMoment{1}_\tVar(\q)   \prob \bigr]
        + \frac{1}{2} \partial_{\q}^2 \bigl[\jumpMoment{2}_\tVar(\q)   \prob  \bigr]  \,.
    \end{equation}
    To apply the procedure from the previous section to this equation, it needs to be brought into the form $\partial_\tVar \prob(\tVar, \q | \cdot) = \transEvoOp_\tVar(\q, \partial_\q) \prob(\tVar, \q | \cdot )$ with a normal-ordered transition operator $\transEvoOp_\tVar(\q, \partial_\q)$. It is readily established that this operator can be expressed in the form of~\eref{eq:PathInts_ForwardKramersMoyal_TransitionOp} with the coefficients
    \begin{eqnarray}
      \alpha_\tVar(\q)  
      \coloneqq -\jumpMoment{1}_\tVar + \partial_{\q} \jumpMoment{2}_\tVar\,,\\
      \beta_\tVar(\q) 
      \coloneqq  \jumpMoment{2}_\tVar  \,,\mathtext{ and}\\
      \perturbationOp_\tVar(\q)
      \coloneqq -\partial_{\q} \jumpMoment{1}_\tVar + \frac{1}{2} \partial_{\q}^2 \jumpMoment{2}_\tVar 
       \,.
    \end{eqnarray}  
    The Fokker-Planck equation now has the same form as the flow equation obeyed by the generating function in \sref{subsec:PathInts_Forward_Derivation}. Therefore, we can follow the steps in that section to represent the solution of the Fokker-Planck equation by the path integral
    \begin{eqnarray}
      \prob(t, \q | t_0, \q_0 )     \label{eq:PathInts_ForwardKramersMoyal_PathInt}
      =  \pathintegral{[t_0}{t)}
        \, \ee^{-\action} 
        \, \delta(\q_0 - \q(t_0))  \big|_{\q(t)=\q}\\
      \mathtext{with }  
      \action  
      \coloneqq  \int_{t_0}^{t} \diff{\tVar}\, 
                  \bigl[
                    \ii\x \partial_{-\tVar} \q 
                    - \transEvoOp_\tVar(\q, \ii\x)  
                  \bigr]\,.                       
    \end{eqnarray} 
    The evaluation of this path integral proceeds analogously to the derivation in \sref{subsec:PathInts_LinearProcesses} and \aref{sec:A_BackwardAlongPaths}. In particular, one can rewrite the probability distribution as
    \begin{equation}
      \prob(t, \q | t_0, \q_0)    \label{eq:PathInts_ForwardKramersMoyal_FeynmanKac}
      = \bigLLangle \ee^{\int_{t_0}^t \diff{\tVar}\, \perturbationOp_\tVar(\q(\tVar))} \delta (\q_0 - \q(t_0)) \bigRRangle_\wienerProcess  \,,
    \end{equation}
    with $\q(\tVar)$ solving the backward-time SDE
    \begin{equation}
      - \diff{\q}(\tVar)     
      = \alpha_\tVar(\q(\tVar)) \diff{\tVar}  + \sqrt{\beta_\tVar(\q(\tVar))}\diff{W}(\tVar)   \,.
    \end{equation} 
    The time evolution of this SDE starts out from the final value $\q(t)=\q$. In a discrete-time approximation, the SDE reads
    \begin{equation}
      \q_{j} - \q_{j+1} = \alpha_{t_{j}}(\q_{j+1}) \Delta t  + \sqrt{\beta_{t_{j}}(\q_{j+1})}\,\Delta \wienerProcess_j  \,.
    \end{equation}
    The increments $\Delta \wienerProcess_j$ are Gaussian distributed with mean $0$ and variance $\Delta t$. Let us note that the two path integral representations~\eref{eq:PathInts_BackwardKramersMoyal_PathIntegral} and~\eref{eq:PathInts_ForwardKramersMoyal_PathInt} of the conditional probability distribution belong to an infinite class of representations~\cite{Haken:1976,Wissel:1979,Risken:1996} (see also \sref{subsubsec:PathInts_BackwardKramersMoyal_Discretization}). We focus on the two representations that follow from the backward and forward Kramers-Moyal expansions via the step-by-step derivations in sections~\ref{subsec:PathInts_Backward_Derivation} and~\ref{subsec:PathInts_Forward_Derivation}, respectively.
    
    To exemplify the validity of the path integral~\eref{eq:PathInts_ForwardKramersMoyal_PathInt} and of the representation~\eref{eq:PathInts_ForwardKramersMoyal_FeynmanKac}, we consider a process with linear drift and no diffusion, i.e.\ a process with jump moments~$\jumpMoment{1}_\tVar = \rateCoeffLinDecay \q$ and~$\jumpMoment{2}_\tVar = 0$. For this process, the representation~\eref{eq:PathInts_ForwardKramersMoyal_FeynmanKac} evaluates to
    \begin{eqnarray}
      \prob(t, \q | t_0, \q_0)  
      &=  \ee^{-\rateCoeffLinDecay (t-t_0)} \delta (\q_0 - \q\, \ee^{-\rateCoeffLinDecay (t-t_0)})\\
      &=  \delta (\q - \q_0\, \ee^{\rateCoeffLinDecay (t-t_0)})  \,.
    \end{eqnarray}
    Thus, the leading exponential in~\eref{eq:PathInts_ForwardKramersMoyal_FeynmanKac} converted the argument of the Dirac delta from the solution of a final value problem to the solution of an initial value problem. It is, of course, no surprise that the probability distribution is given by a Dirac delta function because the process is purely deterministic.
    
    Another simple process that can be solved with the help of~\eref{eq:PathInts_ForwardKramersMoyal_FeynmanKac} is the pure diffusion process with $\jumpMoment{1}_\tVar = 0$ and~$\jumpMoment{2}_\tVar = \rateCoeffDiffCont$. The derivation is performed most easily in the discrete-time approximation and results in the Wiener path integral
    \begin{eqnarray}
      \prob(t, \q | t_0, \q_0) = \\
      \lim_{N\to\infty} \Bigl(\prod_{j=1}^{N} \int_\reals \diff{\tq_{j-1}}\Bigr)   \Bigl(\prod_{j=1}^{N} 
        \gaussianDistribution_{0, \rateCoeffDiffCont\Delta t}(\tq_{j-1} - \tq_{j}) \Bigr)
        \delta(\tq_0 - \q_0)  \nonumber    \,,
    \end{eqnarray} 
    with $\tq_N \coloneqq \q$ and Gaussian distribution $\gaussianDistribution_{\mu, \sigma^2}$ (cf.~\eref{eq:PathInts_BackwardKramersMoyal_Gaussian}). An evaluation of the convolutions of Gaussian distributions shows that the process is solved by $\gaussianDistribution_{0, \rateCoeffDiffCont (t-t_0)}(\q-\q_0)$. Further uses of the path integral representation~\eref{eq:PathInts_ForwardKramersMoyal_PathInt} of the (forward) Fokker-Planck equation remain to be explored. 

  \subsection{R\'esum\'e}\label{subsec:PathInts_Forward_Resume}  

    Here we derived the forward path integral representation
    \begin{eqnarray}
        \ketA{\gen(t | t_0, n_0)}   \label{eq:PathInts_Forward_Resume_PathInt}
          = \pathintegral{[t_0}{t)} 
          \ee^{-\action}    
          \, \ketA{n_0}_{t_0}  
    \end{eqnarray} 
   of the probability generating function $\ketA{\gen(t | t_0, n_0)} = \sum_{n}\ketA{n}_{t} \, \prob(t, n | t_0, n_0)$. The conditional probability distribution is recovered from the generating function via the inverse transformation $\prob(t, n | t_0, n_0)    =   \braketA{n}{\gen(t | t_0, n_0)}$. The path integral representation of the generating function has, for example, been employed to compute rare event probabilities~\cite{Elgart:2004} by a method that we discuss in \sref{sec:StationaryPaths}. Most often, however, the representation has only served as an intermediate step in deriving a path integral representation of averaged observables. Such a representation is considered in the next section. In \sref{subsec:PathInts_LinearProcesses}, we showed how the forward path integral~\eref{eq:PathInts_Forward_Resume_PathInt} can be evaluated along the paths of a differential equation proceeding backward in time. We thereby obtained the generating function of generic linear processes. Besides, we derived a novel path integral representation of processes with continuous state spaces in \sref{subsec:PathInts_ForwardKramersMoyal}, based on the (forward) Kramers-Moyal expansion. The potential use of this representation remains to be explored.
        
\section{Path integral representation of averaged observables}\label{sec:PathInts_Observables}  

  The backward and forward path integral representations  of the conditional probability distribution provide a full characterization of a Markov process. Yet, an intuitive understanding of how a process evolves is often attained more easily by looking at the mean particle number $\langle n \rangle$ and at its variance $\bigl\langle \bigl(n-\langle n \rangle\bigr)^2 \bigr\rangle$. Although both of these averages can in principle be inferred from a given probability distribution, it often proves convenient to bypass the computation of the distribution and to focus directly on the observables. Thus, we now show how one can derive a path integral representation of the average $\bigl\langle A \bigr\rangle$ of an observable $A(n)$. The path integral representation applies to all processes that can be decomposed additively into reactions of the form $k\, A \to l\, A$ in a well-mixed, non-spatial environment. The extension of the path integral to multiple types of interacting particles and to processes with spatial degrees of freedom is straightforward. In the derivation, we assume that the number of particles in the system is initially Poisson distributed with mean $\xAd(t_0)$. This assumption is common in the study of reaction-diffusion master equations and will allow us to focus on the marginalized distribution from \sref{sec:GenFctnl} instead of on the conditional probability distribution. The derived path integral has been applied in various contexts, particularly in the analysis of decay laws of diffusion-limited reactions and in the identification of universality classes~\cite{Taeuber:2005,Taeuber:2014}.
    
  \subsection{Derivation}\label{subsec:PathInts_Observables_Derivation} 
  
    Assuming that the number of particles is initially Poisson distributed with mean $\xAd(t_0)$, the probability of finding $n$ particles at time $t$ is $\prob(t, n) = \sum_{n_0} \prob(t, n | t_0, n_0) \prob(t_0, n_0)$. Here we make use of the initial (single-time) distribution
  \begin{equation}
    \prob(t_0,n_0) = \frac{{\xAd(t_0)}^{n_0} \ee^{-{\xAd(t_0)}}}{n_0!}  \,.
  \end{equation}
   Consequently, the average value of an observable $A(n)$ at time $t$ evaluates to
    \begin{equation}    \label{eq:PathInts_Observables_Derivation_Average_Orig}
      \langle \observable \rangle_{\xAd(t_0)} 
      = \sum_{n=0}^\infty A(n) \prob(t, n)\,.
    \end{equation}
    Here we emphasize that the average depends on the mean of the initial Poisson distribution. As the single-time distribution coincides with the marginalized distribution $\ketA{\prob(t, n | t_0)}$ in~\eref{eq:GenFctnl_MarginalizedDist_Expansion} for the Poisson basis function $\ketA{n_0}_{\xAd(t_0)} = \frac{{\xAd(t_0)}^{n_0} \ee^{-{\xAd(t_0)}}}{n_0!}$, the above average can equivalently be written as
    \begin{equation} 
      \langle \observable \rangle_{\xAd(t_0)}  \label{eq:PathInts_Observables_Derivation_Average}
      = \sum_{n=0}^\infty A(n) \ketA{\prob(t, n | t_0)}_{\xAd(t_0)}   \,.
    \end{equation} 
    The path integral representation of $\langle \observable \rangle_{\xAd(t_0)}$  then follows directly from the backward path integral representation~\eref{eq:PathInts_Backward_ContinuousTime_Solution_MarginalizedDistribution} of the marginalized distribution as
    \begin{eqnarray}
      &\langle \observable \rangle_{\xAd(t_0)}
      =  \pathintegral{(t_0}{t]}
        \ee^{-\actionAd} 
        \, \LLangle \observable \RRangle_{\xAd(t)}                    \label{eq:PathInts_Observables_MeanOfObservables} \\
      & \mathtext{with } 
      \LLangle \observable \RRangle_{\xAd} 
      \coloneqq \sum_{n=0}^\infty  \frac{\xAd^{n} \ee^{-\xAd}}{n!}  \observable(n)     \label{eq:PathInts_Observables_FieldMeanOfObservables}\,.
    \end{eqnarray} 
    With the adjoint transition operator $\transitionOpAd_\tVar(\cre, \ann) = \rateCoeffGeneric_\tVar \cre^k (\ann^l - \ann^k)$ of the reaction $k\, A \to l\, A$ (cf.~\sref{subsec:GenFctnl_BackwardBases_ChemReactions}), the action $\actionAd$ in~\eref{eq:PathInts_Backward_ContinuousTime_Action} reads
    \begin{equation}
      \actionAd  \label{eq:PathInts_Observables_Derivation_Action}
      =  \int_{t_0}^{t} \diff{\tVar}\, 
        \bigl[
          \ii\qAd \partial_\tVar \xAd 
          - \transitionOpAd_\tVar(\xAd, \ii\qAd + 1)  
        \bigr]                \,.             
    \end{equation}
    The ``$+1$'' in the transition operator is called the ``Doi-shift''~\cite{Cardy:2008}. In our above derivation, this shift followed from choosing a Poisson distribution as the basis function. The unshifted version of the path integral can be derived by choosing the basis function of the marginalized distribution~\eref{eq:GenFctnl_MarginalizedDist_Expansion} as $\ketA{n_0}_{\xAd(t_0)} = \frac{{\xAd(t_0)}^{n_0} }{n_0!}$, turning the average~\eref{eq:PathInts_Observables_Derivation_Average} into
    \begin{equation} 
      \langle \observable \rangle_{\xAd(t_0)}  
      = \ee^{-{\xAd(t_0)}} \sum_{n=0}^\infty A(n) \ketA{\prob(t, n | t_0)}_{\xAd(t_0)}   \,.
    \end{equation} 
    Upon rewriting the marginalized distribution in terms of the backward path integral~\eref{eq:PathInts_Backward_ContinuousTime_Solution_MarginalizedDistribution}, the action~\eref{eq:PathInts_Observables_Derivation_Action} acquires the addition summand $\xAd(t_0)-\xAd(t)$ and now involves the unshifted transition operator $\transitionOpAd_\tVar(\xAd, \ii\qAd)$.
    
    The average over a Poisson distribution in~\eref{eq:PathInts_Observables_FieldMeanOfObservables} establishes the link between the particle number $n$ and the path integral variable $\xAd$. For the simplest observable $A(n) \coloneqq n$, i.e.\ for the particle number itself, it holds that $\LLangle n \RRangle_{\xAd} = \xAd$. This relation generalizes to factorial moments of order $k\in\naturals$ for which $\LLangle (n)_k \RRangle_{\xAd}   = \xAd^k$ (recall that $(n)_k \coloneqq n (n-1)\cdots (n-k+1)$). The computation of factorial moments may serve as an intermediate step in obtaining ordinary moments of the particle number. For this purpose, one may use the relation $ n^{k} = \sum_{l=0}^{k} {k \brace l} (n)_l$, where the curly braces represent a Stirling number of the second kind (cf.\ section~26.8 in~\cite{NIST:2010}). An extension of the path integral representation~\eref{eq:PathInts_Observables_MeanOfObservables} to multi-time averages of the form
    \begin{equation}
      \sum_{n_2,n_1=0}^\infty A(n_2,n_1) \prob(\tVar_2, n_2 ; \tVar_1, n_1 | t_0, n_0)
    \end{equation}
     remains open (with $\tVar_2 > \tVar_1 > t_0$). The results of Elderfield~\cite{Elderfield:1985} may prove helpful for this purpose.
    
    In a slightly rewritten form, the Doi-shifted path integral~\eref{eq:PathInts_Observables_MeanOfObservables} was, for example, employed by Lee in his study of the diffusion-controlled annihilation reaction $k\, A \to \emptyset$ with $k\geq 2$~\cite{Lee:1994a,Lee:1994b}. He found that below the critical dimension $d_c=2/(k-1)$, the particle density asymptotically decays as $n\sim A_k t^{-d/2}$ with a universal amplitude $A_k$. At the critical dimension, the particle density instead obeys $n\sim A_k (\ln t/t)^{1/(k-1)}$. Neglecting the diffusion of particles, the above action~\eref{eq:PathInts_Observables_Derivation_Action} of the reaction $k\, A \to \emptyset$ reads
    \begin{equation}
      \actionAd  \label{eq:PathInts_Observables_Derivation_Action_Ann}
      =  \int_{t_0}^{t} \diff{\tVar}\, 
        \Bigl(
          \ii\qAd \partial_\tVar \xAd 
          + \rateCoeffGeneric_\tVar \xAd^k \sum_{j=1}^k \binomCoeff{k}{j} (\ii\qAd)^j
        \Bigr)                         \,.
    \end{equation}
    Besides using different names for integration variables ($\ii\qAd\to\overline{\psi}$ and $\xAd\to\psi$), the action employed by Lee involves an additional boundary term. This term can be introduced by rewriting the path integral representation~\eref{eq:PathInts_Observables_MeanOfObservables} as 
    \begin{eqnarray}
      &\langle \observable \rangle_{\lambda_0}
      =  \pathintegral{[t_0}{t]}
        \ee^{-\actionAd} 
        \, \LLangle \observable \RRangle_{\xAd(t)}       \,.
    \end{eqnarray} 
    This path integral involves integrations over the variables $\qAd(t_0)$ and $\xAd(t_0)$, and the mean of the initial Poisson distribution is denoted by $\lambda_0$. Consequently, one needs to add the boundary term $\ii\qAd(t_0) (\xAd(t_0) - \lambda_0)$ to the action~\eref{eq:PathInts_Observables_Derivation_Action_Ann} to equate $\xAd(t_0)$ with $\lambda_0$. The factor $\ii\qAd(t_0)\xAd(t_0)$ of the new term is, however, often dropped eventually~\cite{Lee:1994a,Lee:1994b}. The convention of Lee is also commonly used, for example, by T\"auber~\cite{Taeuber:2014}. The unshifted version of the path integral with transition operator $\transitionOpAd_\tVar(\xAd, \ii\qAd)$ is recovered via $\ii\qAd + 1 \to \ii\qAd$.
    
    For completeness, let us note that the path integral representation~\eref{eq:PathInts_Observables_MeanOfObservables} can also be derived from the forward path integral~\eref{eq:PathInts_Forward_DiscreteTime_Solution_GeneratingFunction}, provided that the observable $A(n)$ is analytic in $n$. It then suffices to consider the factorial moment $A(n) \coloneqq (n)_k$. To perform the derivation, one may choose the basis function of the generating function~\eref{eq:GenFct_Flow_GenFct} as $\ketA{n} \coloneqq (\ii \q + 1)^n$  (insert $\basisPrefactor\coloneqq\ii$, $\tq\coloneqq 1$ and $\tx\coloneqq 0$ into~\eref{eq:GenFct_Flow_Poisson_BaseFunction}). As the first step, the forward path integral~\eref{eq:PathInts_Forward_ContinuousTime_Solution_GeneratingFunction} is summed over an initial Poisson distribution with mean $\x(t_0)$. The average of the factorial moment is then obtained via $\langle (n)_k \rangle_{\x(t_0)} = \partial_{\ii\q_N}^k   \ketA{\gen(t | t_0; \x(t_0))}_{\q_N} |_{\q_N=0}$. To recover the action~\eref{eq:PathInts_Observables_Derivation_Action}, one may note that the operator $\transitionOp_\tVar(\cre, \ann)$ in~\eref{eq:GenFct_Flow_Poisson_HamiltonianF} and the operator $\transitionOpAd_\tVar(\cre, \ann)$ in~\eref{eq:GenFctnl_Flow_Poisson_HamiltonianB} fulfil $\transitionOp_\tVar(\ii\q + 1, \x) = \transitionOpAd_\tVar(\x, \ii\q + 1)$ for scalar arguments. Thus, both the marginalized distribution approach and the generating function approach result in the same path integral representation of averaged observables. A detailed derivation of the path integral~\eref{eq:PathInts_Observables_MeanOfObservables} from the generating function is, for example, included in the article of Dickman and Vidigal~\cite{Dickman:2003} (see their equations (106) and (108)).
 
  \subsection{Intermezzo: Alternative derivation based on coherent states}\label{subsec:PathInts_Observables_Algebraic} 
  
    We noted previously in \sref{subsec:GenFct_ForwardBases_Algebraic} that the path integral representation~\eref{eq:PathInts_Observables_MeanOfObservables} of the average
    \begin{equation}
      \langle \observable \rangle_{\x(t_0)}  
      =  \sum_{n_0,n=0}^\infty 
          \observable(n)\, \prob(t,n|t_0,n_0) 
          \frac{\x(t_0)^{n_0} \ee^{-\x(t_0)}}{n_0!}   
      \label{eq:PathInts_Observables_Observable}
    \end{equation}
    can be derived without first casting the master equation into a linear PDE~\cite{Lee:1994a,Lee:1994b,Mattis:1998,Cardy:2008,Wiese:2015}. The following section outlines this derivation for the process $k\, A \to l\, A$ with time-independent rate coefficient $\rateCoeffGeneric$. Moreover, we assume $\observable(n)$ to be analytic in $n$. 
    
    The alternative derivation of the path integral representation~\eref{eq:PathInts_Observables_MeanOfObservables} starts out from the exponential solution of the master equation, i.e. from (cf.~\eref{eq:GenFct_ForwardBases_Prob})
    \begin{equation}
      \prob(t, n | t_0, n_0)     \label{eq:PathInts_Observables_Algebraic_Solution}
      = \braA{n}\ee^{\qMatrix(t-t_0)}\ketA{n_0}  \,.
    \end{equation}
    The bras are chosen as the unit row vectors $\braA{n} = \hat{\vect{e}}_n^\transpose$ and the kets as the unit column vectors $\ketA{n} = \hat{\vect{e}}_n$. As before, the transition matrix of the reaction $k\, A \to l\, A$ with rate coefficient $\rateCoeffGeneric$ reads (cf.~\eref{eq:GenFct_ForwardBases_QMatrix})
    \begin{equation}
      \qMatrix(\cre, \ann)          \label{eq:PathInts_Observables_Algebraic_QMatrix}
      = \rateCoeffGeneric (\cre^l - \cre^k) \ann^k   \,.
    \end{equation}
    The creation matrix fulfils $\cre \ketA{n} = \ketA{n+1}$ and $\braA{n} \cre = {\braA{n-1}}$, and the annihilation matrix $\ann \ketA{n} = n \ketA{n-1}$ and $\braA{n} \ann = {(n+1)} {\braA{n+1}}$. Thus, the basis column vectors can be generated incrementally via $\ketA{n} = \cre^n\ketA{0}$ and the basis row vectors via $\braA{n} = \braA{0} \frac{\ann^n}{n!}$. 
    
    Since the bra $\braA{n}$ is a left eigenvector of the number matrix $\numberOperator \coloneqq \cre \ann$ with eigenvalue $n$, one may write $\observable(n) \braA{n} = \braA{n} \observable(\numberOperator)$ for an analytic observable $A$. After inserting the exponential solution~\eref{eq:PathInts_Observables_Algebraic_Solution} into the averaged observable~\eref{eq:PathInts_Observables_Observable}, an evaluation of the sums therefore results in
    \begin{equation}
      \langle \observable \rangle_{\x_0}
      =  \braA{0}
          \ee^\ann 
          \observable(\cre \ann) 
          \ee^{\qMatrix(\cre, \ann) (t-t_0)}   
          \ee^{\x_0(\cre-\unitMatrix)} \ketA{0}  \,.
          \label{eq:PathInts_Observables_Observable_Step1}  
    \end{equation}
    Here we employed the (infinitely-large) unit matrix~$\unitMatrix$. Moreover, we changed the variable $\x(t_0)$ to $\x_0$ in anticipation of a discrete-time approximation.
    
    Following the lecture notes of Cardy~\cite{Cardy:2008}, we now perform the Doi-shift by shifting the first exponential in the above expression to the right. To do so, we require certain relations, which are all based on the commutation relation $[\ann, \cre]=\unitMatrix$. First, it follows from this relation that $[\ann, \cre^n]=n \cre^{n-1} \unitMatrix$ holds for all $n\in\naturals_0$, and more generally that $[\ann, [ \ann \cdots [\ann, \cre^n]]] = (n)_j \cre^{n-j} \unitMatrix$ holds for nested commutators with $j \leq n$ annihilation matrices. Nested commutators of higher order vanish. The Hadamard lemma~\cite{Mansour:2015} can therefore be employed to write (with $z\in\complex$ and $n\in \naturals$)
    \begin{eqnarray}
        \ee^{z \ann}\cre^n 
        %= (\ee^{z \ann}\cre^n \ee^{-z \ann}) \ee^{z \ann}\\ 
        = \bigl( \cre^n + [\ann, \cre^n] z  + \frac{1}{2} [\ann, [\ann, \cre^n]] z^2 + \hldots \bigr)\ee^{z \ann}  \\
        = \bigl( \cre^n + n \cre^{n-1} z\,\unitMatrix+ \binomCoeff{n}{2} \cre^{n-2} z^2\,\unitMatrix+ \hldots \bigr)\ee^{z \ann} \\
        = (\cre+z\unitMatrix)^n \ee^{z \ann}\,.    \nonumber
    \end{eqnarray} 
    This expression generalizes to the following shift operations for an analytic function $f$:
    \begin{eqnarray}
      \ee^{z \ann}f(\cre) 
      = f(\cre+z\unitMatrix) \ee^{z \ann}   \mathtext{ and}        \label{eq:GenFct_ForwardBases_Algebraic_Identity1}\\
      f(\ann) \ee^{z \cre}
      = \ee^{z \cre} f(\ann+z\unitMatrix)  \,.      \label{eq:GenFct_ForwardBases_Algebraic_Identity2}
    \end{eqnarray} 
    Shifting of the first exponential in~\eref{eq:PathInts_Observables_Observable_Step1} to the right therefore results in
    \begin{equation}
      \langle \observable \rangle_{\x_0}
      =
      \braA{0}
      \observable\bigl((\cre+\unitMatrix)\ann\bigr) 
      \ee^{\qMatrix(\cre+\unitMatrix,\ann) (t-t_0)}   
      \ee^{\x_0\cre} \ketA{0}  \,.
      \label{eq:PathInts_Observables_Observable_Step2}  
    \end{equation}

    To proceed, we split the time interval $[t_0,t]$ into $N$ pieces of length $\Delta t \coloneqq (t-t_0)/N$. The Trotter formula~\cite{Trotter:1959} can then be used to express the right hand side of~\eref{eq:PathInts_Observables_Observable_Step2} in terms of the limit
    \begin{equation}
      \lim_{N\to\infty}
      \braA{0}
      \observable\bigl((\cre+\unitMatrix)\ann\bigr) 
      \bigl[1+ \qMatrix(\cre+\unitMatrix, \ann) \Delta t  \bigr]^N
      \ee^{\x_0\cre} \ketA{0}  .
      \label{eq:PathInts_Observables_Observable_Step3}  
    \end{equation}
    This expression is now rewritten by inserting $N$ of the identity matrices~\cite{Peliti:1985}
   \begin{eqnarray}
      \unitMatrix
        &=  \sum_{m,n=0}^\infty  \frac{1}{m!}  
            \Bigl(
              \int_{\reals} \!\diff{\x}\, (\partial_\x^n\, \x^m) \, 
              \int_{\reals} \frac{\diff{\q}}{2\pi}\, \ee^{-\ii\q \x}
            \Bigr)  
            \ketA{m}\braA{n} \\
        &= \int_{\reals^2} 
          \frac{\diff{\x}\diff{\q}}{2\pi}
          \ee^{-(\ii\q-1) \x}
          \ketB{\x}\braB{-\ii\q}    \,.
    \end{eqnarray}
    We obtained the expression in the second line by performing integrations by parts while neglecting any potential boundary terms (note that the derivations in sections~\ref{sec:PathInts_Backward} and~\ref{sec:PathInts_Forward} did not involve integrations by parts). Moreover, we introduced the right eigenvector $\ketB{z} \coloneqq  \ee^{z (\cre-\unitMatrix)} \ketA{0}$ of the annihilation matrix ($\ann\ketB{z} = z \ketB{z}$ with $z\in\complex$), and the left eigenvector $\braB{z^\star} \coloneqq  \braA{0} \ee^{z \ann}$ of the creation matrix ($\braB{z^\star} \cre = z \braB{z^\star}$). These vectors are commonly referred to as ``coherent states''. The eigenvector conditions can be easily verified by rewriting the vectors as
    \begin{equation}
      \ketB{z} = \sum_{m=0}^\infty \frac{z^m \ee^{-z}}{m!}  \ketA{m}
      \mathtext{ and }
      \braB{z^\star} = \sum_{n=0}^\infty z^n  \braA{n}   \,.
    \end{equation}
    Insertion of the identity matrices into~\eref{eq:PathInts_Observables_Observable_Step3} results in
    \begin{eqnarray} 
      \langle \observable \rangle_{\x_0} 
      &= \lim_{N\to\infty}
          \pathintegral{1}{N}  
        \, \braA{0} \observable((\cre+\unitMatrix)\ann) \ee^{\x_N \cre} \ketA{0}      \\
      &\ \cdot \prod_{j=1}^N  \braA{0} 
        \ee^{-\ii\q_j (\x_j - \ann)} 
        \bigl[1+ \qMatrix(\cre+\unitMatrix, \ann) \Delta t  \bigr]
        \ee^{\x_{j-1} \cre} 
        \ketA{0}      \nonumber  \,.                        
    \end{eqnarray} 
    This expression can be simplified with the help of the shift operations in~\eref{eq:GenFct_ForwardBases_Algebraic_Identity1} and~\eref{eq:GenFct_ForwardBases_Algebraic_Identity2}. Since the $\qMatrix$-matrix~\eref{eq:PathInts_Observables_Algebraic_QMatrix} is a normal-ordered polynomial in $\cre$ and $\ann$ (i.e.\ all the $\cre$ are to the left of all the $\ann$), and both $\braA{0}\cre$ and $\ann\ketA{0}$ vanish, the above expression evaluates to
    \begin{eqnarray} 
      \langle \observable \rangle_{\x_0}
      &= \lim_{N\to\infty}
          \pathintegral{1}{N}
          \, \braA{0} \observable((\cre+\unitMatrix)(\ann+\x_N \unitMatrix)) \ketA{0}     \label{eq:PathInts_Observables_Observable_Step4}   \\
            &\ \cdot\prod_{j=1}^N  
            \ee^{-\ii\q_j (\x_j - \x_{j-1})}  
             \bigl[1+ \qMatrix(\ii\q_j+1, \x_{j-1})  \Delta t   \bigr]
              \nonumber  \,.                                   
    \end{eqnarray} 

    The factor $\braA{0} \observable((\cre+\unitMatrix)(\ann+\x_N \unitMatrix)) \ketA{0}$ could be evaluated by normal-ordering the observable with respect to $\cre$ and $\ann$ before employing $\braA{0}\cre=0$ and $\ann\ketA{0}=0$ again. The resulting object is sometimes called a ``normal-ordered observable''~\cite{Wiese:2015}. In the following, we show that this object agrees with the average over a Poisson distribution in~\eref{eq:PathInts_Observables_FieldMeanOfObservables}. The proof of this assertion is based on the observation that $f_j(\xAd) \coloneqq  \braA{0}((\cre+\unitMatrix)(\ann+\xAd \unitMatrix))^j \ketA{0} $ fulfils the following defining relation of Touchard polynomials (see~\cite{Touchard:1939,Peccati:2011} for information on these polynomials):
    \begin{equation} 
      f_{j+1}(\xAd) = \xAd \sum_{i=0}^j \binomCoeff{j}{i} f_i(\xAd) \mathtext{ with } f_0(\xAd)=1   \,.
    \end{equation} 
    Since the $j$-th Touchard polynomial $f_{j}(\xAd)$ agrees with the $j$-th moment of a Poisson distribution with mean $\xAd$, i.e.\ with $\LLangle n^j \RRangle_{\xAd}$ as defined in~\eref{eq:PathInts_Observables_FieldMeanOfObservables}, our above assertion holds true. 
    
    The path integral representation~\eref{eq:PathInts_Observables_MeanOfObservables} of averaged observables is recovered as the continuous-time limit of~\eref{eq:PathInts_Observables_Observable_Step4} (upon rewriting $1+ \qMatrix\Delta t$ as an exponential). The action~\eref{eq:PathInts_Observables_Derivation_Action} is also recovered because the transition matrix $\qMatrix(\cre, \ann)$ in~\eref{eq:PathInts_Observables_Algebraic_QMatrix} and the adjoint transition operator $\transitionOpAd_\tVar(\cre, \ann)$ in~\eref{eq:GenFctnl_Flow_Poisson_HamiltonianB} fulfil $\qMatrix(\ii\q + 1, \x) = \transitionOpAd(\x, \ii\q + 1)$ for scalar arguments.

  \subsection{Perturbation expansions}\label{subsec:PathInts_Observables_Perturbation} 
  
    The path integral representation~\eref{eq:PathInts_Observables_MeanOfObservables} of factorial moments can be rewritten in terms of a $(\QAd, \XAd)$-generating functional analogously to \sref{subsec:PathInts_BackwardAlongPaths}. The resulting expression may serve as the starting point for a perturbative~\cite{Peliti:1985,Taeuber:2014} or a non-perturbative analysis~\cite{Canet:2011} of the path integral. Here we focus on the perturbative approach. To derive the representation, we assume that the adjoint transition operator can be split into drift, diffusion, and perturbation parts as in~\eref{eq:PathInts_BackwardAlongPaths_Hamiltonian}, so that
    \begin{equation}
       \transitionOpAd_\tVar(\xAd, \partial_\xAd + 1) \label{eq:PathInts_Observables_Perturbation_Op}
        = \alpha_\tVar(\xAd) \partial_\xAd + \beta_\tVar(\xAd) \frac{\partial_\xAd^2}{2}  + \perturbationOpAd_\tVar(\xAd, \partial_\xAd)    \,.
    \end{equation} 
    By following the steps in \aref{sec:A_BackwardAlongPaths}, the path integral representation of a factorial moment can be written as
    \begin{eqnarray}
      \langle (n)_j \rangle_{\xAd(t_0)}
      =
      \frac{\delta^j}{\delta \QAd(t)^j}
         \ee^{\int_{t_0}^{t} \diff{\tVar}\, \perturbationOpAd_\tVar (\frac{\delta}{\delta \QAd(\tVar)}, \frac{\delta}{\delta \XAd(\tVar)})}
         \fieldGenFctAd^0_{\QAd, \XAd}   
        \Big|_{\QAd=\XAd=0} \,,   \nonumber\\ \label{eq:PathInts_Observables_Perturbation_Factorial}
    \end{eqnarray}
    where $\fieldGenFctAd^0_{\QAd, \XAd} = \bigLLangle \ee^{  \int_{t_0}^t\! \diff{\tVar}  \QAd(\tVar) \xAd(\tVar)} \bigRRangle_{\wienerProcess}$ represents the $(\QAd, \XAd)$-generating functional~\eref{eq:PathInts_BackwardAlongPaths_PhiXGeneratingFunction_0}. Moreover, $\xAd(\tVar)$ solves the It\^{o} SDE
    \begin{equation}
      \diff{\xAd}(\tVar) 
      = \bigl[\alpha_\tVar(\xAd) + \XAd(\tVar)\bigr]\diff{\tVar}  + \sqrt{\beta_\tVar(\xAd)}\diff{\wienerProcess}(\tVar)     
    \end{equation} 
    with initial condition $\xAd(t_0)$. If both the diffusion and perturbation parts of the transition operator~\eref{eq:PathInts_Observables_Perturbation_Op} vanish, the factorial moment at time $t$ obeys the statistics of a Poisson distribution, i.e.\ $\langle (n)_j \rangle = \xAd(t)^j$. In the general case, the representation~\eref{eq:PathInts_Observables_Perturbation_Factorial} may be evaluated perturbatively by expanding its exponential as~\cite{Peliti:1985}
    \begin{equation}
       \ee^{\int_{t_0}^{t} \diff{\tVar}\, \perturbationOpAd_\tVar}     \label{eq:PathInts_Observables_Perturbation_Expansion}
       =  \sum_{m=0}^{\infty}
         \int_{t_0}^t \diff{\tVar_{m}}\, \perturbationOpAd_{\tVar_{m}}
        \cdots  
        \int_{t_0}^{\tVar_2} \diff{\tVar_{1}}\, \perturbationOpAd_{\tVar_1}     \,.
    \end{equation}
    Note that this expression involves only a single sum and not a double sum as the expansion~\eref{eq:PathInts_BackwardSolutions_DiffusionAndDecay_Expansion} did.
    
  \subsection{Coagulation}\label{subsec:PathInts_Observables_Binary} 
    
    Let us exemplify the perturbation expansion of~\eref{eq:PathInts_Observables_Perturbation_Factorial} for particles that diffuse and coagulate in the reaction $2\, A \to A$. This reaction exhibits the same asymptotic particle decay as the binary annihilation reaction $2\, A \to \emptyset$ and thus belongs to the same universality class~\cite{Kang:1984,Peliti:1986}. Only a pre-factor differs between the perturbation operators of the two reactions (see below). A full renormalization group analysis of general annihilation reactions $k\, A\to l\, A$ with $l<k$ was performed by Lee~\cite{Lee:1994a,Lee:1994b} (see also~\cite{Taeuber:2005}). The procedure of Lee differs slightly from the one presented below, but it involves the same Feynman diagrams. We restrict ourselves to the ``tree level'' of the diagrams. The coagulation reaction $2\, A \to A$ and the annihilation reaction $2\, A \to \emptyset$ eventually cause all but possibly one particle to vanish from a system. The transition into an absorbing steady state can be prevented; for example, by allowing for the creation of particles through the reaction $\emptyset \to A$. The fluctuations around the ensuing non-trivial steady state have been explored with the help of path integrals in~\cite{Droz:1993,Rey:1997}.
        
    In the following, we consider particles that diffuse on an arbitrary network $\lattice$ and that coagulate in the reaction $2\, A \to A$. Hence, all results from \sref{subsubsec:PathInts_BackwardSolutions_DiffusionAndDecay} apply here as well, except that the transition operator of the coagulation reaction now acts as a perturbation. With the local coagulation rate coefficient $\rateCoeffCoagulation_{\tVar,i}$ at lattice site $i\in\lattice$, this perturbation reads (cf.~\eref{eq:GenFctnl_Flow_Poisson_HamiltonianB})
    \begin{eqnarray}
      \perturbationOpAd_\tVar(\vect{\xAd}, \ii\vect{\qAd} )
      & =  \sum_{i\in\lattice} \rateCoeffCoagulation_{\tVar,i} 
        \xAd_i^2  \bigl[ (\ii\qAd_i + 1) - (\ii\qAd_i+1)^2  \bigr]  \\
        &=  \sum_{i\in\lattice}   \bigl[
         (-\rateCoeffCoagulation_{\tVar,i}) \xAd_i^2 \ii \qAd_i  
         + (-\rateCoeffCoagulation_{\tVar,i}) \xAd_i^2 (\ii \qAd_i)^2  
       \bigr]    \,.
    \end{eqnarray}
    For the binary annihilation reaction $2\, A\to\emptyset$, the first rate coefficient $\rateCoeffCoagulation_{\tVar,i}$ in this expression differs by an additional pre-factor $2$. Note that we allow the rate coefficient to depend both on time and on the node $i\in\lattice$. 
    
    As explained in \sref{subsubsec:PathInts_BackwardSolutions_DiffusionAndDecay}, summands of the expansion of~\eref{eq:PathInts_Observables_Perturbation_Factorial} can be represented by Feynman diagrams. For the mean local particle number $\langle n_i \rangle$, each of the diagrams is composed of a sink with one incoming leg, and possibly of the vertices
    \begin{center}    
      \begin{equation}    \label{eq:PathInts_Observables_Binary_Vertices}
        \begin{tikzpicture}[node distance=7.071mm and 10mm,baseline=(current  bounding  box.center)]
          \coordinate[vertex,label=right:$-\rateCoeffCoagulation_{\tVar,i}$] (tripleVertex);
          \coordinate[left=of tripleVertex] (tripleSink);
          \draw[lineWithArrow1] (tripleVertex) -- (tripleSink);
          \coordinate[above right=of tripleVertex,xshift=-2.93mm] (tripleSource1);
          \coordinate[below right=of tripleVertex,xshift=-2.93mm] (tripleSource2);
          \draw[lineWithArrow1] (tripleSource1) -- (tripleVertex);
          \draw[lineWithArrow1] (tripleSource2) -- (tripleVertex);
          \coordinate[right=of tripleVertex,xshift=3.0mm,label=right:and] (centerCoord);
          \coordinate[vertex,right=of centerCoord,xshift=7.0mm,label=right:$-\rateCoeffCoagulation_{\tVar,i}$] (quadVertex);
          \coordinate[above left=of quadVertex,xshift=2.93mm] (quadSink1);
          \coordinate[below left=of quadVertex,xshift=2.93mm] (quadSink2);
          \draw[lineWithArrow1] (quadVertex) -- (quadSink1);
          \draw[lineWithArrow1] (quadVertex) -- (quadSink2);
          \coordinate[above right=of quadVertex,xshift=-2.93mm] (quadSource1);
          \coordinate[below right=of quadVertex,xshift=-2.93mm] (quadSource2); %,label=right:$\quad.$
          \draw[lineWithArrow1] (quadSource1) -- (quadVertex);
          \draw[lineWithArrow1] (quadSource2) -- (quadVertex);
        \end{tikzpicture}
        \hspace{2mm}  .
      \end{equation}
    \end{center}
   According to the functional derivative~\eref{eq:PathInts_BackwardSolutions_DiffusionAndDecay_Diff2}, contracted lines, either between two vertices or between a vertex and the sink, are associated to the propagator $\propagator(\tVar|\tVar^\prime)$. Dangling incoming lines, on the other hand, introduce a factor $\vect{\xAd}_h(\tVar)$, which represents the homogeneous solution of $\partial_\tVar \xAd_i = \discreteLaplaceOp \rateCoeffDiffDisc_{\tVar,i} \xAd_i + \XAd_i(\tVar)$ (cf.~\eref{eq:PathInts_BackwardSolutions_DiffusionAndDecay_DiffEqSol} and~\eref{eq:PathInts_BackwardSolutions_DiffusionAndDecay_Diff1}). 
   
   In general, diagrams constructed from the above building blocks exhibit internal loops. The simplest connected diagram with such a loop and with a single sink is given by
    \begin{center}     
      \begin{equation}
        \begin{tikzpicture}[node distance=7.071mm and 10mm,baseline=(current  bounding  box.center)]
          \coordinate[vertex] (sink);
          \coordinate[vertex, right=of sink] (tripleVertex);
          \draw[lineWithArrow1] (tripleVertex) -- (sink);
          \coordinate[vertex, right=of tripleVertex,xshift=4.5mm] (quadVertex);
          \upperloop[linePlain]{quadVertex}{tripleVertex};
          \lowerloop[linePlain]{quadVertex}{tripleVertex};
          \coordinate[above right=of quadVertex,xshift=-2.93mm] (upperSource);
          \coordinate[below right=of quadVertex,xshift=-2.93mm] (lowerSource); %,label=right:$\quad.$
          \draw[lineWithArrow1] (upperSource) -- (quadVertex);
          \draw[lineWithArrow1] (lowerSource) -- (quadVertex);
          %
          % The following lines put Arrowheads on the loops.
          % Its ugly. But the automatic placement performed an ugly rotation of the arrows.
          %
          \coordinate (centerOfUpperLoopTip) at (17.9mm, 7.74mm);
          \coordinate (centerOfUpperLoopTail) at (17.9mm+0.0001mm, 7.74mm);
          \draw[black, thick, decoration={markings, mark=at position 1 with {\arrow[scale=1.75]{stealth}}}, postaction={decorate}, draw opacity=0] (centerOfUpperLoopTail) -- (centerOfUpperLoopTip);
          \coordinate (centerOfLowerLoopTip) at (18.0mm, -7.74mm);
          \coordinate (centerOfLowerLoopTail) at (18.0mm+0.0001mm, -7.74mm);
          \draw[black, thick, decoration={markings, mark=at position 1 with {\arrow[scale=1.75]{stealth}}}, postaction={decorate}, draw opacity=0] (centerOfLowerLoopTail) -- (centerOfLowerLoopTip);
        \end{tikzpicture}
        \hspace{2mm}  .
      \end{equation}  
    \end{center}  
    This loop is part of the $m=2$ summand of~\eref{eq:PathInts_Observables_Perturbation_Expansion}. Its mathematical expression is obtained by tracing the diagram from right to left and reads
    \begin{eqnarray}
      \int_{t_0}^t \diff{\tVar_2} \sum_{j\in\lattice} \propagator(t,i|\tVar_2,j)
      (-\rateCoeffCoagulation_{\tVar_2,j})\\
      \cdot \int_{t_0}^{\tVar_2} \diff{\tVar_1} \sum_{k\in\lattice}
      2 \bigl[\propagator(\tVar_2,j|\tVar_1,k)\bigr]^2
      (-\rateCoeffCoagulation_{\tVar_1,k})  
      \bigl[\xAd_{h,k}(\tVar_1)\bigr]^2      \,.\nonumber
    \end{eqnarray}
    The combinatorial factor $2$ stems from the two possible ways of connecting the two vertices (either outgoing leg can connect to either incoming leg).  Note that the sink, which corresponds to the final derivative $\delta/\delta \QAd(t)$, is not associated to an additional pre-factor (unlike in \sref{subsubsec:PathInts_BackwardSolutions_DiffusionAndDecay}). 
    
    In the following, we focus on the contribution of ``tree diagrams'' to the mean particle number. These diagrams do not exhibit internal loops. Thus, upon removing all the diagrams containing loops from the expansion of~\eref{eq:PathInts_Observables_Perturbation_Factorial}, one can define the ``tree-level average''
    \begin{center}
        \begin{tikzpicture}[node distance=0.707cm and 1.cm]
          \coordinate[vertex,label={[xshift=-0.9cm,yshift=-0.325cm]$ \bar{n}_i(t)\,  \coloneqq $}] (sink1);
          \coordinate[right=of sink1,label={[xshift=0.28cm,yshift=-0.23cm]$+$}] (source1);
          \draw[lineWithArrow1] (source1) -- (sink1);
          \coordinate[vertex,right=of source1,xshift=-0.4cm] (sink2);
          \coordinate[vertex,right=of sink2,label={[xshift=0.65cm,yshift=-0.31cm]$+\,\,2$}] (tripleVertex2);
          \draw[lineWithArrow1] (tripleVertex2) -- (sink2);
          \coordinate[above right=of tripleVertex2,xshift=-0.293cm] (source12);
          \coordinate[below right=of tripleVertex2,xshift=-0.293cm] (source22);
          \draw[lineWithArrow1] (source12) -- (tripleVertex2);
          \draw[lineWithArrow1] (source22) -- (tripleVertex2);
          \coordinate[vertex,right=of tripleVertex2,xshift=-0.025cm] (sink3);
          \coordinate[vertex,right=of sink3,label={[xshift=0.9cm,yshift=-0.31cm]$+\hspace{1mm}\hldots \,\,.$}] (tripleVertex13);
          \draw[lineWithArrow1] (tripleVertex13) -- (sink3);
          \coordinate[vertex,above right=of tripleVertex13,xshift=-0.293cm] (tripleVertex23);
          \coordinate[right=of tripleVertex23,xshift=-0.2cm] (source13);
          \coordinate[above=of tripleVertex23,yshift=0.093cm] (source23);
          \draw[lineWithArrow1] (tripleVertex23) -- (tripleVertex13);
          \draw[lineWithArrow1] (source13) -- (tripleVertex23);
          \draw[lineWithArrow1] (source23) -- (tripleVertex23);
          \coordinate[below right=of tripleVertex13,xshift=-0.293cm] (source33);
          \draw[lineWithArrow1] (source33) -- (tripleVertex13);
          % the next two lines are causing the mess
        \end{tikzpicture}
        \begin{equation} \label{eq:PathInts_Observables_Binary_SumOfTrees}\end{equation}
    \end{center}
    The corresponding mathematical expression reads
    \begin{eqnarray}
       \bar{n}_i(t)        \label{eq:PathInts_Observables_Binary_SumOfTrees_Math}
      &= \xAd_{h,i}(t)   \\
      &+ \int_{t_0}^t \diff{\tVar_1} \sum_{j\in\lattice} \propagator(t,i|\tVar_1,j) (-\rateCoeffCoagulation_{\tVar_1,j}) \bigl[\xAd_{h,j}(\tVar_1)\bigr]^2   \nonumber\\
      &+ 2 \int_{t_0}^t \diff{\tVar_2} \sum_{j\in\lattice} \propagator(t,i|\tVar_2,j) (-\rateCoeffCoagulation_{\tVar_2,j}) \xAd_{h,j}(\tVar_2)   \nonumber\\
      &\phantom{{}+{}} \cdot \int_{t_0}^{\tVar_2} \diff{\tVar_1} \sum_{k\in\lattice} \propagator(\tVar_2,j|\tVar_1,k) (-\rateCoeffCoagulation_{\tVar_1,k}) \bigl[\xAd_{h,k}(\tVar_1)\bigr]^2 \nonumber\\
      &+ \hldots \nonumber \,.
    \end{eqnarray}
    The inclusion of diagrams with loops would correct $\bar{n}_i$ to the true mean $\langle n_i \rangle$. For the treatment of loops, see for example~\cite{Lee:1994a,Lee:1994b,Cardy:1997,Taeuber:2014}.

    The tree-level average $\bar{n}_i$ defined above fulfils the deterministic rate equation of the coagulation process, i.e. it fulfils
    \begin{equation}
      \partial_\tVar  \bar{n}_i    \label{eq:PathInts_Observables_Binary_RateEq}
      = \discreteLaplaceOp \rateCoeffDiffDisc_{\tVar,i}  \bar{n}_i - \rateCoeffCoagulation_{\tVar,i} \bar{n}_i^2 \,.
    \end{equation}
    If both the rates $\rateCoeffDiffDisc_\tVar$ and $\rateCoeffCoagulation_\tVar$, and also the mean $\xAd(t_0)$ of the initial Poisson distribution are spatially homogeneous, the well-mixed analogue of \eref{eq:PathInts_Observables_Binary_RateEq} can be derived by direct resummation of~\eref{eq:PathInts_Observables_Binary_SumOfTrees_Math}. In the general case, the validity of the  rate equation~\eref{eq:PathInts_Observables_Binary_RateEq} can be established by exploiting a self-similarity of $\bar{n}_i$. For that purpose, we assume that $\bar{n}_i(\tVar)$ is known for $\tVar < t$ and we want to extend its validity up to time~$t$. To do so, we remove all the sinks from the diagrams in~\eref{eq:PathInts_Observables_Binary_SumOfTrees} and connect their now dangling outgoing legs with the incoming leg of another three-vertex at time $\tVar$. The second incoming leg of this vertex is contracted with every other tree diagram and its outgoing leg with a sink at time $t$. The corresponding expression reads $\sum_{j\in\lattice} \propagator(t,i|\tVar,j) (-\rateCoeffCoagulation_{\tVar,j}) \bar{n}_j(\tVar)^2$ and it contributes to $\bar{n}_i(t)$ for every time $\tVar$. To respect also the initial condition $\bar{n}_i(t_0) = \xAd_{i}(t_0)$, we introduce the first diagram of~\eref{eq:PathInts_Observables_Binary_SumOfTrees} by hand so that in total
    \begin{equation}
      \bar{n}_i(t)   \label{eq:PathInts_Observables_Binary_Solution}
      = \xAd_{h,i}(t) + \int_{t_0}^t \diff{\tVar} \sum_{j\in\lattice} \propagator(t,i|\tVar,j) (-\rateCoeffCoagulation_{\tVar,j}) \bar{n}_j(\tVar)^2 \,.
    \end{equation}
    Differentiation of this equation with respect to $t$ confirms the validity of the rate equation~\eref{eq:PathInts_Observables_Binary_RateEq} (recall the definition of the propagator in \sref{subsubsec:PathInts_BackwardSolutions_DiffusionNetwork}). 
    
  \subsection{R\'esum\'e}\label{subsec:PathInts_Observables_Resume}  
  
    Path integral representations of averaged observables have proved useful in a variety of contexts. Their use has deepened our knowledge about the critical behaviour of diffusion-limited annihilation and coagulation reactions~\cite{Peliti:1986,Lee:1994b,HowardMJ:1997,Hinrichsen:1999,Cardy:1997,Hnatich:2000,Vernon:2003,Taeuber:2005,Hnatic:2013,Taeuber:2014}, of branching and annihilating random walks and percolation processes~\cite{Cardy:1996,Cardy:1998,Taeuber:1998,Goldschmidt:1999,Janssen:2005,Taeuber:2005,Benitez:2012,Benitez:2013,Taeuber:2014}, and of elementary multi-species reactions~\cite{Lee:1994c,Howard:1995,Lee:1995,Howard:1996a,Howard:1996b,Oerding:1996,Sasamoto:1997,Wijland:1998,Rey:1999,Konkoli:1999,Konkoli:2000,Hilhorst:2004}. In the present section, we derived a path integral representation of averaged observables for processes that can be decomposed additively into reactions of the form $k\, A \to l\, A$. Provided that the number of particles $n$ in the system is initially Poisson distributed with mean $\xAd(t_0)$, we showed that the average of an observable $A(n)$ can be represented by the path integral 
    \begin{eqnarray}
      &\langle \observable \rangle_{\xAd(t_0)}    \label{eq:PathInts_Observables_Resume_PathInt}
      =  \pathintegral{(t_0}{t]}
        \ee^{-\actionAd} 
        \, \LLangle \observable \RRangle_{\xAd(t)}                     \\
      & \mathtext{with } 
      \LLangle \observable \RRangle_{\xAd}    \label{eq:PathInts_Observables_Resume_Av}
      = \sum_{n=0}^\infty  \frac{\xAd^{n} \ee^{-\xAd}}{n!}  \observable(n)      \,.
    \end{eqnarray}     
    We derived this representation in \sref{subsec:PathInts_Observables_Derivation} by summing the backward path integral representation~\eref{eq:PathInts_Backward_ContinuousTime_Solution_MarginalizedDistribution} of the marginalized distribution over the observable $A(n)$ (using the Poisson basis function $\ketA{n}=\frac{\xAd^{n} \ee^{-\xAd}}{n!}$). The generalization of the above path integral to reactions with multiple types of particles and spatial degrees of freedom is straightforward. The above path integral representation was found to be equivalent to Doi-shifted path integrals used in the literature~\cite{Lee:1994b,Lee:1995,Cardy:1996,Cardy:1998,Janssen:2005,Taeuber:2014}. Its unshifted version is obtained for a redefined basis function. Unlike path integral representations encountered in the literature, our representation~\eref{eq:PathInts_Observables_Resume_PathInt} does not involve a so-called ``normal-ordered observable''. By using a defining relation of Touchard polynomials, we could show that this object agrees with the average~\eref{eq:PathInts_Observables_Resume_Av} of $A(n)$ over a Poisson distribution (cf.\ \sref{subsec:PathInts_Observables_Algebraic}).
  
    As shown in \sref{subsec:PathInts_Observables_Perturbation}, the path integral representation~\eref{eq:PathInts_Observables_Resume_PathInt} can be rewritten in terms of a perturbation expansion. We demonstrated the evaluation of this expansion in \sref{subsec:PathInts_Observables_Binary} for diffusing particles that coagulate according to the reaction $2\, A \to A$. In doing so, we restricted ourselves to the tree-level of the Feynman diagrams associated to the perturbation expansion. Information on perturbative renormalization group techniques for the treatment of diagrams with loops can be found in~\cite{Taeuber:2005,Taeuber:2014}. Recently, non-perturbative renormalization group techniques have been developed for the evaluation of stochastic path integrals~\cite{Canet:2004a,Canet:2011}. These techniques have proved particularly useful in studying branching and annihilating random walks~\cite{Canet:2004a,Canet:2004b} and annihilation processes~\cite{Winkler:2012,Homrighausen:2013}.

\section{Stationary paths}\label{sec:StationaryPaths}  

  In the previous sections, we outlined how path integrals can be expressed in terms of averages over the paths of stochastic differential equations (SDEs). Corrections to those paths were treated in terms of perturbation expansions. In the following, we formulate an alternative method in which the variables of the path integrals act as deviations from ``stationary'', or ``extremal'', paths. The basic equations of the method have the form of Hamilton's equations from classical mechanics. Their application to stochastic path integrals goes back at least to the work of Mikhailov~\cite{Mikhailov:1981a}. 
  
  More recently, Elgart and Kamenev extended the method for the study of rare event probabilities~\cite{Elgart:2004} and for the classification of phase transitions in reaction-diffusion models with a single type of particles~\cite{Elgart:2006}. These studies are effectively based on the forward path integral representation~\eref{eq:PathInts_Forward_DiscreteTime_Solution_GeneratingFunction} of the generating function. After reviewing how this approach can be used for the approximation of probability distributions, we extend it to the backward path integral representation~\eref{eq:PathInts_Backward_DiscreteTime_Solution_MarginalizedDistribution} of the marginalized distribution. Whereas the generating function approach requires an auxiliary saddle-point approximation to extract probabilities from the generating function, the backward approach provides direct access to probabilities. A proper normalization of the resulting probability distribution is, however, only attained beyond leading order. The generating function technique respects the normalization of the distribution even at leading order, but this normalization may be violated by the subsequent saddle-point approximation.

  The methods discussed in the following all apply to the chemical master equation~\eref{eq:Intro_Mescoscopic_ChemicalMasterEq} and employ the basis functions that we introduced in sections~\ref{subsec:GenFct_ForwardBases_ChemReactions} and~\ref{subsec:GenFctnl_BackwardBases_ChemReactions}. Moreover, we assume the number of particles to be initially Poisson distributed with mean $\xAd(t_0)$. This assumption proves to be convenient in the analysis of explicit stochastic processes but it can be easily relaxed.

  \subsection{Forward path integral approach}\label{subsec:StationaryPaths_GenFct}  
  
    Our goal lies in the approximation of the marginalized probability distribution 
    \begin{equation} 
      \ketA{\prob(t, n | t_0)}_{\xAd(t_0)}    \label{eq:StationaryPaths_GenFct_Marginalized}
      = \sum_{n_0=0}^\infty \prob(t, n|t_0,n_0)  \frac{\xAd(t_0)^{n_0} \ee^{-\xAd(t_0)}}{n_0!}   \,.      
    \end{equation} 
    For this purpose, we employ the forward path integral~\eref{eq:PathInts_Forward_DiscreteTime_Solution_GeneratingFunction} to formulate an alternative representation of the ordinary probability generating function
    \begin{equation}
      \ketA{\gen(t | t_0, n_0)}_{\q(t)}   \label{eq:StationaryPaths_GenFct_GenFct}
      \coloneqq \sum_{n = 0}^\infty \q(t)^n \prob(t, n|t_0,n_0)  \,.
    \end{equation}
    As the first step, we rewrite the argument $\q(t)$ of this function in terms of a deviation $\Delta\q(t)$ from a ``stationary path'' $\tq(t)$, i.e.\ $\q(t)=\tq(t)+\Delta\q(t)$. The path $\tq(t)$ and an auxiliary path $\tx(\tVar)$ are chosen so that the action of the resulting path integral is free of terms that are linear in the path integral variables $\Delta\q(\tVar)$ and $\Delta\x(\tVar)$ (with $\tVar\in[t_0,t]$). Thus, the approach bears similarities with the stationary phase approximation of oscillatory integrals~\cite{Bleistein:1986}. In the next section, we apply the method to the binary annihilation process $2\, A\to\emptyset$. 
    
    In order to implement the above steps, we define the basis function
    \begin{equation}
      \ketA{n}_{\tVar,\Delta\q(\tVar)} \coloneqq (\basisPrefactor\Delta\q(\tVar) + \tq(\tVar))^n \ee^{-\tx(\tVar) (\basisPrefactor\Delta\q(\tVar) + \tq(\tVar))}
    \end{equation}
    for yet to be specified paths $\tq(\tVar)$ and $\tx(\tVar)$, and a free parameter $\basisPrefactor$. The basis function has the same form as the basis function~\eref{eq:GenFct_Flow_Poisson_BaseFunction} but with its second argument being written as $\Delta\q(\tVar)$. Provided that the path $\tq(\tVar)$ fulfils the final condition $\tq(t)=\q(t)$, one can trivially rewrite the generating function~\eref{eq:StationaryPaths_GenFct_GenFct} in terms of the above basis function as
    \begin{eqnarray}
        \ketA{\gen}_{\q(t)}
        =
        \ee^{\tx(t) \tq(t)}
        \Bigl(
        \sum_{n = 0}^\infty 
        \ketA{n}_{t,\Delta\q(t)} \,
        \prob(t, n | \cdot)   
        \Bigr)
        \Big|_{\Delta \q(t) = 0}    \,.  
    \end{eqnarray} 
    The term in brackets has the form of the generalized generating function~\eref{eq:GenFct_Flow_GenFct}, and thus it can be rewritten in terms of the forward path integral~\eref{eq:PathInts_Forward_ContinuousTime_Solution_GeneratingFunction} by following the steps in~\sref{subsec:PathInts_Forward_Derivation}. Using $\Delta\q(\tVar)$ and $\Delta\x(\tVar)$ as labels for the path integral variables, one arrives at the representation
    \begin{equation}
        \ketA{\gen}_{\q(t)}
        =
      \ee^{\tx(t) \tq(t)}  \label{eq:StationaryPaths_GenFct_DerivationStep1}
        \Bigl(
          \pathintegral{[t_0}{t)}
          \, \ee^{-\action}
          \, \ketA{n_0}_{t_0, \Delta\q(t_0)}  
          \Bigr)
          \Big|_{\Delta \q(t) = 0}      \,.
    \end{equation} 
    In analogy with \eref{eq:PathInts_Forward_DiscreteTime_Measure}, the integral sign is defined as the continuous-time limit
   \begin{equation}
        \pathintegral{[t_0}{t)} = 
        \lim_{N \to \infty} 
        \prod_{j=0}^{N-1}  \int_{\reals^2}\!\frac{\diff\Delta{\x_j}\diff{\Delta\q_{j}}}{2\pi}    \,.   
    \end{equation} 
    The action~\eref{eq:PathInts_Forward_ContinuousTime_Solution_Action} inside the generating function~\eref{eq:StationaryPaths_GenFct_DerivationStep1} reads
    \begin{eqnarray}
      \action   \label{eq:StationaryPaths_GenFct_Action}
      =  \int_{t_0}^{t}\!\diff{\tVar}\, \bigl[
                  \ii\Delta\x \partial_{-\tVar} \Delta\q - \transEvoOp_\tVar(\Delta\q, \ii\Delta\x) 
                  \bigr] \,.               
    \end{eqnarray} 

    As emphasized above, our interest lies in a system whose initial number of particles is Poisson distributed with mean $\x(t_0)$. Thus, instead of dealing with the ordinary generating function~\eref{eq:StationaryPaths_GenFct_GenFct}, it proves convenient to work in terms of the generating function
    \begin{eqnarray}
      \ketA{\gen(t | t_0; \x(t_0))}_{\q(t)} 
      \coloneqq \sum_{n_0=0}^\infty \ketA{\gen(t | t_0, n_0)}_{\q(t)}  \frac{\xAd(t_0)^{n_0} \ee^{-\xAd(t_0)}}{n_0!}    \nonumber\\
      = \ee^{\tx(t) \tq(t)-\x(t_0)}  
          \pathintegral{[t_0}{t)}
          \, \ee^{-\action}
          \Big|_{\Delta \q(t) = 0}    \label{eq:StationaryPaths_GenFct_GenFct_PathInt1}   \,.
    \end{eqnarray}
    To arrive at the second line, we made use of the path integral representation~\eref{eq:StationaryPaths_GenFct_DerivationStep1} while requiring that the path $\tx(\tVar)$ meets the initial condition $\tx(t_0) = \x(t_0)$. Note that the marginalized distribution~\eref{eq:StationaryPaths_GenFct_Marginalized} is recovered from the redefined generating function via
    \begin{equation}
      \ketA{\prob(t, n | t_0)}_{\x(t_0)} 
      = \frac{1}{n!} \partial_{\q(t)}^n \ketA{\gen(t | t_0; \x(t_0))}_{\q(t)}  \Big|_{\q(t)=0}      \,.
    \end{equation}
    
    Thus far, we have not yet specified the paths $\tq(\tVar)$ and $\tx(\tVar)$, besides requiring that they fulfil the boundary conditions $\tq(t)=\q(t)$ and $\tx(t_0) = \x(t_0)$. We specify these paths in such a way that the action~\eref{eq:StationaryPaths_GenFct_Action} becomes free of terms that are linear in the deviations $\Delta\q$ and $\Delta\x$. For this purpose, let us recall the definitions of the creation operator $\cre=\basisPrefactor\Delta\q + \tq$ and of the annihilation operator $\ann=\partial_{\basisPrefactor\Delta\q} + \tx$ from section \sref{subsec:GenFct_ForwardBases_ChemReactions}, as well as the definition of the transition operator
    \begin{equation}
       \transEvoOp_\tVar(\Delta\q, \partial_{\Delta\q}) 
       =   \transitionOp_\tVar(\cre, \ann) 
           + \evolutionOp_\tVar(\cre, \ann)  \,.
    \end{equation}
    The basis evolution operator is specified in~\eref{eq:GenFct_Flow_Poisson_HamiltonianT} as
    \begin{equation}
        \evolutionOp_\tVar(\cre, \ann)    
        =   (\partial_\tVar \tq) (\ann - \tx)
                    - (\partial_\tVar \tx) \cre  \,.     
      \end{equation}
     Upon performing a Taylor expansion of the operator $\transEvoOp_\tVar(\Delta\q, \ii\Delta\x)$ in the action~\eref{eq:StationaryPaths_GenFct_Action} with respect to the deviations $\Delta\q$ and $\Delta\x$, one observes that terms that are linear in the deviations vanish from the action if the paths $\tx(\tVar)$ and $\tq(\tVar)$ fulfil%    
    \footnote{
    If Hamilton's equations are fulfilled, it holds that
    \begin{eqnarray*}
      \transEvoOp_\tVar(\Delta\q, \ii\Delta\x)  
      &=  \bigl(\transitionOp_{\tVar}(\tq, \tx) 
          + \frac{\partial \transitionOp_{\tVar}}{\partial\tq} \basisPrefactor\Delta\q
          + \frac{\partial \transitionOp_{\tVar}}{\partial\tx} \basisPrefactor^{-1} \ii\Delta\x   
          + \Delta \transitionOp_\tVar  \bigr)\\
          &\phantom{{}={}}
          + \bigl((\partial_\tVar\tq)\basisPrefactor^{-1} \ii\Delta\x
          - (\partial_\tVar\tx)(\basisPrefactor\Delta\q + \tq)    \bigr)\\
      &=   -\partial_\tVar(\tx\tq)
          - \bigl(\txAd\partial_{-\tVar}\tq - \transitionOp_{\tVar}(\tq, \tx)\bigr)
          + \Delta \transitionOp_\tVar        \,.
    \end{eqnarray*}
    Here, $\Delta \transitionOp_\tVar$ represents all terms of the Taylor expansion of $\transitionOp_\tVar$ that are of second or higher order in the deviations.}
    \begin{eqnarray}
      \partial_\tVar \tx        \label{eq:StationaryPaths_GenFct_1}
       = \frac{\partial\transitionOp_\tVar(\tq, \tx)}{\partial \tq} 
      \mathtext{ with }
      \tx(t_0) = \x(t_0)                
      \mathtext{ and } \\
      \partial_{-\tVar} \tq          \label{eq:StationaryPaths_GenFct_2}
      =  \frac{\partial \transitionOp_\tVar(\tq, \tx)}{\partial \tx} 
      \mathtext{ with }
      \tilde{\q}(t) = \q(t)     \,.
    \end{eqnarray}
    These equations resemble Hamilton's equations from classical mechanics. Just as in classical mechanics, $\transitionOp_\tVar$ is conserved along solutions of the equations if it does not depend on time itself. This property follows from $\frac{\diff{}}{\diff{\tVar}} \transitionOp_\tVar(\tq, \tx) = \partial_\tVar \transitionOp_\tVar(\tq, \tx)$, with $\frac{\diff{}}{\diff{\tVar}}$ being the total time derivative. Since the ordinary generating function obeys the flow equation $\partial_\tVar \ketA{\gen}_\q = \transitionOp_\tVar(\q,\partial_\q)\ketA{\gen}_\q$, the conservation of total probability requires that $\transitionOp_\tVar(1,\partial_\q)=0$. This condition is, for example, fulfilled by the transition operator $\transitionOp_\tVar(\q, \partial_\q) = \rateCoeffGeneric_\tVar (\q^l - \q^k) \partial_\q^k $ of the generic reaction $k\, A \to l\, A$ (cf.~\eref{eq:GenFct_Flow_Poisson_HamiltonianF}). For the final value $\q(t)=1$, Hamilton's equations~\eref{eq:StationaryPaths_GenFct_1} and~\eref{eq:StationaryPaths_GenFct_2} are solved by $\tq(\tVar)=1$ with $\tx(\tVar)$ solving the rate equation $\partial_\tVar \tx = \rateCoeffGeneric_\tVar (l - k) \tx^k$ of the process. Elgart and Kamenev analysed the topology of $\transitionOp_\tVar(\tq, \tx)=0$ lines of reaction-diffusion models with a single type of particles to classify the phase transitions of these models~\cite{Elgart:2006} (they use the label $p$ instead of $\tq$, and $q$ instead of $\tx$).
         
    Provided that Hamilton's equations~\eref{eq:StationaryPaths_GenFct_1} and~\eref{eq:StationaryPaths_GenFct_2} are fulfilled, the action~\eref{eq:StationaryPaths_GenFct_Action} evaluates to
    \begin{eqnarray}
       \action 
       = \tx(t) \tq(t) - \x(t_0) 
         + \tilde{\action}  + \Delta \action 
    \end{eqnarray}
    with the definitions
    \begin{eqnarray} 
      \tilde{\action}     
        \coloneqq  \x(t_0)(1-\tq(t_0))
                    + \int_{t_0}^{t}\!\diff{\tVar}\, \bigl[
                      \tx \partial_{-\tVar} \tq - \transitionOp_\tVar(\tq, \tx) 
                    \bigr] \\
      \mathtext{and }  
      \Delta \action 
      \coloneqq  \int_{t_0}^{t}\!\diff{\tVar}\, \bigl(
                    \ii\Delta\x \partial_{-\tVar} \Delta\q - \Delta\transitionOp_\tVar    
                  \bigr) \,.    
    \end{eqnarray} 
    The transition operator $\Delta\transitionOp_\tVar$ absorbs all terms of the Taylor expansion of $\transitionOp_\tVar$ that are of second or higher order in the deviations. Combined with~\eref{eq:StationaryPaths_GenFct_GenFct_PathInt1}, the above action results in the following path integral representation of the generating function:
    \begin{eqnarray}
      \ketA{\gen(t | t_0; \x(t_0))}_{\q(t)}  \label{eq:StationaryPaths_GenFct_PathIntegral}
      = \ee^{-\tilde{\action}}
        \pathintegral{[t_0}{t)}
        \ee^{-\Delta \action}\big|_{\Delta \q(t)=0}       \,.
    \end{eqnarray} 
    Although this path integral representation may seem daunting, our primary interest lies only in its leading-order approximation $\ketA{\gen} \approx \ee^{-\tilde{\action}}$. This approximation is exact if $\Delta \transitionOp_\tVar$ vanishes. That is, for example, the case for the simple growth process $\emptyset \to A$. For the binary annihilation reaction $2\, A \to \emptyset$, the leading-order approximation was evaluated by Elgart and Kamenev up to a pre-exponential factor~\cite{Elgart:2004}. The pre-exponential factor was later approximated by Assaf and Meerson for large times~\cite{Assaf:2006a}. In the next section, we evaluate the leading-order approximation of the binary annihilation reaction and evaluate its pre-exponential factor for arbitrary times.
    
  \subsection{Binary annihilation}\label{subsec:StationaryPaths_GenFct_BinaryAnnihilation}  
  
    Let us demonstrate the use of the representation~\eref{eq:StationaryPaths_GenFct_PathIntegral} for the binary annihilation reaction $2\, A \to \emptyset$ with rate coefficient $\rateCoeffAnnihilation_\tVar$. First, we evaluate the generating function in leading order. Afterwards, the generating function is cast into a probability distribution using the inverse transformation~\eref{eq:GenFct_Flow_InverseTransformation}. The derivatives involved in the inverse transformation are expressed by Cauchy's differentiation formula, which is evaluated in terms of a saddle-point approximation.
    
    \begin{figure}[tb] 
        \centering
        \includegraphics{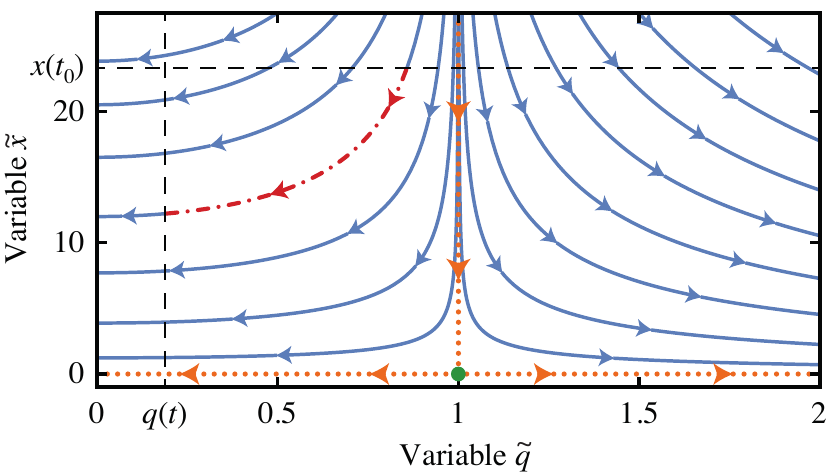}
        \caption{\label{fig:StationaryPaths_GenFct_BinaryAnnihilation_Flow}
          Phase portrait of Hamilton's equations~\eref{eq:StationaryPaths_GenFct_BinaryAnnihilation_XEq} and~\eref{eq:StationaryPaths_GenFct_BinaryAnnihilation_QEq} for the rate coefficient $\rateCoeffAnnihilation_\tVar=1$. The transition operator $\transitionOp_{\tVar}(\tq, \tx)=\rateCoeffAnnihilation_\tVar (1 - \tq^2)\tx^2$ of the binary annihilation process vanishes for $\tx=0$ and $\tq=1$ (orange dotted lines). The green disk represents the fixed point $(\tq,\tx)=(1,0)$ of Hamilton's equations. The red dash-dotted line exemplifies a particular solution of Hamilton's equations for a given time interval $(t_0,t)$, a given initial value $\x(t_0)$ and a given final value $\q(t)$ (black dashed lines).
        }
      \end{figure} 
      
    The transition operator of the binary annihilation reaction follows from~\eref{eq:GenFct_Flow_Poisson_HamiltonianF} as $\transitionOp_\tVar(\cre_\tVar, \ann_\tVar)    = \rateCoeffAnnihilation_\tVar (1 - \cre_\tVar^2) \ann_\tVar^2$ with the annihilation rate coefficient $\rateCoeffAnnihilation_\tVar$. Therefore, Hamilton's equations~\eref{eq:StationaryPaths_GenFct_1} and~\eref{eq:StationaryPaths_GenFct_2} read
    \begin{eqnarray}
      \partial_\tVar \tx          \label{eq:StationaryPaths_GenFct_BinaryAnnihilation_XEq}
       = - 2 \rateCoeffAnnihilation_\tVar \tq \tx^2
      \mathtext{ with }
      \tx(t_0) = \x(t_0)      
      \mathtext{ and } \\
      \partial_{-\tVar} \tq          \label{eq:StationaryPaths_GenFct_BinaryAnnihilation_QEq}      
      =  2 \rateCoeffAnnihilation_\tVar(1-\tq^2) \tx
      \mathtext{ with }
      \tq(t) = \q(t)              \,.
    \end{eqnarray}
    A phase portrait of these equations is shown in \fref{fig:StationaryPaths_GenFct_BinaryAnnihilation_Flow}. Equation~\eref{eq:StationaryPaths_GenFct_BinaryAnnihilation_QEq} is solved by $\tq=1$, for which the previous equation simplifies to the rate equation $\partial_\tVar \tx = - 2 \rateCoeffAnnihilation_\tVar \tx^2$ of the process. This rate equation is solved by
    \begin{equation}
      \bar{\x}(t)   \label{eq:StationaryPaths_GenFct_BinaryAnnihilation_RateEq}
      \coloneqq \tx(t)
      = \frac{\x(t_0)}{1+\x(t_0)/\bar{\x}_\infty(t)}  
    \end{equation}
    with the asymptotic limit $\bar{\x}_\infty(t) \coloneqq (2\int_{t_0}^t \diff{\tVar}\, \rateCoeffAnnihilation_\tVar)^{-1} \to 0$. Since Hamilton's equations conserve $(\tq^2-1) \tx^2$, one can rewrite the equation~\eref{eq:StationaryPaths_GenFct_BinaryAnnihilation_QEq} as
    \begin{equation}
      \partial_\tVar \tq          
      =  - 2 \rateCoeffAnnihilation_\tVar\x(t_0) \sqrt{1-\tq(t_0)^2}\sqrt{1-\tq^2}       \,.
    \end{equation}
    Here we assume $\tq(t)=\q(t) < 1$ but the derivation can also be performed for $\q(t) > 1$  (these inequalities are preserved along the flow). The above equation only allows for an implicit solution that provides $\tq(t_0)$ for a given $\q(t)$, namely
    \begin{equation}
      \arccos{\tq(t_0)}    \label{eq:StationaryPaths_GenFct_BinaryAnnihilation_Eq1}      
        + \frac{\x(t_0)}{\bar{\x}_\infty(t)} \sqrt{1-\tq(t_0)^2}  
      = \arccos{\q(t)}      \,.      
    \end{equation}
    The first term of this equation was neglected in previous studies~\cite{Elgart:2004,Assaf:2006a} (for large times $t$, $\tq(t_0)\approx 1$). Using the conservation of $(\tq^2-1) \tx^2$ once again, the action in the leading-order approximation $\ketA{\gen}_{\q(t)}  = \ee^{-\tilde{\action}(\q(t))}$ can be written as
    \begin{equation}
      \tilde{\action}(\q(t))
      =  \bigl[1-\tq(t_0)\bigr]\x(t_0)
        +\frac{\bigl[1-\tq(t_0)^2\bigr] \x(t_0)^2}{2 \bar{\x}_\infty(t)}   \,.
    \end{equation} 
    We have thus fully specified the generating function in terms of its argument $\q(t)$ and the mean $\x(t_0)$ of the initial Poisson distribution. The leading-order approximation $\ketA{\gen}_{\q(t)}  = \ee^{-\tilde{\action}(\q(t))}$ respects the normalization of the underlying probability distribution because 
    \begin{equation}
      \sum_{n = 0}^\infty \prob(t, n | t_0, n_0)  =\ketA{\gen}_{\q(t)=1} = 1  \,.
    \end{equation}
    Here we used that for $\q(t)=1$, Hamilton's equation~\eref{eq:StationaryPaths_GenFct_BinaryAnnihilation_QEq} is solved by $\tq(\tVar)=1$, implying that $\tilde{\action}(\q(t)=1)=0$.
    
    The probability distribution follows from the generating function via the inverse transformation $\ketA{\prob(t, n | t_0)}_{\x(t_0)} = \frac{1}{n!} \partial_{\q(t)}^n \ketA{\gen}_{\q(t)} \big|_{\q(t)=0}$. This transformation is trivial for $n=0$, for which one obtains the probability of observing an empty system. For other values of $n$, the derivatives can be expressed by Cauchy's differentiation formula so that in leading order
    \begin{equation}
      \ketA{\prob(t, n | t_0)}_{\x(t_0)}  \label{eq:StationaryPaths_GenFct_BinaryAnnihilation_Cauchy} 
      =  \frac{1}{2\pi\ii}  \oint_\complexPath  \frac{\diff{\q(t)}}{\q(t)} \ee^{-\tilde{\action}(\q(t)) - n\ln\q(t)}\,.
    \end{equation}
    Here, $\complexPath$ represents a closed path around zero in the complex domain and is integrated over once in counter-clockwise direction. 
    
    The contour integral in~\eref{eq:StationaryPaths_GenFct_BinaryAnnihilation_Cauchy} can be evaluated in a saddle-point approximation as
    \begin{equation}
      \ketA{\prob(t, n | t_0)}_{\x(t_0)}   \label{eq:StationaryPaths_GenFct_BinaryAnnihilation_SaddleptApprox}    
      \approx \frac{\q_s^{-n}  \ee^{-\tilde{\action}(\q_s)}}{\sqrt{2\pi (n-\q_s^2 \tilde{S}^{\prime\prime}(\q_s))}}    \,.      
    \end{equation} 
    The saddle-point $\q_s$ is found by solving the equation $n/\q_s = -\tilde{S}^{\prime}(\q_s)$. By differentiating the implicit solution~\eref{eq:StationaryPaths_GenFct_BinaryAnnihilation_Eq1} with respect to $\q(t)$, the saddle-point condition can be rewritten as
    \begin{equation}
      \frac{n}{\q_s}   \label{eq:StationaryPaths_GenFct_BinaryAnnihilation_FixPt}
      = \x(t_0)  \frac{\sqrt{1-\tq(t_0)^2}}{\sqrt{1-\q_s^2}}\,.
    \end{equation}
    A closed equation for $\q_s$ is obtained by combining this equation with the implicit solution~\eref{eq:StationaryPaths_GenFct_BinaryAnnihilation_Eq1}. The resulting equation is solved by $\q_s=1$ for $n=\bar{\x}(t)$, i.e.\ if $n$ lies on the trajectory of the rate equation. For other values of $n$, the equation has to be solved numerically. 
    Once $\q_s$ has been obtained, the value $\tq(t_0)$ can be inferred from~\eref{eq:StationaryPaths_GenFct_BinaryAnnihilation_FixPt}.
        
    One piece is still missing for the numeric evaluation of the probability distribution~\eref{eq:StationaryPaths_GenFct_BinaryAnnihilation_SaddleptApprox}, namely its denominator. It evaluates to
    \begin{eqnarray}
      n-\q_s^2 \tilde{\action}^{\prime\prime}(\q_s)  \label{eq:StationaryPaths_GenFct_BinaryAnnihilation_Denominator}
      = \bar{\x}_\infty(t)
        \frac{\q_s^2 \alpha - n / \bar{\x}_\infty(t)}{\q_s^2 -1}      \\
      \mathtext{with } \alpha \coloneqq 1 - \Bigl( 1+\frac{\x(t_0) \tq(t_0)}{\bar{\x}_\infty(t)} \Bigr)^{-1} \,.
    \end{eqnarray} 
    The pre-exponential factor of the distribution~\eref{eq:StationaryPaths_GenFct_BinaryAnnihilation_SaddleptApprox} computed by Assaf and Meerson~\cite{Assaf:2006a} is recovered for large times for which $\bar{\x}_\infty(t) \to 0$ and thus $\alpha \to 1$. The above expressions hold both for $\q_s<1$ and $\q_s>1$. Care has to be taken in evaluating the limit $\q_s \to 1$. A rather lengthy calculation employing L'H\^{o}pital's rule shows that the left-hand side of~\eref{eq:StationaryPaths_GenFct_BinaryAnnihilation_Denominator} evaluates to
    \begin{equation}
      \frac{2}{3} \bar{\x}(t) \biggl(1+\frac{1}{2}\Bigl[1+\frac{\x(t_0)}{\bar{\x}_\infty(t)}\Bigr]^{-3}\biggr)    \,.
    \end{equation} 
    The pre-factor of this expression matches the normalization constant that Elgart and Kamenev inserted by hand~\cite{Elgart:2004}. 

    \begin{figure*}[tb] 
      \centering
      \includegraphics{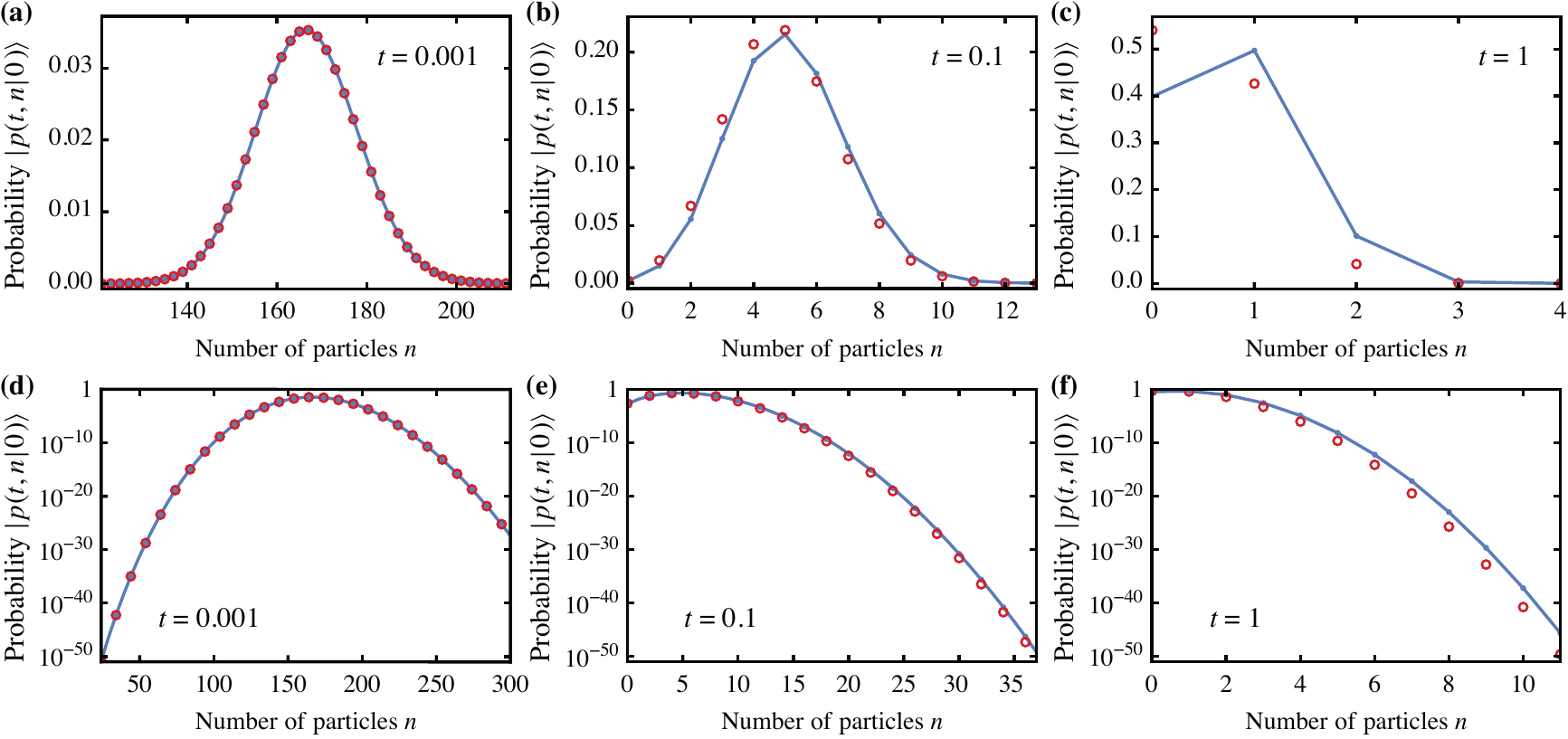}
      \caption{\label{fig:SP_Forward}
        Solution of the binary annihilation reaction $2\, A \to \emptyset$ via the numerical integration of the master equation (blue lines) and approximation of the process using the stationary path method in~\sref{subsec:StationaryPaths_GenFct_BinaryAnnihilation} (red circles). The figures in the upper row show the marginalized probability distribution $\ketA{\prob(t, n | 0)}$ at times (\textit{a}) $t=0.001$, (\textit{b}) $t=0.1$, and (\textit{c}) $t=1$ on a linear scale. The figures in the lower row show the distribution at times (d) $t=0.001$, (e) $t=0.1$, and (f) $t=1$ on a logarithmic scale. The rate coefficient of the process was set to $\rateCoeffAnnihilation_\tVar=1$. For the numerical integration, the master equation was truncated at $n=450$. The marginalized distribution was initially of Poisson shape with mean $\xAd(0) = 250$. Its leading-order approximation is given in~\eref{eq:StationaryPaths_GenFct_BinaryAnnihilation_SaddleptApprox}.
      }
    \end{figure*}    
    
    After putting all of the above pieces together, the probability distribution~\eref{eq:StationaryPaths_GenFct_BinaryAnnihilation_SaddleptApprox} provides a decent approximation of the binary annihilation process. \Fref{fig:SP_Forward} compares the distribution with a distribution that was obtained through a numerical integration of the master equation. For very large times, the quality of the approximation deteriorates. In particular, the approximation does not capture the final state of the process in which it is equally likely to find a single surviving particle or none at all. Limiting cases of the approximation~\eref{eq:StationaryPaths_GenFct_BinaryAnnihilation_SaddleptApprox} are provided in~\cite{Elgart:2004,Assaf:2006a}. 
    
    Although the above evaluation of the saddle-point approximation is rather elaborate, it is still feasible. It becomes infeasible if one goes beyond the leading-order term of the generating function. In the section after the next, we show how higher order terms can be included in a dual approach, which is based on the backward path integral~\eref{eq:PathInts_Backward_ContinuousTime_Solution_MarginalizedDistribution}.

  \subsection{Backward path integral approach}\label{subsec:StationaryPaths_GenFctnl}  

    The above method can be readily extended to a novel path integral representation of the marginalized distribution
    \begin{equation} 
      \ketA{\prob(t, n | t_0)}_{\xAd(t_0)} \label{eq:StationaryPaths_GenFct_MarginalDist}
      = \sum_{n_0=0}^\infty \prob(t, n|t_0,n_0) \frac{\xAd(t_0)^{n_0} \ee^{-\xAd(t_0)}}{n_0!}         \,.
    \end{equation} 
    Unlike in the previous section, no saddle-point approximation is required to evaluate $\ketA{\prob(t, n | t_0)}_{\xAd(t_0)}$. The resulting path integral can also be evaluated beyond leading order, at least numerically. In the next section, we outline such an evaluation for the binary annihilation reaction $2\, A \to \emptyset$. On the downside, the leading order of the ``backward'' approach does not respect the normalization of $\ketA{\prob(t, n | t_0)}_{\xAd(t_0)}$. The normalization of the distribution has to be implemented by hand or by evaluating higher order terms.
    
    To represent the marginalized distribution~\eref{eq:StationaryPaths_GenFct_MarginalDist} by a path integral, we require that for a given particle number $n$ and for a given initial mean $\xAd(t_0)$, there exist functions $\txAd(\tVar)$ and $\tqAd(\tVar)$ fulfilling
    \begin{eqnarray}
      \partial_\tVar \txAd          
      =  \frac{\partial \transitionOpAd_\tVar(\txAd, \tqAd)}{\partial \tqAd} 
      \mathtext{ with }
      \tilde{\xAd}(t_0) = \xAd(t_0)      \label{eq:StationaryPaths_GenFctnl_1}
      \mathtext{ and } \\
      \partial_{-\tVar} \tqAd          
       = \frac{\partial \transitionOpAd_\tVar(\txAd, \tqAd)}{\partial \txAd} 
      \mathtext{ with }
      \tqAd(t) = \frac{n}{\txAd(t)}          \,.            \label{eq:StationaryPaths_GenFctnl_2}
    \end{eqnarray}
    Since the transition operators $\transitionOpAd_\tVar$ in~\eref{eq:GenFctnl_Flow_Poisson_HamiltonianB} and $\transitionOp_\tVar$ in~\eref{eq:GenFct_Flow_Poisson_HamiltonianF} are connected via $\transitionOpAd_\tVar(\txAd, \tqAd)=\transitionOp_\tVar(\tq, \tx)$, the above equations differ from the previous equations~\eref{eq:StationaryPaths_GenFct_1} and~\eref{eq:StationaryPaths_GenFct_2} only in the final condition on $\tqAd$. As shown below, the marginalized distribution~\eref{eq:StationaryPaths_GenFct_MarginalDist} can then be expressed as
    \begin{eqnarray}
        \ketA{\prob(t, n | t_0)}_{\xAd(t_0)}      \label{eq:StationaryPaths_GenFctnl_PathIntegral}
      = \frac{\txAd(t)^{n} \ee^{-\tilde{\actionAd}}}{n!}
        \pathintegral{(t_0}{t]}
        \ee^{-\Delta \actionAd}
        \big|_{\Delta \xAd(t_0)=0}      \\
        \mathtext{with }
        \tilde{\actionAd}  \label{eq:StationaryPaths_GenFctnl_Action}
        \coloneqq  \xAd(t_0) +
                    \int_{t_0}^{t}\!\diff{\tVar}\, \bigl(
                      \tqAd \partial_{\tVar} \txAd - \transitionOpAd_\tVar(\txAd, \tqAd) 
                    \bigr)  \\
      \mathtext{and }  
      \Delta \actionAd   \label{eq:StationaryPaths_GenFctnl_DeltaAction}
      \coloneqq  n \Bigl( \frac{\basisPrefactor\Delta \xAd(t)}{\txAd(t)} - \ln\Bigl[1 + \frac{\basisPrefactor\Delta \xAd(t)}{\txAd(t)}\Bigr] \Bigr)
                  \\\phantom{\mathtext{and } t\Delta \actionAd \coloneqq}
                  + \int_{t_0}^{t}\!\diff{\tVar}\, \bigl(
                    \ii\Delta\qAd \partial_{\tVar} \Delta\xAd - \Delta\transitionOpAd_\tVar    
                  \bigr) \,.        \nonumber
    \end{eqnarray} 
    The variables $\Delta\xAd$ and $\Delta\qAd$ again represent deviations from the stationary paths $\txAd$ and $\tqAd$. The transition operator $\Delta\transitionOpAd_\tVar$ encompasses all the terms of an expansion of $\transitionOpAd_\tVar(\basisPrefactor\Delta\xAd + \txAd, \basisPrefactor^{-1} \ii\Delta\qAd + \tqAd)$ that are of second or higher order in the deviations. A Taylor expansion of the action shows that it is free of terms that are linear in $\Delta\xAd$ and $\Delta\qAd$.
    
    The derivation of the above representation proceeds analogously to \sref{subsec:StationaryPaths_GenFct}. It begins by using $\tilde{\xAd}(t_0) = \xAd(t_0)$ and the basis function $\ketA{n}_{\tVar,\Delta\xAd} = \frac{1}{n!} (\basisPrefactor\Delta\xAd + \txAd)^n \ee^{-\tqAd (\basisPrefactor\Delta\xAd + \txAd)}$ from~\eref{eq:GenFctnl_Flow_Poisson_BaseFunction} to rewrite the right-hand side of the marginalized distribution~\eref{eq:StationaryPaths_GenFct_MarginalDist} as
    \begin{equation}
      \ee^{(\tqAd(t_0)-1) \txAd(t_0)}  \!
          \sum_{n_0 = 0}^\infty \prob(t, n | t_0, n_0)   
          \ketA{n_0}_{t_0,\Delta \xAd(t_0)}   
          \big|_{\Delta \xAd(t_0) = 0}
            \,.
    \end{equation} 
    The sum in this expression can be represented by the backward path integral~\eref{eq:PathInts_Backward_DiscreteTime_Solution_MarginalizedDistribution}, turning the expression into
    \begin{equation}
        \ee^{(\tqAd(t_0)-1) \txAd(t_0)}  
          \pathintegral{(t_0}{t]}
          \ee^{-\actionAd}
          \, \ketA{n}_{t, \Delta\xAd(t)}  
          \big|_{\Delta \xAd(t_0) = 0}      \label{eq:StationaryPaths_GenFctnl_Derivation1}      \,.
    \end{equation} 
    Recalling the definition of $\evolutionOpAd_\tVar$ in~\eref{eq:GenFctnl_Flow_Poisson_HamiltonianT}, the transition operator
    \begin{eqnarray}
       \transEvoOpAd_\tVar(\Delta\xAd, \ii\Delta\qAd) 
       &= \transitionOpAd_\tVar(\basisPrefactor\Delta\xAd + \txAd, \basisPrefactor^{-1} \ii\Delta\qAd + \tqAd) \\
       &  -  \evolutionOpAd_\tVar(\basisPrefactor\Delta\xAd + \txAd, \basisPrefactor^{-1} \ii\Delta\qAd + \tqAd)    \nonumber
    \end{eqnarray}
    can now be expanded in the deviations. We thereby obtain the action
    \begin{eqnarray}
       \actionAd 
       = & \tqAd(t_0) \txAd(t_0) - \tqAd(t) \txAd(t) 
         + \int_{t_0}^{t}\!\diff{\tVar}\, \bigl[
            \tqAd \partial_{\tVar} \txAd - \transitionOpAd_\tVar(\txAd, \tqAd) 
          \bigr] \nonumber\\
      &+  \int_{t_0}^{t}\!\diff{\tVar}\, \bigl(
            \ii\Delta\qAd \partial_{\tVar} \Delta\xAd - \Delta\transitionOpAd_\tVar    
          \bigr)  \,.
    \end{eqnarray}
    The path integral representation~\eref{eq:StationaryPaths_GenFctnl_PathIntegral} follows upon inserting this action into~\eref{eq:StationaryPaths_GenFctnl_Derivation1} and employing the final condition $\tqAd(t) = n/\txAd(t)$.

  \subsection{Binary annihilation}\label{subsec:StationaryPaths_GenFctnl_BinaryAnnihilation}  
    
    To complement the approximation of the binary annihilation reaction $2\, A \to \emptyset$ with rate coefficient $\rateCoeffAnnihilation_\tVar$ in \sref{subsec:StationaryPaths_GenFct_BinaryAnnihilation}, we now perform an approximation of the process using the backward path integral representation~\eref{eq:StationaryPaths_GenFctnl_PathIntegral} of the marginalized distribution. We first perform the approximation in leading order, which amounts to the evaluation of the pre-factor of~\eref{eq:StationaryPaths_GenFctnl_PathIntegral}. For the simple growth process $\emptyset\to A$ and for the linear decay process $A\to\emptyset$, the pre-factor provides exact solutions.
    
    \subsubsection{Leading order.}\label{subsec:StationaryPaths_GenFctnl_BinaryAnnihilation_Leading}      
      
      The evaluation of the leading-order approximation
      \begin{equation}
        \ketA{\prob(t, n | t_0)}_{\xAd(t_0)}        \label{eq:StationaryPaths_GenFctnl_BinaryAnnihilation_Leading_Approx}
        \approx \frac{\txAd(t)^{n} \ee^{-\tilde{\actionAd}}}{n!}   
      \end{equation} 
      proceeds very much like the derivation in \sref{subsec:StationaryPaths_GenFct_BinaryAnnihilation}. Using the adjoint transition operator $\transitionOpAd_\tVar(\cre_\tVar, \ann_\tVar) = \rateCoeffAnnihilation_\tVar \cre_\tVar^2 (1 - \ann_\tVar^2)$ from~\eref{eq:GenFctnl_Flow_Poisson_HamiltonianB}, one finds that Hamilton's equations
      \begin{eqnarray}
        \partial_\tVar \txAd          
        =  - 2 \rateCoeffAnnihilation_\tVar \tqAd \txAd^2
        \mathtext{ with }
        \tilde{\xAd}(t_0) = \xAd(t_0)      
        \mathtext{ and } \\
        \partial_{-\tVar} \tqAd        
         = 2 \rateCoeffAnnihilation_\tVar (1-\tqAd^2) \txAd
        \mathtext{ with }
        \tqAd(t) = n/\txAd(t)                
      \end{eqnarray}
      agree with~\eref{eq:StationaryPaths_GenFct_BinaryAnnihilation_XEq} and~\eref{eq:StationaryPaths_GenFct_BinaryAnnihilation_QEq}, apart from the final condition on $\tqAd$. The conservation of $(\tqAd^2-1) \txAd^2$ and the asymptotic deterministic limit $\bar{\x}_\infty(t) = (2\int_{t_0}^t \diff{\tVar}\, \rateCoeffAnnihilation_\tVar)^{-1}$ can be used to rewrite the action~\eref{eq:StationaryPaths_GenFctnl_Action} as
      \begin{equation}
        \tilde{\actionAd}
        =  n + \Bigl( 1- \sqrt{1 - \frac{\txAd(t)^2 - n^2}{\xAd(t_0)^2}}\Bigr) \xAd(t_0) 
            + \frac{\txAd(t)^2-n^2}{2 \bar{\x}_\infty(t)}    \,.
      \end{equation} 
      In addition, the conservation law can be used to infer the flow equation
      \begin{equation}
        \partial_{\tVar} \tqAd      
        =  -2 \rateCoeffAnnihilation_\tVar\sqrt{1-\tqAd^2}\sqrt{\txAd(t)^2-n^2}    \,.
      \end{equation}
      Here we assume $\tq(t) < 1$ but the derivation can also be performed for $\tq(t) > 1$. The equation is implicitly solved by
      \begin{eqnarray}
        \arccos{\sqrt{1 - \frac{\txAd(t)^2 - n^2}{\xAd(t_0)^2}}}      
            + \frac{\sqrt{\txAd(t)^2-n^2}}{\bar{\x}_\infty(t)}     \label{eq:StationaryPaths_GenFctnl_BinaryAnnihilation_Leading_FixedPoint} \\
          = \arccos{\frac{n}{\txAd(t)}}    \,.    \nonumber
      \end{eqnarray}
      For a given initial mean $\xAd(t_0)$, this equation can be solved numerically for $\txAd(t)$, which is then inserted into~\eref{eq:StationaryPaths_GenFctnl_BinaryAnnihilation_Leading_Approx}. Upon normalizing the distribution by hand, it provides a reasonable approximation of the process. The quality of the approximation comes close to the quality of the approximation discussed in \sref{subsec:StationaryPaths_GenFct_BinaryAnnihilation}.  

    \subsubsection{Beyond leading order.}\label{subsubsec:StationaryPaths_GenFctnl_HigherOrder}  
      
      \begin{figure*}[tb] 
        \centering
        \includegraphics{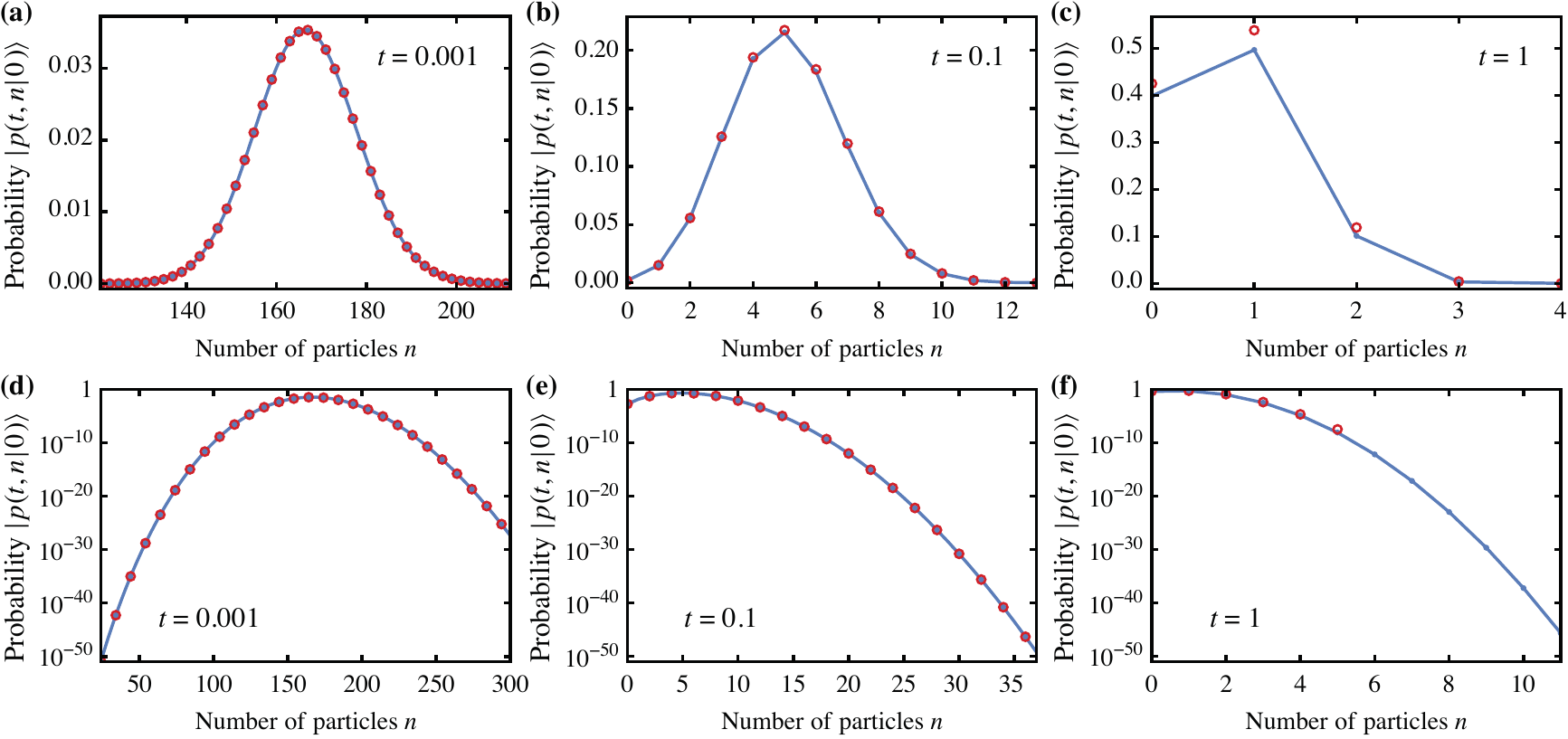}
        \caption{\label{fig:SP_Backward}
          Solution of the binary annihilation reaction $2\, A \to \emptyset$ via the numerical integration of the master equation (blue lines) and approximation of the process using the stationary path method in~\sref{subsec:StationaryPaths_GenFctnl_BinaryAnnihilation} (red circles). The figures in the upper row show the marginalized probability distribution $\ketA{\prob(t, n | 0)}$ at times (\textit{a}) $t=0.001$, (\textit{b}) $t=0.1$, and (\textit{c}) $t=1$ on a linear scale. The figures in the lower row show the distribution at times (d) $t=0.001$, (e) $t=0.1$, and (f) $t=1$ on a logarithmic scale. The rate coefficient of the process was set to $\rateCoeffAnnihilation_\tVar=1$. For the numerical integration, the master equation was truncated at $n=450$. The marginalized distribution was initially of Poisson shape with mean $\xAd(0) = 250$. Its leading-order approximation~\eref{eq:StationaryPaths_GenFctnl_BinaryAnnihilation_Leading_Approx} was corrected by the factor $1/\sqrt{\det{A}}$ (cf.\ \sref{subsubsec:StationaryPaths_GenFctnl_HigherOrder}). The numerical solution of the fixed point equation~\eref{eq:StationaryPaths_GenFctnl_BinaryAnnihilation_Leading_FixedPoint} failed for large times (missing circles in (f)).  
        }
      \end{figure*} 
      
      Instead of normalizing the function~\eref{eq:StationaryPaths_GenFctnl_BinaryAnnihilation_Leading_Approx} by hand, it can be normalized by evaluating the path integral~\eref{eq:StationaryPaths_GenFctnl_PathIntegral}. In the following, we perform this evaluation, but only after restricting the action $\Delta \actionAd$ in~\eref{eq:StationaryPaths_GenFctnl_DeltaAction} to terms that are of second order in the deviations $\Delta\xAd$ and $\Delta\qAd$, i.e. to
      \begin{equation}
        \Delta \actionAd 
        \approx  - \frac{(\Delta \xAd(t))^2}{2} \frac{n}{\txAd(t)^2}
            + \int_{t_0}^{t}\!\diff{\tVar}\, \bigl(
                \ii\Delta\qAd \partial_{\tVar} \Delta\xAd - \Delta\transitionOpAd_\tVar    
              \bigr) \,.        
      \end{equation} 
      The corresponding transition operator 
      \begin{equation}
        \Delta\transitionOpAd_\tVar    
          \approx  \ii\Delta\qAd \Delta\xAd \, \alpha_\tVar  - \frac{(\Delta\qAd)^2}{2} \beta_\tVar + \frac{(\Delta\xAd)^2}{2} c_\tVar    
      \end{equation} 
      is obtained as explained in \sref{subsec:StationaryPaths_GenFctnl} (with $\basisPrefactor\coloneqq\ii$). Its coefficients read
      \begin{equation}
        \alpha_\tVar \coloneqq - 4 \rateCoeffAnnihilation_\tVar \txAd\tqAd    \,,\quad
        \beta_\tVar \coloneqq 2 \rateCoeffAnnihilation_\tVar \txAd^2  \,,  \quad
        c_\tVar \coloneqq 2\rateCoeffAnnihilation_\tVar (\tqAd^2 - 1) \,.
      \end{equation} 
      The path integral can now be evaluated in one of the following two ways. 
      
      On the one hand, the transition operator $\Delta\transitionOpAd_\tVar$ has the same form as the one in \sref{subsec:PathInts_BackwardAlongPaths} and thus can be treated in the same way. Following \aref{sec:A_BackwardAlongPaths}, the path integral~\eref{eq:StationaryPaths_GenFctnl_PathIntegral} can be expressed in terms of the following average over a Wiener process $\wienerProcess$:
      \begin{eqnarray}
        \pathintegral{(t_0}{t]}    \label{eq:StationaryPaths_GenFctnl_HigherOrder_PathIntegralV1}
          \, \ee^{-\Delta \actionAd}
          \big|_{\Delta \xAd(t_0)=0}    \\
        =  \BigLLangle 
              \exp{\Bigl(  \frac{(\Delta \xAd(t))^2}{2} \frac{n}{\txAd(t)^2} + \int_{t_0}^t \diff{\tVar}\, \frac{(\Delta\xAd)^2}{2}c_\tVar\Bigr)}
            \BigRRangle_{\wienerProcess}    \nonumber    \,.
      \end{eqnarray}   
      Here, $\Delta \xAd(\tVar)$ solves the It\^{o} SDE
      \begin{equation}
        \diff{\Delta \xAd} 
        = \alpha_\tVar \Delta \xAd \diff{\tVar}  + \sqrt{\beta_\tVar}\,\diff{\wienerProcess}(\tVar)    \,,
      \end{equation} 
      with initial value $\Delta \xAd(t_0) = 0$. Unfortunately, the computation of a sufficient number of sample paths to approximate the average~\eref{eq:StationaryPaths_GenFctnl_HigherOrder_PathIntegralV1} was found to be rather slow.    
    
      As an alternative to the above way, one can discretize the action $\Delta \actionAd$ and cast it into the quadratic form $\Delta \actionAd=\frac{1}{2}\xi^\transpose A \xi$ with $\xi \coloneqq \{\Delta \qAd_j, \Delta \xAd_j\}_{j=1,\hldots,N}$. The matrix $A$ is symmetric and tridiagonal. If $A$ is also positive-definite, the path integral~\eref{eq:StationaryPaths_GenFctnl_PathIntegral} evaluates to $1/\sqrt{\det{A}}$. Curiously, we found that this factor even matches the average~\eref{eq:StationaryPaths_GenFctnl_HigherOrder_PathIntegralV1} when $A$ is not positive-definite. Therefore, we used the factor $1/\sqrt{\det{A}}$ to correct the approximation~\eref{eq:StationaryPaths_GenFctnl_BinaryAnnihilation_Leading_Approx} and evaluated the resulting expression for various times $t$. A comparison with distributions obtained via a numeric integration of the master equation is provided in \fref{fig:SP_Backward}. Apparently, the quality of the approximation is even better than the quality of the approximation discussed in \sref{subsec:StationaryPaths_GenFct_BinaryAnnihilation}, at least for early times. However, upon approaching the asymptotic limit of the process, it becomes increasingly difficult to numerically solve~\eref{eq:StationaryPaths_GenFctnl_BinaryAnnihilation_Leading_FixedPoint} for a fixed point $\txAd(t)$.

  \subsection{R\'esum\'e}\label{subsec:StationaryPaths_Resume}  
  
    The solution of a master equation can be approximated by expanding its forward or backward path integral representations around stationary paths of their respective actions. Such an expansion proves particularly useful in the approximation of ``rare-event'' probabilities in a distribution's tail. Algorithms such as the stochastic simulation algorithm (SSA) of Gillespie typically perform poorly for this purpose. Whereas the expansion of the forward path integral representation provides the ordinary probability generating function~\eref{eq:StationaryPaths_GenFct_GenFct} as an intermediate step, the expansion of the backward path integral representation provides the marginalized distribution~\eref{eq:StationaryPaths_GenFct_MarginalDist}. In both cases, the stationary paths obey differential equations resembling Hamilton's equations from classical mechanics. In sections~\ref{subsec:StationaryPaths_GenFct_BinaryAnnihilation} and~\ref{subsec:StationaryPaths_GenFctnl_BinaryAnnihilation}, we demonstrated the two approaches in approximating the binary annihilation reaction $2\, A \to \emptyset$. Our approximation based on the forward path integral amends an earlier computation of Elgart and Kamenev~\cite{Elgart:2004}. The computation requires a saddle-point approximation of Cauchy's differentiation formula to extract probabilities from the generating function. The advantage of the  approach lies in the fact that its leading order term respects the normalization of the underlying probability distribution. The backward approach does not require the auxiliary saddle-point approximation but its leading order term is not normalized. In the approximation of the binary annihilation reaction, we demonstrated how the expansion of the backward path integral can be evaluated beyond leading order. Future studies are needed to explore whether the two approaches are helpful in analysing processes with multiple types of particles and spatial degrees of freedom. The efficient evaluation of contributions beyond the leading order approximations also remains a challenge.
  
\section{Summary and outlook}\label{sec:Summary}  

  On sufficiently coarse time and length scales, many complex systems appear to evolve through finite jumps. Such a jump may be the production of an mRNA or protein in a gene regulatory network~\cite{Elowitz:2000,Thattai:2001,Shahrezaei:2008}, the step of a molecular motor along a cytoskeletal filament~\cite{Lipowsky:2001,Parmeggiani:2003,Parmeggiani:2004}, or the flipping of a spin~\cite{Glauber:1963a,Kawasaki:1966a,Kawasaki:1966b,Kawasaki:1966c}. The stochastic evolution of such processes is commonly modelled in terms of a master equation~\cite{Kolmogoroff:1931,Nordsieck:1940}. This equation provides a generic description of a system's stochastic evolution under the following three conditions: First, time proceeds continuously. Second, the system's future is fully determined by the system's present state. Third, changes of the system's state proceed via discontinuous jumps. 
  
  In this work, we reviewed the mathematical theory of master equations and discussed analytical and numerical methods for their solution. Special attention was paid to methods that apply even when stochastic fluctuations are strong and to the representation of master equations by path integrals. In the following, we provide brief summaries of the discussed methods, which all have their own merits and limitations. Information on complementary approximation methods can be found in the text books~\cite{vanKampen:2007,Gardiner:2009} and in the review~\cite{Schnoerr:2016}.
  
  \paragraph{The stochastic simulation algorithm} (\sref{subsec:Intro_MasterEqSolution})\\
  The SSA~\cite{Gillespie:1976,Gillespie:1977,Gillespie:1992,Gillespie:2007} and its variations~\cite{Gibson:2000,Gillespie:2007,Slepoy:2008,Gillespie:2013} come closest to being all-purpose tools in solving the (forward) master equation numerically. The SSA enables the computation of sample paths with the correct probability of occurrence. Since these paths are statistically independent of one another, their computation can be easily distributed to individual processing units. Besides providing insight into a system's ``typical'' dynamics, the sample paths can be used to compute a histogram approximation of the master equation's solution or to approximate observables. In principle, the SSA can be applied to arbitrarily complex systems with multiple types of particles and possibly spatial degrees of freedom. Consequently, the algorithm is commonly used in biological studies~\cite{Elowitz:2000,Thattai:2001,Shahrezaei:2008} and extensions of it have been implemented in various simulation packages~\cite{Tomita:1999,Adalsteinsson:2004,Ander:2004,Hattne:2005,Sanft:2011,Hepburn:2012,Drawert:2012,StochSS:2015}. A fast but only approximate alternative to the SSA for the simulation of processes evolving on multiple time scales is $\tVar$-leaping~\cite{Gillespie:2001,Gillespie:2003,Rathinam:2003,Tian:2004,Chatterjee:2005,Cao:2005,Cao:2006,Auger:2006,Gillespie:2007,Cao:2007,Anderson:2008,Gillespie:2013}. Algorithms for the simulation of processes with time-dependent transition rates are discussed in~\cite{Lu:2004,Anderson:2007}. Let us note, however, that for small systems, a direct numerical integration of the master equation may be more efficient than the use of the SSA. 
  
  \paragraph{Alternative path summation algorithms} (\sref{subsec:Intro_MasterEqSolution})\\
  Given enough time, the SSA could be used to generate every possible sample path of a process (at least if the state space is finite). If every path is recorded just once, the corresponding histogram approximation of the conditional probability distribution $\prob(\tVar, n| \tVar_0, n_0)$ would agree with the ``path summation representation''~\eref{eq:Intro_Mescoscopic_Solution}. Various alternatives to the SSA have been proposed for the numerical evaluation of this representation, mostly based on its Laplace transform~\cite{Empacher:1992,Helbing:1994,Helbing:1996,Sun:2006,Harland:2007,Jackson:2009}. Thus far, these algorithms have remained restricted to rather simple systems. The analytical evaluation of the sums and convolutions that are involved proves to be a challenge.
  
  \paragraph{Exponentiation and uniformization} (\sref{subsec:Intro_MasterEqSolution})\\
  If the state space of a system is finite, the (forward) master equation $\partial_\tVar \prob(\tVar | t_0) = \qMatrix \prob(\tVar | t_0)$ is solved by the matrix exponential $\prob(\tVar| \tVar_0) = \ee^{\qMatrix(\tVar-\tVar_0)}\unitMatrix$. Here, $\qMatrix$ represents the transition matrix of a process and $\prob(\tVar | t_0)$ the matrix of the conditional probabilities $\prob(\tVar, n| \tVar_0, n_0)$. If the state space of the system is countably infinite but the exit rates from all states are bounded (i.e.\ $\sup_{m} |\qMatrix(m,m)| < \infty$), the conditional probability distribution can be represented by the ``uniformization'' formula~\eref{eq:Intro_MasterEqSolution_Uniform}. As  the evaluation of this formula involves the computation of an infinite sum of matrix powers, its use is restricted to small systems and to systems whose transition matrices exhibit special symmetries. The same applies to the evaluation of the above matrix exponential (note, however, the projection techniques proposed in~\cite{Munsky:2006,Burrage:2006}).
 
  \paragraph{Flow equations} (sections~\ref{sec:GenFct} and~\ref{sec:GenFctnl})\\
  For the chemical master equation~\eref{eq:Intro_Mescoscopic_ChemicalMasterEq} of the reaction $k\,A \to l\,A$ with rate coefficient $\rateCoeffGeneric_\tVar$, it is readily shown that the probability generating function
  \begin{equation}
    \ketA{\gen(\tVar | t_0, n_0)} = \sum_{n} \ketA{n} \, p(\tVar, n | t_0, n_0)
  \end{equation}
  with basis function $\ketA{n}_{\q} = \q^n$ obeys the linear partial differential equation, or ``flow equation''
  \begin{equation}
    \partial_\tVar \ketA{\gen}
    = \rateCoeffGeneric_\tVar (\q^l - \q^k) \partial_{\q}^k \ketA{\gen} \,.
  \end{equation}
  Analogously, the marginalized distribution
  \begin{equation}
    \ketA{\prob(t, n | \tVar)} = \sum_{n_0} \prob(t, n|\tVar,n_0) \ketA{n_0}
  \end{equation}
  with Poisson basis function $\ketA{n_0}_{\xAd}   = \frac{\xAd^{n_0} \ee^{-\xAd}}{n_0!}$ obeys the backward-time flow equation
  \begin{equation}
    \partial_{-\tVar} \ketA{\prob}
    = \rateCoeffGeneric_\tVar \xAd^k \bigl[ (\partial_{\xAd}+1)^l - (\partial_{\xAd}+1)^k\bigr] \ketA{\prob}  \,.
  \end{equation}
  After one has solved one of these equations, the conditional probability distribution can be recovered via the inverse transformations $\prob(\tVar, n | t_0, n_0) = \braketA{n}{\gen(\tVar | t_0, n_0)}$ or $\prob(t, n | \tVar, n_0) =  \braketA{n_0}{\prob(t,n|\tVar)}$. In sections~\ref{sec:GenFct} and~\ref{sec:GenFctnl}, we generalized the above approaches and formulated conditions under which the forward and backward master equations can be transformed into flow equations. Besides the flow equations obeyed by the generating function and the marginalized distribution, we also derived a flow equation obeyed by the generating functional
  \begin{equation}
    \braA{\gen(\tVar | t_0, n_0)}  = \sum_{n} \braA{n} \prob(\tVar, n | t_0, n_0)     \,.
  \end{equation} 
  For the Poisson basis function $\ketA{n}_{\xAd}   = \frac{\xAd^{n} \ee^{-\xAd}}{n!}$, the inverse transformation $\prob(\tVar, n | t_0, n_0) =  \braketA{\gen(\tVar|t_0,n_0)}{n}$ of this functional recovers the Poisson representation of Gardiner and Chaturvedi~\cite{Gardiner:1977,Chaturvedi:1978}. The Poisson representation can, for example, be used for the computation of mean extinction times~\cite{Drummond:2010}, but we found the marginalized distribution to be more convenient for this purpose (cf. \sref{subsec:GenFctnl_ExtTimes}). 
  
  Thus far, most methods that have been proposed for the analysis of the generating function's flow equation have only been applied to systems without spatial degrees of freedom. These methods include a variational approach~\cite{Eyink:1996, Sasai:2003}, spectral formulations and WKB approximations~\cite{Assaf:2006a,Assaf:2006b,Assaf:2007,Assaf:2008,Assaf:2009,Walczak:2009,Mugler:2009,Assaf:2010b,Walczak:2012,Assaf:2016}. We outlined some of these methods in \sref{subsec:GenFct_Spectral}. ``Real-space'' WKB approximations, which employ an exponential ansatz for the probability distribution rather than for the generating function, were not discussed in this review (information on these approximations can be found in~\cite{Kubo:1973,Gang:1987,Dykman:1994,Kessler:2007,Meerson:2008,Escudero:2009,Assaf:2010,Mobilia:2010b,Assaf:2010c,Meerson:2011,Black:2011,Black:2012b,Bressloff:2014,Bressloff:2014b,Smith:2016,Assaf:2016}). WKB approximations often prove helpful in computing a mean extinction time or a (quasi)stationary probability distribution. Future studies could explore whether the flow equation obeyed by the marginalized distribution can be evaluated in terms of a WKB approximation or using spectral methods. Moreover, further research is needed to investigate whether the above methods may be helpful in studying processes with spatial degrees of freedom and multiple types of particles. For that purpose, it will be crucial to specify satisfactory boundary conditions for the flow equations obeyed by the generating function and the marginalized distribution (see~\cite{Assaf:2007} for a discussion of ``lacking'' boundary conditions in a study of the branching-annihilation reaction $A \to 2\, A$ and $2\, A \to \emptyset$). 
  
  Recently, we have been made aware of novel ways of treating the generating function's flow equation based on duality relations~\cite{Ohkubo:2010,Ohkubo:2010b,Ohkubo:2013,Ohkubo:2015,Benitez:2016}. These approaches are not yet covered here.

  \paragraph{Stochastic path integrals} (sections~\ref{sec:PathInts_Backward}--\ref{sec:PathInts_Observables})\\
  Path integral representations of the master equation have proved invaluable tools in gaining analytical and numerical insight into the behaviour of stochastic processes, particularly in the vicinity of phase transitions~\cite{Mikhailov:1985,Peliti:1986,Droz:1993,Lee:1994b,Lee:1994c,Howard:1995,Lee:1995,Howard:1996a,Howard:1996b,Oerding:1996,Cardy:1996,Rey:1997,HowardMJ:1997,Sasamoto:1997,Cardy:1997,Cardy:1998,Wijland:1998,Taeuber:1998,Hinrichsen:1999,Rey:1999,Konkoli:1999,Goldschmidt:1999,Hnatich:2000,Konkoli:2000,Bettelheim:2001,Dickman:2002,Vernon:2003,Canet:2004a,Canet:2004b,Hilhorst:2004,Janssen:2004,Janssen:2005,Whitelam:2005,Taeuber:2005,Canet:2005,Buice:2007,Taeuber:2011,Winkler:2012,Benitez:2012,Taeuber:2012,Buice:2013,Hnatic:2013,Benitez:2013,Homrighausen:2013,Taeuber:2014}. A classical example of such a phase transition is the transition between an active and an absorbing state of a system~\cite{Henkel:2008}. In \sref{sec:PathInts_Observables}, we showed that for a process whose initial number of particles is Poisson distributed with mean $\xAd(t_0)$, the average of an observable $A(n)$ at time $t$ can be represented by the path integral
    \begin{eqnarray}
      &\langle \observable \rangle_{\xAd(t_0)}
      =  \pathintegral{(t_0}{t]}
        \ee^{-\actionAd} 
        \, \LLangle \observable \RRangle_{\xAd(t)}  \label{eq:Summary_Observable}     \\
      & \mathtext{with } 
      \LLangle \observable \RRangle_{\xAd} 
      \coloneqq \sum_{n=0}^\infty  \frac{\xAd^{n} \ee^{-\xAd}}{n!}  \observable(n)    \,.
    \end{eqnarray} 
    The above path integral can be derived both from the novel \textit{backward path integral representation}
    \begin{equation}
      \prob(t, n | t_0, n_0)    \label{eq:Summary_BwdPI}
      = \braA{n_0}_{t_0} 
        \pathintegral{(t_0}{t]}
        \, \ee^{-\actionAd} 
        \, \ketA{n}_{t} \,,
    \end{equation} 
    of the conditional probability distribution  (cf.\ \sref{sec:PathInts_Backward}), and from the \textit{forward path integral representation}
    \begin{equation}
      \prob(t, n | t_0, n_0)    \label{eq:Summary_FwdPI}
      =  \braA{n}_{t} 
         \pathintegral{[t_0}{t)} 
        \ee^{-\action}    
        \, \ketA{n_0}_{t_0}         
    \end{equation}
    (cf.\ \sref{sec:PathInts_Forward}). The meanings of the integral signs and of the actions $\action$ and $\actionAd$ were explained in the respective sections. We derived both of the above representations of the conditional probability distribution from the flow equations discussed in the previous paragraph. An extension of the path integrals to systems with multiple types of particles or spatial degrees of freedom is straightforward. We demonstrated the use of the integrals in solving various elementary processes, including the pair generation process and a process with linear decay of diffusing particles. Although we did not discuss the application of renormalization group techniques, we showed how the above path integrals can be evaluated in terms of perturbation expansions using Feynman diagrams. Information on perturbative renormalization group techniques can be found in~\cite{Taeuber:2005,Taeuber:2014}, information on non-perturbative techniques in~\cite{Canet:2011}. Besides the above path integrals, we showed how one can derive path integral representations for processes with continuous state spaces. These path integrals were based on Kramers-Moyal expansions of the respective backward and forward master equations. Upon truncating the expansion of the backward master equation at the level of a diffusion approximation, we recovered a classic path integral representation of the (backward) Fokker-Planck equation~\cite{Martin:1973,deDominicis:1976,Janssen:1976,Bausch:1976}. We hope that our exposition of the path integrals helps in developing new methods for the analysis of stochastic processes. Future studies may focus directly on the backward or forward path integral representations~\eref{eq:Summary_BwdPI} and~\eref{eq:Summary_FwdPI} of the conditional probability distribution, or use the representation~\eref{eq:Summary_Observable} to explore how (arbitrarily high) moments of the particle number behave in the vicinity of phase transitions.
    
 \paragraph{Stationary path approximations} (\sref{sec:StationaryPaths})\\
  Elgart and Kamenev recently showed how the forward path integral representation~\eref{eq:Summary_FwdPI} can be approximated by expanding its action around ``stationary paths''~\cite{Elgart:2004}. The paths obey Hamilton's equations of the form 
  \begin{eqnarray}
    \partial_\tVar \tx       \label{eq:Summary_Hamiltons}
     = \frac{\partial\transitionOp_\tVar}{\partial \tq}            
    \mathtext{ and } 
    \partial_{-\tVar} \tq   
    =  \frac{\partial \transitionOp_\tVar}{\partial \tx}        \,,
  \end{eqnarray}
  with the transition operator $\transitionOp_\tVar$ acting as the ``Hamiltonian''. In \sref{sec:StationaryPaths}, we reviewed this ``stationary path method'' and showed how it can be extended to the backward path integral representation~\eref{eq:Summary_BwdPI}. We found that this backward approach does not require an auxiliary saddle-point approximation if the number of particles is initially Poisson distributed, but that a proper normalization of the probability distribution is only attained beyond leading order. Future work is needed to apply the method to systems with spatial degrees of freedom and multiple types of particles. The latter point also applies to the classification of phase transitions based on phase-space trajectories of the equations~\eref{eq:Summary_Hamiltons} as has been proposed in~\cite{Elgart:2006}.\\

  We hope that our review of the above methods inspires future research on master equations and that it helps researchers who are new to the field of stochastic processes to become acquainted with the theory of stochastic path integrals.

% END - Main Text

\ack{For critical reading of a pre-submission manuscript and for providing valuable feedback, we would like to thank Michael Assaf, L\'{e}onie Canet, Alexander Dobrinevski, Nigel Goldenfeld, Peter Grassberger, Baruch Meerson, Ralf Metzler, Mauro Mobilia, Harold P. de Vladar, Herbert Wagner, and Royce Zia. Moreover, we would like to thank Johannes Knebel, Isabella Kr\"amer, and Cornelius Weig for helpful discussions during the preparation of the manuscript. This research was financially supported by the German Excellence Initiative via the program `NanoSystems Initiative Munich' (NIM). }

% BEGIN - Appendix

\appendix
\renewcommand\thesection{\Alph{section}}%  remove the "Appendix" prefix... makes problems in the TOC

  \section{Proof of the Feynman-Kac formula in \sref{subsec:Intro_Mesoscopic}}\label{sec:A_FeynmanKacProof}      

    Here we provide a brief proof of a special case of the Feynman-Kac, or Kolmogorov, formula (see section~4.3.5 in~\cite{Gardiner:2009}). In particular, we assume that for $\tVar\in[t_0,t]$, a function $u(\tVar, \xAd)$ obeys the linear PDE
    \begin{eqnarray}  
      \partial_{-\tVar} u(\tVar, \xAd)       \label{eq:A_FeynmanKacProof_PDE}    
        =   \alpha_\tVar(\xAd) \partial_{\xAd} u
          +\frac{1}{2} \beta_\tVar(\xAd)\partial_{\xAd}^2 u \\
      \mathtext{with \mathtextit{final} value } u(t,\xAd) = G(\xAd)   \label{eq:A_FeynmanKacProof_PDE_Initial}  \,.  
    \end{eqnarray} 
    The function $\alpha_\tVar$ is called a drift coefficient and $\beta_\tVar$ a diffusion coefficient. According to the Feynman-Kac formula, the above PDE is solved by
    \begin{equation}
      u(\tVar, \xAd)     \label{eq:A_FeynmanKacProof_FK}
      = \bigLLangle G (\XAd(t)) \bigRRangle_\wienerProcess    
    \end{equation}
    with $\LLangle \cdot \RRangle_{\wienerProcess}$ representing an average over realizations of the Wiener process $\wienerProcess$. The function $\XAd(\sVar)$ with $s \in [\tVar,t]$ and $\tVar \geq t_0$ represents a solution of the It\^{o} stochastic differential equation (SDE)
    \begin{eqnarray}
      \diff{\XAd(\sVar)}      \label{eq:A_FeynmanKacProof_SDE}
      = \alpha_\sVar(\XAd(\sVar)) \diff{\sVar}  + \sqrt{\beta_\sVar(\XAd(\sVar))}\diff{\wienerProcess}(\sVar)\\
      \mathtext{with \mathtextit{initial} value } \XAd(\tVar) = \xAd  \,.     \label{eq:A_FeynmanKacProof_SDE_Initial}
    \end{eqnarray} 
    If the initial value $\xAd$ of the SDE is chosen as a real number, the drift coefficient $\alpha_\sVar$ is a real function, and the diffusion coefficient $\beta_\sVar$ as a real non-negative function, the ``sample path'' $\XAd(\sVar)$ assumes only real values along its temporal evolution. In a multivariate extension of the Feynman-Kac formula, one requires a matrix $\sqrt{\beta_s} \coloneqq \gamma_s$ fulfilling $\gamma_s \gamma_s^\transpose = \beta_s$. If the matrix $\beta_\sVar$ is symmetric and positive-semidefinite, one may choose $\gamma_s$ as its unique symmetric and positive-semidefinite square root~\cite{Harville:1997}.
    
    The solution~\eref{eq:Intro_Mescoscopic_FP_BW_Solution} of the backward Fokker-Planck equation~\eref{eq:Intro_Mescoscopic_FP_BW} constitutes a special case of the above formula. There, the independent variable is $\xAd_0$ instead of $\xAd$, and $u(\tVar, \xAd_0)$ is the conditional probability distribution $\prob(t, \xAd | \tVar, \xAd_0)$ (with $\xAd$ being an arbitrary parameter). The final value of the distribution is $G(\xAd_0) = \delta(\xAd - \xAd_0)$. The Feynman-Kac formula~\eref{eq:A_FeynmanKacProof_FK} then implies that $\bigLLangle \delta(\xAd - \XAd(t)) \bigRRangle_\wienerProcess$ solves the backward Fokker-Planck equation~\eref{eq:Intro_Mescoscopic_FP_BW}. Note that in the main text, we use a small letter to denote the sample path.

    To prove the Feynman-Kac formula~\eref{eq:A_FeynmanKacProof_FK}, we assume that $\XAd(\sVar)$ solves the SDE~\eref{eq:A_FeynmanKacProof_SDE}. By It\^{o}'s Lemma and the PDE~\eref{eq:A_FeynmanKacProof_PDE}, it then holds that
    \begin{equation}
      \diff{u(\sVar,\XAd(\sVar))}
      = \frac{\partial u(\sVar,\XAd(\sVar))}{\partial \XAd(\sVar)} \sqrt{\beta_\sVar(\XAd(\sVar))}\diff{\wienerProcess}(\sVar)   \,.
    \end{equation}
    As the next step, we integrate this differential from $\sVar=\tVar$ to $s=t$ and average the result over realizations of the Wiener process $W$. Since $\LLangle\diff{\wienerProcess}\RRangle_\wienerProcess = 0$, it follows that $\LLangle u(\tVar,\XAd(\tVar)) \RRangle_\wienerProcess = \LLangle u(t,\XAd(t)) \RRangle_\wienerProcess$. This expression coincides with the Feynman-Kac formula~\eref{eq:A_FeynmanKacProof_FK} because the initial condition~\eref{eq:A_FeynmanKacProof_SDE_Initial} implies that $u(\tVar,\XAd(\tVar)) = u(\tVar,\xAd)$ does not depend on the Wiener process, and because the final condition~\eref{eq:A_FeynmanKacProof_PDE_Initial} implies that $u(t,\XAd(t))=G(\XAd(t))$.

  \section{Proof of the path summation representation in \sref{subsec:Intro_MasterEqSolution}}\label{sec:A_PathSummationProof}      
    
    In the following, we prove that the path summation representation~\eref{eq:Intro_Mescoscopic_Solution} solves the forward master equation~\eref{eq:Intro_Mescoscopic_MasterEqGainLoss} if the transition rate $\transitionRate(n, m)$ is independent of time. Hence, the process is homogeneous in time and we may choose $t_0 \coloneqq 0$. After defining $d(n,m) \coloneqq \delta_{n,m} \exitRate(m) = \delta_{n,m} \sum_{k} \transitionRate(k, m)$, we first rewrite the master equation as
    \begin{equation}
      \partial_\tVar \vect{\prob}(\tVar | 0, n_0 )    
      = (w-d) \vect{\prob}(\tVar | \cdot ) 
    \end{equation} 
    with the probability vector $\vect{\prob}(\tVar|t_0,n_0)$. Note that the matrix notation assumes some mapping between the state space of $n$ and an index set $I\subset \naturals$. Following~\cite{Helbing:1994}, the Laplace transform
    \begin{equation}
      \legendreTransform f(s) \coloneqq \int_{0}^\infty \diff{\tVar}\, \ee^{-s \tVar} f(\tVar)     
    \end{equation}
    of the above master equation is given by
    \begin{equation}
      s \legendreTransform \vect{\prob}(s | \cdot ) - \hat{\vect{e}}_{n_0}
      = (w-d) \legendreTransform\vect{\prob}(s | \cdot ) \,.
    \end{equation} 
    Here, $\vect{\prob}(0 | 0, n_0 ) = \hat{\vect{e}}_{n_0}$ represents a unit vector pointing in direction $n_0$. Note that $s$ is a scalar but that $w$ and $d$ are matrices. Making use of the unit matrix $\unitMatrix$, the above expression can be solved for $\legendreTransform \vect{\prob}(s | \cdot )$ and be rewritten as
    \begin{equation}
      \legendreTransform \vect{\prob}(s | \cdot )\label{eq:A_PathSummationProof_Step1}  
      = \bigl[\unitMatrix - (s\unitMatrix + d)^{-1} w\bigr]^{-1} (s\unitMatrix + d)^{-1} \hat{\vect{e}}_{n_0}     \,.
    \end{equation} 
    The value of the real part of $s$ is determined by the parameter $\varepsilon$ in the inverse Laplace transformation
    \begin{equation}
      f(t) = \frac{1}{2\pi\ii} \int_{\varepsilon - \ii \infty}^{\varepsilon + \ii \infty} \diff{s}\, \ee^{s t} \legendreTransform f(s)    \,.
    \end{equation}
    We assume that $\varepsilon$ can be chosen so large that the first factor in the solution~\eref{eq:A_PathSummationProof_Step1} can  be rewritten as a geometric series, i.e. as
    \begin{equation}
      \legendreTransform \vect{\prob}(s | \cdot )  \label{eq:A_PathSummationProof_Step2}
      = \sum_{J=0}^\infty \bigl[(s\unitMatrix + d)^{-1} w\bigr]^J  (s\unitMatrix + d)^{-1} \hat{\vect{e}}_{n_0}      \,.
    \end{equation} 
    This expression may be simplified by noting that
    \begin{equation}
      \bigl[(s\unitMatrix + d)^{-1} w\bigr]_{n,m}
      = \frac{\transitionRate(n,m)}{s+\exitRate(n)}     \,.
    \end{equation}
    By defining $\sum_{\{\stochPath_J\}}\coloneqq \sum_{n_1} \cdots \sum_{n_{J-1}}$, \eref{eq:A_PathSummationProof_Step2} becomes
    \begin{eqnarray}
      \legendreTransform \prob(s, n | \cdot ) =  \label{eq:A_PathSummationProof_Step3}\\
      \sum_{J=0}^\infty \sum_{\{\stochPath_J\}}\biggl( \prod_{j=0}^{J-1} \transitionRate(n_{j+1},n_j) \biggr) \prod_{j=0}^{J} \frac{1}{s+\exitRate(n_j)} \bigg|_{n_J=n}    \,.\nonumber
    \end{eqnarray} 
    Since an exponential function $f(\tVar) \coloneqq \ee^{-\alpha \tVar}\HeavisideStep(\tVar)$ transforms as $\legendreTransform f(s) = (s+\alpha)^{-1}$ and since the Laplace transform converts convolutions into products, we thus find that~\eref{eq:A_PathSummationProof_Step3} agrees with the Laplace transform of the path summation representation~\eref{eq:Intro_Mescoscopic_Solution}.
      
  \section{Solution of the random walk in sections~\ref{subsec:GenFct_ForwardBases_RandomWalk} and~\ref{subsec:GenFctnl_BackwardBases_RandomWalk}}\label{sec:A_RandomWalk}       
    
    We here provide the conditional probability distribution $\prob(\tVar, n | t_0, n_0)$ solving the random walk from sections~\ref{subsec:GenFct_ForwardBases_RandomWalk} and~\ref{subsec:GenFctnl_BackwardBases_RandomWalk}. In the first of these sections, we showed that this distribution is obtained by applying the functional $\braA{n}f  = \int_{-\pi}^\pi \frac{\diff{\q}}{2\pi} \ee^{-\ii n \q} f(\q)$ to the generating function $\ketA{\gen(\tVar | t_0, n_0)}$, which we derived as
    \begin{equation*}
      \exp\Bigl((\ee^{\ii\q} - 1) \int_{t_0}^{\tVar}\diff{s}\, r_s+   (\ee^{-\ii\q} - 1) \int_{t_0}^{\tVar}\diff{s}\,l_s  +  \ii n_0 \q  \Bigr)    \,.
    \end{equation*}
    Multiple series expansions show that this expression can be rewritten as $\ee^{-\int_{t_0}^{\tVar}\diff{s}\,(r_s+l_s) }$ times
    \begin{equation*}
        \sum_{k=0}^\infty \sum_{m=0}^k 
        \frac{\bigl(\int_{t_0}^{\tVar}\diff{s}\,r_s\bigr)^m}{m!} 
        \frac{\bigl(\int_{t_0}^{\tVar}\diff{s}\,l_s\bigr)^{k-m}}{(k-m)!}  
        \ee^{\ii(2m-k+n_0) \q}   \,.
    \end{equation*}
    We ignore the constant pre-factor $\ee^{-\int_{t_0}^{\tVar}\diff{s}\,(r_s+l_s) }$ for now and apply the functional $\braA{n}$ to this expression. After carefully noting the restrictions imposed by Kronecker deltas, the resulting expression reads
    \begin{eqnarray}
       \sum_{k=0}^\infty \frac{\bigl(\int_{t_0}^{\tVar}\diff{s}\,r_s \cdot \int_{t_0}^{\tVar}\diff{s}\,l_s\bigr)^k}{k!(k+|n-n_0|)!}
      \cdot
      %\begin{cases}
        %  \bigl(\int_{t_0}^{\tVar}\diff{s}\,r_s\bigr)^{n-n_0}  \,\mathtext{ for $n \geq n_0$}
        %  \\
        %  \bigl(\int_{t_0}^{\tVar}\diff{s}\,l_s\bigr)^{n_0-n}  \,  \mathtext{ for $n \leq n_0$ .}
       %\end{cases}      \nonumber
    \end{eqnarray}
    This sum can be expressed by a modified Bessel function of the first kind (10.25.2 in~\cite{NIST:2010}), resulting in
    \begin{equation*}
      \Bigl(\frac{\int_{t_0}^{\tVar}\diff{s}\,r_s}{\int_{t_0}^{\tVar}\diff{s}\,l_s}\Bigr)^{\frac{n-n_0}{2}}
      I_{n-n_0}\Bigl(2 \Bigl(\int_{t_0}^{\tVar}\diff{s}\,r_s \int_{t_0}^{\tVar}\diff{s}\,l_s\Bigr)^{\frac{1}{2}}\Bigr)  
      \,. 
    \end{equation*}
    Multiplied with $\ee^{-\int_{t_0}^{\tVar}\diff{s}\,(r_s+l_s) }$, this expression corresponds to a Skellam distribution with mean $\rateCoeffLinDecay = n_0 + \int_{t_0}^{\tVar} \diff{s}\, (r_s - l_s)$ and variance $\sigma^2 = \int_{t_0}^{\tVar} \diff{s}\, (r_s + l_s)$.

  \section{Details on the evaluation of the backward path integral representation in \sref{subsec:PathInts_BackwardAlongPaths}}\label{sec:A_BackwardAlongPaths}       
      
     In the following, we fill out the missing steps in \sref{subsec:PathInts_BackwardAlongPaths} and rewrite the backward path integral representation~\eref{eq:PathInts_Backward_ContinuousTime_Solution_MarginalizedDistribution} of the marginalized distribution in terms of the $(\QAd, \XAd)$-generating functional~\eref{eq:PathInts_BackwardAlongPaths_PhiXGeneratingFunction}. Upon comparing the discrete-time approximation of the marginalized distribution~\eref{eq:PathInts_Backward_DiscreteTime_Solution_MarginalizedDistribution} with its representation in~\eref{eq:PathInts_BackwardAlongPaths_MarginalizedDistribution}, one observes that the corresponding $(\QAd, \XAd)$-generating functional should read
    \begin{eqnarray}
      \fieldGenFctAd_{\QAd, \XAd,N}  \coloneqq    \label{eq:A_BackwardAlongPaths_QXGenFct}
        \pathintegral{1}{N-1} 
        \ee^{  \ii\qAd_N \xAd_N - \actionAd_N } \ee^{ Z_N}    \\
      \mathtext{with } Z_{N} \coloneqq 
        \sum_{j=1}^{N} \Delta t\, \QAd_{j}  \xAd_{j-1} 
        + \sum_{j=1}^{N-1} \Delta t\,\XAd_{j-1}  \ii \qAd_{j}      \,.
    \end{eqnarray} 
    A differentiation of~\eref{eq:A_BackwardAlongPaths_QXGenFct} with respect to $\Delta t\, \QAd_j$ generates a factor $\xAd_{j-1}$, a differentiation with respect to $\Delta t\, \XAd_{j-1}$ a factor $\ii\qAd_j$. Upon recalling the definition of the action $\actionAd_N$ in~\eref{eq:PathInts_Backward_DiscreteTime_Solution_Action}, the first exponential in~\eref{eq:A_BackwardAlongPaths_QXGenFct} can be rewritten via
    \begin{eqnarray}
      &\ii\qAd_N \xAd_N - \actionAd_N \\
      &=-\sum_{j=1}^{N-1} \ii\qAd_{j}  \Bigl(   
             \xAd_{j} - \bigl[\xAd_{j-1} + \alpha_{t_{j-1}}(\xAd_{j-1}) \Delta t \bigr]  
          \Bigr)   
         \label{eq:A_BackwardAlongPaths_Step1}
         \\&\phantom{{}={}}   \nonumber
        - \sum_{j=1}^{N-1}  \frac{\qAd^2_{j}}{2}  \beta_{t_{j-1}}(\xAd_{j-1})\Delta t    
        + \sum_{j=1}^{N-1} \Delta t\, \perturbationOpAd_{t_{j-1}}(\xAd_{j-1}, \ii\qAd_{j})       
         \nonumber\\&\phantom{{}={}}
        + \ii\qAd_N  \xAd_{N-1}   \nonumber  \,.
    \end{eqnarray} 
    The equality holds up to corrections of $\bigO(\Delta t)$ (because of the sums, all remaining terms are of $\bigO(1)$). Note that the right hand side of~\eref{eq:A_BackwardAlongPaths_Step1} is independent of $\xAd_N$. One can linearize the terms that are quadratic in $\qAd_j$ by the completion of a square. In particular, we write
    \begin{eqnarray}
      \exp{\Bigl( - \sum_{j=1}^{N-1}  \frac{\qAd^2_{j}}{2}  \beta_{t_{j-1}}(\xAd_{j-1}) \Delta t \Bigr)}   \\
      =  \BigLLangle \exp{\Bigl( \sum_{j=1}^{N-1} \ii\qAd_{j}  \sqrt{\beta_{t_{j-1}}(\xAd_{j-1})}{\Delta \wienerProcess_j}  \Bigr)}  \BigRRangle_{\wienerProcess}      \nonumber
    \end{eqnarray} 
    with the average being defined as
    \begin{equation}
      \bigLLangle\,\cdot\,\bigRRangle_{\wienerProcess}  
      \coloneqq \Bigl(
        \prod_{j=1}^{N-1} \int_{\reals} \diff{\Delta \wienerProcess_j}
        \, \gaussianDistribution_{0,\Delta t}(\Delta \wienerProcess_j)
      \Bigr) \bigl(\,\cdot\,\bigr)            \,.
    \end{equation} 
    The average employs the Gaussian distribution
    \begin{equation}
      \gaussianDistribution_{0, \Delta t}(\Delta \wienerProcess_j)  
        = \frac{\ee^{-(\Delta \wienerProcess_j)^2/(2\Delta t)}}{\sqrt{2\pi\Delta t}}
    \end{equation}
     with zero mean and variance $\Delta t$ (cf.~\eref{eq:PathInts_BackwardKramersMoyal_Gaussian}). The sequence $\Delta \wienerProcess_1, \hldots, \Delta \wienerProcess_{N-1}$ can be interpreted as the steps of a discretized Wiener process. To proceed with the derivation, we now move the perturbation operator $\perturbationOpAd$ to the front of the $(\QAd, \XAd)$-generating functional~\eref{eq:A_BackwardAlongPaths_QXGenFct} by writing
    \begin{eqnarray}
      \ee^{ \sum_{j=1}^{N-1}\Delta t\, \perturbationOpAd_{t_{j-1}}(\xAd_{j-1}, \ii\qAd_{j})}
        \ee^{ Z_N} \\
      = \ee^{\sum_{j=1}^{N-1}\Delta t\, 
          \perturbationOpAd_{t_{j-1}}
          \bigl(
            \frac{1}{\Delta t}\frac{\partial}{\partial \QAd_{j}}, \frac{1}{\Delta t}\frac{\partial}{\partial \XAd_{j-1}}
          \bigr)
        }
        \ee^{ Z_N}     \nonumber\,.
    \end{eqnarray} 
    Upon combining all of the above steps, one finds that
    \begin{eqnarray}
      \fieldGenFctAd_{\QAd, \XAd,N}    \label{eq:A_BackwardAlongPaths_GenFctnl}\\
      =  \ee^{
              \ii\qAd_N  \frac{1}{\Delta t}\frac{\partial}{\partial \QAd_{N}}
              + \sum_{j=1}^{N-1} \Delta t\, \perturbationOpAd_{t_{j-1}}\bigl(\frac{1}{\Delta t}\frac{\partial}{\partial \QAd_{j}}, \frac{1}{\Delta t}\frac{\partial}{\partial \XAd_{j-1}}\bigr)
            } \fieldGenFctAd^0_{\QAd, \XAd,N}  \nonumber
    \end{eqnarray}
    with the definition
    \begin{eqnarray}
      \fieldGenFctAd^0_{\QAd, \XAd,N}  \coloneqq \BigLLangle
          \Bigl(\prod_{j=1}^{N-1}
          \int_{\reals} \diff{\xAd_{j}}  
          \,\delta\Bigl( \xAd_{j} - \xAd_{j-1}    \\
      -  \Bigl\{
            \bigl[\alpha_{t_{j-1}}(\xAd_{j-1}) + \XAd_{j-1}\bigr] \Delta t  + \sqrt{\beta_{t_{j-1}}(\xAd_{j-1})}\,\Delta \wienerProcess_j
          \Bigr\}\Bigr)\Bigr)    \nonumber  \\
      \cdot \,\ee^{  \sum_{j=1}^{N} \Delta t\, \QAd_{j}  \xAd_{j-1} + \bigO(\Delta t)}
        \BigRRangle_{\wienerProcess}    \,.\nonumber
    \end{eqnarray}
    The sequence of Dirac delta functions implies that $\xAd_j$ depends on $\XAd_i$ only for $i < j$. This property has to be kept in mind when the functional derivatives in~\eref{eq:A_BackwardAlongPaths_GenFctnl} are evaluated in continuous-time. In the continuous-time limit, i.e.\ for $N \to \infty$ and $\Delta t \to 0$, one recovers the marginalized distribution~\eref{eq:PathInts_BackwardAlongPaths_MarginalizedDistribution} with generating functional~\eref{eq:PathInts_BackwardAlongPaths_PhiXGeneratingFunction} and It\^{o} SDE~\eref{eq:PathInts_BackwardAlongPaths_SDE}.

 % END - Appendix
  
\ \\
\bibliography{frey_bibliography}

\end{document}